\documentclass[prb,showpacs,floatfix,superscriptaddress,twocolumn,aps,10,groupedaddress]{revtex4-2}
\usepackage[english]{babel} 
\usepackage{bm}
\usepackage{graphicx}
\usepackage{amsmath}
\usepackage{amssymb}
\usepackage{wasysym}
\usepackage{comment}
\usepackage[dvipsnames]{xcolor}
\usepackage{tikz}
\usepackage{adjustbox}
\usepackage{booktabs}
\usepackage{array}
\usepackage{multirow}
\usepackage{lipsum}   
\usepackage{makecell}
\usepackage{placeins}

\def\genbox#1#2#3#4#5#6{
    \leavevmode\raise#4bp\hbox to#5bp{\vrule height#5bp depth0bp width0bp
    \pdfliteral{q .5 w \csname #2COLOR\endcsname\space RG
                       \csname #3PDF\endcsname{#5}{#6} S Q
             \ifx1#1 q \csname #2COLOR\endcsname\space rg 
                       \csname #3PDF\endcsname{#5}{#6} f Q\fi}\hss}}

\def\sqbox      #1#2{\genbox{#1}{#2}  {sq}       {0}   {4.5}  {2.25}}
\def\circbox    #1#2{\genbox{#1}{#2}  {circ}     {0}   {5}    {2.5}}
 
\usepackage[hidelinks]{hyperref}

\begin{document}

\title{Decorated cluster lattices: a natural framework for classical spin liquids and flat bands}

    \author{Naïmo Davier}
	\email[]{naimo.davier@u-bordeaux.fr}
    \affiliation{CNRS, Universit\'e de Bordeaux, LOMA, UMR 5798, 33400 Talence, France}
		
	\date{\today}

\begin{abstract} 
    Classical spin liquids are frustrated magnetic phases characterized by local constraints, flat bands in reciprocal space, and emergent gauge structures with distinctive signatures such as pinch points. These arise generally in \emph{cluster systems}, where spin interactions can be expressed as constraints on clusters of spins. In this work we present the different generic rules allowing to build such cluster systems together with a few tools allowing to quickly characterize it. We show that based on these rules, it is possible to conceive a tunable recipe for generating such models by decorating a parent lattice on its bonds and/or vertices with symmetry-compatible clusters. This approach highlights a key design trade-off: using fewer cluster types increases the number of flat bands and enhances spin-liquid behavior, but produces denser connectivity that is harder to realize experimentally. The framework is highly tunable, extends naturally to two and three dimensions, and provides a versatile toolbox for engineering new classical spin-liquid candidates with targeted features such as higher-rank pinch points or pinch lines. Importantly, decorated cluster lattices constitute an ideal playground for producing multiple flat bands—not only at the spectral bottom but also embedded within the spectrum. This proliferation of flat bands can be understood either from general connectivity arguments or, equivalently, by explicitly constructing compact localized eigenmodes in an associated hopping description.
\end{abstract}
\maketitle

\section{Introduction}

Classical spin liquids (CSLs) are frustrated magnetic phases in which geometric frustration suppresses conventional long-range order, even at zero temperature. Rather than selecting an ordered state, these systems exhibit highly degenerate ground-state manifolds governed by local constraints, which often manifest as flat bands in reciprocal space. When these constraints take a divergence-free form—analogous to a Gauss law—they give rise to an emergent gauge structure reminiscent of classical electromagnetism \cite{Henley2005, henley2010coulomb}. This leads to characteristic signatures such as algebraically decaying correlations and pinch-point singularities in the static structure factor, which are directly accessible through neutron scattering experiments \cite{castelnovo08a, Gauss_pyro}. Prototypical examples include nearest-neighbor Heisenberg antiferromagnets on the pyrochlore and kagome lattices, where frustration and geometry stabilize disordered yet strongly correlated Coulomb phases \cite{Anderson_pyro, Garanin_LargeN}.

More recently, it has been shown that classical spin liquids arise naturally within a broad class of models known as \emph{cluster systems} \cite{Davier_2023, Yan_2024_short, Yan_2024_long, Davier_2025_interacting_clusters}. In these models, interactions are encoded as constraints acting on clusters of spins, so that the number of independent conditions is controlled by the number of clusters rather than by the number of bonds. This reorganization of constraints typically reduces their effective number, enhancing the likelihood of extensive degeneracy and flat-band formation. As a result, cluster systems provide a particularly fertile framework for realizing and classifying classical spin liquids.

The primary focus of the present work is to present how such cluster systems can be \emph{constructed} in a systematic and geometric manner, with the aim to enlarge the catalog of cluster systems. We begin by defining the general class of cluster Hamiltonians and by introducing a set of compact criteria that allow one to rapidly assess whether a given construction is likely to host a classical spin liquid. These indicators are then illustrated across a broad collection of explicit examples.
Building on these ingredients, we propose a simple and highly tunable construction scheme that upgrades a standard parent lattice into a cluster lattice through bond and/or vertex decoration, effectively promoting edges and/or sites to clusters. This framework applies equally well in two and three dimensions and naturally generates a large variety of cluster systems. Their basic properties can be efficiently analyzed within the large-$\mathcal{N}$ approximation \cite{Henley2005, Garanin_LargeN, Canals_2001_LargeN, Garanin_SCGA}, providing rapid access to their low-temperature structure factors and approximate band structures.

Decorated cluster systems often host additional flat bands in their spectrum, not necessarily at its bottom. Flat bands have been investigated extensively in a variety of condensed-matter settings \cite{Kim_Rhim_2023_flat_bands, Yan_2024_long, Rhim_2019_flat_bands, Sutherland_1986_FB, Dias_2015_FB, Morales-Inostroza_2016_FB, Miyahara_2005_FB, Graf_Piechon_2021_flat_bands, Biplab_2018_FB}, since they enable a range of remarkable phenomena. On the one hand, band flatness implies a macroscopic degeneracy of eigenstates. On the other hand, the vanishing bandwidth means that even weak perturbations can act effectively nonperturbatively, strongly reshaping the low-energy physics and potentially stabilizing exotic phases of matter \cite{Graf_Piechon_2021_flat_bands}. This motivates a detailed understanding of why decorated lattices so generically produce flat bands—especially because this mechanism is rooted in lattice-graph properties and is therefore not specific to spin models. We analyze the origin of these flat bands using two complementary approaches: (i) graph-connectivity arguments, and (ii) an explicit construction of compact localized states in an associated hopping problem. Both viewpoints, developed in Secs.~\ref{sec: additional_bfb}--\ref{sec: meta_spin}, show that in decorated systems flat bands can be traced to a characteristic graph structure built from fully connected clusters. In particular, flat bands occurring \emph{within} the spectrum are linked to the appearance of this structure in the system’s bidual lattice.

\section{Cluster Hamiltonian}

Cluster spin systems are defined by Hamiltonians that can be written as a sum over interacting clusters of spins
\begin{equation}
    \mathcal{H} = \sum_{n} \sum_X |\bm{\mathcal{C}}_{n,X}|^2 ,
    \label{Eq: general cluster H}
\end{equation}
where $\bm{\mathcal{C}}_{n,X}$ denotes the \emph{constrainer} associated with unit cell $n$ and cluster type $X$. For example in the case of the kagome lattice, see Table \ref{tab: 2d cluster systems}, $X$ would stand for up and down triangles, the unit cell being composed of two triangles.
Each constrainer is a weighted local magnetization built from the spins belonging to the cluster,
\begin{equation}
    \bm{\mathcal{C}}_{n,X} = \sum_{i \in n \cap X} \gamma_i^X \mathbf{S}_i ,
    \label{Eq: general constrainer}
\end{equation}
with coefficients $\gamma_i^X$ fixing the relative strength of spin–spin interactions inside the cluster. We consider translation invariant systems where the coefficients $\gamma_i^X$ are only cluster dependent and repeat over all unit cells. Two spins $i$ and $j$ that both belong to cluster $X$ thus interact with an effective exchange constant $2 \gamma_i^X \gamma_j^X$. Note that a spin $i$ interacts with itself among a cluster $X$ to produce an energy term $(\gamma_i^X)^2 |\mathbf{S}_i|^2$. These additional onsite terms serve as Lagrange constraints enforcing the spin length constraint $|\mathbf{S}_i| = 1$, assuring ground states obtained by 
minimizing of Eq.~(\ref{Eq: general cluster H}) and fulfilling the conditions
\begin{equation}
    \bm{\mathcal{C}}_{n,X} = 0 \qquad \forall \; n, X,
\end{equation}
are physical ground states. In practice, these zero energy ground states are accessible provided the coefficients $\gamma_i^X$ do not deviate too strongly from unity; otherwise, the fixed spin-length condition $|\mathbf{S}_i|=1$ may prevent zero energy solutions from existing. Since our goal is to construct cluster systems that can support classical spin liquids in some region of parameter space, we will not dwell on this restriction.
Each cluster $(n,X)$ generates a constraint $\bm{\mathcal{C}}_{n,X} = 0$, so the number of independent ground-state constraints scales with the number of clusters rather than the number of bonds. Because the number of clusters is typically smaller than the number of spins, an extensive number of degrees of freedom remain unconstrained. These surviving modes are responsible for the appearance of flat bands at the bottom of the excitation spectrum. 
To obtain a first view of these systems, one can compute the zero-temperature structure factor together with the band structure within the large-$\mathcal{N}$ approximation\cite{Henley2005, Garanin_LargeN, Canals_2001_LargeN, Garanin_SCGA}. Inspired by the Luttinger--Tisza method \cite{LT1, LT2, kaplan_2007}, this approximation considers $\mathcal{N}$-component spins and relaxes the fixed-length constraint $|\mathbf{S}_i|=1$, enforcing it only on average over the ensemble. This amounts to treating the spin components as independent scalars, which allows one to compute an effective band structure for the spin system, and has been shown to be a good approximation for many (though not all) Heisenberg candidate CSLs\cite{Yan_2024_long}.
Within this framework, the spectrum is obtained by diagonalizing the Hamiltonian matrix resulting from a direct Fourier transform of Hamiltonian (\ref{Eq: general cluster H}). This matrix corresponds to the adjacency matrix of the lattice graph, shifted by an appropriate set of Lagrange multipliers fixed by the cluster structure (\ref{Eq: general cluster H}) of the Hamiltonian\cite{Davier_2025_interacting_clusters}. Indeed, expressing the Hamiltonian as a sum of squared constrainers generates onsite terms $(\gamma_i^X)^2 |\mathbf{S}_i|^2$, which---after summing over all clusters containing site $i$---can be interpreted as the Lagrange terms enforcing $|\mathbf{S}_i|=1$ on average. The zero-temperature structure factor is then obtained by computing the projector onto the flat-band ground-state manifold; see Appendix \ref{Appendix:Large N} for details.
The number of bottom flat bands $n_\text{b.f.b}$ can be simply estimated within this framework\cite{Yan_2024_long, Davier_2025_interacting_clusters} as $n_\text{b.f.b} = n_s - n_c$ where $n_s$ is the number of spin sublattices and $n_c$ is the number of distinct cluster types. This can be simply understood as a $\mathbf{k}$-space counting of the zero modes, with $n_s$ representing the number of degrees of freedom and $n_c$ the number of constraints per unit cell. Note that this counting argument only gives a lower bound for the number of bottom flat bands, and that they may exists systems where the number of bottom flat bands is accidentally larger than this estimate if some of the cluster constraints appear to be trivial.
A sufficient condition for realizing a classical spin liquid therefore seems to be $n_{\text{b.f.b.}} > 0$. 
Going beyond large $\mathcal{N}$ approximation, one must also ensure that real-space superpositions of flat-band states can satisfy the spin-length constraint $|\mathbf{S}_i|=1$. This requirement is met\cite{Davier_2025_interacting_clusters} when the number of flat bands is at least
\begin{equation}
    n_{\text{b.f.b.}} \geq \frac{n_s}{\mathcal{N}},
    \label{Eq: Criterion Classical Spin liquid}
\end{equation}
where $\mathcal{N}$ is the spin dimensionality (e.g., $\mathcal{N}=1$ for Ising spins, $\mathcal{N}=2$ for XY spins and $\mathcal{N}=3$ for Heisenberg spins).
Even if this criterion is a sufficient and not a necessary condition to observe spin liquids, it serves as a practical guideline for identifying promising cluster systems: candidates with $n_{\text{b.f.b.}} \geq n_s/\mathcal{N}$ are good contenders for hosting classical spin-liquid phases. 

Having established the generic structure and constraints of cluster Hamiltonians, we now turn to constructive principles for building cluster lattices capable of realizing such phases.

\begin{table*}[t]
    \centering
    \renewcommand{\arraystretch}{1.5} 
    \resizebox{\textwidth}{!}{
    \begin{tabular}{c c c c c c c c c c}
        \hline
        \textbf{\makecell{2D Cluster\\ Lattice}} & Kagome & Checkerboard & Hexagonal & \makecell{Kagome\\ hexagonal} & \makecell{Square\\ hexagonal} & \makecell{Square\\ octagon} & Ruby & Square kagome & \makecell{Decorated\\ square kagome} \\ 
        \hline
        \textbf{Parent Lattice} & Honeycomb & Square & Triangular & Triangular & Square & Square $J_1-J_2$ & Kagome & Square octagon & Square \\ 
        \hline
        \noalign{\vskip 1mm}
        \makecell{ \vspace{-2.5cm} \\ \textbf{Lattice}  \\ \textbf{Scheme} } & 
        \includegraphics[width=2.5cm]{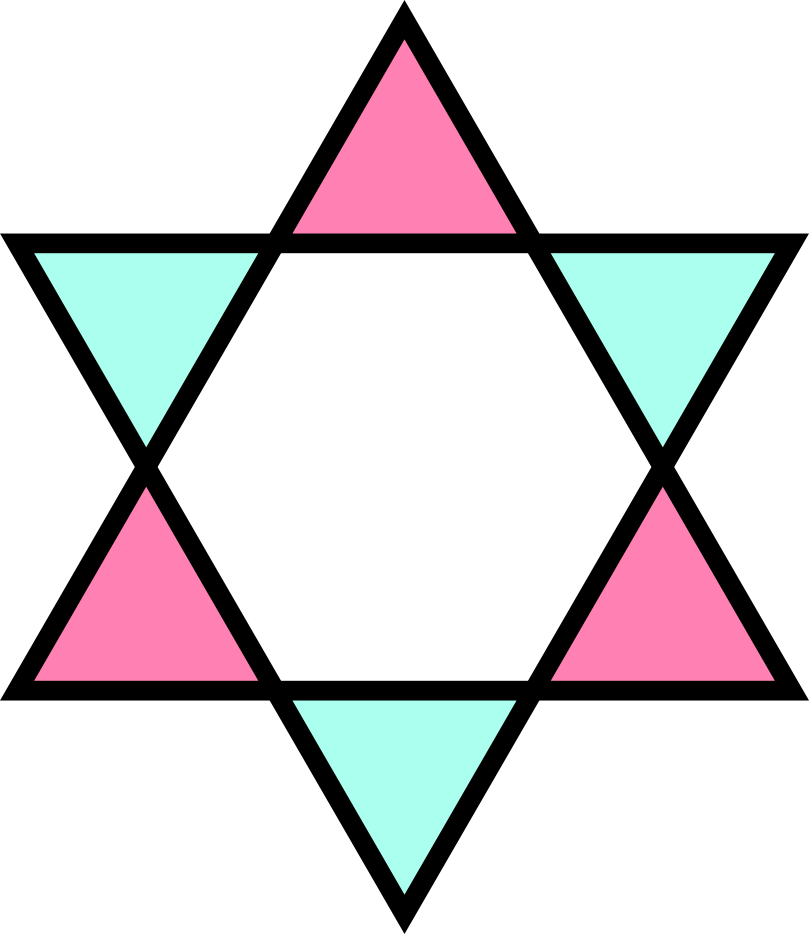} & 
        \includegraphics[width=2.5cm]{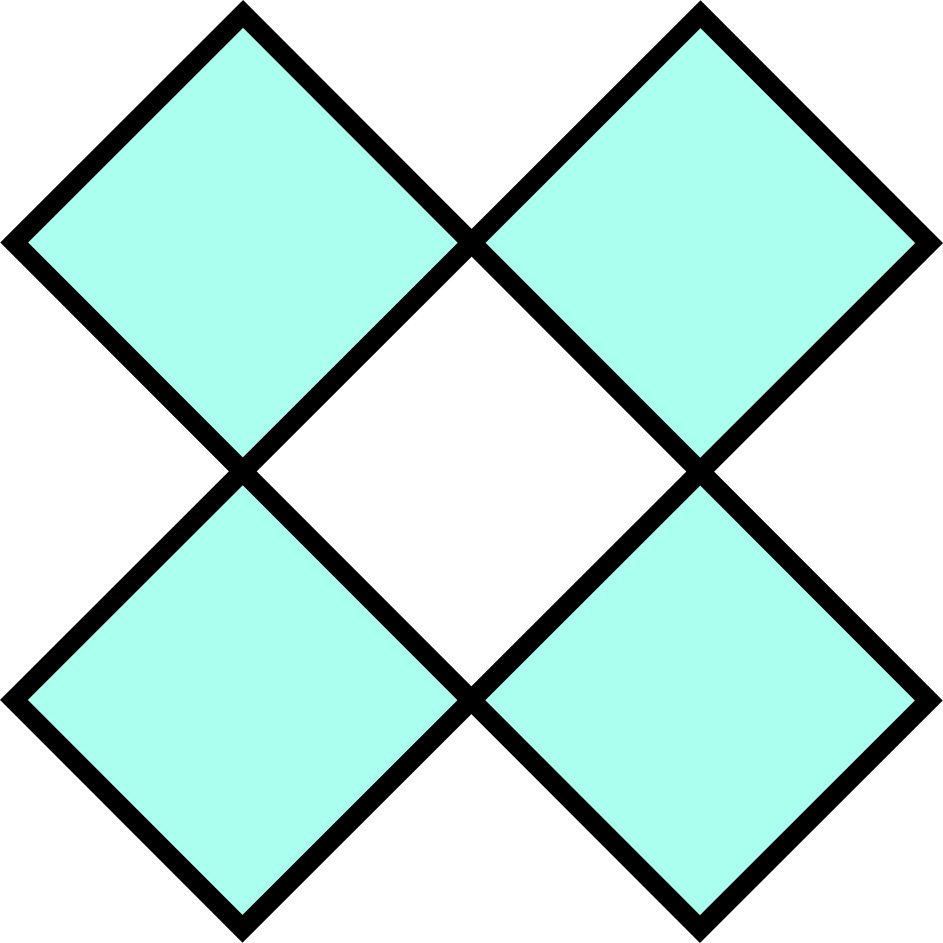} & 
        \includegraphics[width=2.5cm]{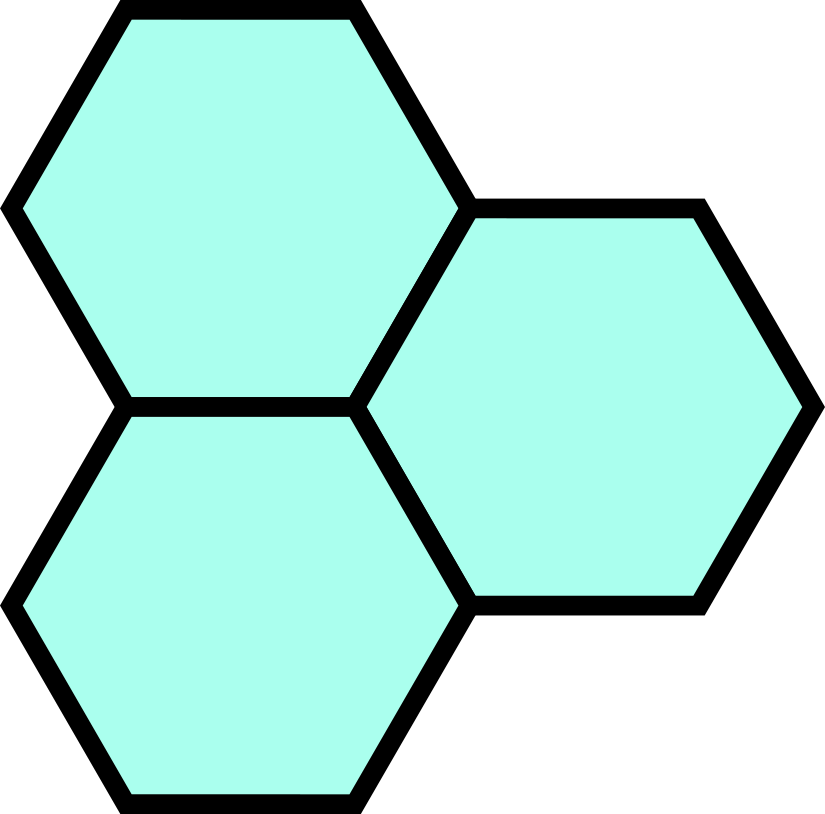} & 
        \includegraphics[width=2.5cm]{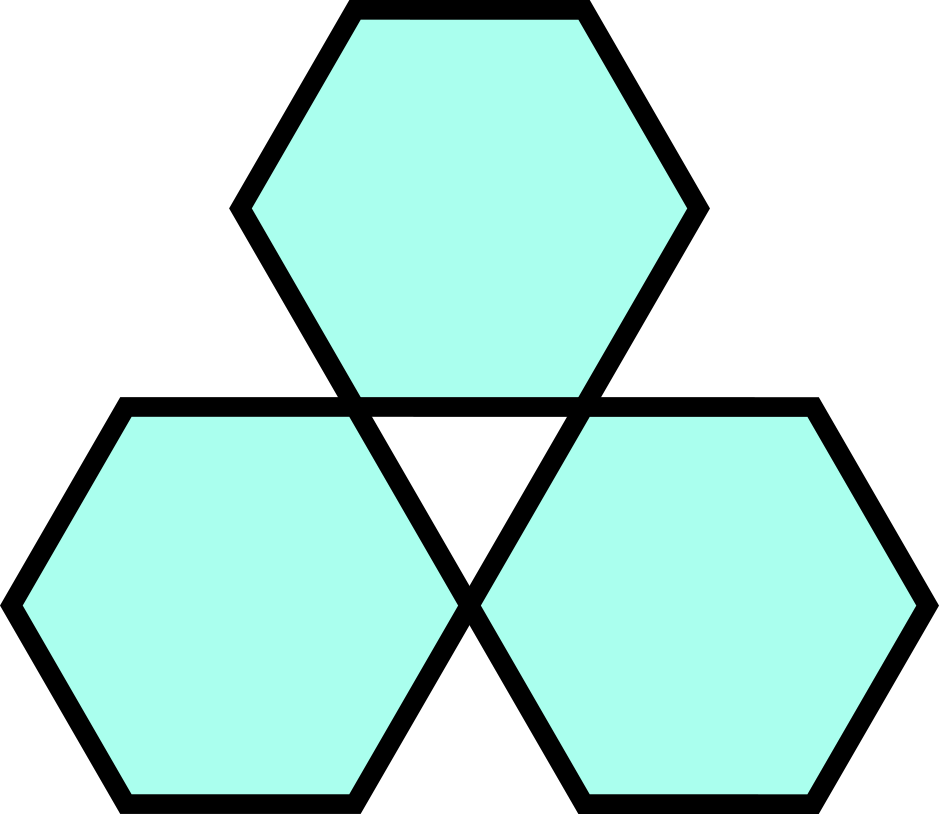} & 
        \includegraphics[width=2.5cm]{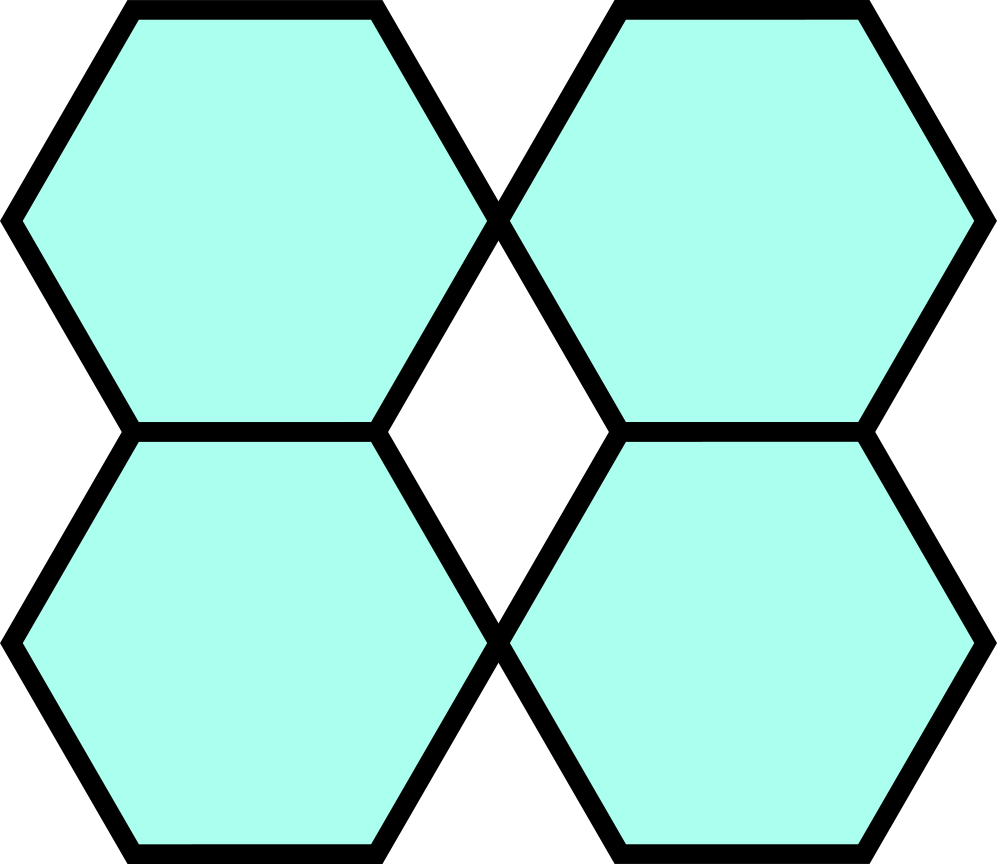} & 
        \includegraphics[width=2.5cm]{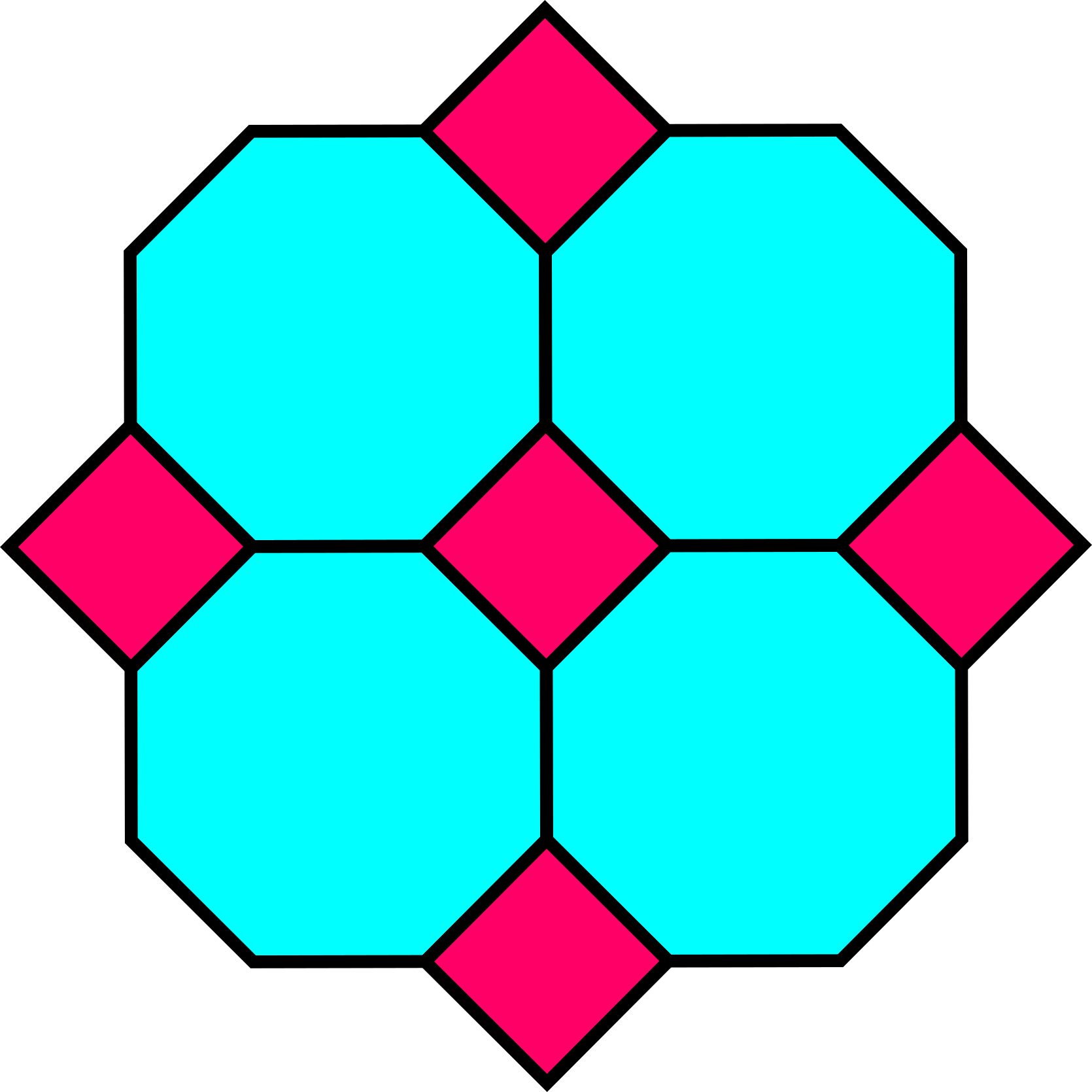} & 
        \includegraphics[width=2.5cm]{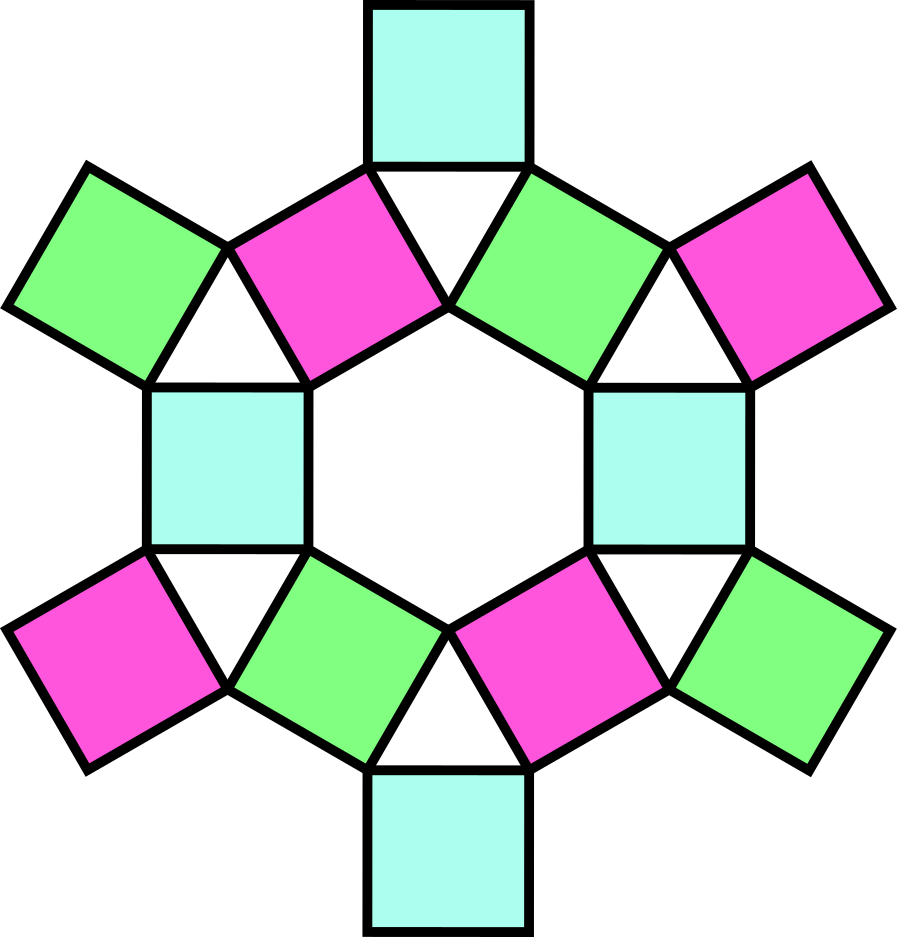} & 
        \includegraphics[width=2.5cm]{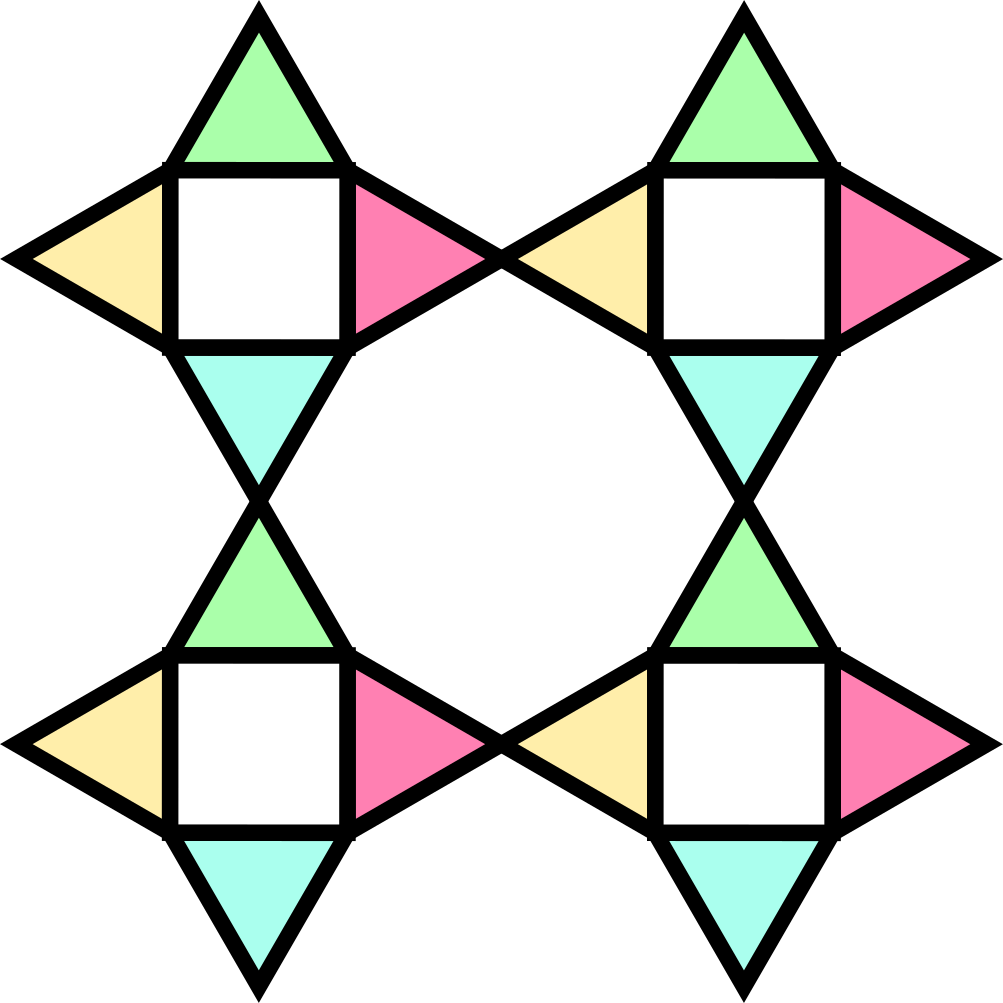} & 
        \includegraphics[width=2.5cm]{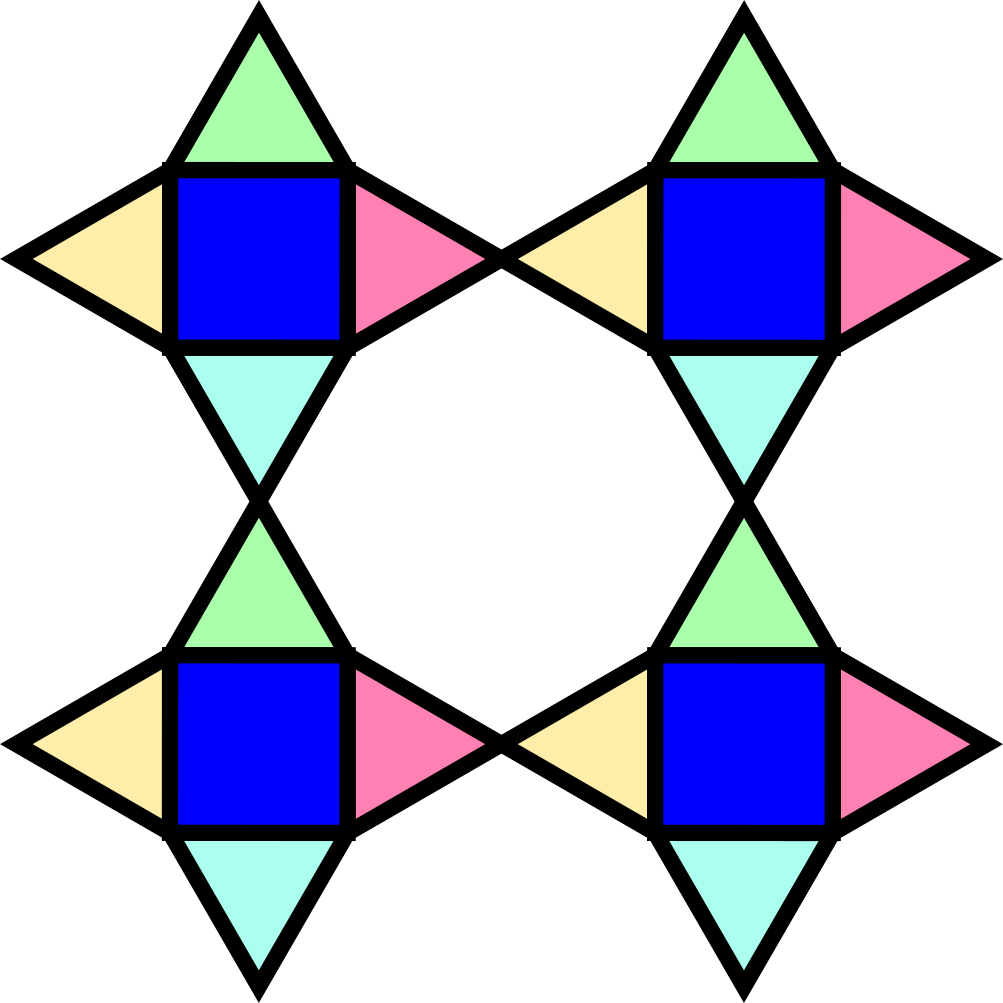} \\
              
        \makecell{ \vspace{-2.5cm} \\ \textbf{Structure}  \\ \textbf{Factor} } & 
        \includegraphics[width=2.5cm,trim={1cm 1cm 2cm 0},clip]{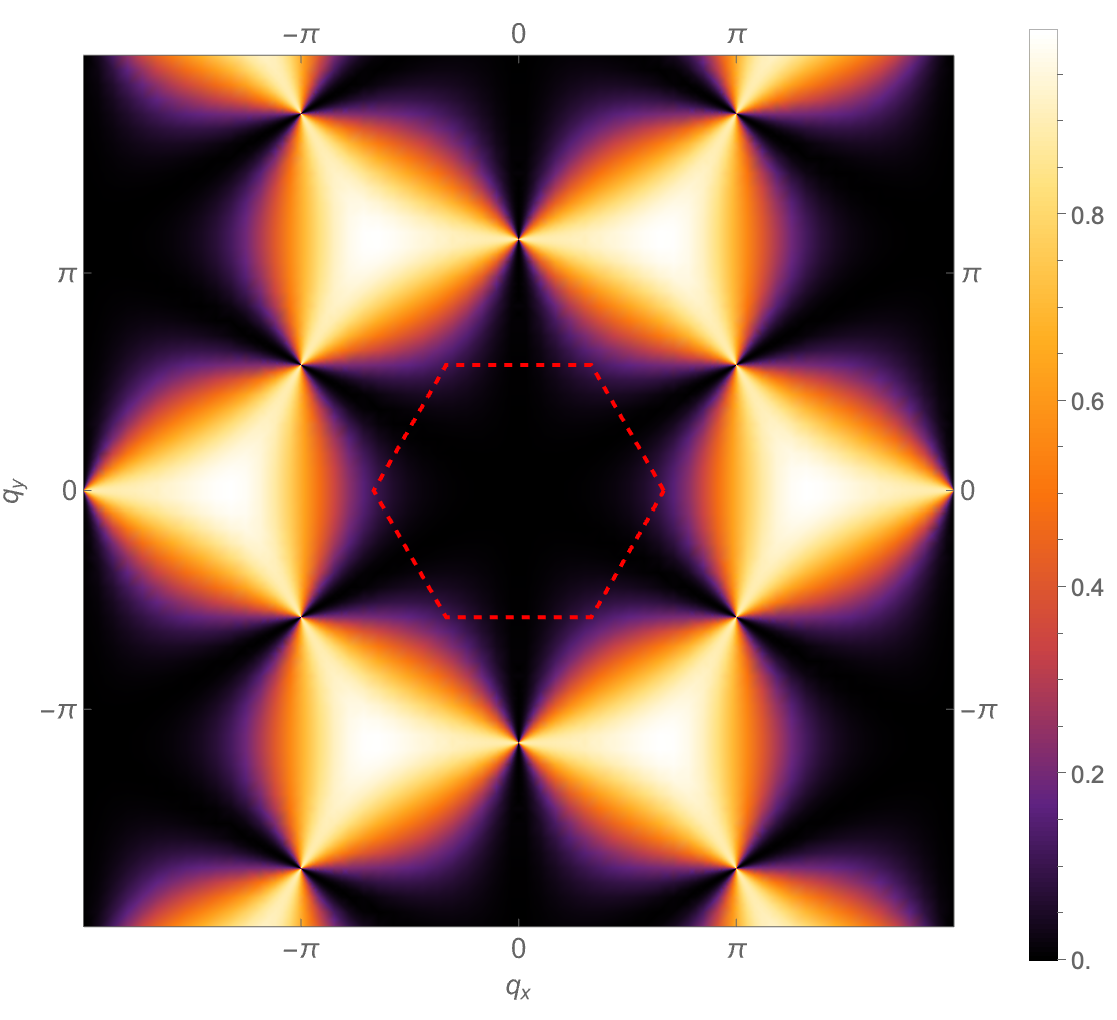} & 
        \includegraphics[width=2.5cm,trim={1cm 1cm 2cm 0},clip]{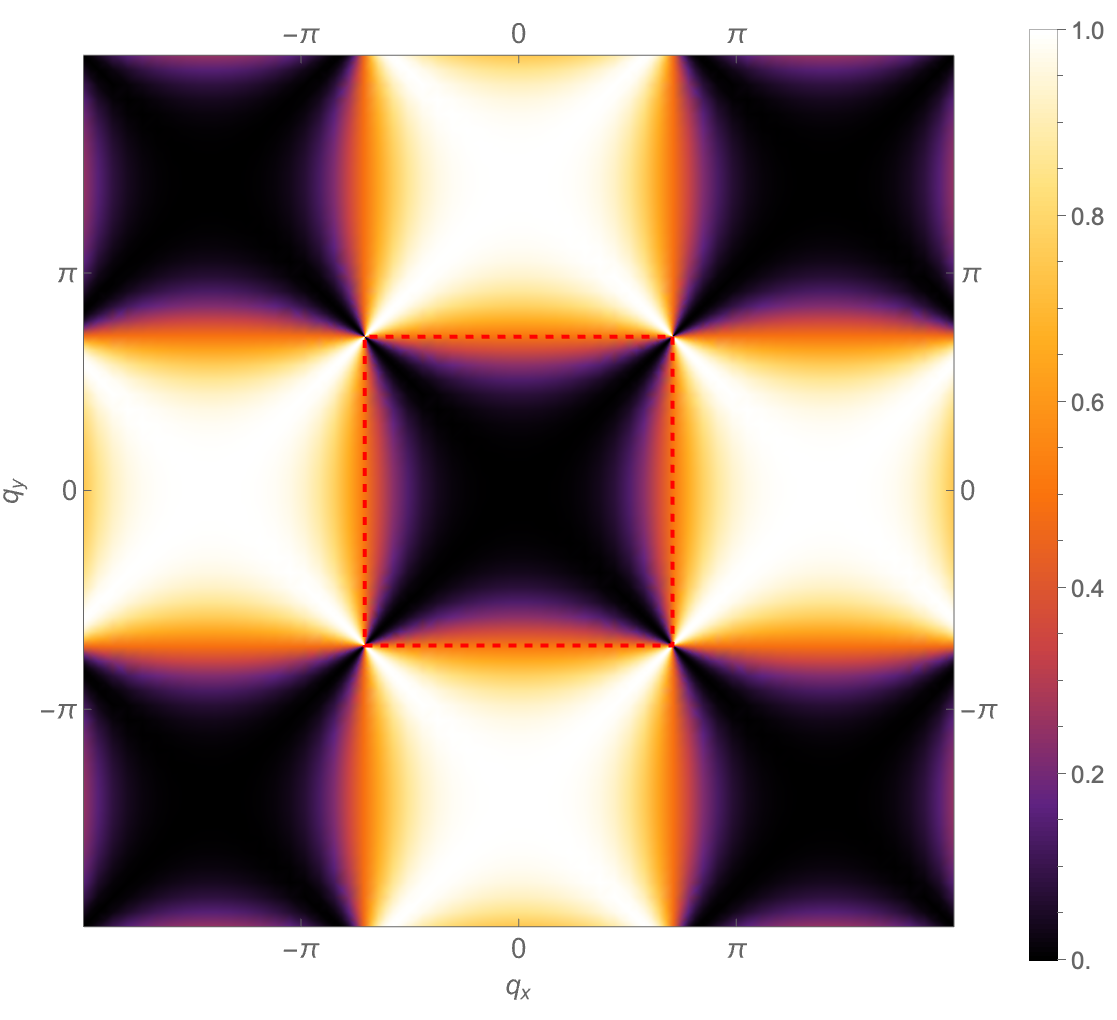} & 
        \includegraphics[width=2.5cm,trim={1cm 1cm 2cm 0},clip]{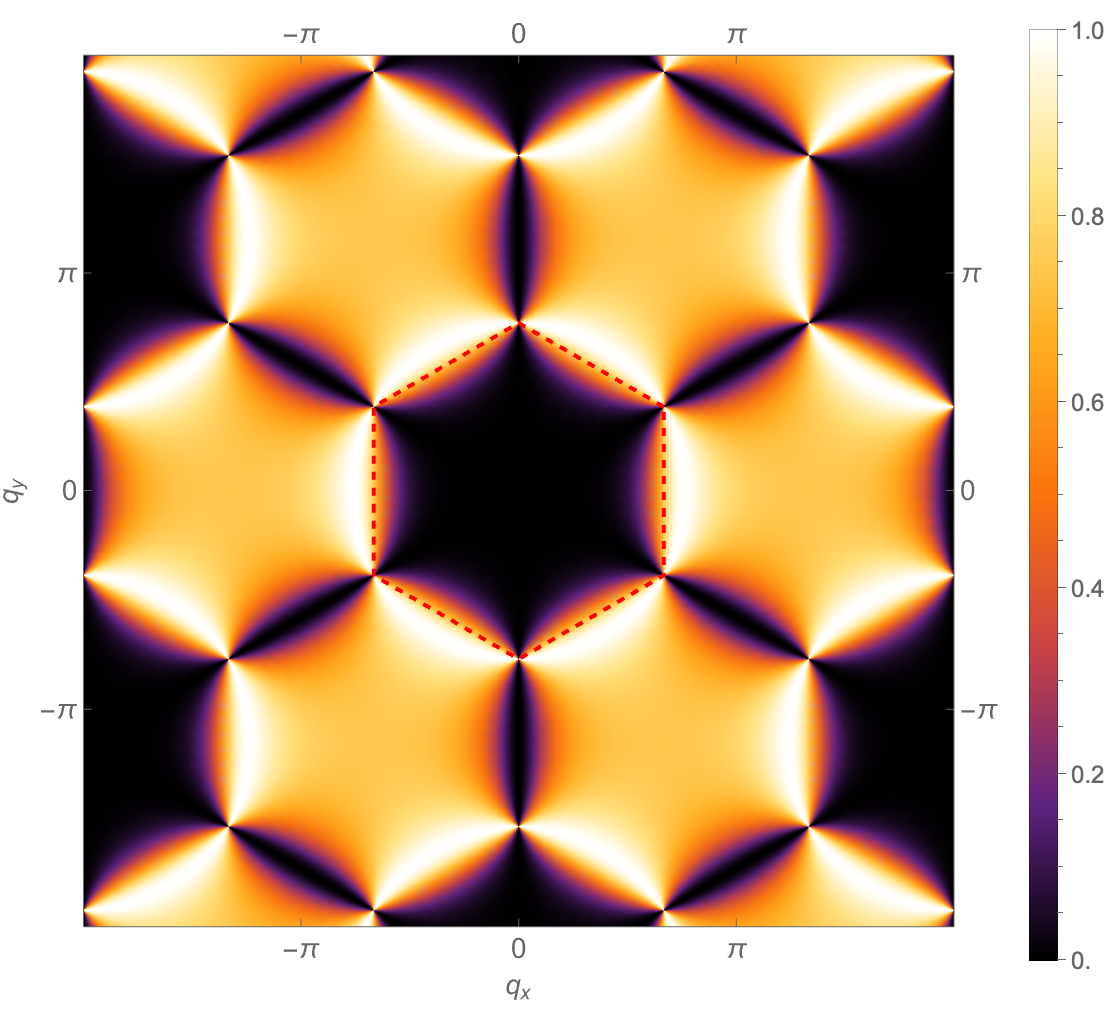} & 
        \includegraphics[width=2.5cm,trim={1cm 1cm 2cm 0},clip]{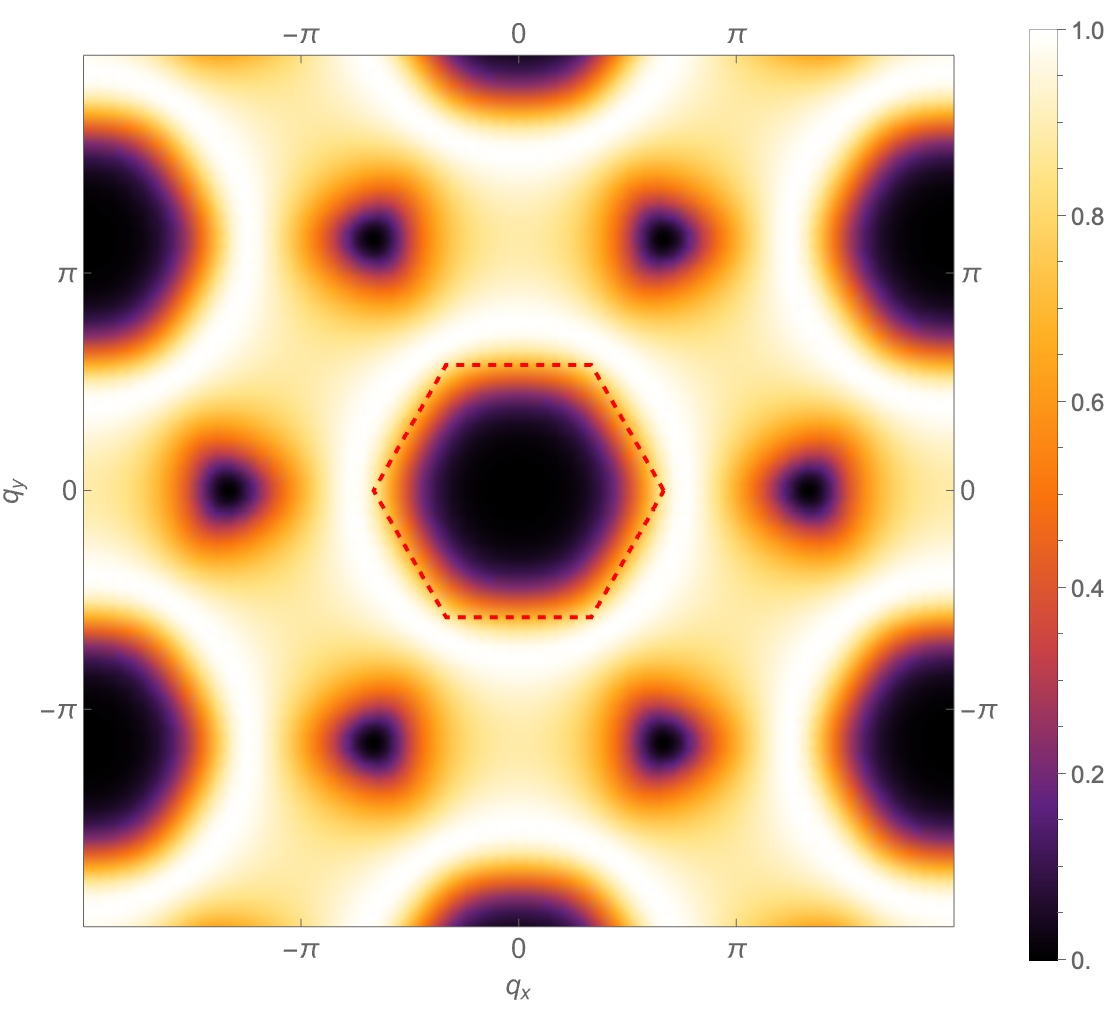} & 
        \includegraphics[width=2.5cm,trim={1cm 1cm 2cm 0},clip]{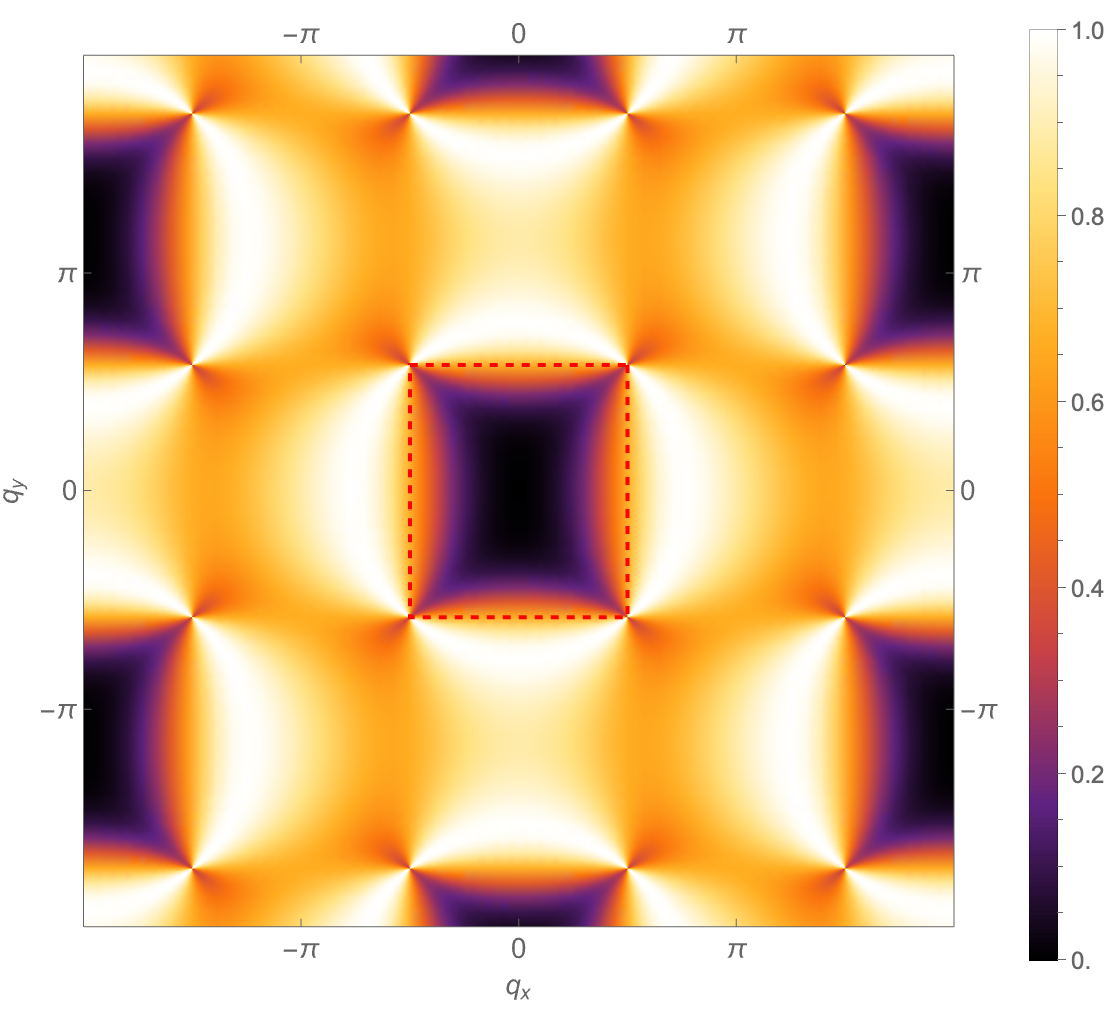} & 
        \includegraphics[width=2.5cm,trim={1cm 1cm 2cm 0},clip]{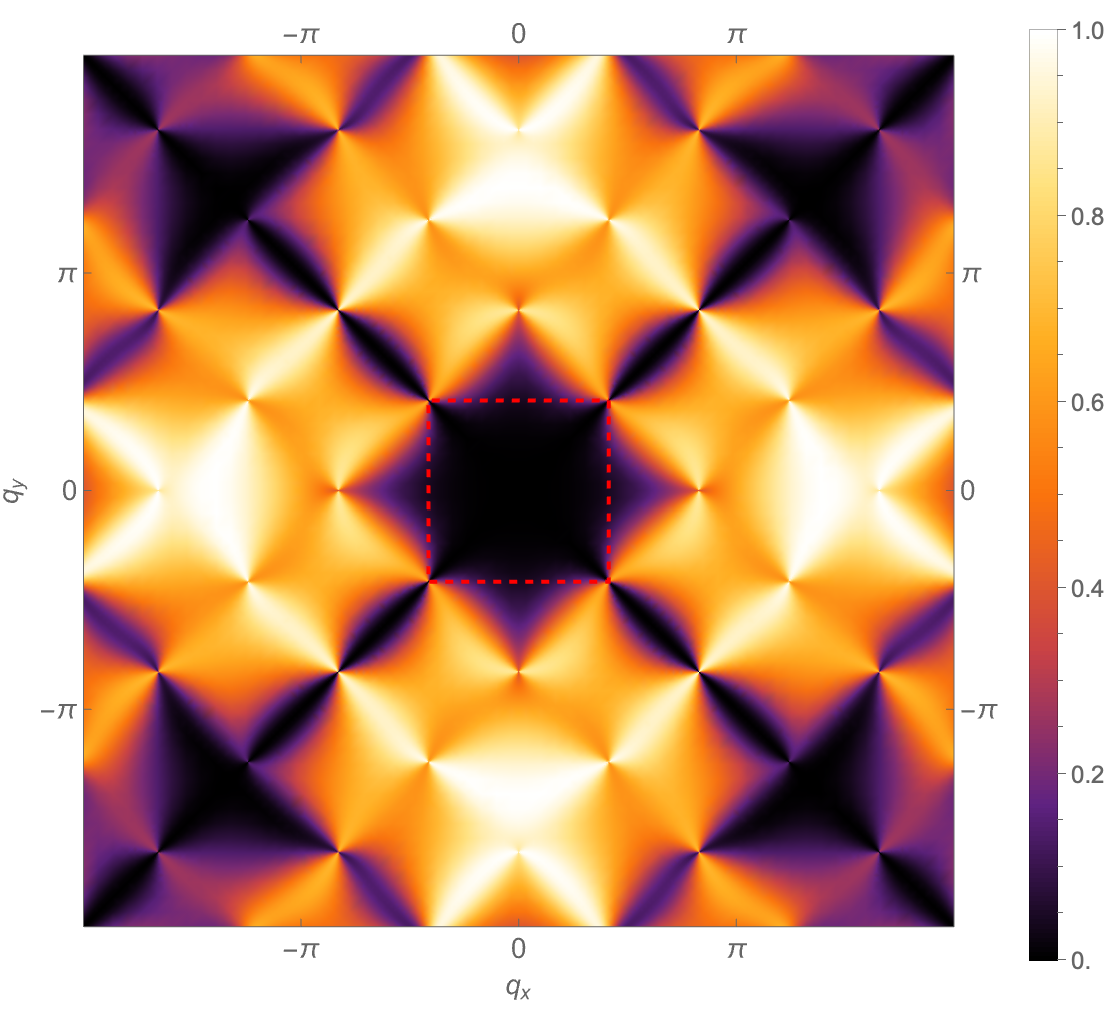} & 
        \includegraphics[width=2.5cm,trim={1cm 1cm 2cm 0},clip]{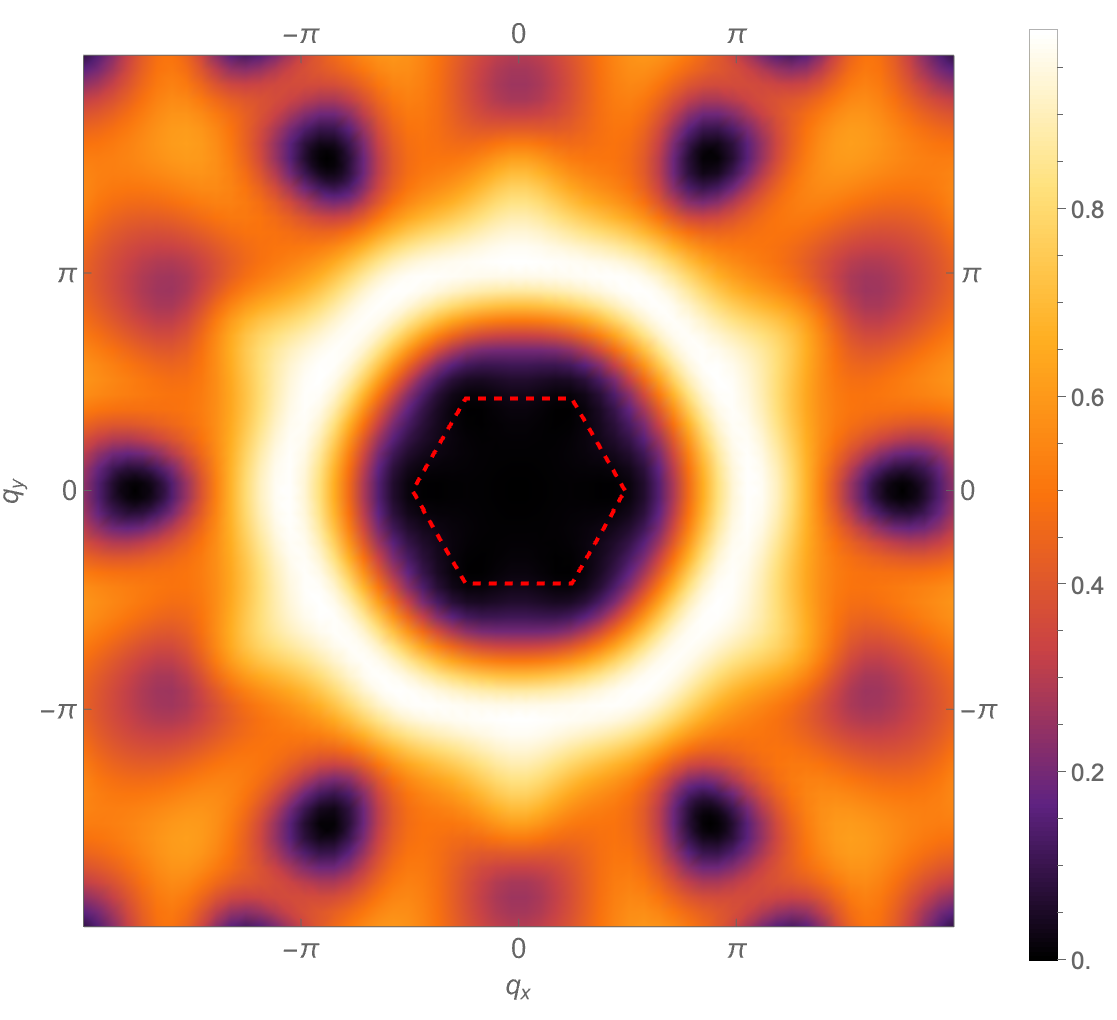} & 
        \includegraphics[width=2.5cm,trim={1cm 1cm 2cm 0},clip]{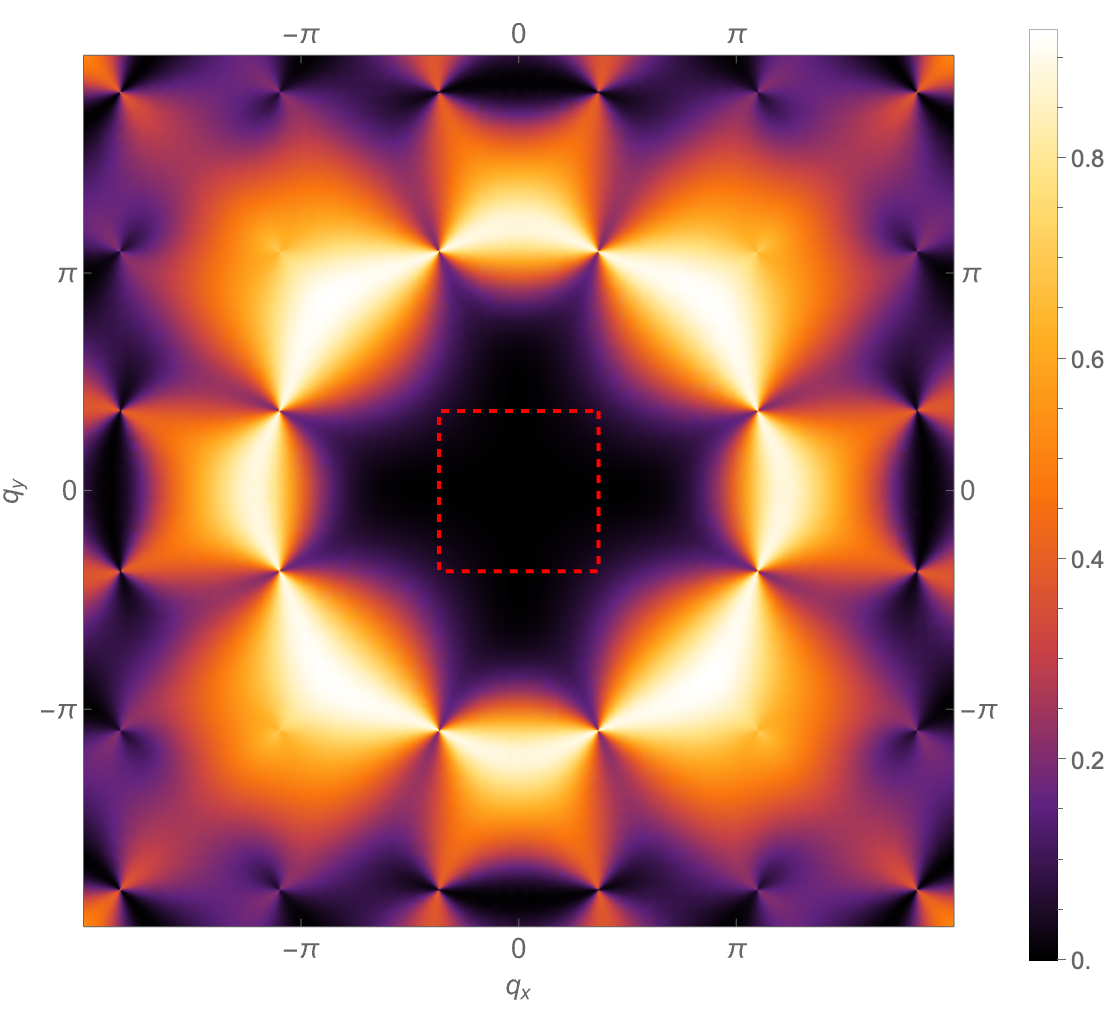} & 
        \includegraphics[width=2.5cm,trim={1cm 1cm 2cm 0},clip]{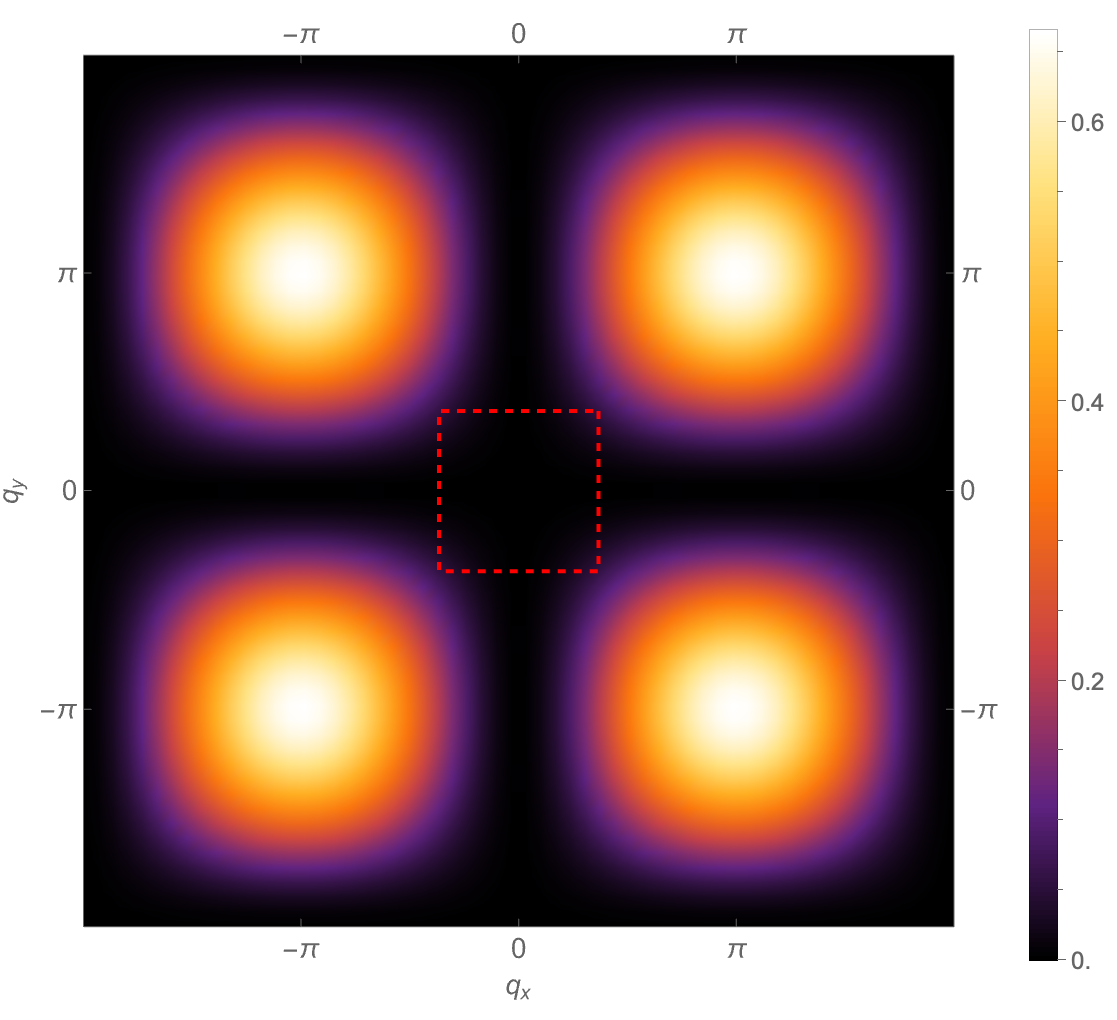} \\
        
        \makecell{ \vspace{-2.2cm} \\ \textbf{3D Band}  \\ \textbf{Structure} } & 
        \includegraphics[width=2.5cm]{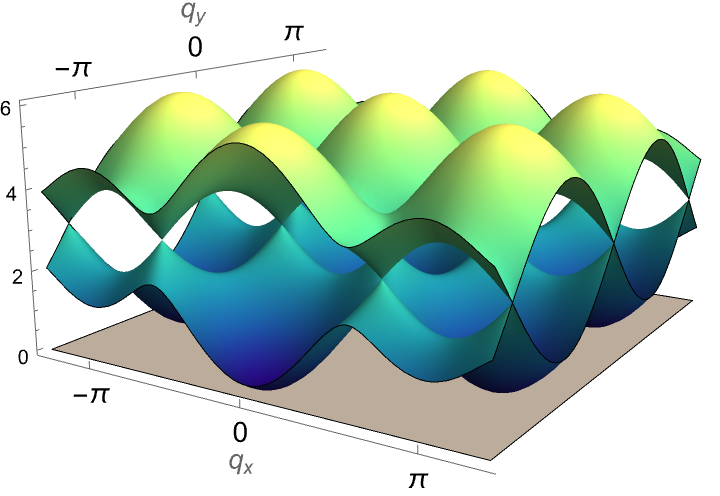} & 
        \includegraphics[width=2.5cm]{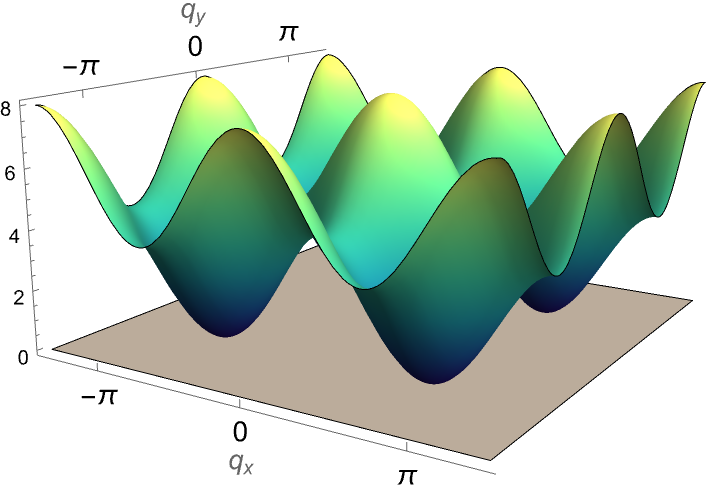} & 
        \includegraphics[width=2.5cm]{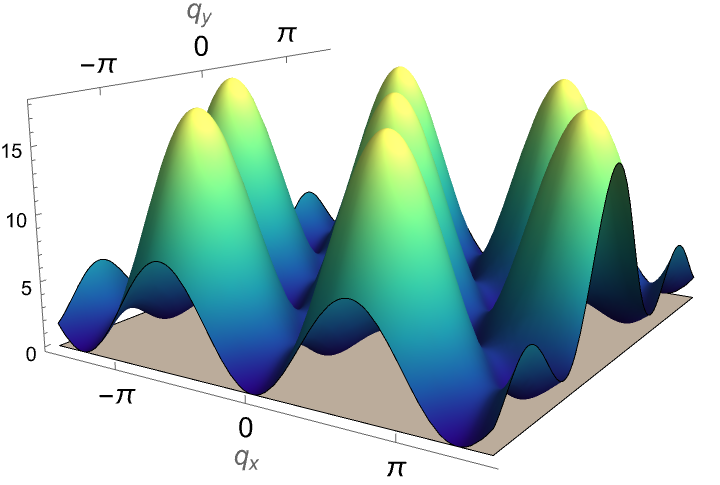} & 
        \includegraphics[width=2.5cm]{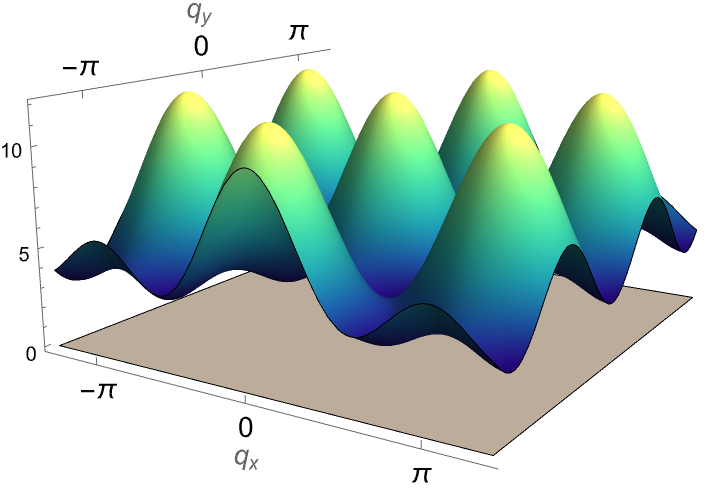} & 
        \includegraphics[width=2.5cm]{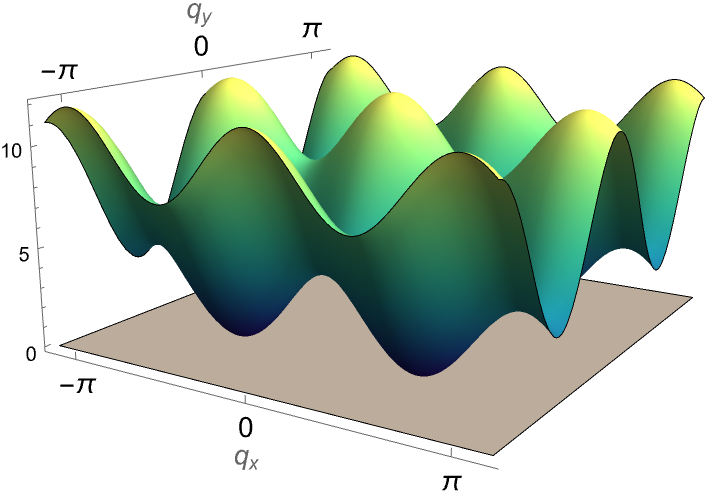} & 
        \includegraphics[width=2.5cm]{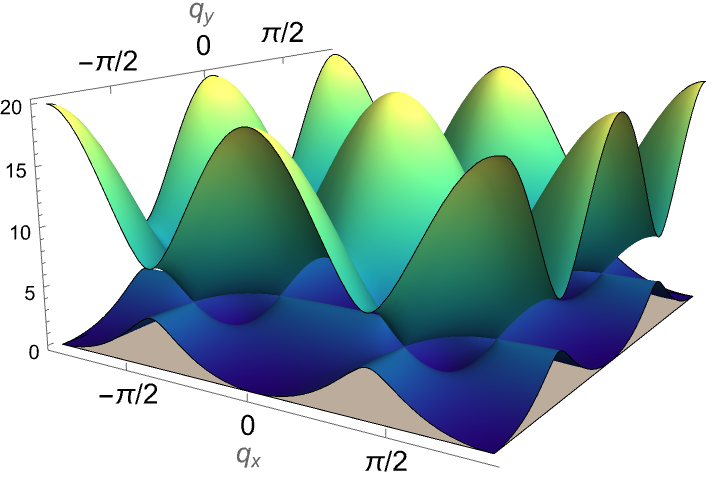} & 
        \includegraphics[width=2.5cm]{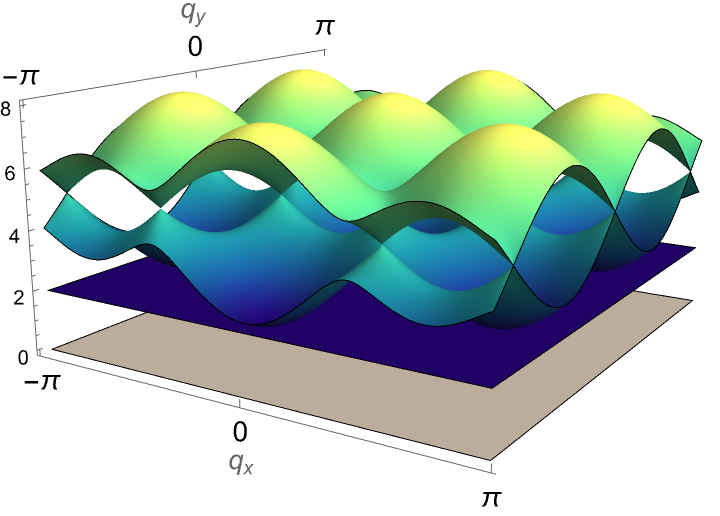} & 
        \includegraphics[width=2.5cm]{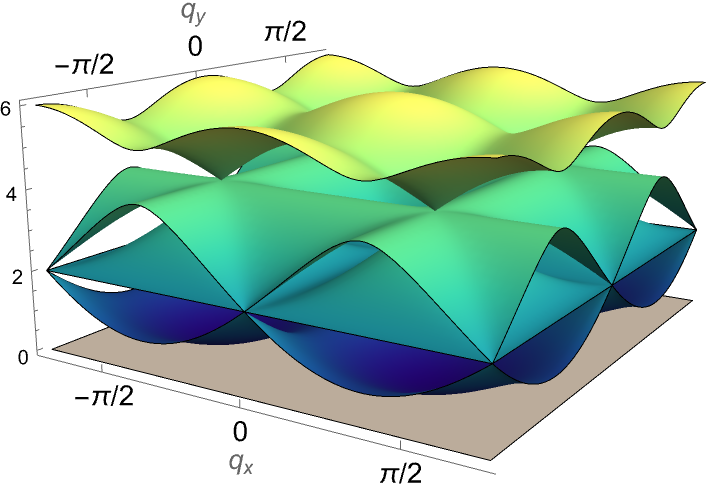} & 
        \includegraphics[width=2.5cm]{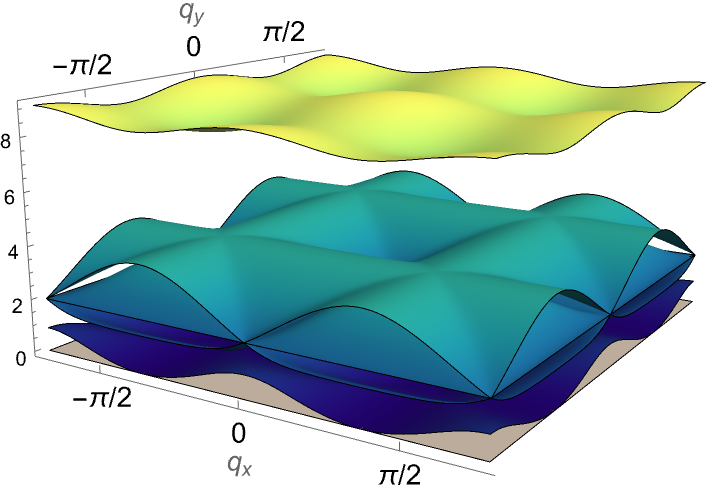} \\

        \makecell{ \vspace{-2.cm} \\ \textbf{Band}  \\ \textbf{Structure} } & 
        \includegraphics[width=2.5cm]{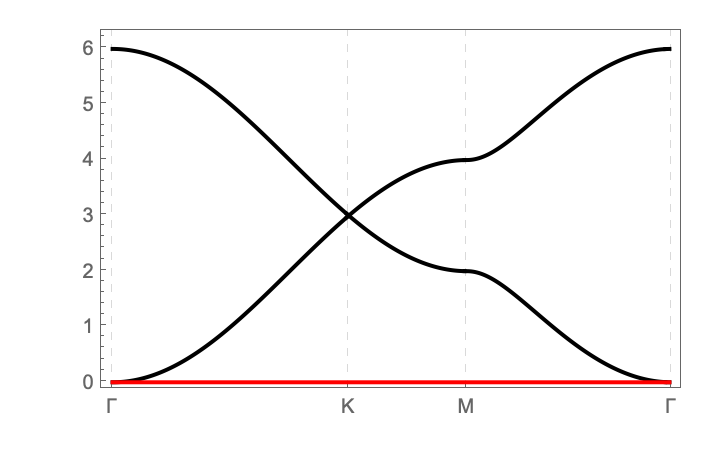} & 
        \includegraphics[width=2.5cm]{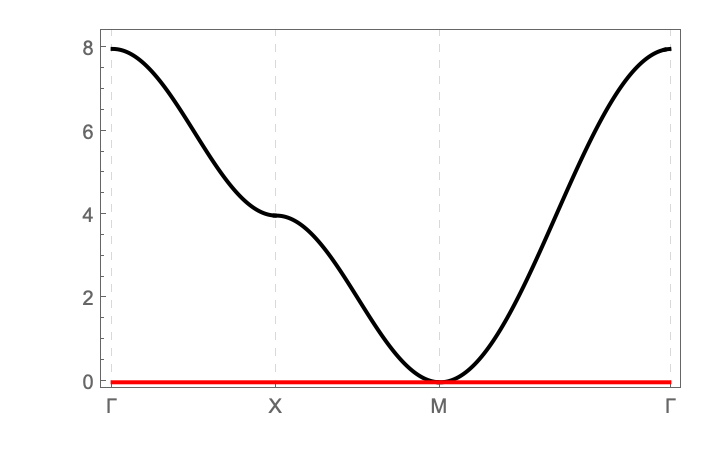} & 
        \includegraphics[width=2.5cm]{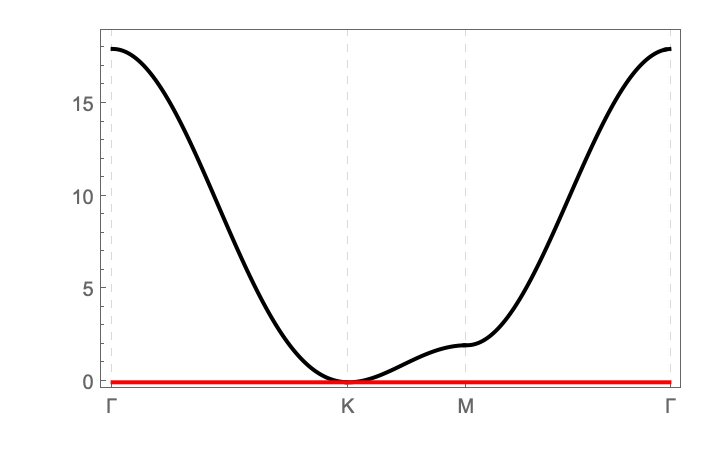} & 
        \includegraphics[width=2.5cm]{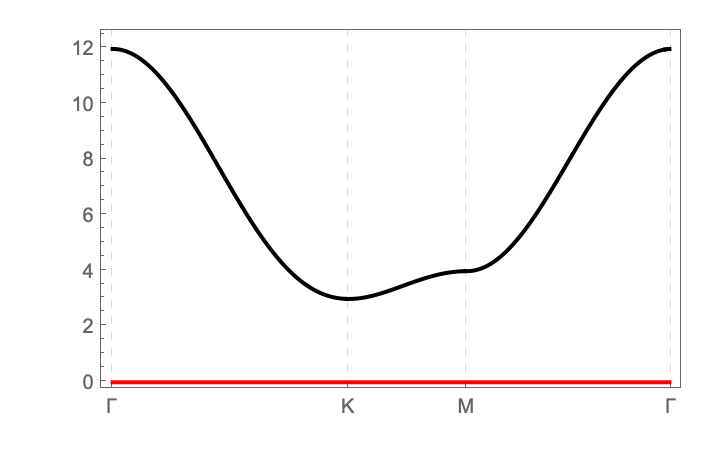} & 
        \includegraphics[width=2.5cm]{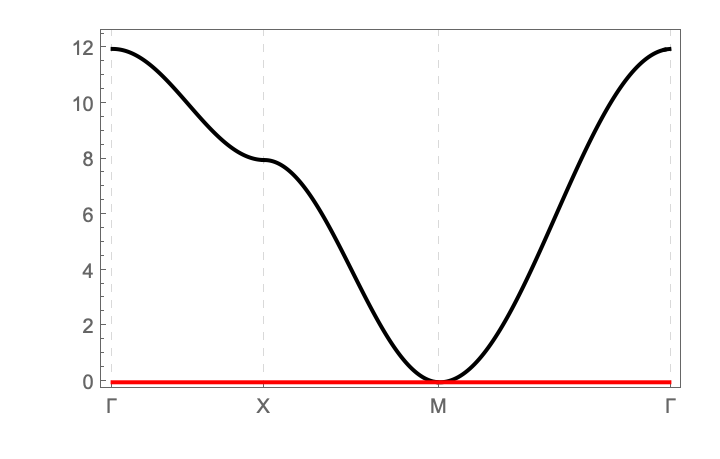} & 
        \includegraphics[width=2.5cm]{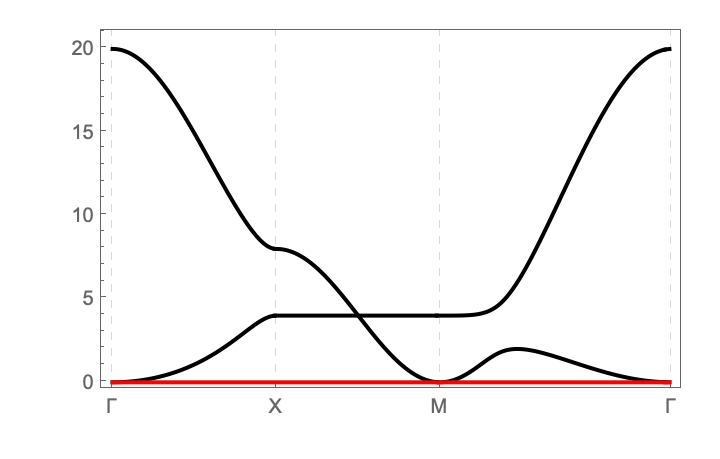} & 
        \includegraphics[width=2.5cm]{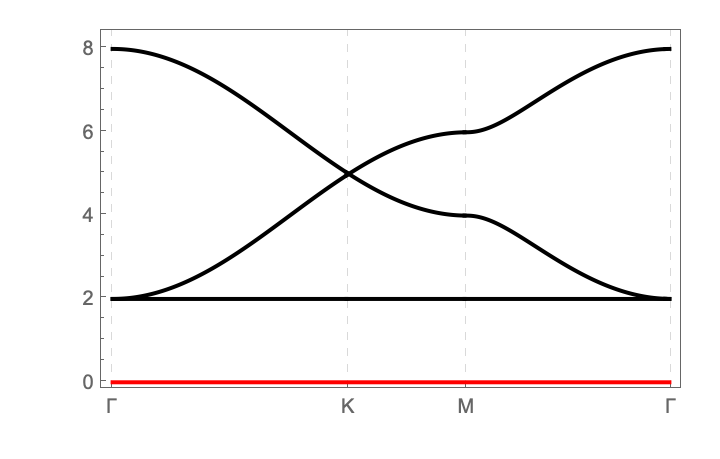} & 
        \includegraphics[width=2.5cm]{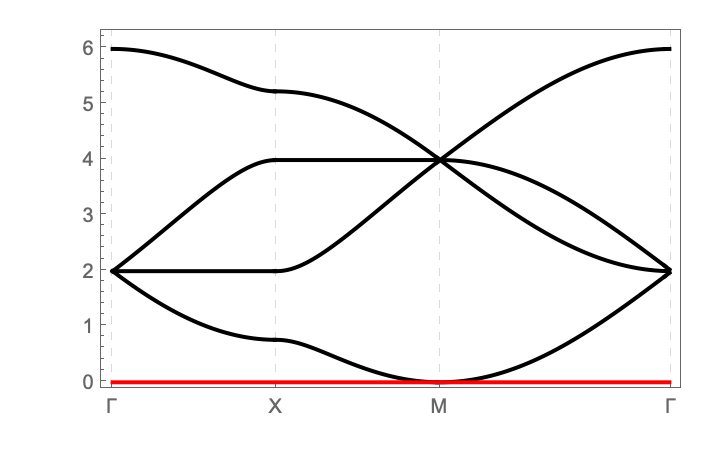} & 
        \includegraphics[width=2.5cm]{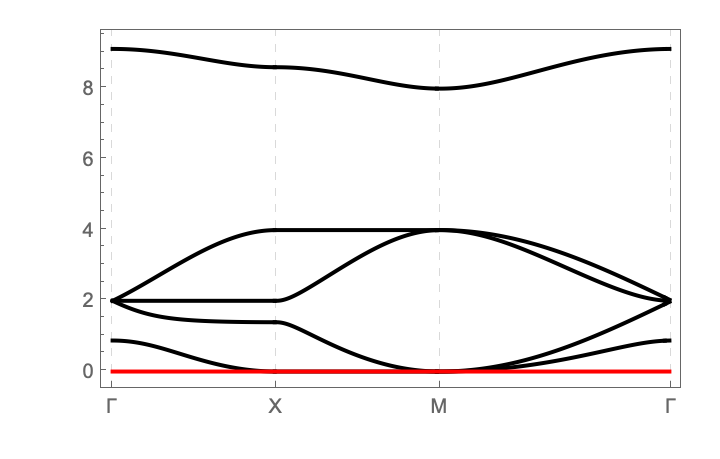} \\

        \textbf{$n_s$}     & 3 & 2 & 2 & 3 & 3 & 4 & 6 & 6 & 6 \\
        \textbf{$n_c$}     & 2 & 1 & 1 & 1 & 1 & 2 & 3 & 4 & 5 \\ 
        \textbf{$n_\text{b.f.b}$} & 1 & 1 & 1 & 2 & 2 & 2 & 3 & 2 & 1 \\
        
        \textbf{References} & \cite{OBDkagome, Davier_2023, Yan_2024_long} & \cite{Davier_2023, Yan_2024_long} & \cite{Benton_Moessner_2021, Yan_2024_long} & \cite{Davier_2023, Yan_2024_long} & & \cite{Codello_2010_square_octagon} & \cite{Rehn_2017_Ruby_lattice, Verresen_2021_ruby_lattice} & \cite{Siddharthan_2001_square_kagome} & \cite{Gonzalez_2025_decorated_square_kagome} \\
        \hline
    \end{tabular}}
    \caption{Examples of common two-dimensional cluster systems. Each system can be viewed as a decoration (bond, vertex, or bond--vertex) of an underlying parent lattice, whose name is indicated for each case. The $n_c$ distinct cluster types within a given lattice are shown in different colors; a colored cluster indicates that the spins within that cluster are fully connected. The number of bottom flat bands, $n_{\mathrm{b.f.b}}$, can be estimated from the number of sublattices $n_s$ and the number of cluster types $n_c$ as $n_{\mathrm{b.f.b}} = n_s - n_c$, in perfect agreement with large-$\mathcal{N}$ band-structure calculations. The values of $n_s$ and $n_{\mathrm{b.f.b}}$ reported here apply to clusters hosting only spins located at their vertices. All structure factors are obtained by projecting onto the flat-band subspace. Reciprocal-space axes are given in units of $a^{-1}$, where $a$ is the nearest-neighbor distance, and all clusters are regular (equilateral triangles, squares, etc.). The first Brillouin-zone contour (see Fig.~\ref{fig:BZ}) is indicated by a dashed red line in each structure-factor panel.
    The band structure is displayed both in a 3D perspective fashion and along the first BZ special path depicted on Fig. \ref{fig:BZ}.}
    \label{tab: 2d cluster systems}
\end{table*}

\section{Building cluster models}
\label{Section : Building cluster systems}

\begin{figure}[ht]
    \centering
    \includegraphics[height=1.1cm]{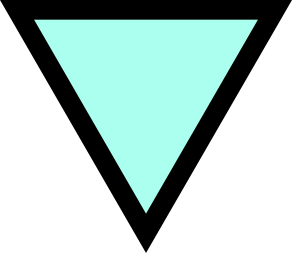} \hspace{0.3 mm}
    \includegraphics[height=1.15cm]{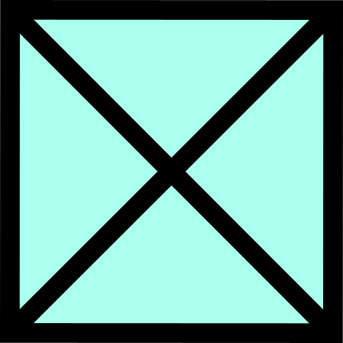} \hspace{1 mm}
    \includegraphics[height=1.2cm]{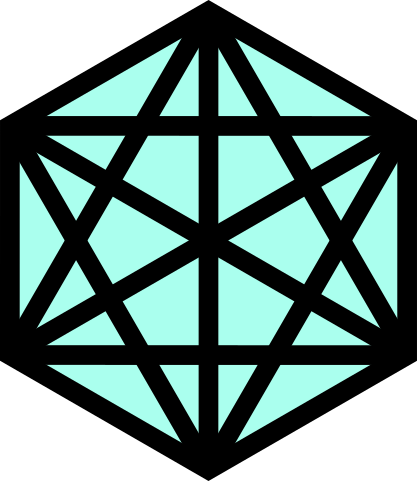} \hspace{1 mm}
    \includegraphics[height=1.2cm]{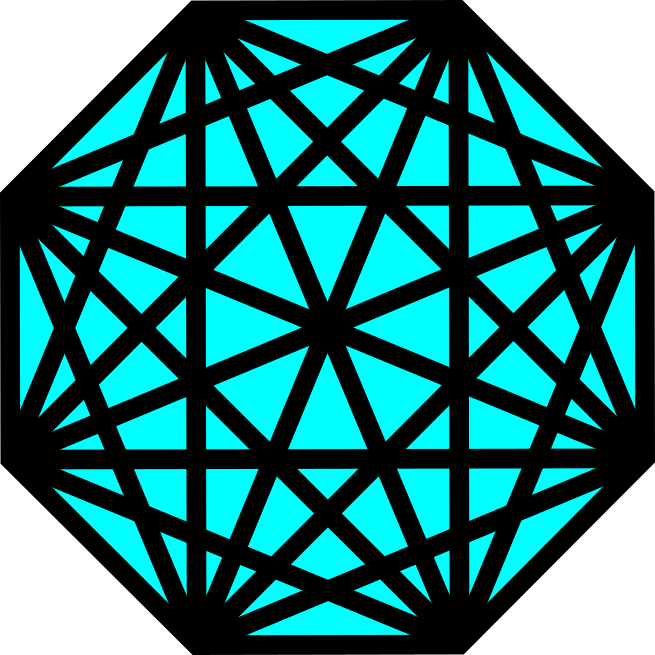} \hspace{1 mm}
    \includegraphics[height=1.2cm]{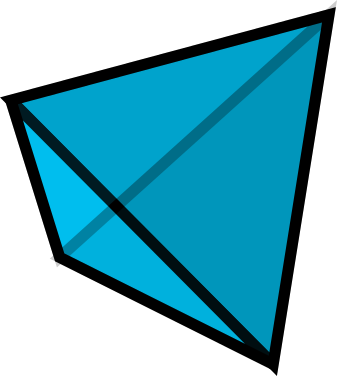} \hspace{0. mm}
    \includegraphics[height=1.2cm]{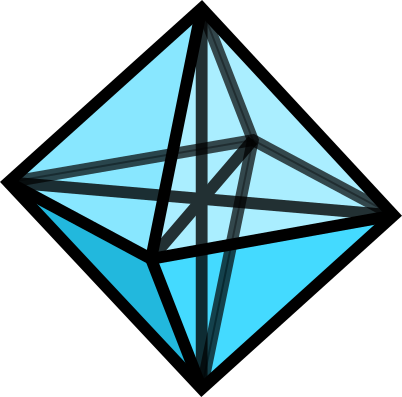}
    \caption{Common choices of clusters: triangles, squares, hexagons, octagons, tetrahedra, or octahedra.}
    \label{fig:clusterexamples}
\end{figure}

The construction of a cluster Hamiltonian begins with the choice of cluster type. By definition, every spin within a cluster interacts with all other spins in the cluster. Beyond this condition, considerable freedom remains in selecting the geometry and size of the clusters. Common examples of clusters include triangles, squares, hexagons, octagons, tetrahedra, and octahedra, as illustrated in Fig.~\ref{fig:clusterexamples}.
To form an extended system, clusters are connected by sharing vertices, edges, or faces.
Representative two-dimensional and three-dimensional realizations are listed in Tables~\ref{tab: 2d cluster systems} and~\ref{tab: 3d cluster systems}.
Notably, the dimensionality of the lattice need not coincide with that of the constituent clusters. For instance, the hyperkagome lattice (Table~\ref{tab: 3d cluster systems}) is constructed from two-dimensional triangular clusters arranged in a three-dimensional network.

\begin{table*}[t]
    \centering
    \renewcommand{\arraystretch}{1.5} 
    \resizebox{\textwidth}{!}{
    \begin{tabular}{c c c c c c}
        \hline
        \textbf{\makecell{3D Cluster \\ Lattice}} & Pyrochlore & Octahedral & \makecell{Kagome\\ bipyramidal} & Quadrupahedral & Hyperkagome\\
        \hline
        \textbf{\makecell{Parent \\ Lattice}} & Diamond & Cubic & \makecell{Staggered\\ Honeycomb} & \makecell{Staggered\\ Honeycomb} & \makecell{Hyperkagome\\ bidual}\\
        \hline
        \noalign{\vskip 0.5 mm}
        \makecell{ \vspace{-3cm} \\ \textbf{Lattice}  \\ \textbf{Scheme} } & 
        \includegraphics[height=3cm]{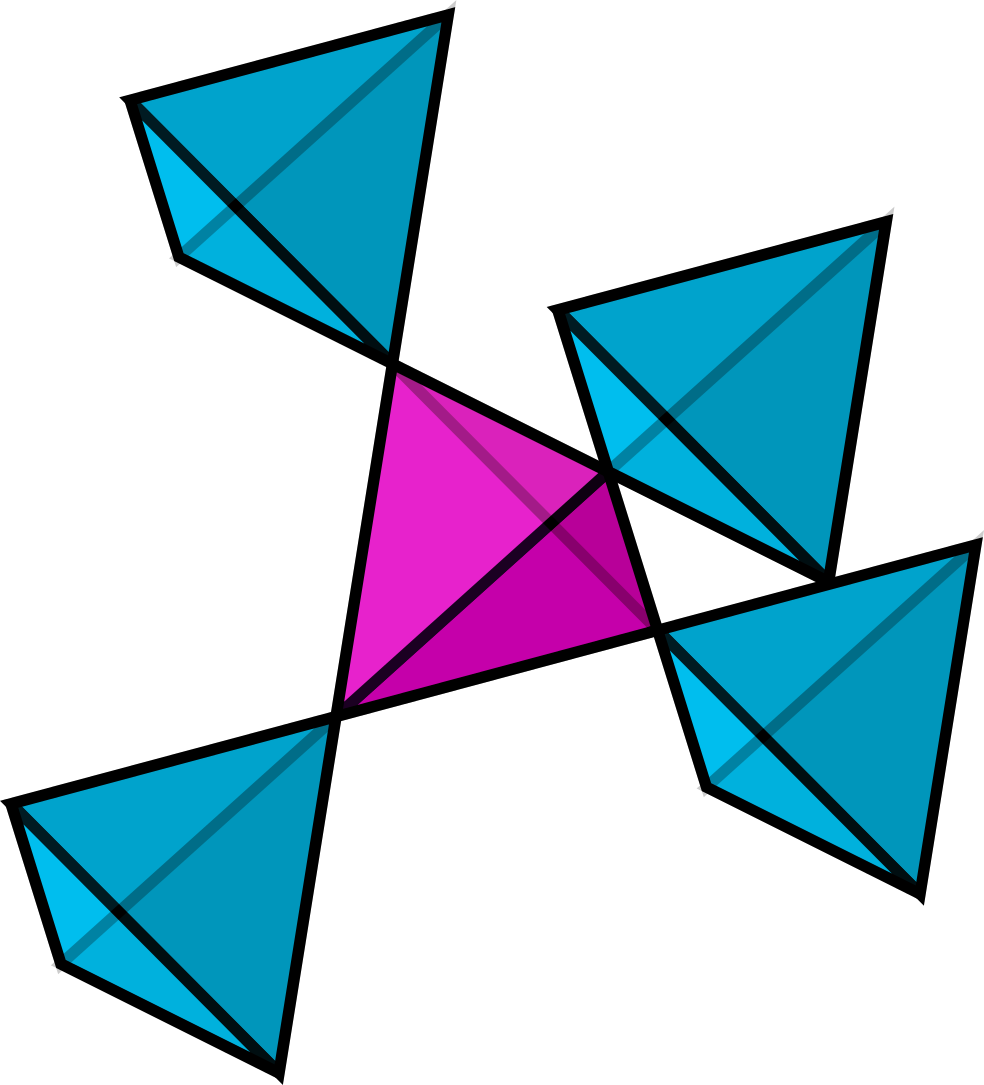} &
        \includegraphics[height=3cm]{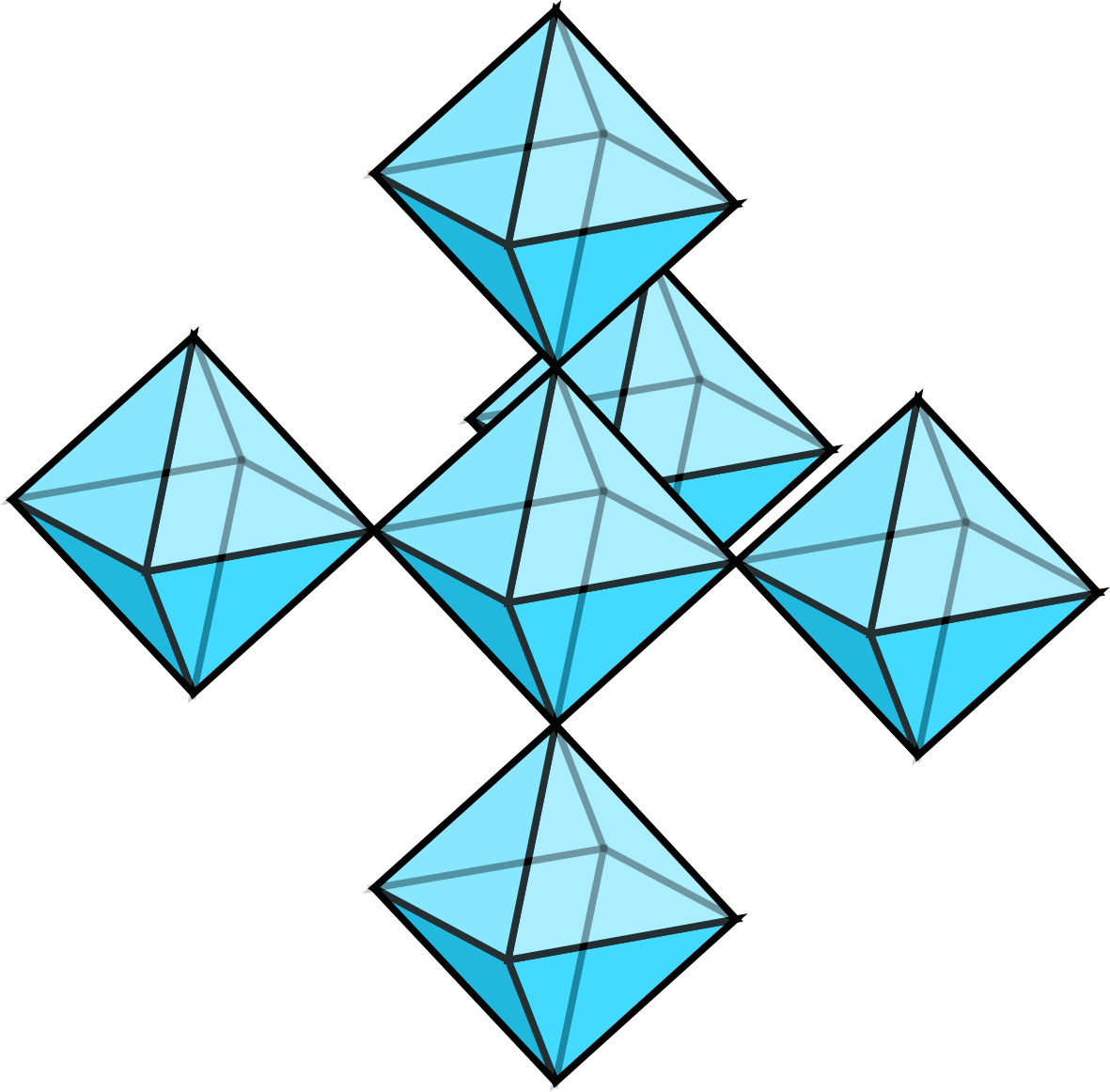} & 
        \includegraphics[height=3cm]{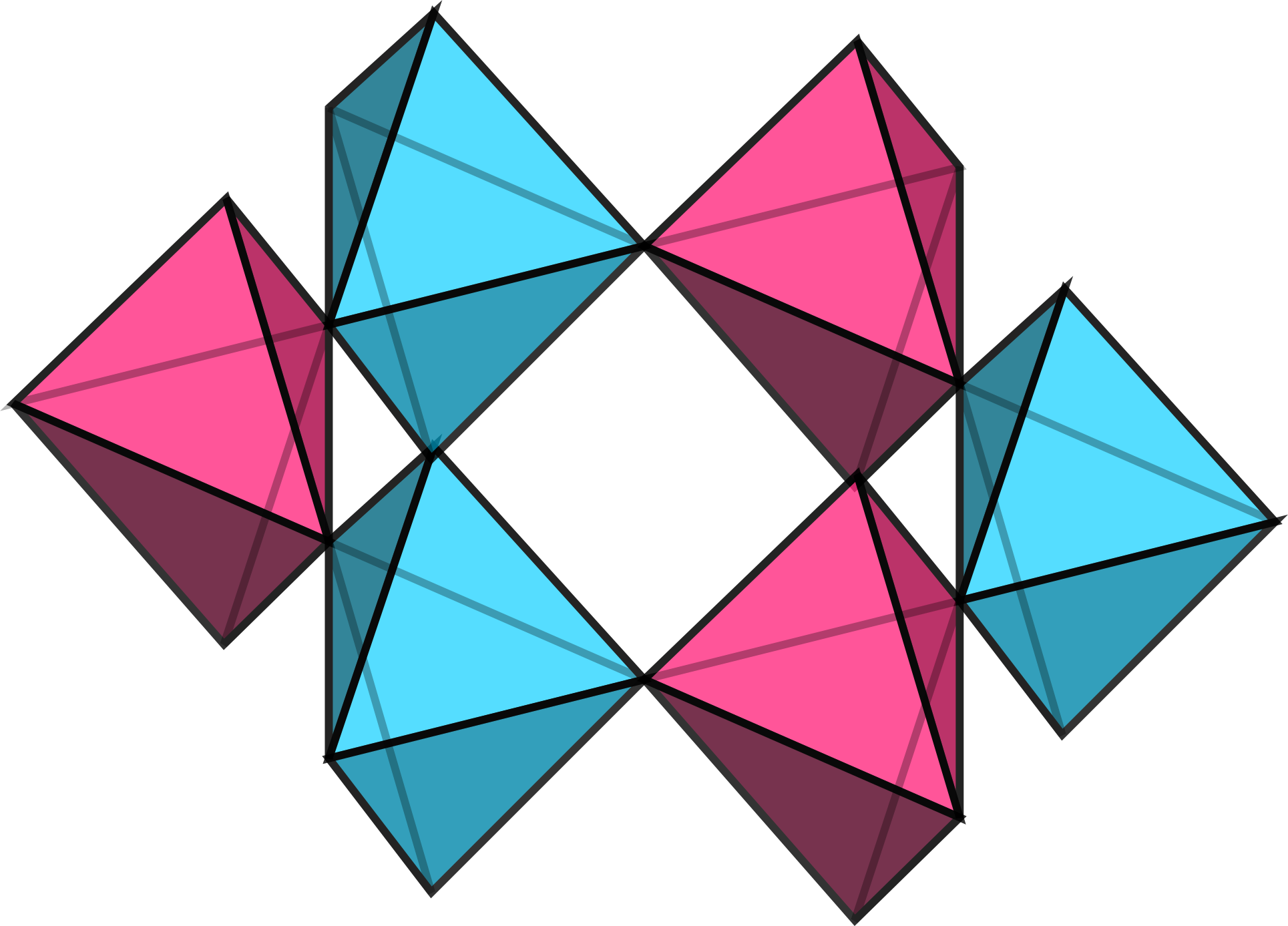} & 
        \includegraphics[height=3cm]{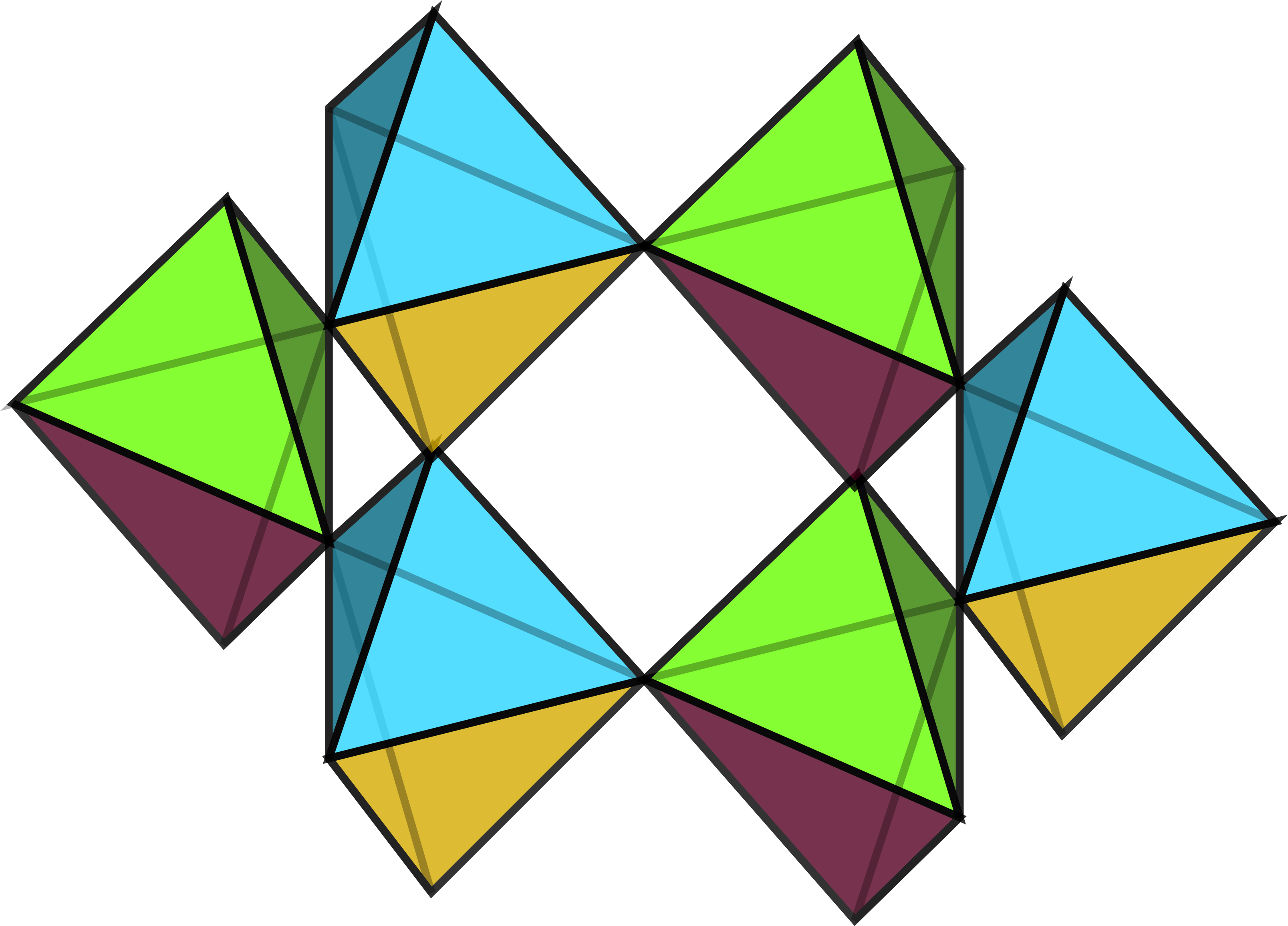} & 
        \includegraphics[height=3cm]{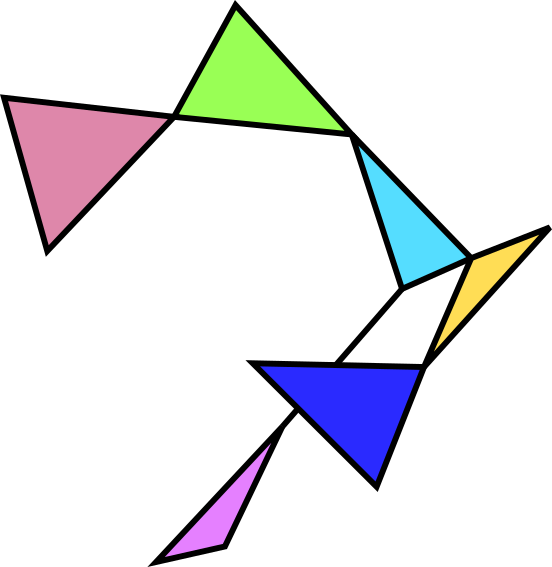} \\
        \textbf{$n_s$} & 4 & 3 & 5 & 5 & 12 \\
        \textbf{$n_c$} & 2 & 1 & 2 & 4 & 6 \\
        \textbf{$n_\text{b.f.b}$} & 2 & 2 & 3 & 1 & 6 \\
        \textbf{References} & \cite{Henley2005, Yan_2024_long} & \cite{Benton_Moessner_2021, Niggemann_2023} & & & \cite{Hopkinson_2007_Hyperkagome, Okamoto_2007_hyperkagome} \\
        \hline
    \end{tabular}}
    \caption{Non-exhaustive list of three-dimensional cluster systems, built from either 2D or 3D clusters. The distinct cluster types within a given lattice are shown in different colors. For each system, the number of sublattices $n_s$, cluster types $n_c$, and bottom flat bands $n_{\mathrm{b.f.b}}$ are reported in the simplest case where spins occupy only the cluster vertices. Note that the kagome bipyramidal lattice is composed of bipyramidal clusters with six faces, whereas the quadrupahedral lattice consists of tetrahedra sharing a face. }
    \label{tab: 3d cluster systems}
\end{table*}

\begin{figure}[h]
    \centering
    \includegraphics[width=0.6\linewidth]{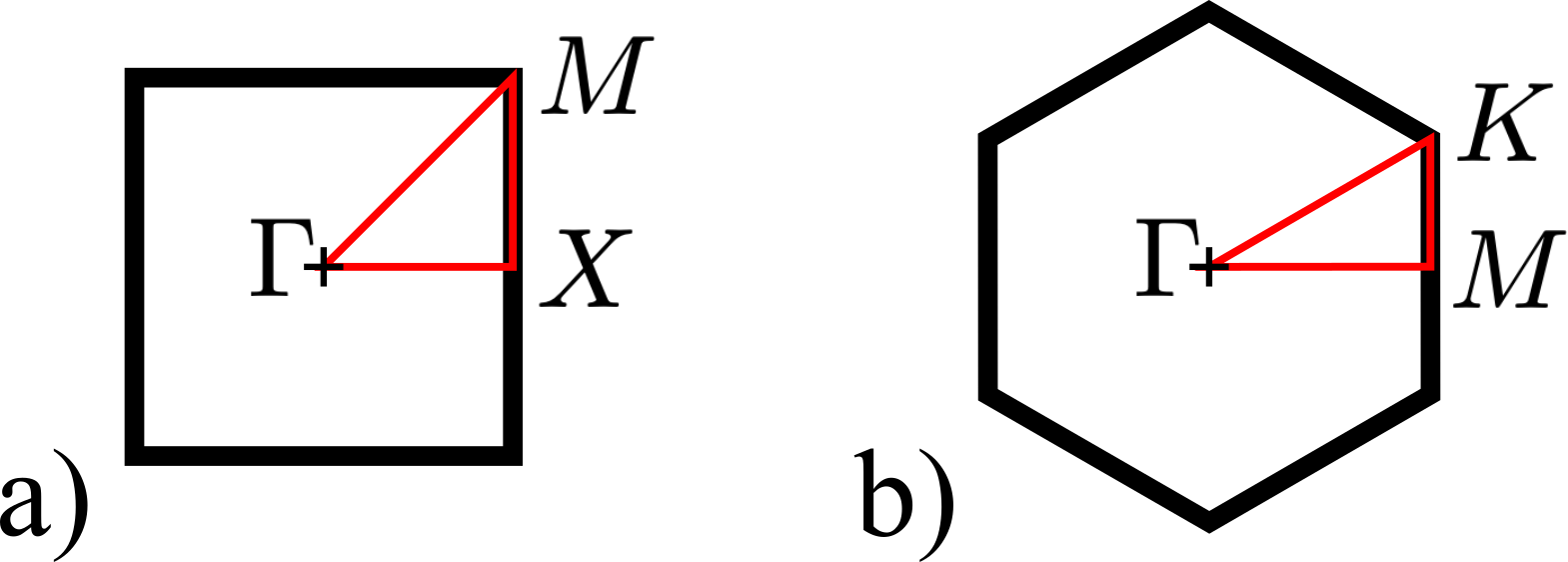}
    \caption{First Brillouin zones of the a) square lattice and b) triangular lattice. }
    \label{fig:BZ}
\end{figure}

Most of the cluster lattice presented in this work are displayed with their zero temperature structure factor and their band structure, calculated using the large $\mathcal{N}$ approximation formalism detailed in Appendix. \ref{Appendix:Large N}. Almost all the lattices are based on the triangular or square lattices, which first Brillouin zone are depicted on Fig.~\ref{fig:BZ}. 

Spins are not restricted to cluster vertices: they may also occupy other sites provided the cluster topology and unit cell are preserved. A particularly simple modification is the addition of a central spin at the geometric center of each cluster. For such centered clusters, each central spin adds an extra sublattice, thus increasing the number of flat bands that can be expressed as $n_\text{b.f.b}^\text{ctrd} = n_\text{b.f.b}^0 + n_c$ where $n_\text{f.b.}^0$ is the number of flat bands in the vertex-only case. More generally, if $m$ additional non-vertex sites are included per cluster, the flat-band count becomes $n_\text{b.f.b}^m = n_\text{b.f.b}^0 + m \times n_c$. For clarity, the different tables presenting examples of cluster systems all along this work report the baseline values $n_\text{f.b.}^0$ for vertex-only systems.

\begin{figure}[ht]
    \centering
    \includegraphics[width = \linewidth]{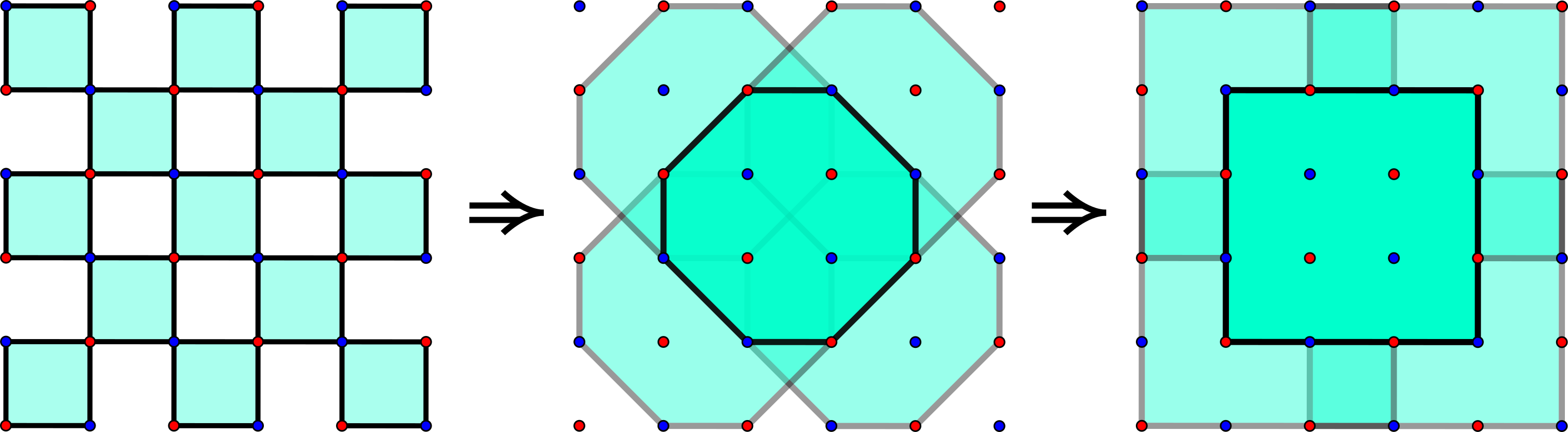}
    \caption{Example of possible cluster extension on the checkerboard lattice.}
    \label{fig: checkerboiard extended cluster}
\end{figure}

Another systematic way to generate new cluster systems is by enlarging existing clusters to include further-neighbor sites, while preserving translational symmetry and the unit cell. For example, in the checkerboard lattice, clusters can be extended to incorporate the eight or even twelve nearest neighbors~\cite{Davier_2023}, see Fig.~\ref{fig: checkerboiard extended cluster}.
This operation leaves the number of flat bands unchanged, so the existence of a spin-liquid phase is maintained if it was already present. However, the precise nature of the spin liquid can be significantly modified, as demonstrated in checkerboard, kagome, hexagonal, and octochlore systems~\cite{Benton_Moessner_2021, Davier_2023, Yan_2024_long, Niggemann_2023}.

These basic ingredients—choice of cluster, connectivity, spin placement, and possible cluster extension—enable the construction of a wide range of cluster systems. Moreover, they can be combined with more systematic procedures~\cite{Fang_2024}, which allow almost any lattice to be upgraded into a cluster lattice in a controlled fashion as we will now see.

\section{Decorated cluster lattices}
\label{Section : Decorated cluster lattices}

\begin{table*}[ht]
    \centering
    \renewcommand{\arraystretch}{1.5} 
    \resizebox{\textwidth}{!}{
    \begin{tabular}{c c c c c c c}
        \hline    
        \textbf{2D Lattice} & \makecell{Square \\ monodiamond } & \makecell{Hexagonal \\ monodiamond } & \makecell{Kagome \\ monodiamond } & \makecell{Triangular \\ monodiamond }  & \makecell{Square \\ monocracker}  & \makecell{Hexagonal\\ monocracker} \\
        \hline
        \noalign{\vskip 1mm}
        \makecell{ \vspace{-2.5cm} \\ \textbf{Lattice}  \\ \textbf{Scheme} } & 
        \includegraphics[height=2.5cm]{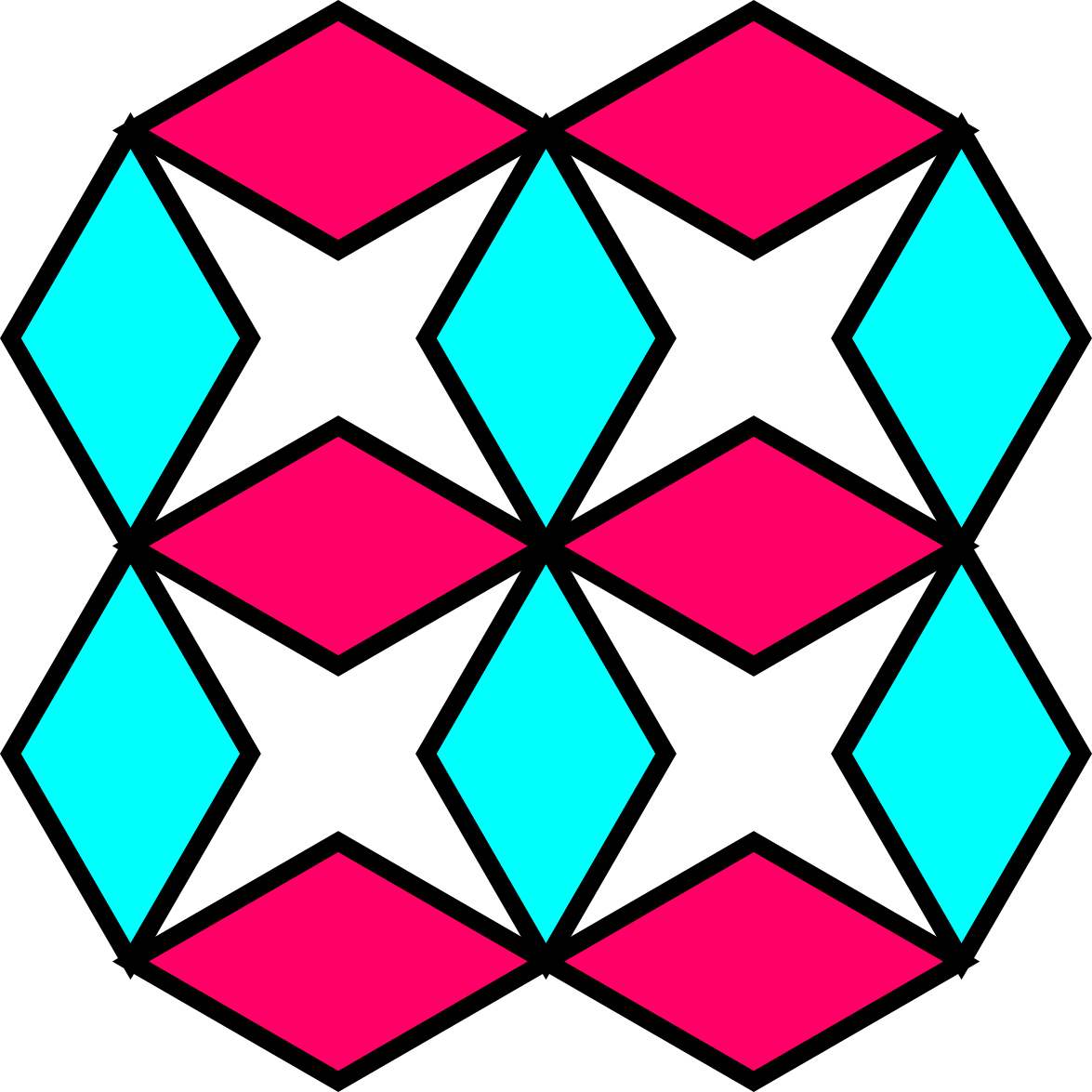} & 
        \includegraphics[height=2.5cm]{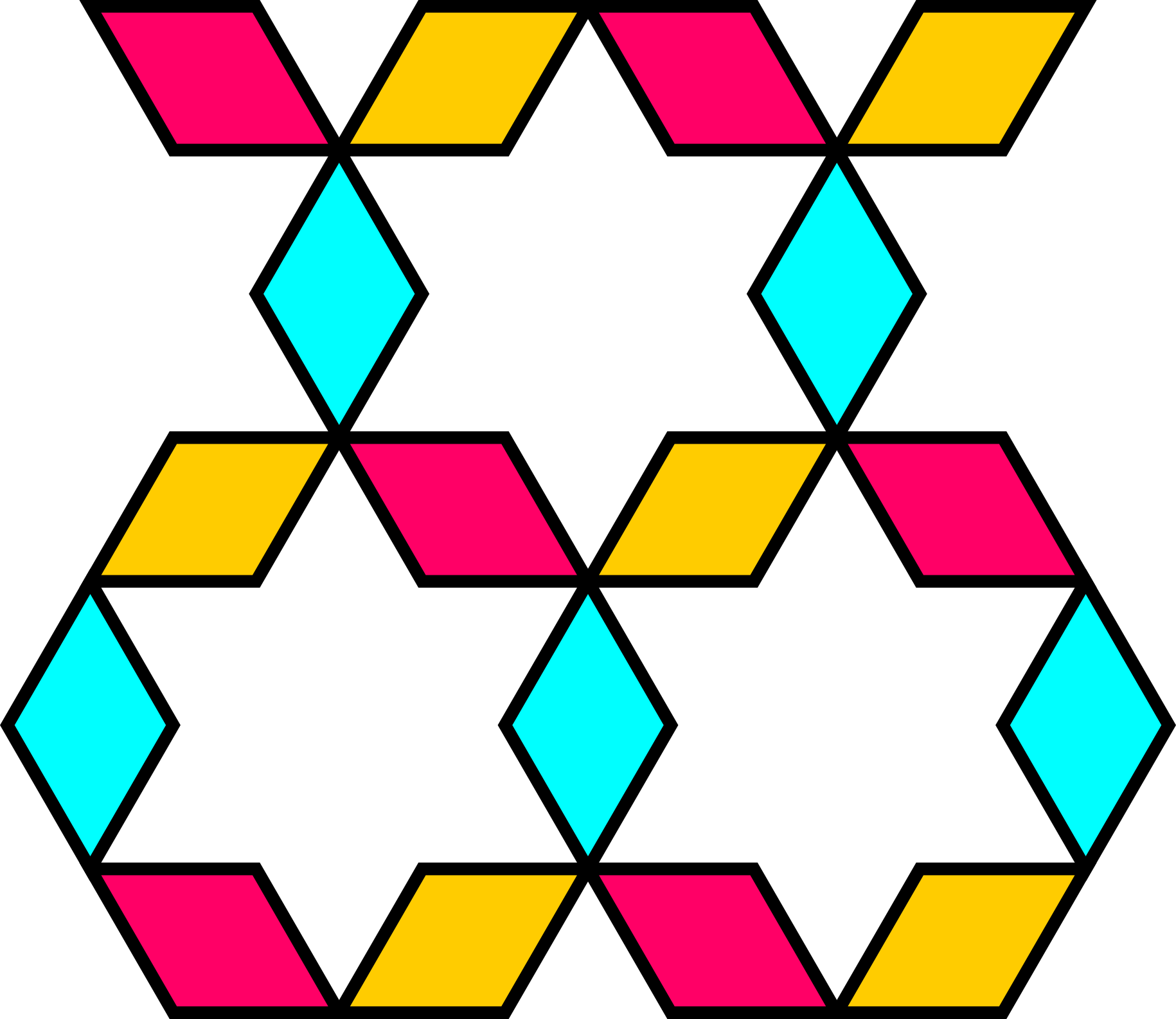} &
        \includegraphics[height=2.5cm]{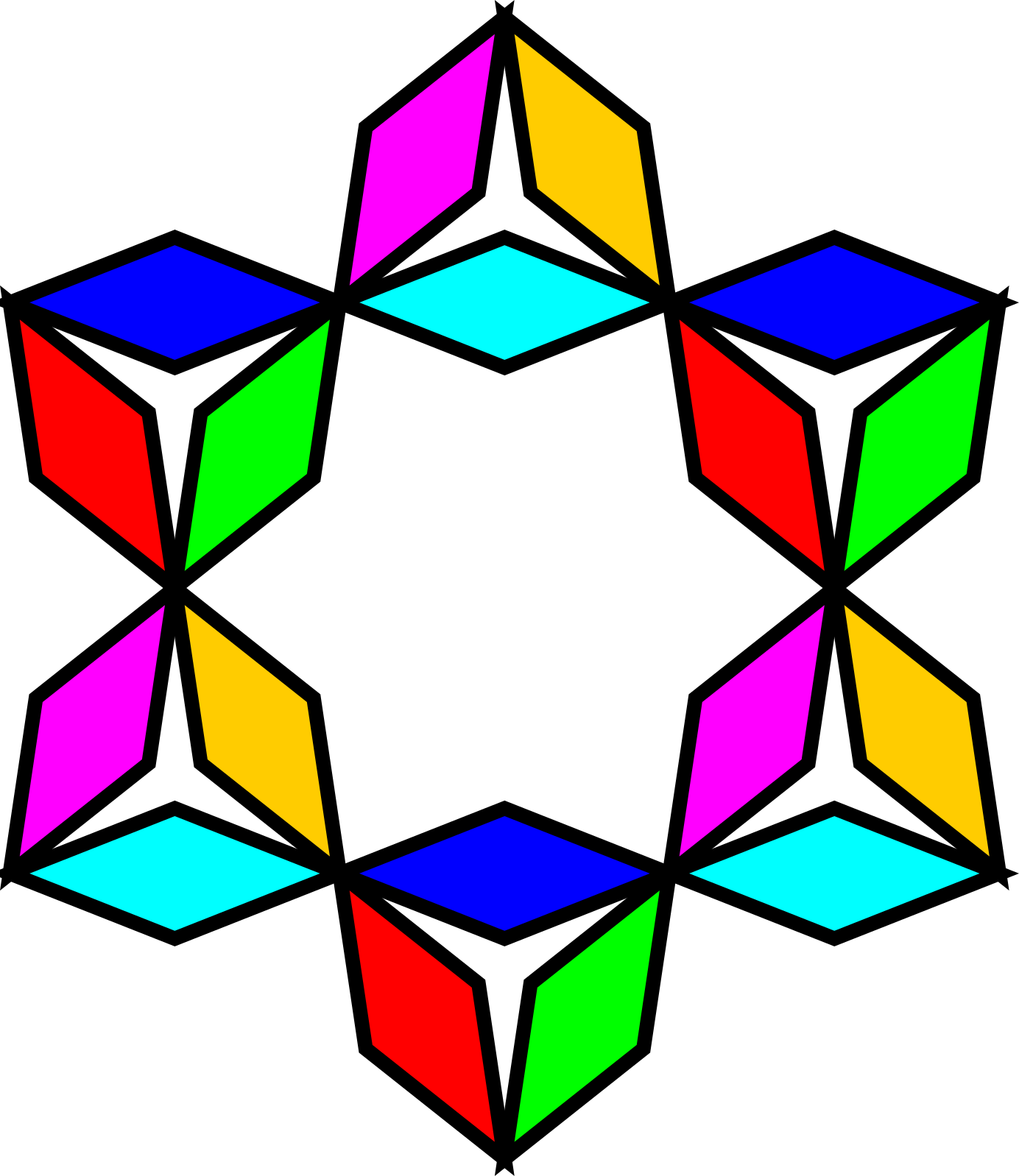} & 
        \includegraphics[height=2.5cm]{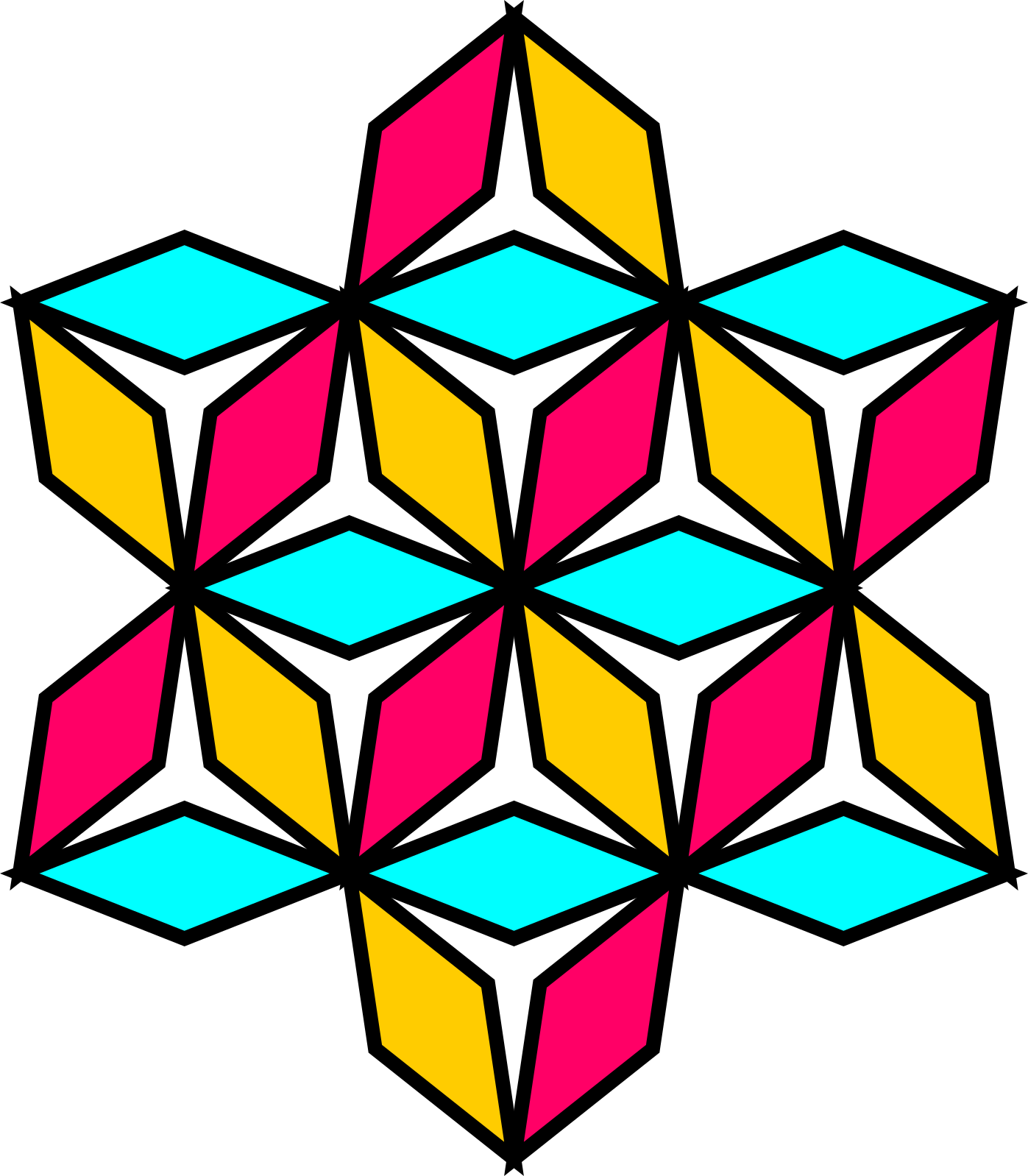} & 
        \includegraphics[height=2.5cm]{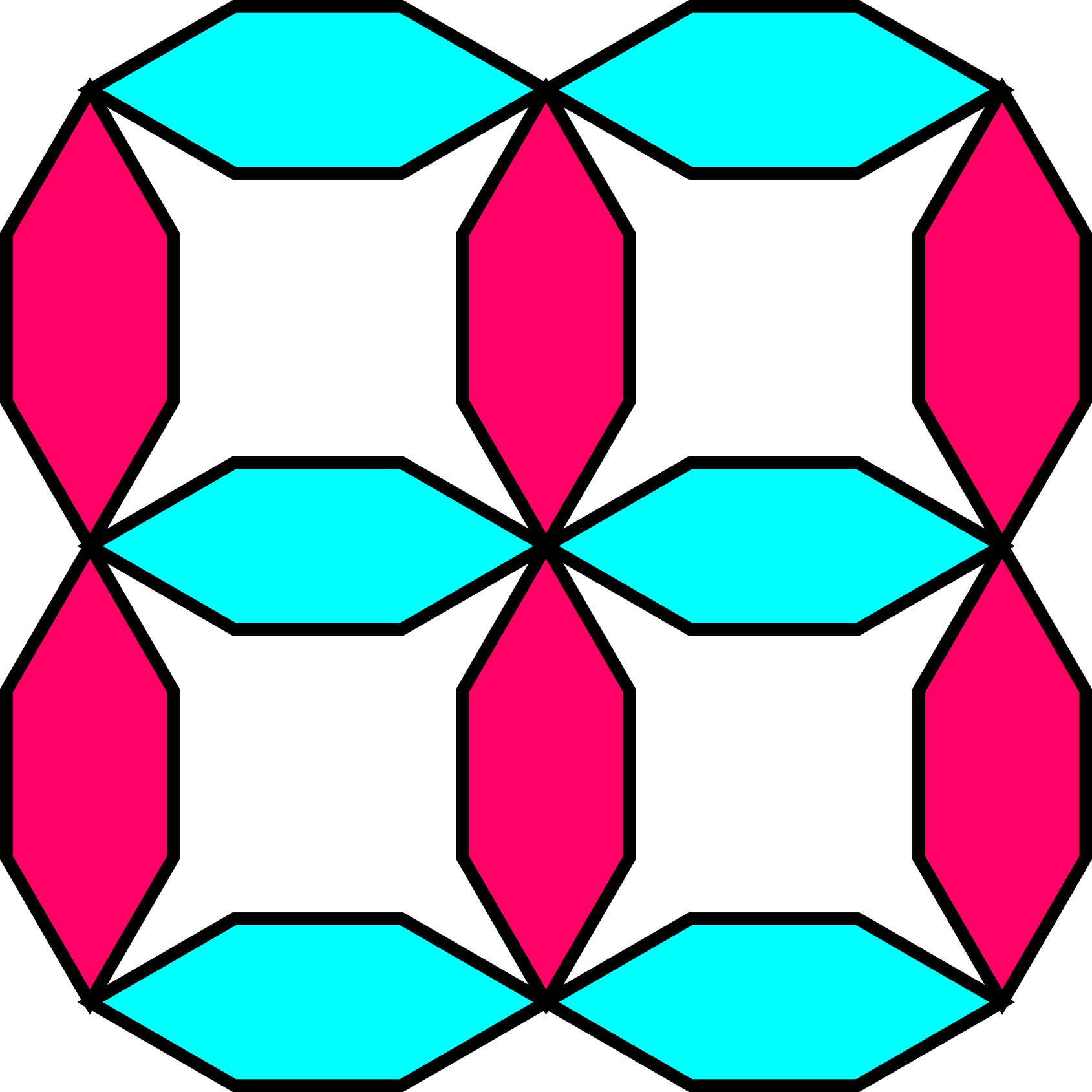} & 
        \includegraphics[height=2.5cm]{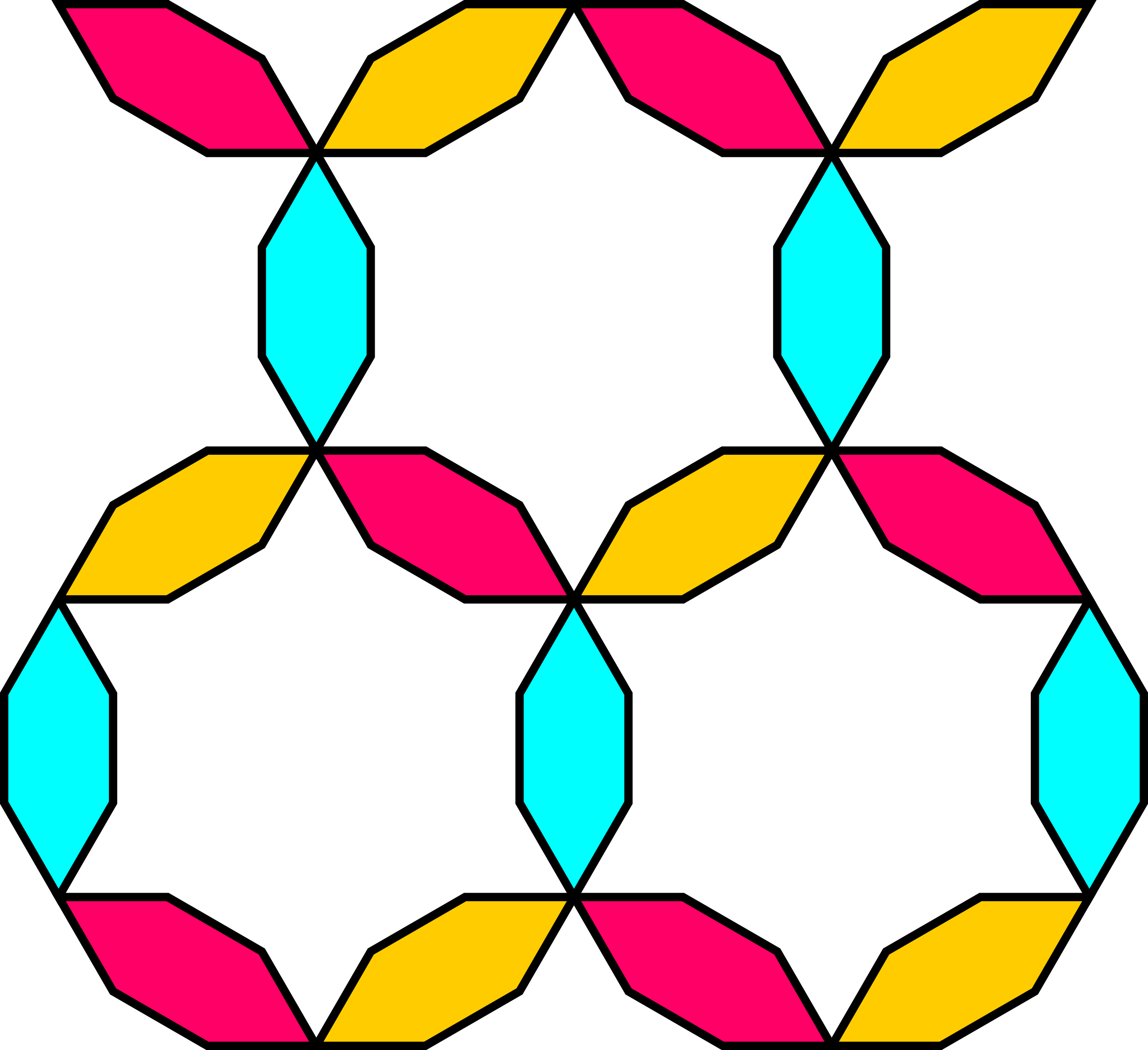} \\

        \makecell{ \vspace{-2.5cm} \\ \textbf{3D Band}  \\ \textbf{Structure} } & 
        \includegraphics[height=2.5cm]{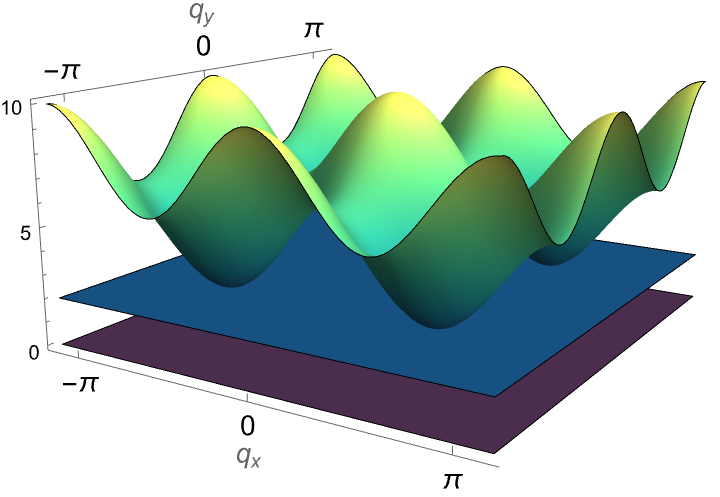} & 
        \includegraphics[height=2.5cm]{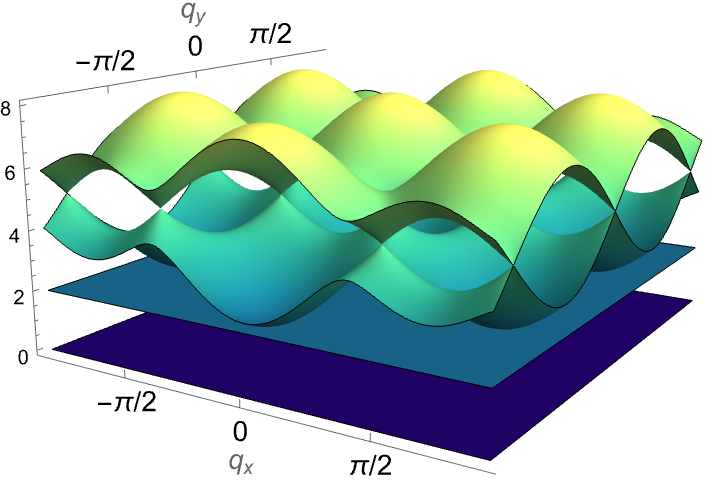} & 
        \includegraphics[height=2.5cm]{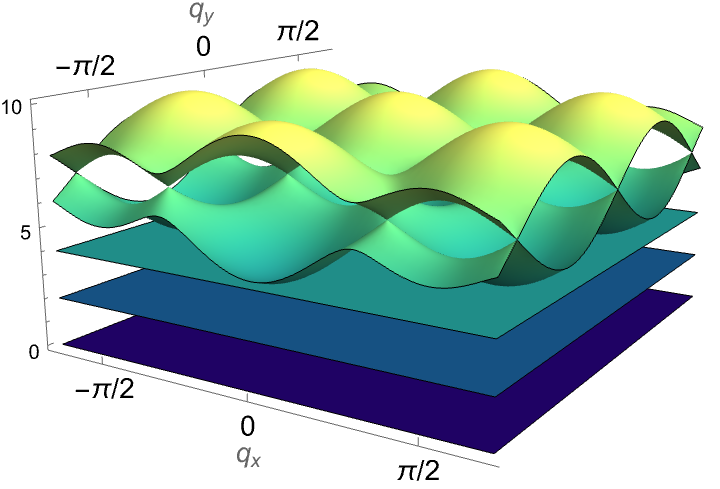} & 
        \includegraphics[height=2.5cm]{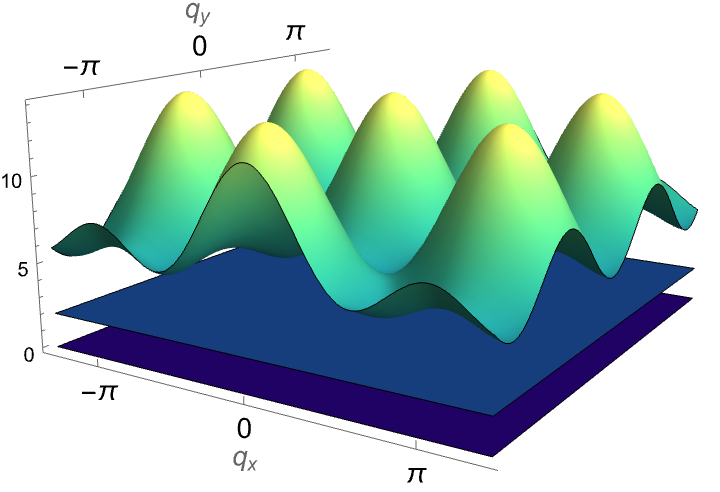} & 
        \includegraphics[height=2.5cm]{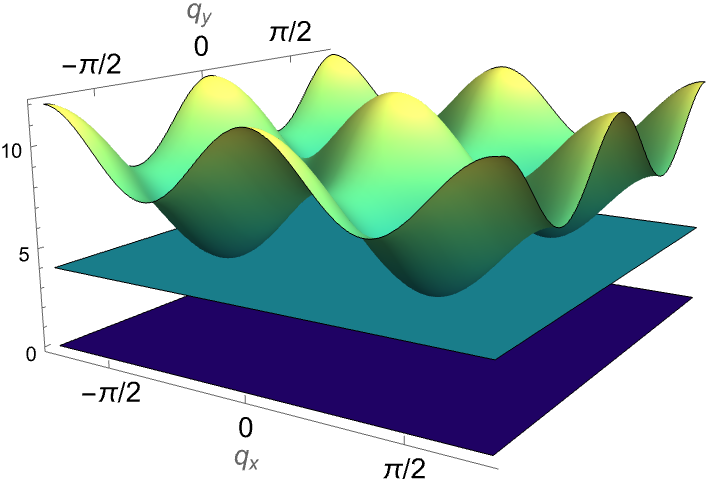} & 
        \includegraphics[height=2.5cm]{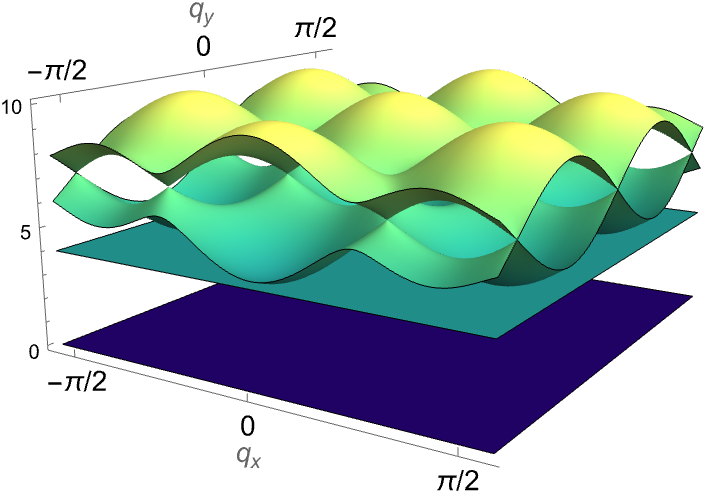} \\

        \makecell{ \vspace{-2.8cm} \\ \textbf{Band}  \\ \textbf{Structure} } & 
        \includegraphics[height=2.5cm]{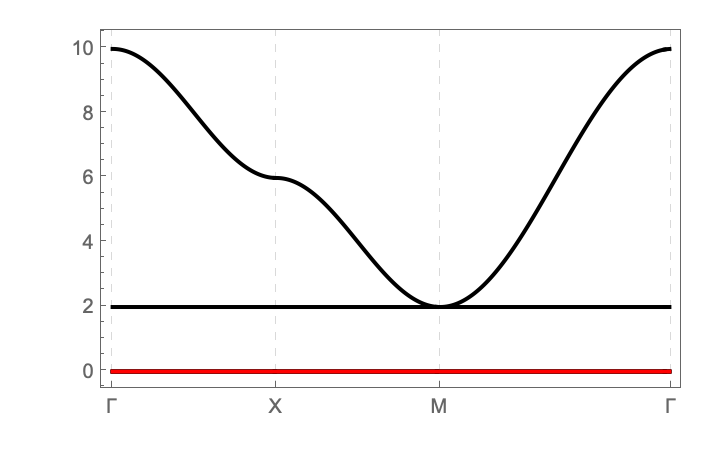} & 
        \includegraphics[height=2.5cm]{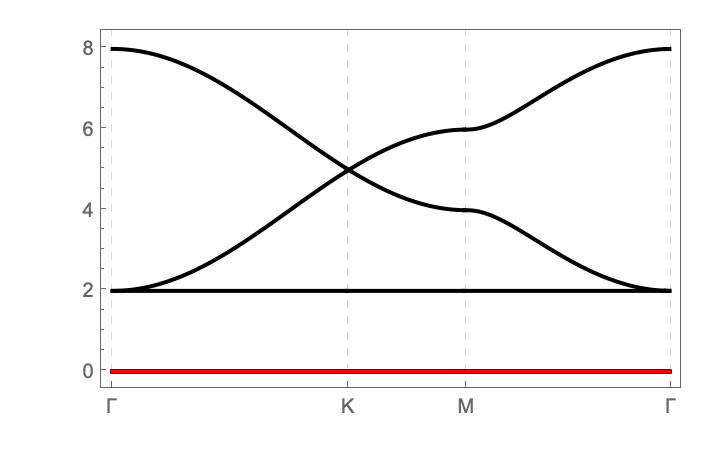} & 
        \includegraphics[height=2.5cm]{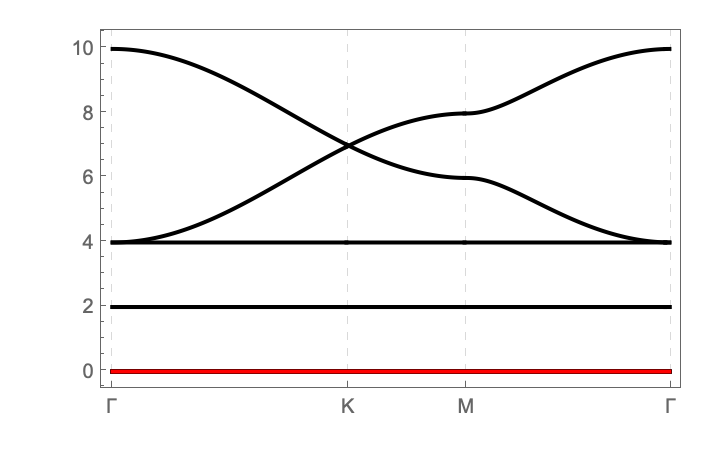} & 
        \includegraphics[height=2.5cm]{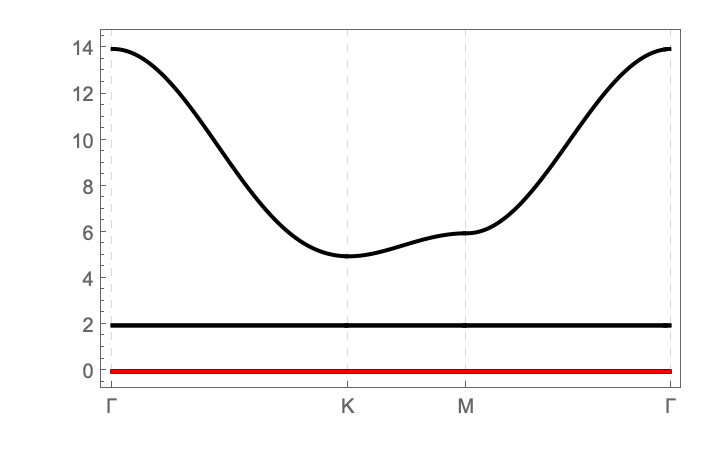} & 
        \includegraphics[height=2.5cm]{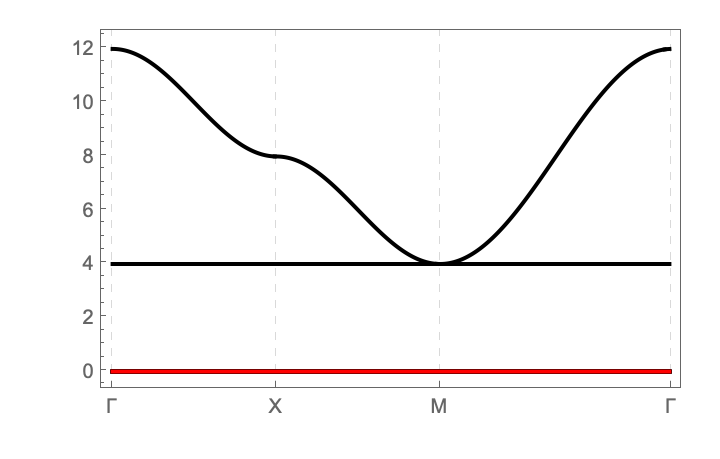} & 
        \includegraphics[height=2.5cm]{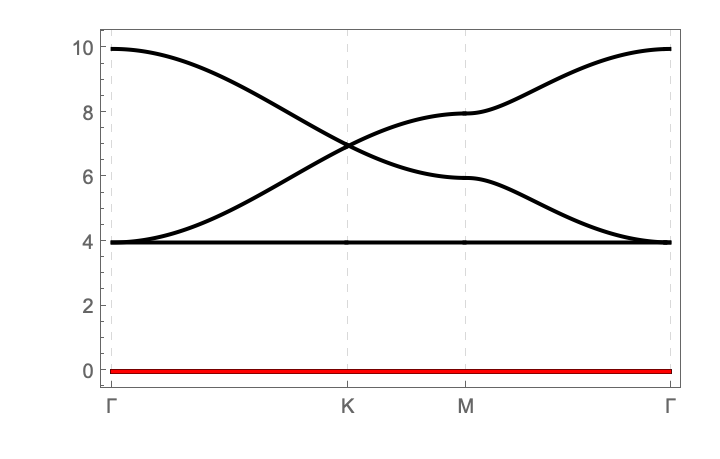} \\
        
        \textbf{$n_s$}            & 5 & 8 & 15  & 7 & 9 & 14 \\
        \textbf{$n_c$}            & 2 & 3 &  6  & 3 & 2 & 3  \\
        \textbf{$n_\text{b.f.b}$} & 3 & 5 &  9  & 4 & 7 & 11 \\
        \textbf{$n_\text{i.f.b}$} & 1 & 1 & 3+1 & 2 & 1 & 1  \\

        \hline
    \end{tabular}}
    \caption{\textbf{Monoblock bond-decorated systems in two dimensions}. 
    Each lattice is constructed from cluster-bonds acting as extended bonds. Identical clusters with the same orientation may nevertheless generate inequivalent local constraints when attached to different sublattices, resulting in distinct effective bonds which are thus depicted with distinct colors. The number of bottom flat bands $n_{\text{b.f.b}}$ is compared to the expected value $n_s - n_c$, assuming spins reside only on cluster vertices. 
    The number of the in-spectrum flat bands $n_{\text{i.f.b}}$ is indicated as a sum to precise the degeneracy of each flat band.
    The structure factors and band structures are shown in units of $a^{-1}$, where $a$ is the nearest-neighbor distance (for kagome- and triangular-based lattices, $a$ denotes the second-nearest-neighbor distance). The first Brillouin zone is indicated by a red contour in the structure-factor plots. 
    Mono-block bond-decorated systems are systematically gapped and therefore realize \emph{fragile topological} classical spin liquids; as their structure factors exhibit no pinch points, they are not shown. 
    The band structure is displayed both in a 3D perspective fashion and along the first BZ special path depicted on Fig. \ref{fig:BZ}.
    }
    \label{tab: 2d monobond decorated lattices}
\end{table*}

\begin{table*}[ht]
    \centering
    \renewcommand{\arraystretch}{1.5} 
    \resizebox{\textwidth}{!}{
    \begin{tabular}{c c c c c c c}    
        \hline    
        \textbf{2D Lattice} & \makecell{Square \\ diamond \cite{Morita_2016_Heisenberg_Diamond_Spin_Lattices, Caci_2023_decorated_square}}  &  \makecell{Hexagonal \\ diamond \cite{Morita_2016_Heisenberg_Diamond_Spin_Lattices}} & \makecell{Kagome \\ diamond} & \makecell{Triangular \\ diamond} & \makecell{Square \\ cracker} & \makecell{Hexagonal\\ cracker} \\
        \hline
        \noalign{\vskip 1mm}
        \makecell{ \vspace{-2.5cm} \\ \textbf{Lattice}  \\ \textbf{Scheme} } & 
        \includegraphics[height=2.5cm]{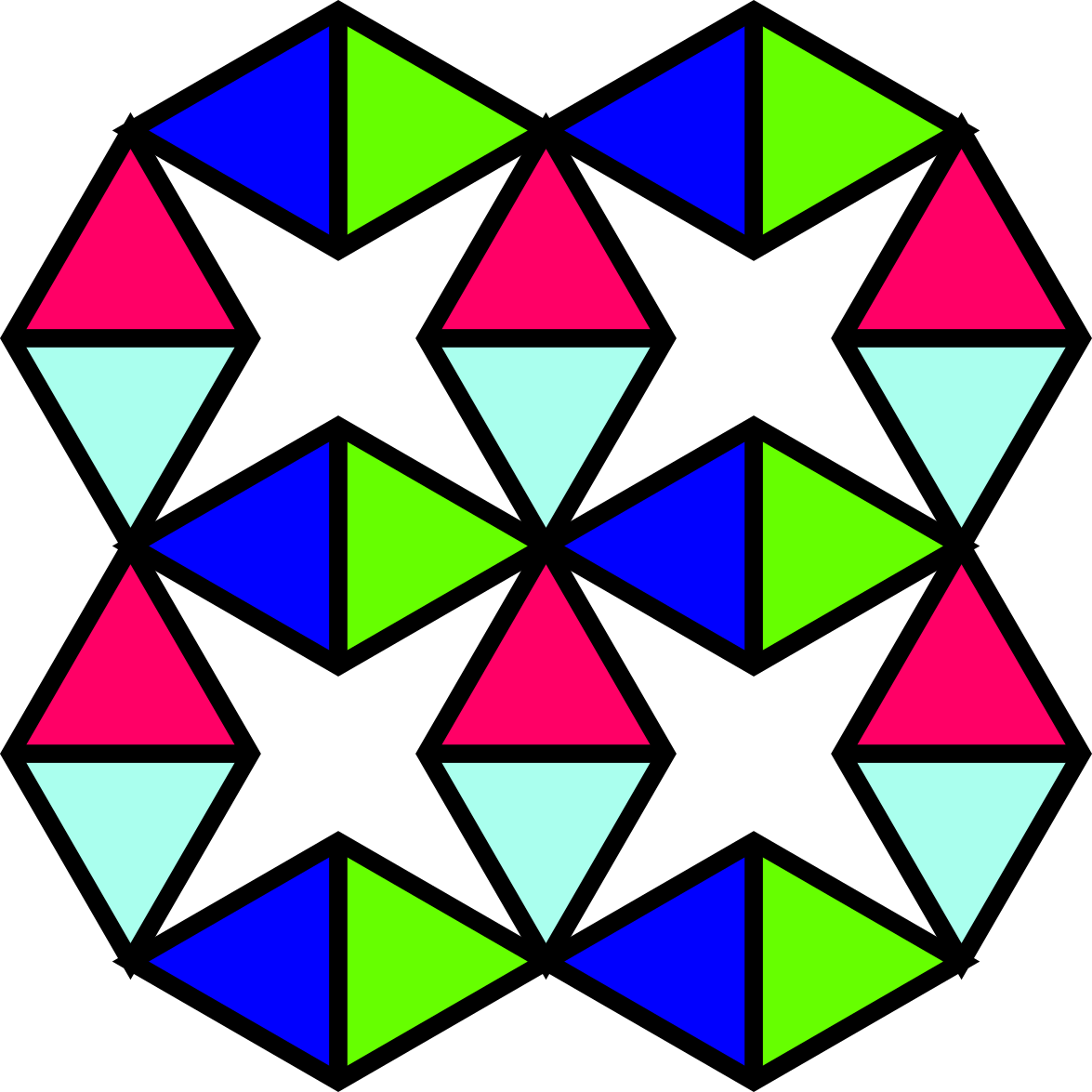} & 
        \includegraphics[height=2.5cm]{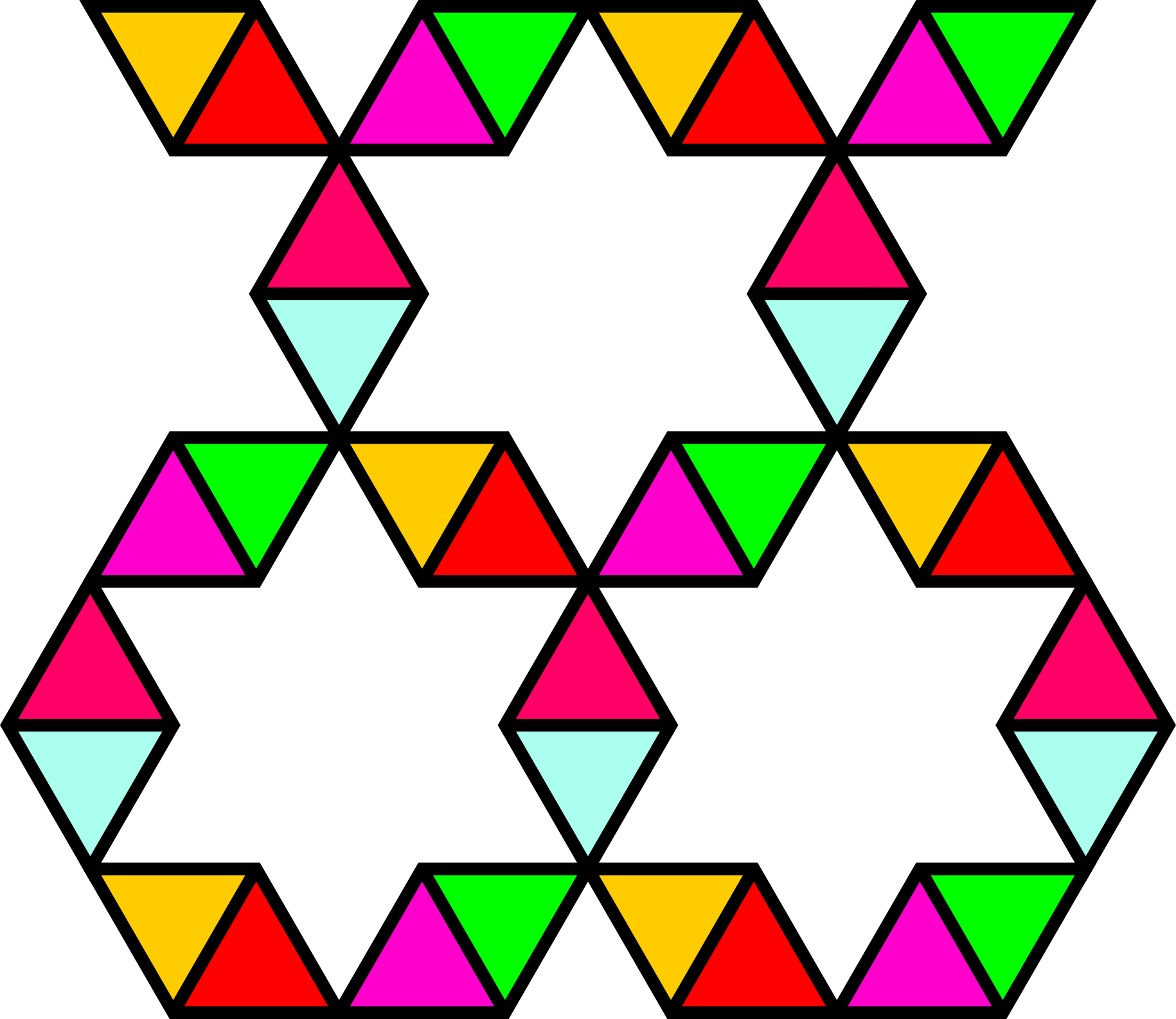} & 
        \includegraphics[height=2.5cm]{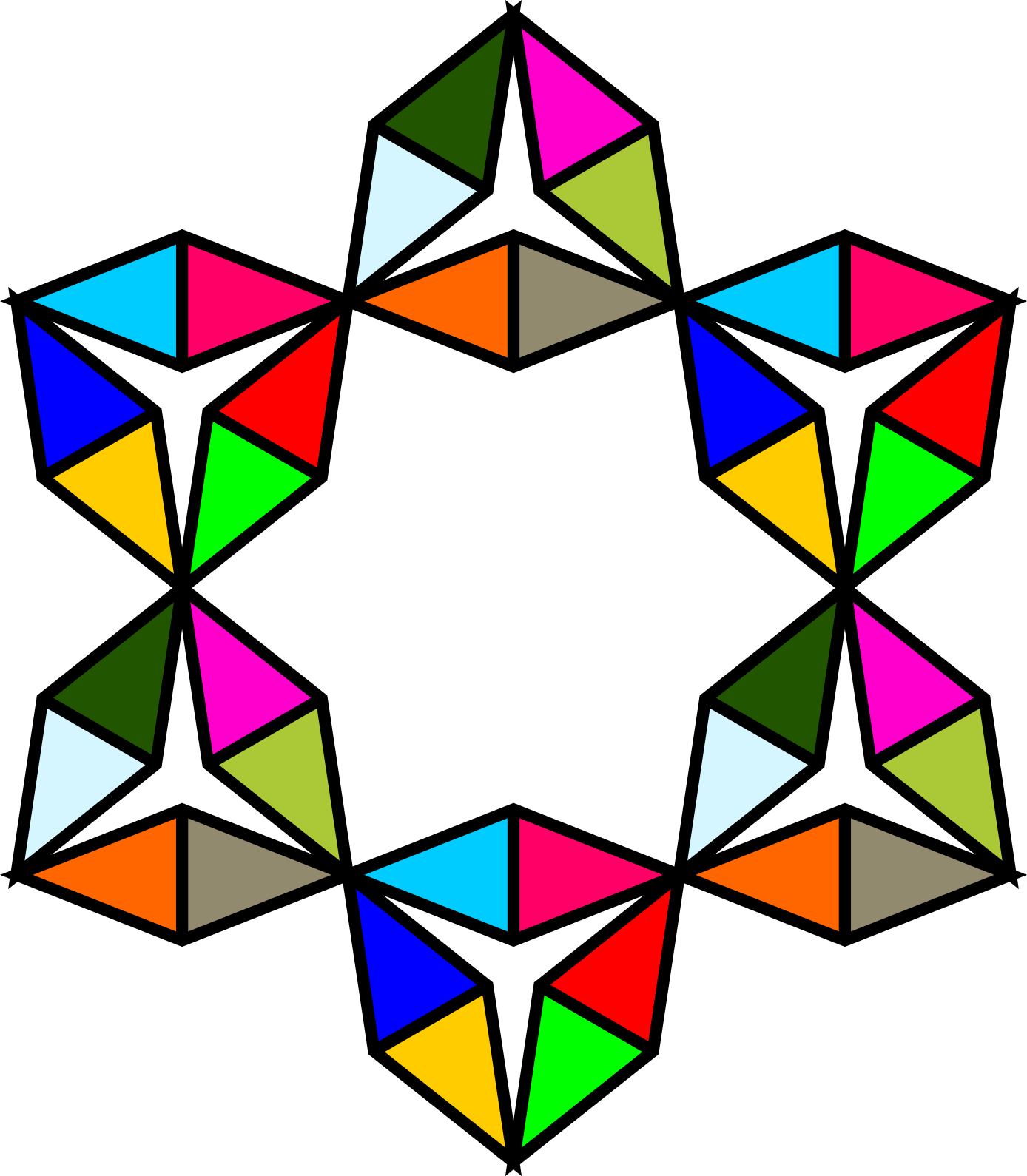} &
        \includegraphics[height=2.5cm]{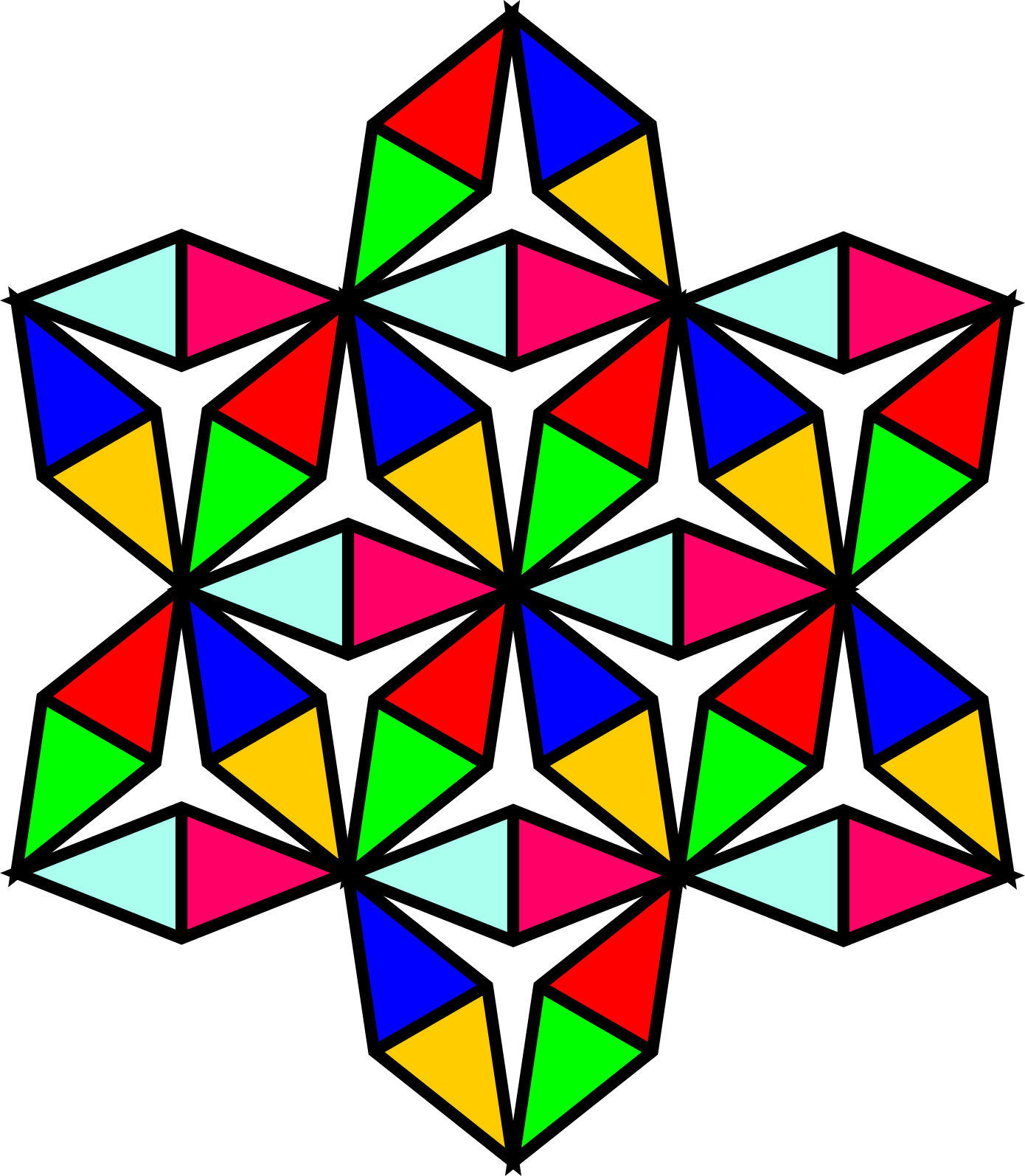} &
        \includegraphics[height=2.5cm]{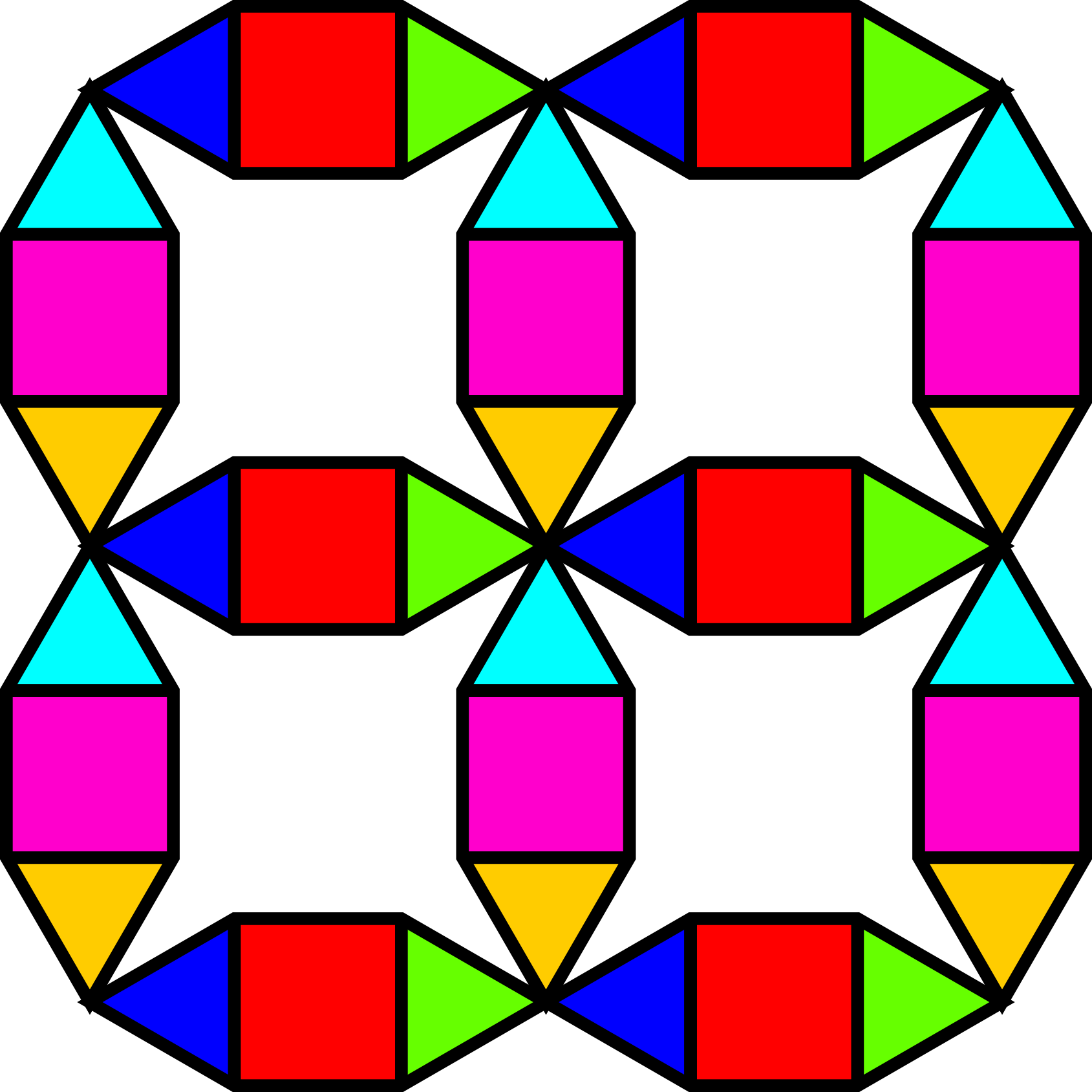} & 
        \includegraphics[height=2.5cm]{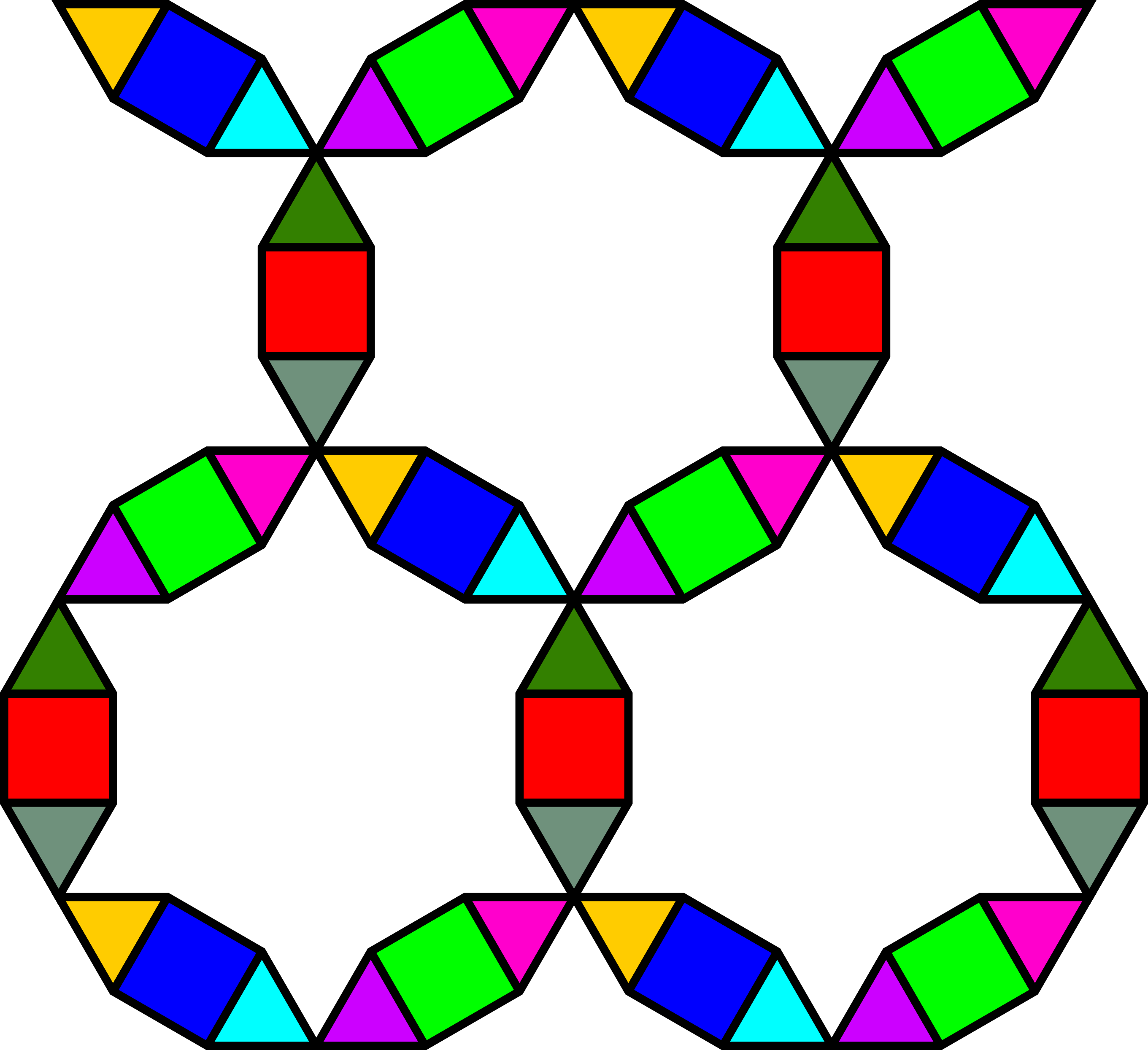} \\

        \makecell{ \vspace{-2.5cm} \\ \textbf{Structure}  \\ \textbf{Factor} } & 
        \includegraphics[width=2.5cm,trim={1cm 1cm 2cm 0},clip]{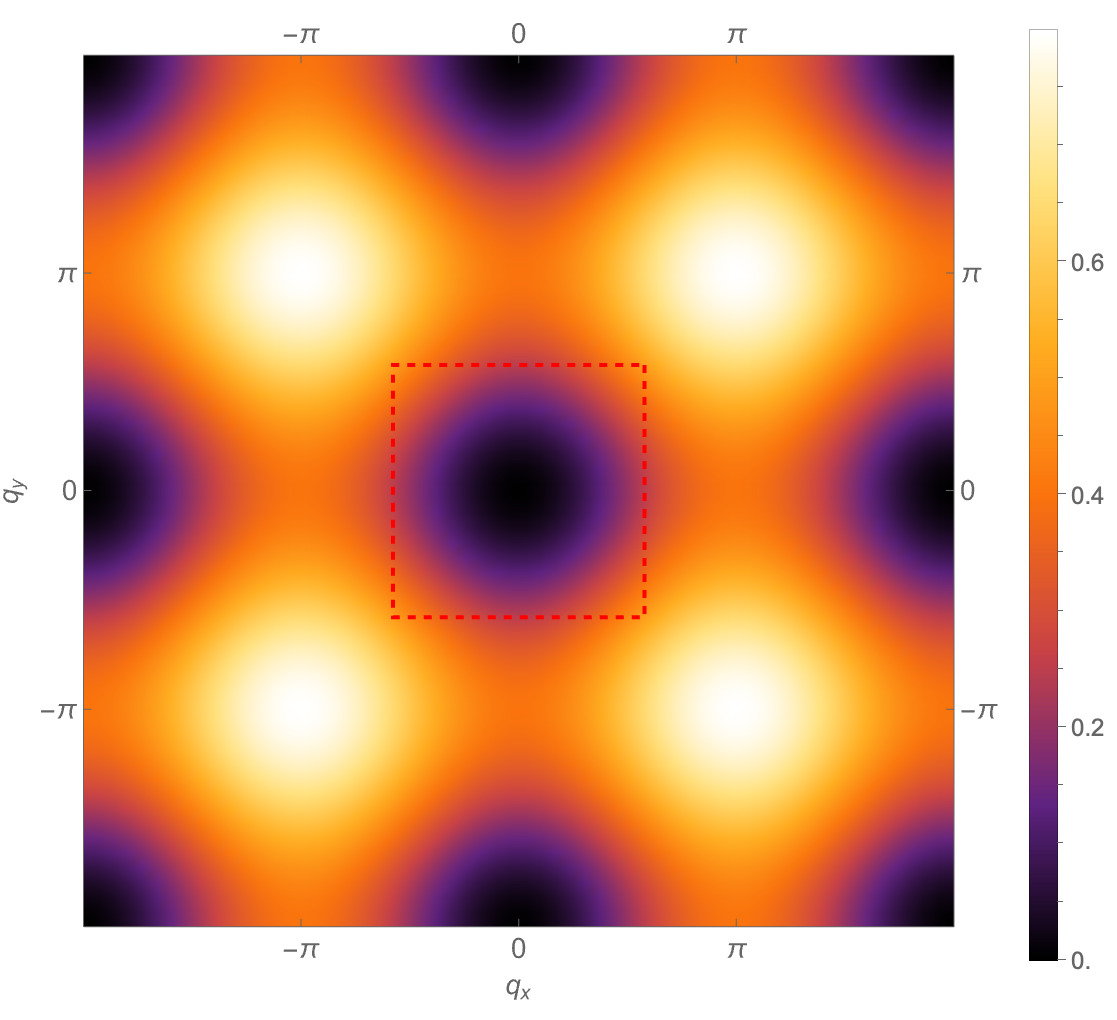} & 
        \includegraphics[width=2.5cm,trim={1cm 1cm 2cm 0},clip]{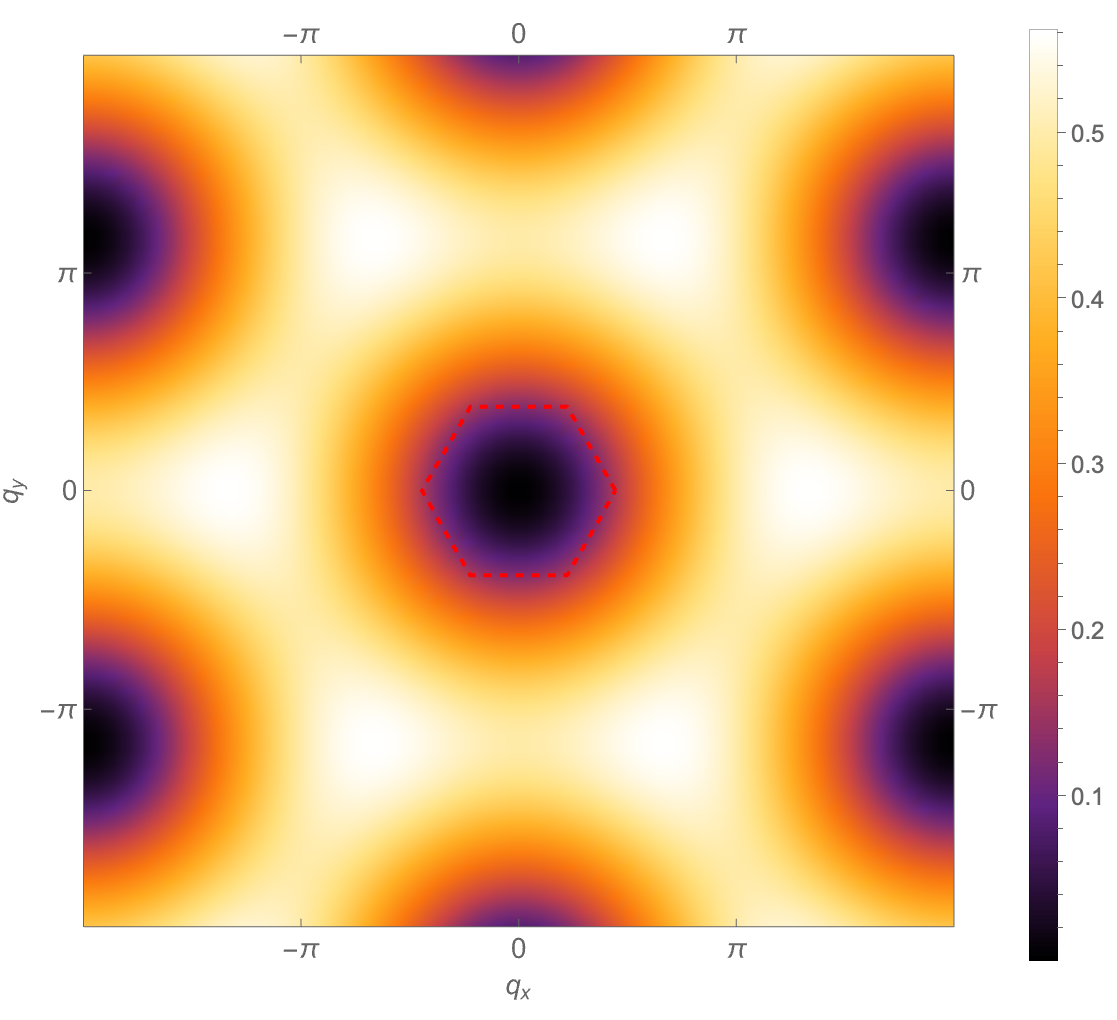} &
        \includegraphics[width=2.5cm,trim={1cm 1cm 2cm 0},clip]{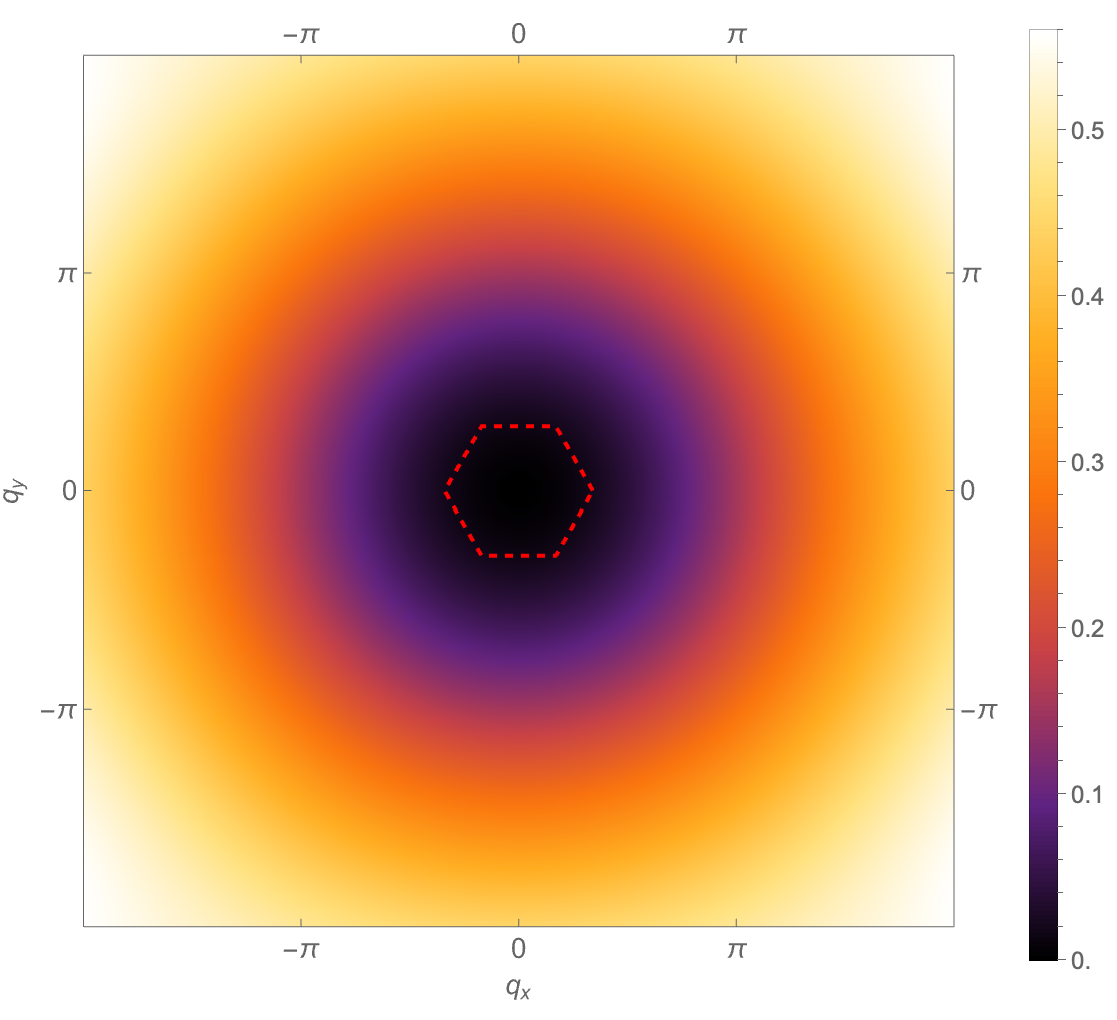} &
        \includegraphics[width=2.5cm,trim={1cm 1cm 2cm 0},clip]{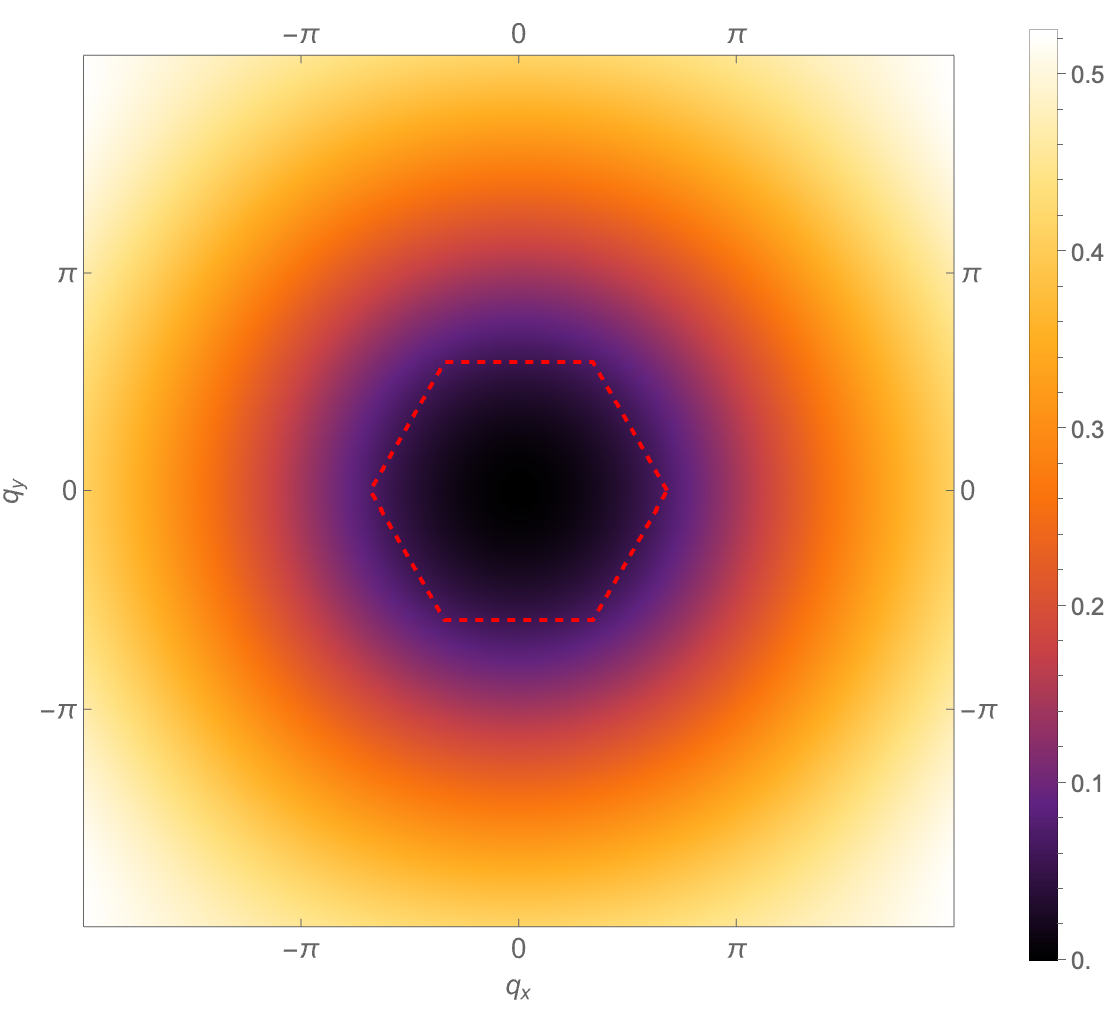} &
        \includegraphics[width=2.5cm,trim={1cm 1cm 2cm 0},clip]{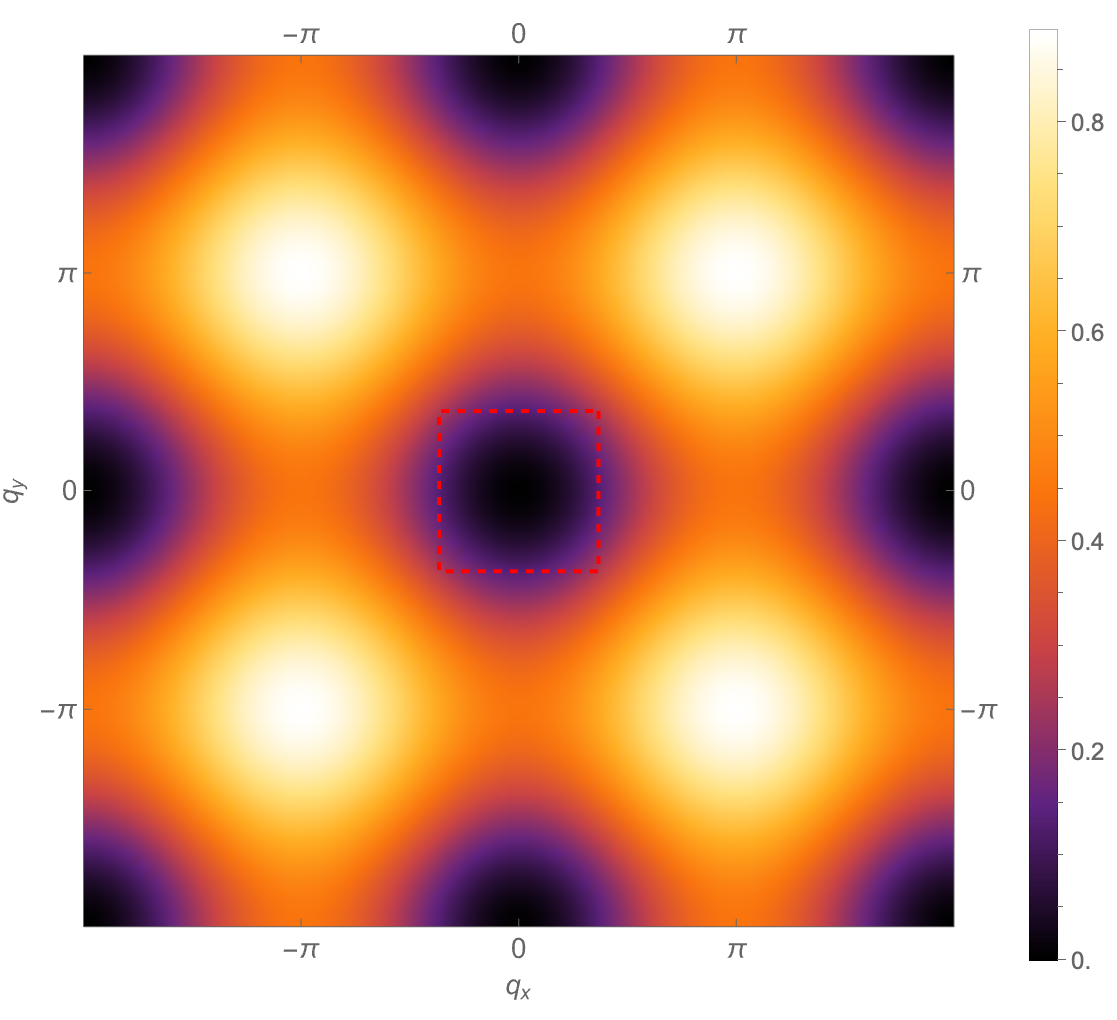} & 
        \includegraphics[width=2.5cm,trim={1cm 1cm 2cm 0},clip]{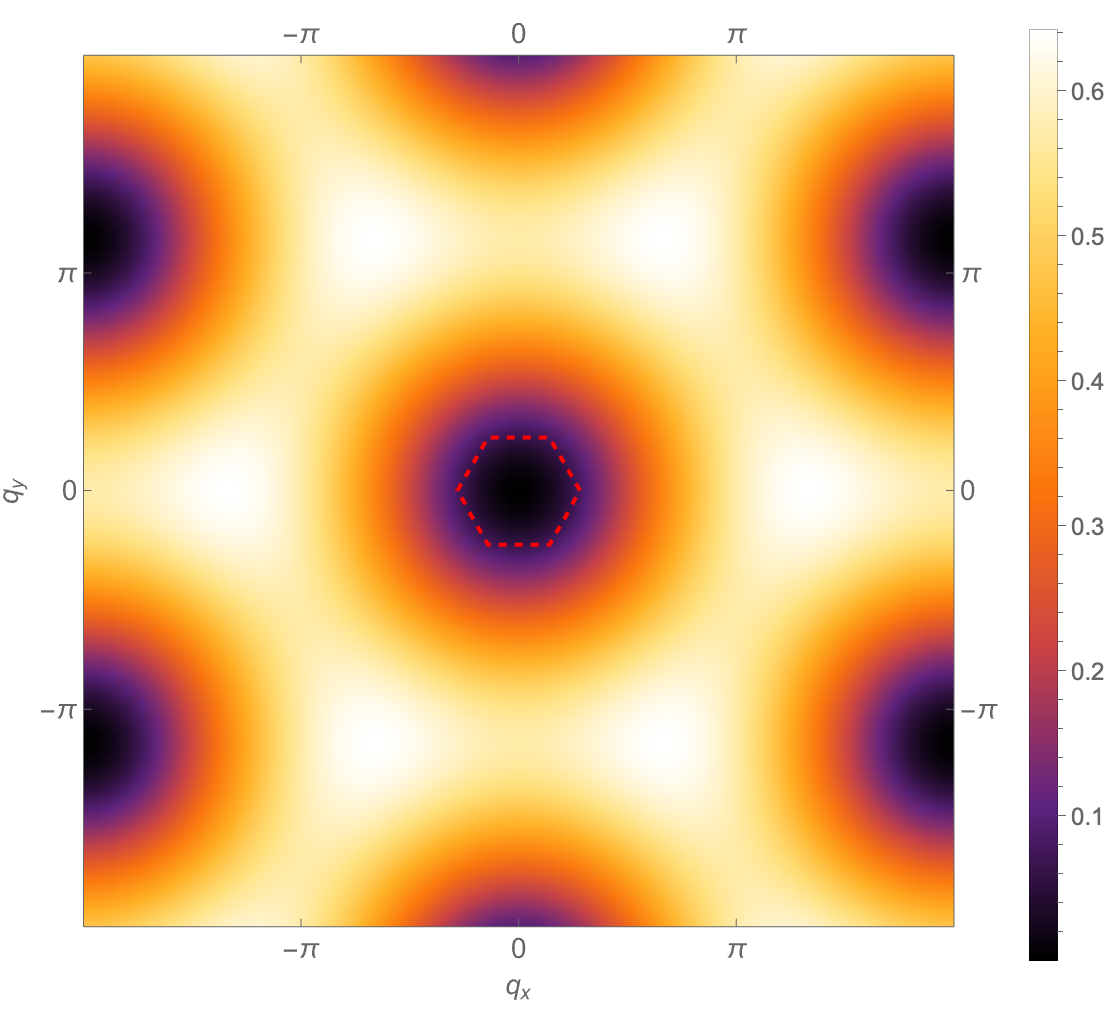} \\

        \makecell{ \vspace{-2.2cm} \\ \textbf{3D Band}  \\ \textbf{Structure} } & 
        \includegraphics[width=2.5cm]{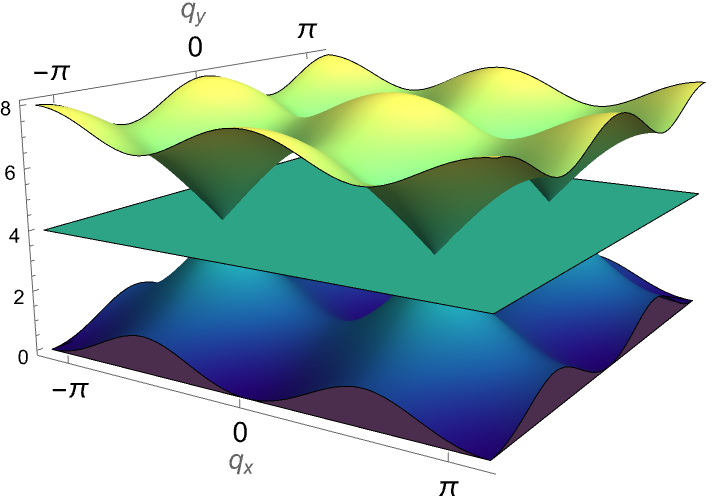} & 
        \includegraphics[width=2.5cm]{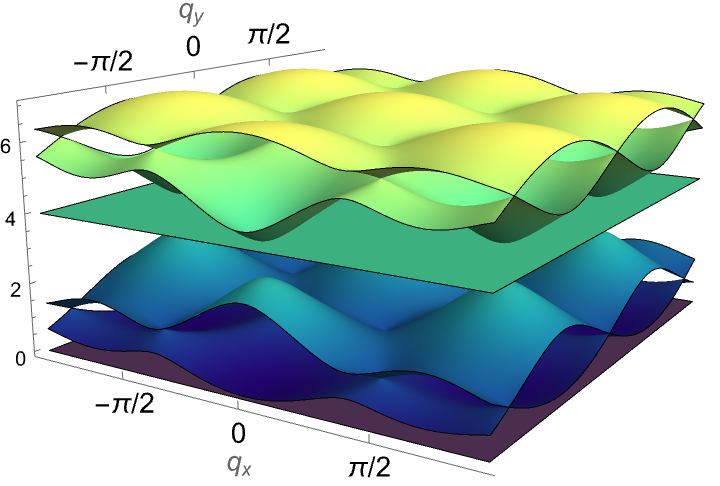} &
        \includegraphics[width=2.5cm]{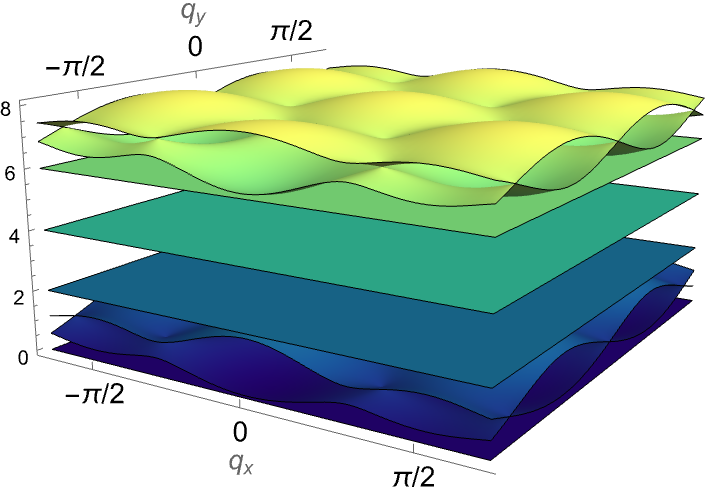} &
        \includegraphics[width=2.5cm]{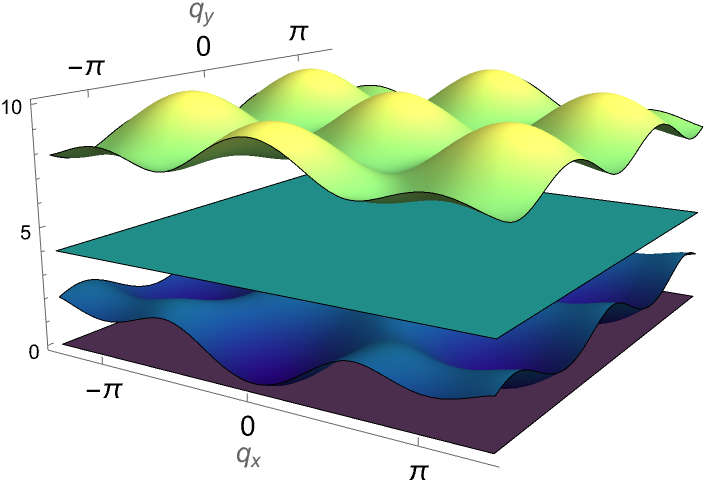} &
        \includegraphics[width=2.5cm]{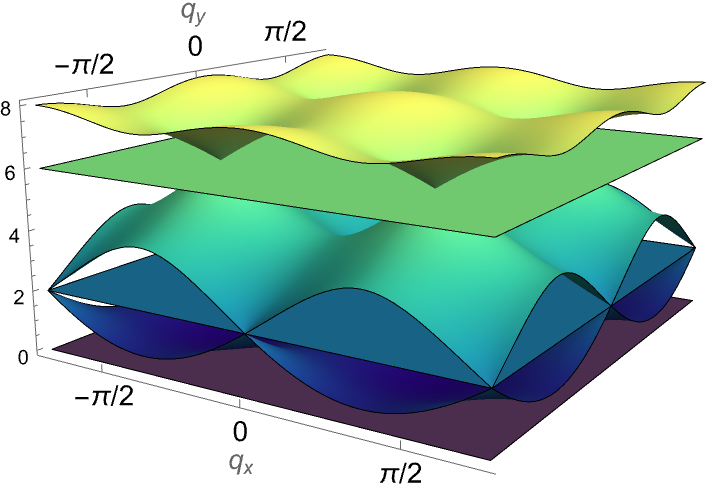} & 
        \includegraphics[width=2.5cm]{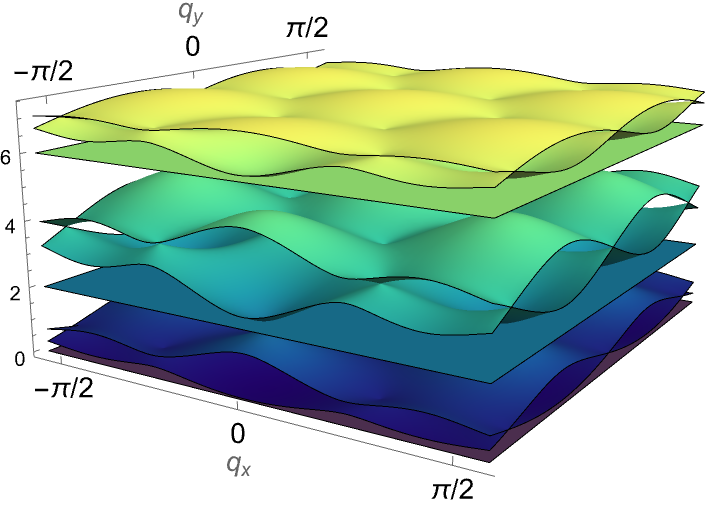} \\

         \makecell{ \vspace{-2.cm} \\ \textbf{Band}  \\ \textbf{Structure} } & 
        \includegraphics[width=2.5cm]{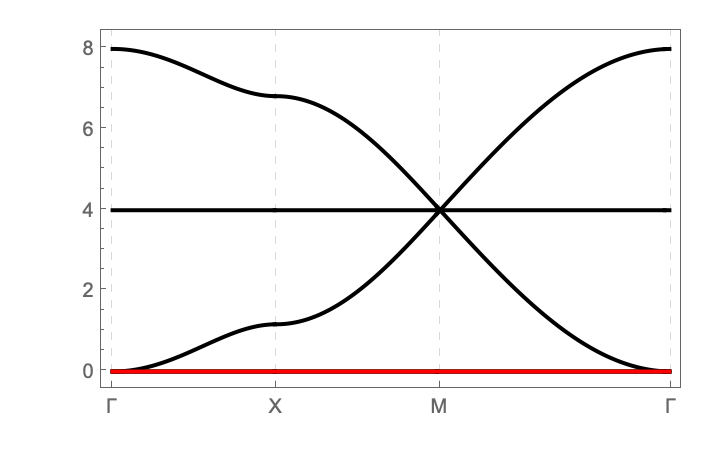} & 
        \includegraphics[width=2.5cm]{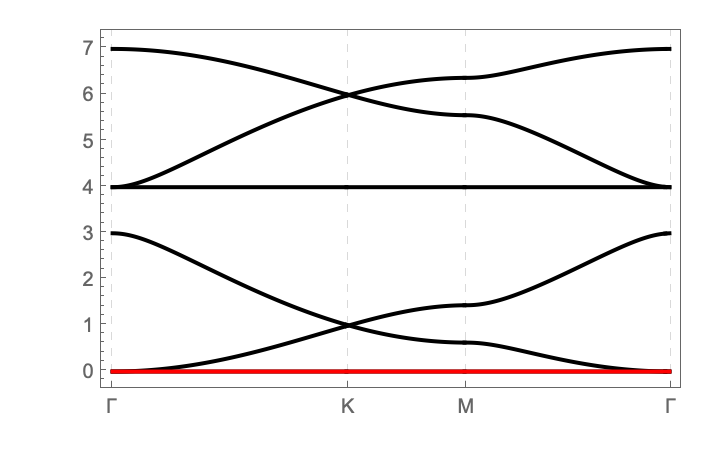} &
        \includegraphics[width=2.5cm]{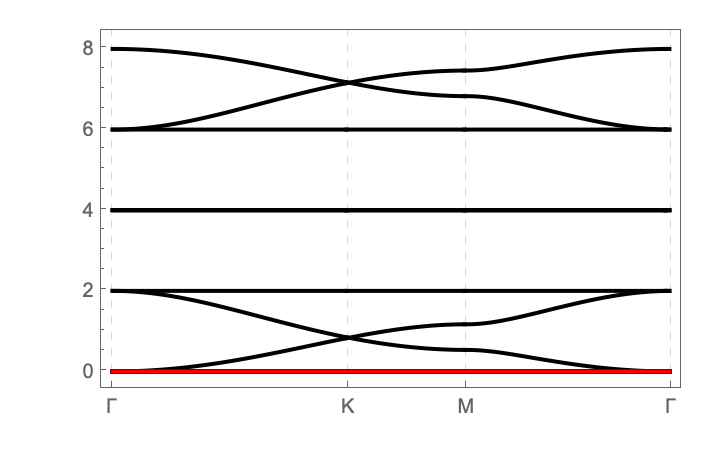} &
        \includegraphics[width=2.5cm]{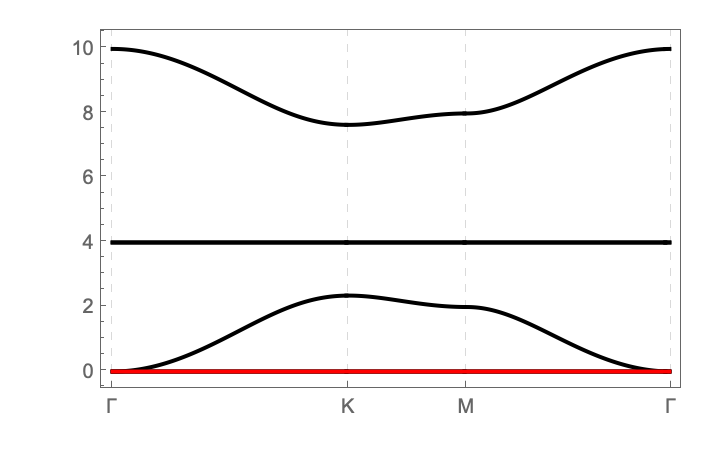} &
        \includegraphics[width=2.5cm]{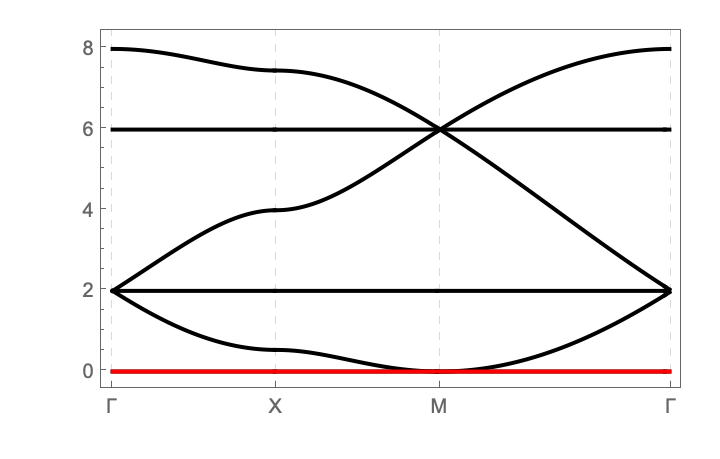} & 
        \includegraphics[width=2.5cm]{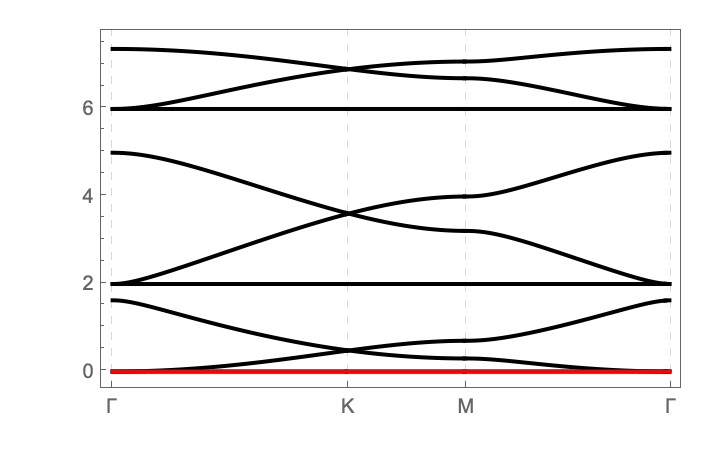} \\
        
        \textbf{$n_s$}     & 5 & 8 & 15 & 7 & 9 & 14 \\
        \textbf{$n_c$}     & 4 & 6 & 12  & 6 & 6 & 9 \\
        \textbf{$n_\text{b.f.b}$} & \textcolor{red}{2} & \textcolor{red}{3} & \textcolor{red}{6} & \textcolor{red}{3}  & \textcolor{red}{4} & \textcolor{red}{6} \\
        \textbf{$n_\text{i.f.b}$} & 1 & 1 & 1+3+1 & 2 & 1+1 & 1+1 \\

        \hline
    \end{tabular}}
    \caption{\textbf{Composite-bond-decorated systems in two dimensions}. The number of bottom flat bands $n_\text{b.f.b}$ is compared to the difference $n_s-n_c$, assuming spins sit only on cluster vertices, indicating it in red when there is a mismatch. 
    The structure factors and band structures are shown in units of $a^{-1}$, where $a$ is the nearest-neighbor distance (for kagome- and triangular-based lattices, $a$ denotes the second-nearest-neighbor distance). The first Brillouin zone is indicated by a red contour in the structure-factor plots. 
    Being all gapless, composite cluster-bonds decorated systems still never display any pinch point in their structure factor. This is because the existence of additional accidental bottom flat bands destroys the singular nature of the band touchings, see main text. All bond decorated systems presented here appear to host intermediate flat bands located inside the spectrum and not at its bottom. As these may be degenerate their number is indicated as a sum over the number of distinct set of intermediate flat bands.
    }
    \label{tab: 2d decorated lattices}
\end{table*}

A particularly elegant way to construct cluster systems is by decorating an existing lattice: replacing each bond with a cluster, replacing each vertex by a cluster, or replacing both bonds and vertices with clusters. We distinguish these three cases below. Since these procedures correspond visually to bond or vertex decorations, we will collectively refer to the resulting structures as \emph{decorated systems}.

\subsection{Bond-decorated systems}
\begin{figure}[ht]
    \centering
    \includegraphics[height=0.65cm]{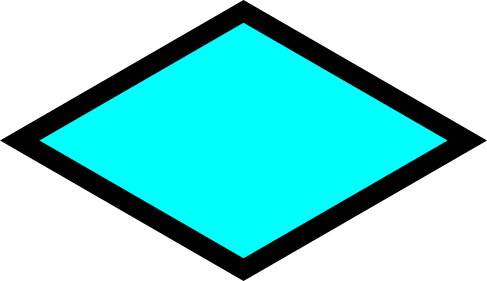} \hspace{0.2 mm}
    \includegraphics[height=0.65cm]{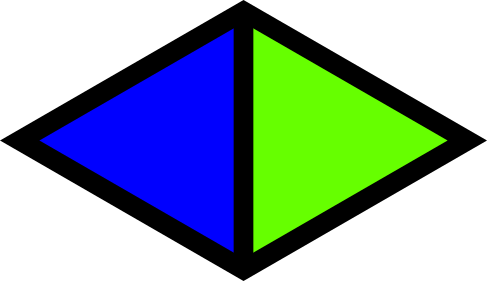} \hspace{0.2 mm}
    \includegraphics[height=0.65cm]{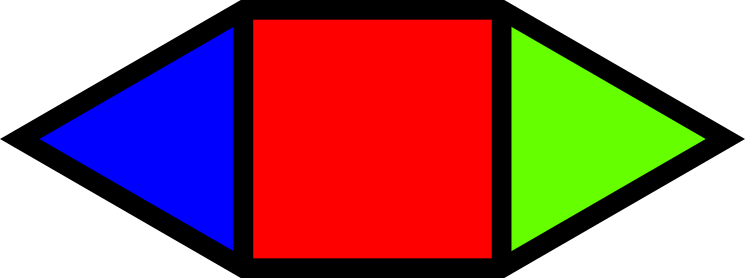}  \hspace{0.2 mm}
    \includegraphics[height=0.65cm]{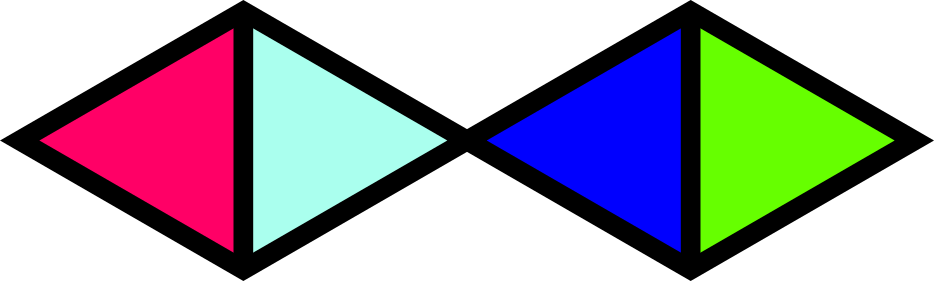}  \hspace{0.2 mm}
    \includegraphics[height=0.65cm]{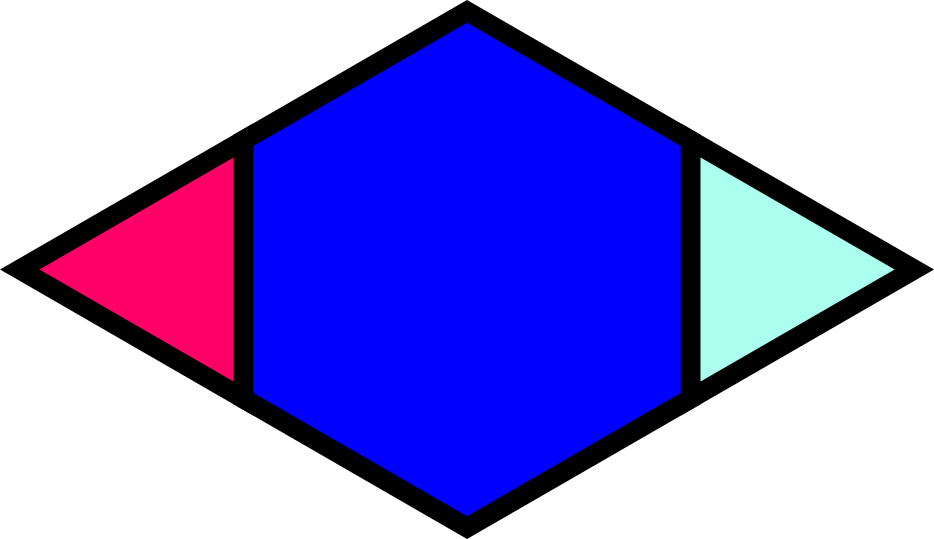}
    \caption{Examples of 2D cluster-bonds: monodiamond, diamond, cracker, double-diamond, and hexagonal-diamond.}
    \label{fig: cluster-links}
\end{figure}

In this construction, each bond of the parent lattice is replaced by a \emph{cluster-bond}, i.e.\ a composite unit acting as an extended bond. Such clusters must possess double axial symmetry and terminate at two vertices that connect naturally to their neighbors. 
The simplest example is the diamond block, built either as a single unit or as two joined triangles\cite{Morita_2016_Heisenberg_Diamond_Spin_Lattices, Caci_2023_decorated_square}. 
More elaborate structures are possible, such as a square inserted between two triangles, yielding a “Christmas cracker”. 
Several examples of 2D cluster-bonds are shown in Fig.~\ref{fig: cluster-links}.

\begin{figure}[ht]
    \centering
    \includegraphics[height=0.63cm]{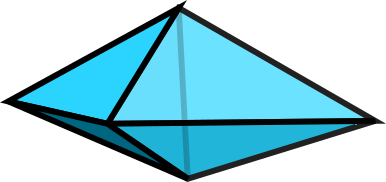}  \hspace{0.2 mm}
    \includegraphics[height=0.63cm]{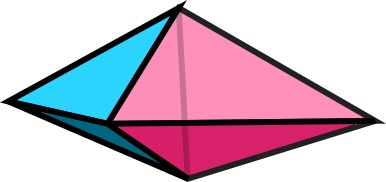}  \hspace{0.2 mm}
    \includegraphics[height=0.63cm]{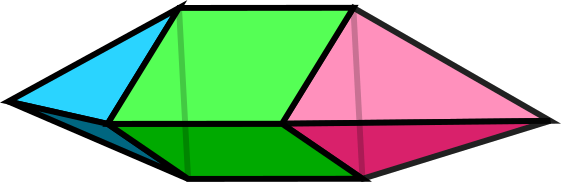}  \hspace{0.2 mm}
    \includegraphics[height=0.63cm]{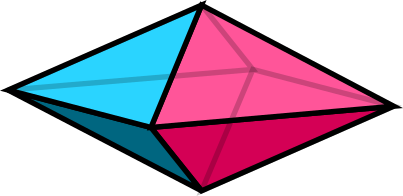}  \hspace{0.2 mm}
    \includegraphics[height=0.63cm]{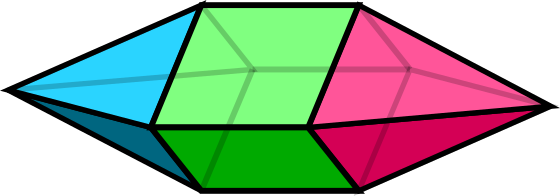} 
    \caption{Examples of 3D cluster-bonds: mono-bloc or double-bloc bi-triangular pyramid, elongated bi-triangular pyramid, bi-pyramid or elongated bi-pyramid.}
    \label{fig: 3D cluster-links}
\end{figure}

These cluster-bonds can be substituted for the bonds of a \emph{parent lattice} to create decorated variants of this lattice, as in decorated versions of the square~\cite{Morita_2016_Heisenberg_Diamond_Spin_Lattices, Caci_2023_decorated_square}, honeycomb~\cite{Morita_2016_Heisenberg_Diamond_Spin_Lattices}, triangular, or kagome lattices (Tables~\ref{tab: 2d monobond decorated lattices} and \ref{tab: 2d decorated lattices}). Additional clusters may also be inserted consistently with the lattice geometry, as presented in Table. \ref{tab: 2d decorated lattices with inter-clusters}. 
The same idea extends to 3D: joining two triangular pyramids or two square pyramids produces cluster-bonds such as those shown in Fig.~\ref{fig: 3D cluster-links}, which can be used to assemble fully three-dimensional networks.

Mono-block cluster-bonds usually reduce the number of distinct cluster types, thereby increasing the number of flat bands—often making them the most promising candidates for classical spin liquids from a theoretical perspective. The drawback, however, is that all sites within such bonds are interconnected, leading to a more intricate real-space lattice that is harder to realize experimentally.

\begin{table*}[ht]
    \centering
    \renewcommand{\arraystretch}{1.5} 
    \resizebox{\textwidth}{!}{
    \begin{tabular}{c c c c c c c}
        \hline    
        \makecell{\textbf{2D parent}\\ \textbf{lattice}} & Square & Square & Hexagonal & Hexagonal & Triangular & Triangular \\
        \textbf{Cluster-bond} & Square & Square & Square & Butterfly & Square & Butterfly \\
        \textbf{Cluster-vertex} & Square & Octagon & Hexagon & Hexagon & Hexagon & Hexagon \\
        \hline
        \noalign{\vskip 1mm}
        \makecell{ \vspace{-2.5cm} \\ \textbf{Lattice}  \\ \textbf{Scheme} } & 
        \includegraphics[width=2.5cm]{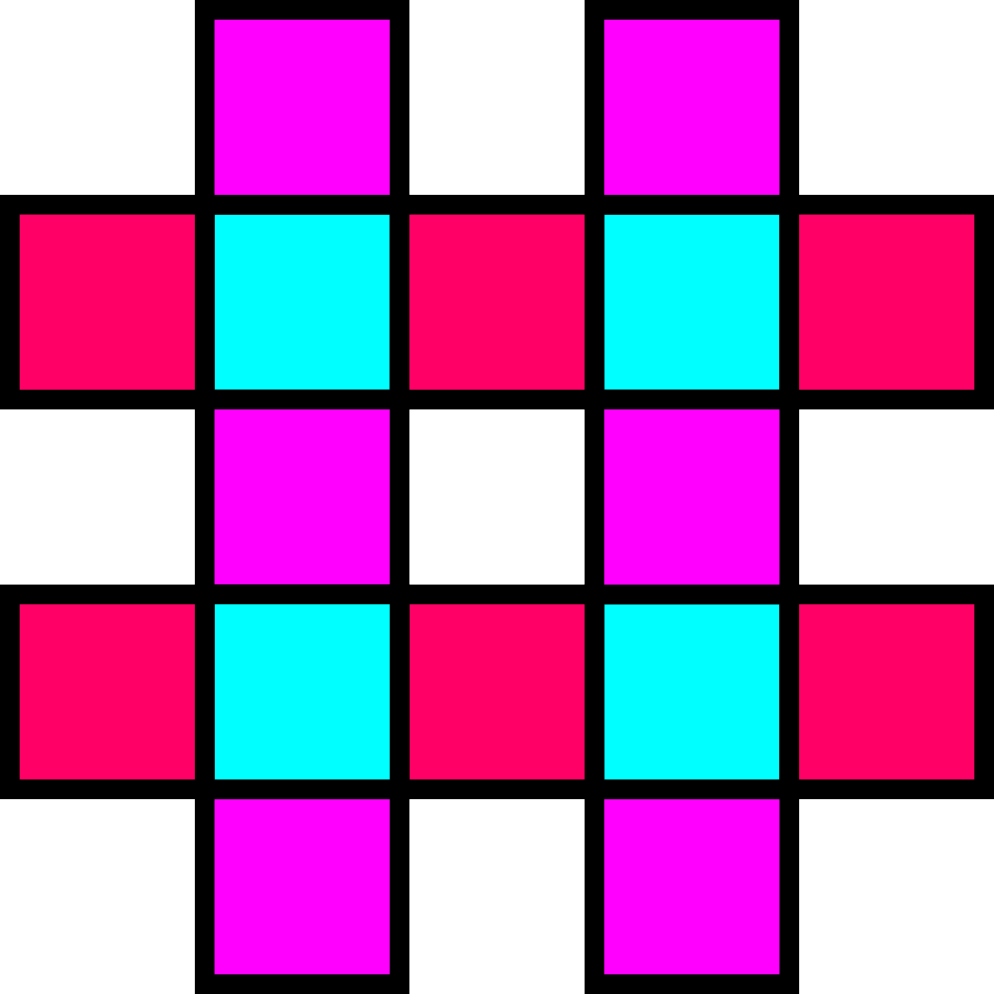} & 
        \includegraphics[width=2.5cm]{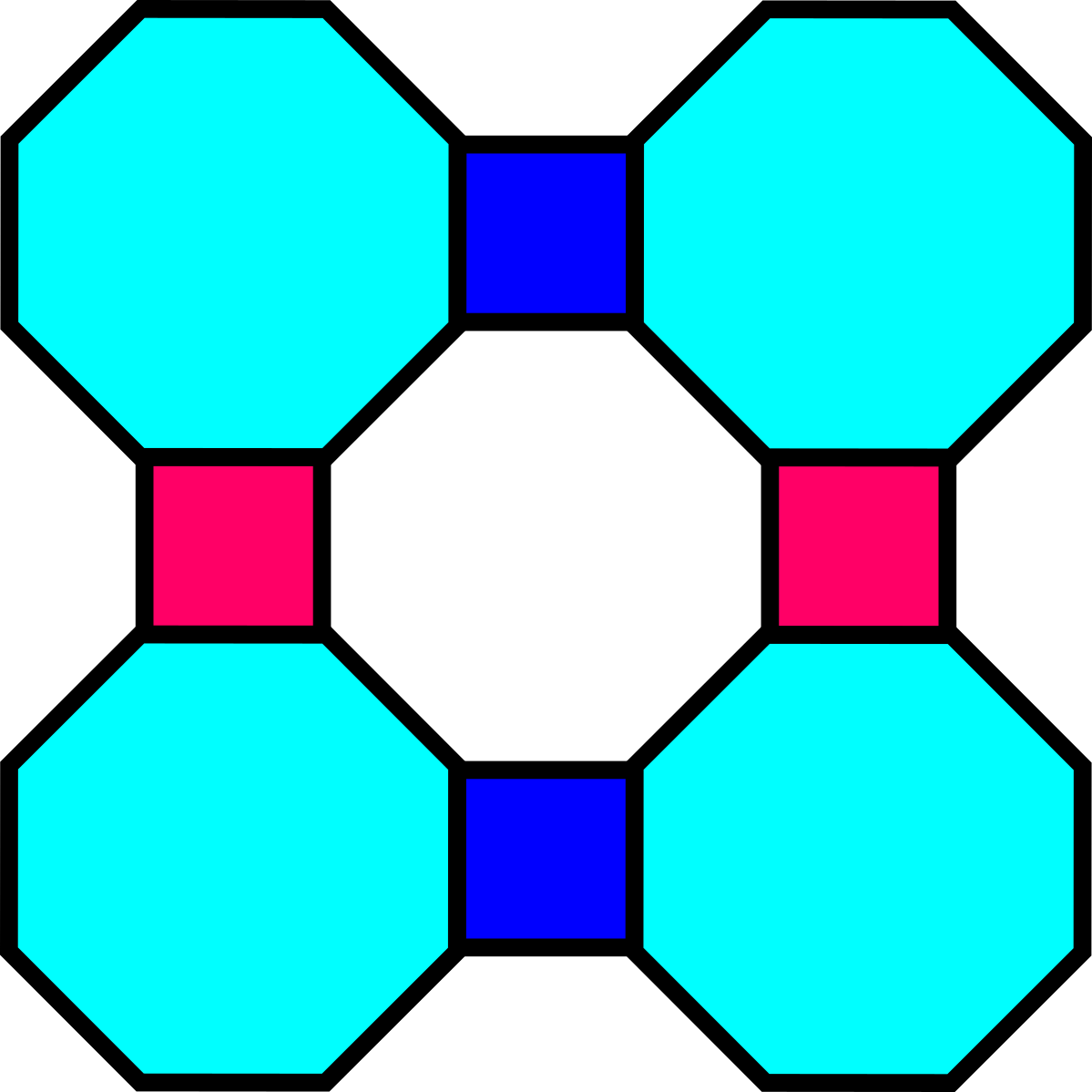} & 
        \includegraphics[width=2.5cm]{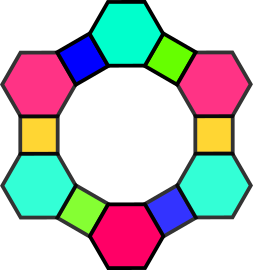} & 
        \includegraphics[width=2.5cm]{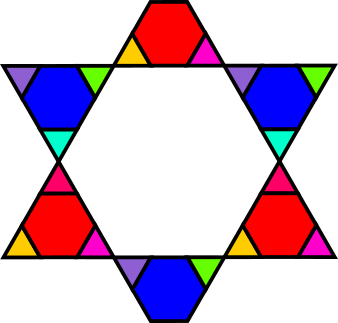} & 
        \includegraphics[width=2.5cm]{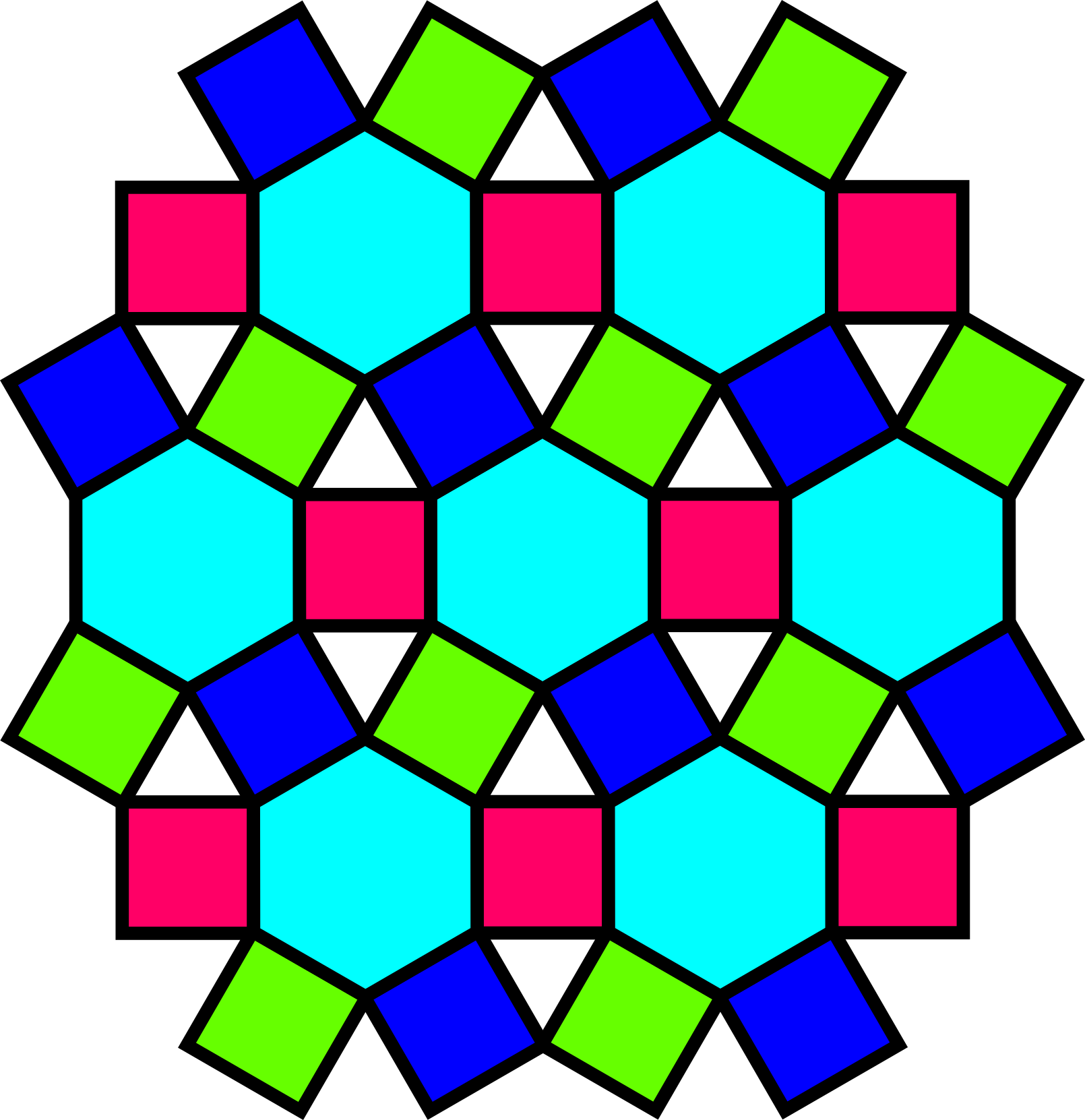} & 
        \includegraphics[width=2.5cm]{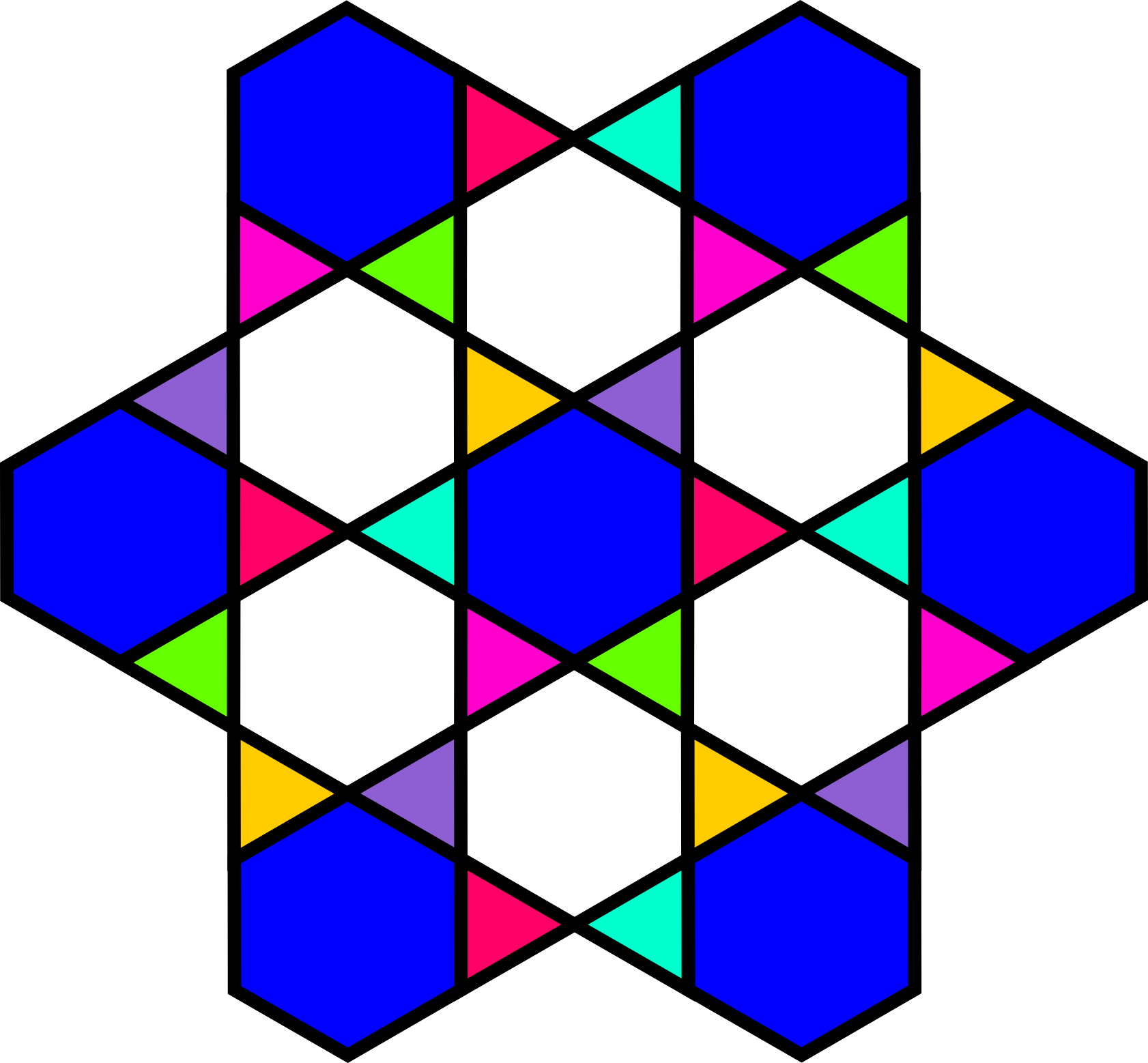} 
        \\
        \noalign{\vskip 1mm}
        \makecell{ \vspace{-2.5cm} \\ \textbf{Structure}  \\ \textbf{Factor} } & 
        \includegraphics[width=2.5cm,trim={1cm 1cm 2cm 0},clip]{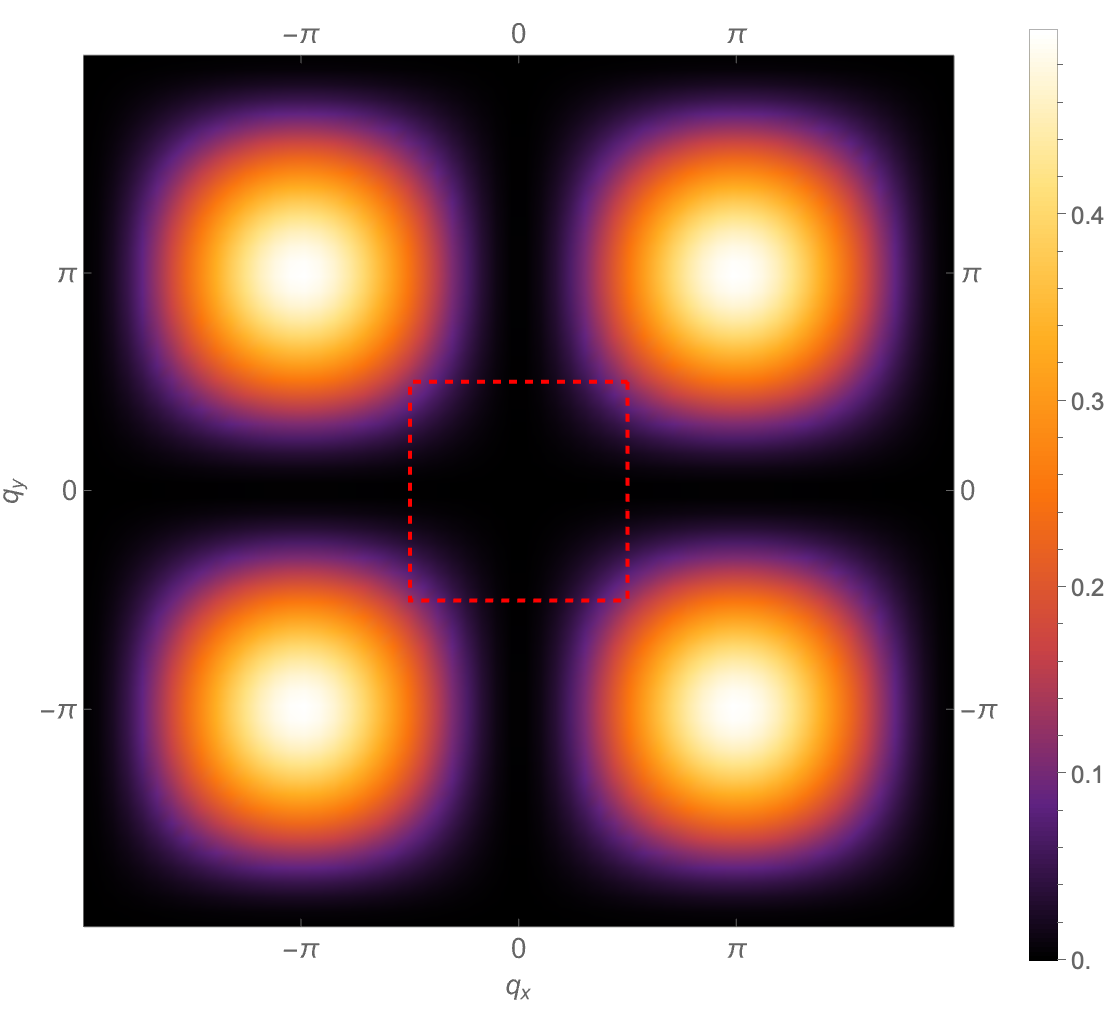} & 
        \includegraphics[width=2.5cm,trim={1cm 1cm 2cm 0},clip]{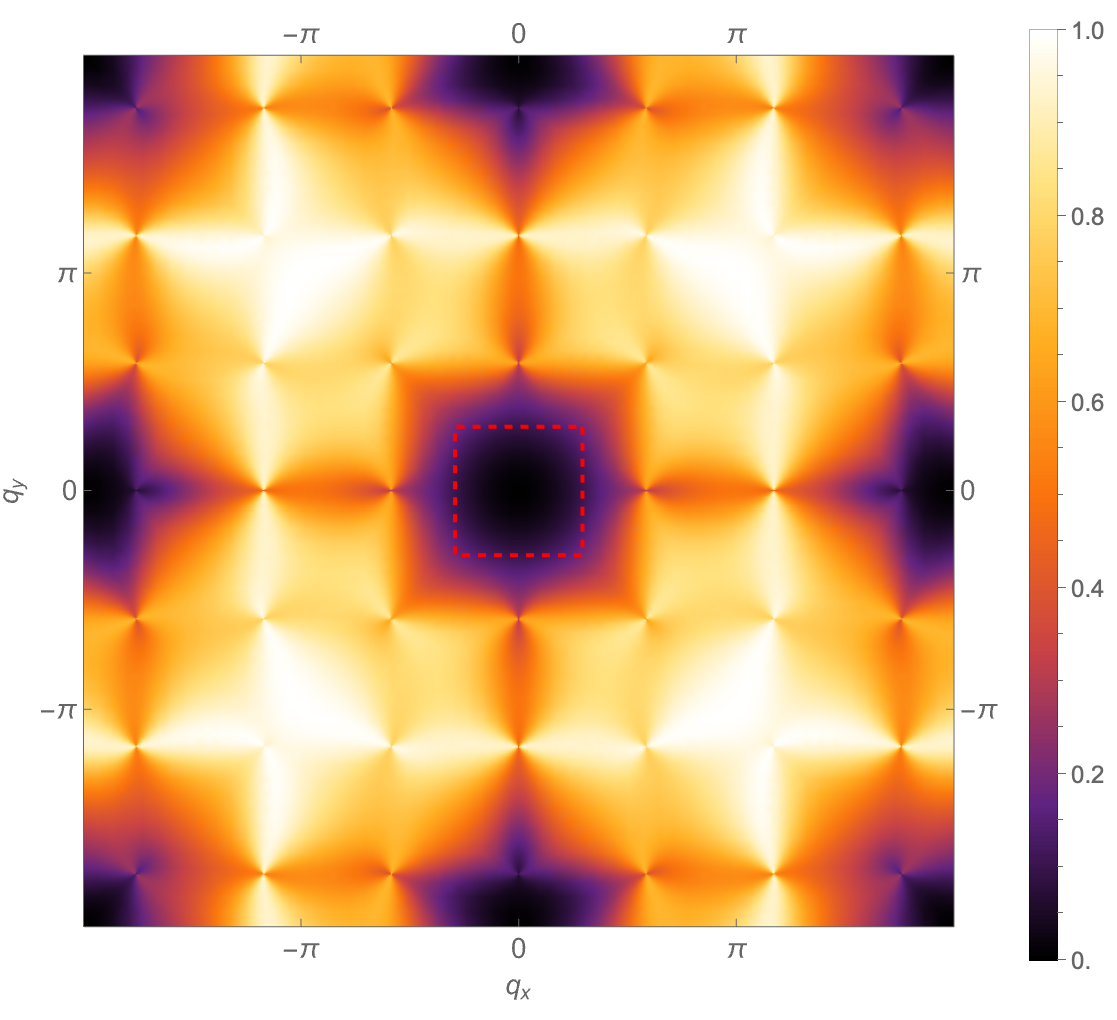} & 
        \includegraphics[width=2.5cm,trim={1cm 1cm 2cm 0},clip]{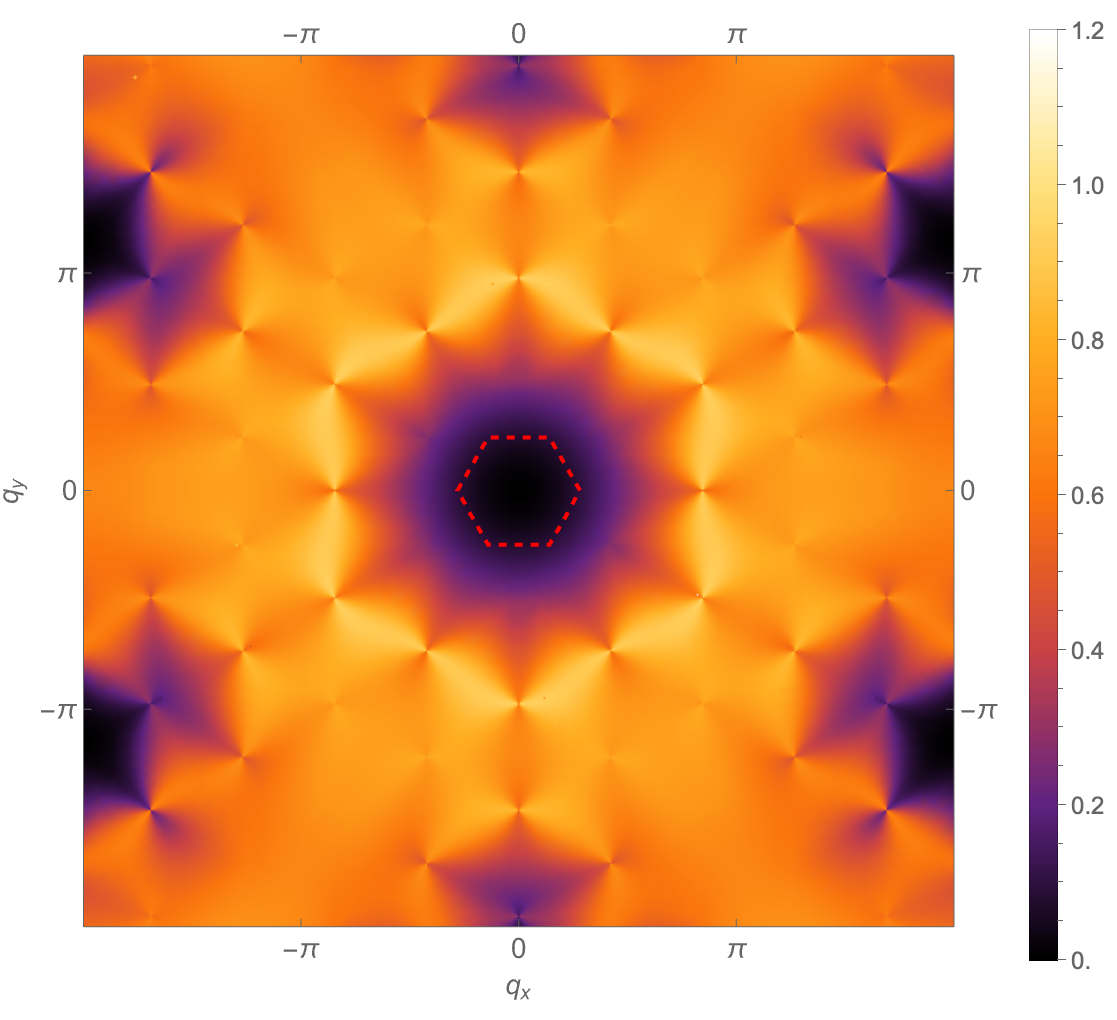} & 
        \includegraphics[width=2.5cm,trim={1cm 1cm 2cm 0},clip]{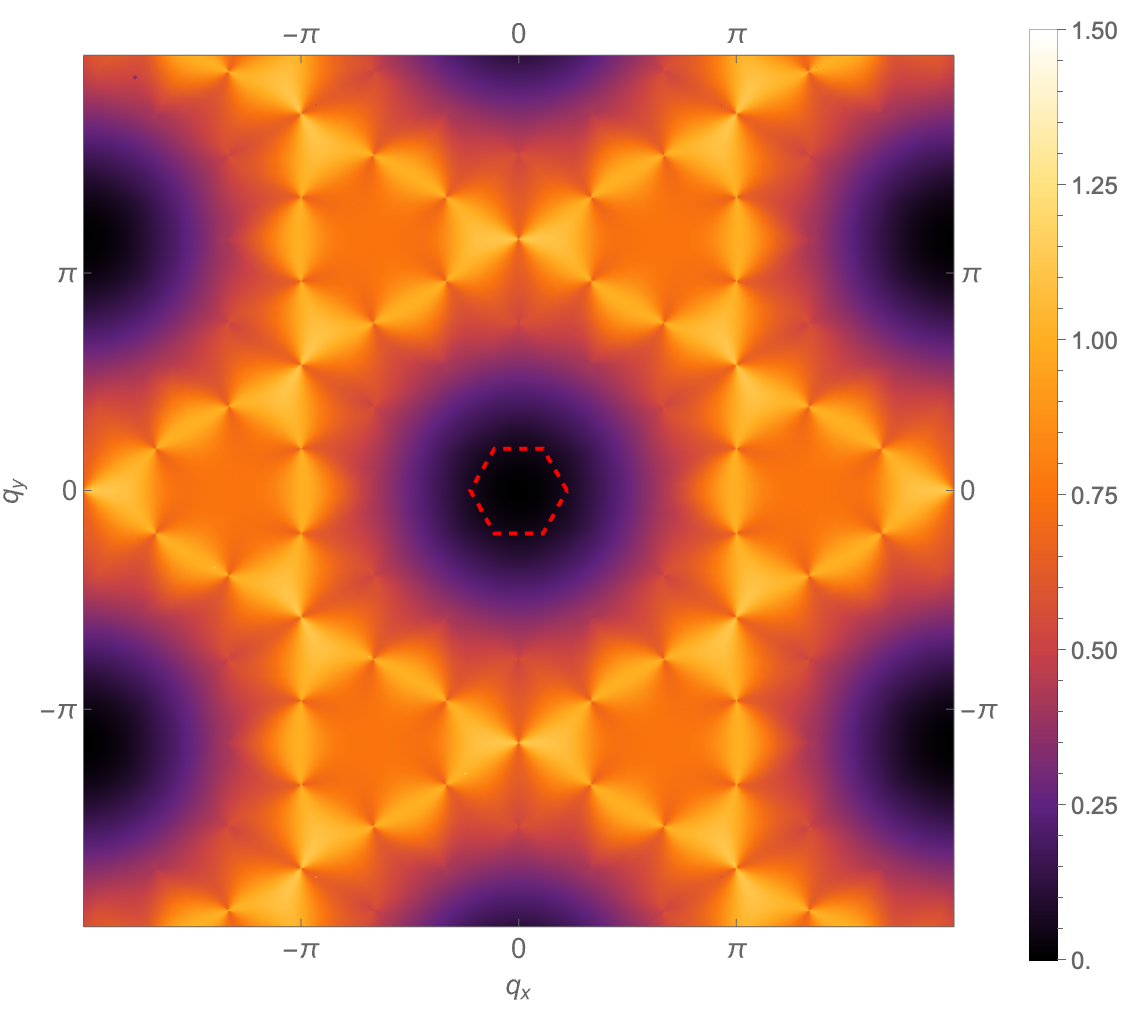} & 
        \includegraphics[width=2.5cm,trim={1cm 1cm 2cm 0},clip]{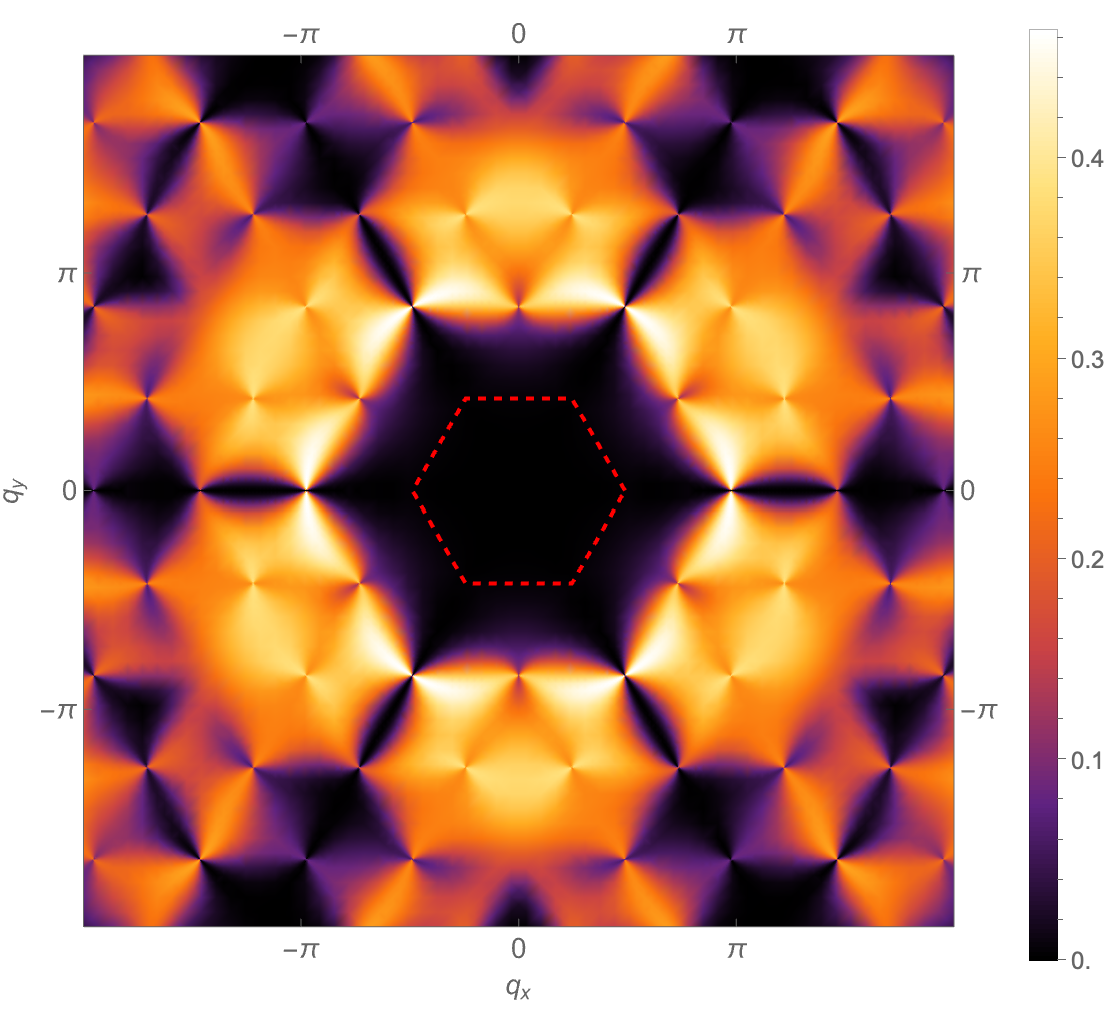} & 
        \includegraphics[width=2.5cm,trim={1cm 1cm 2cm 0},clip]{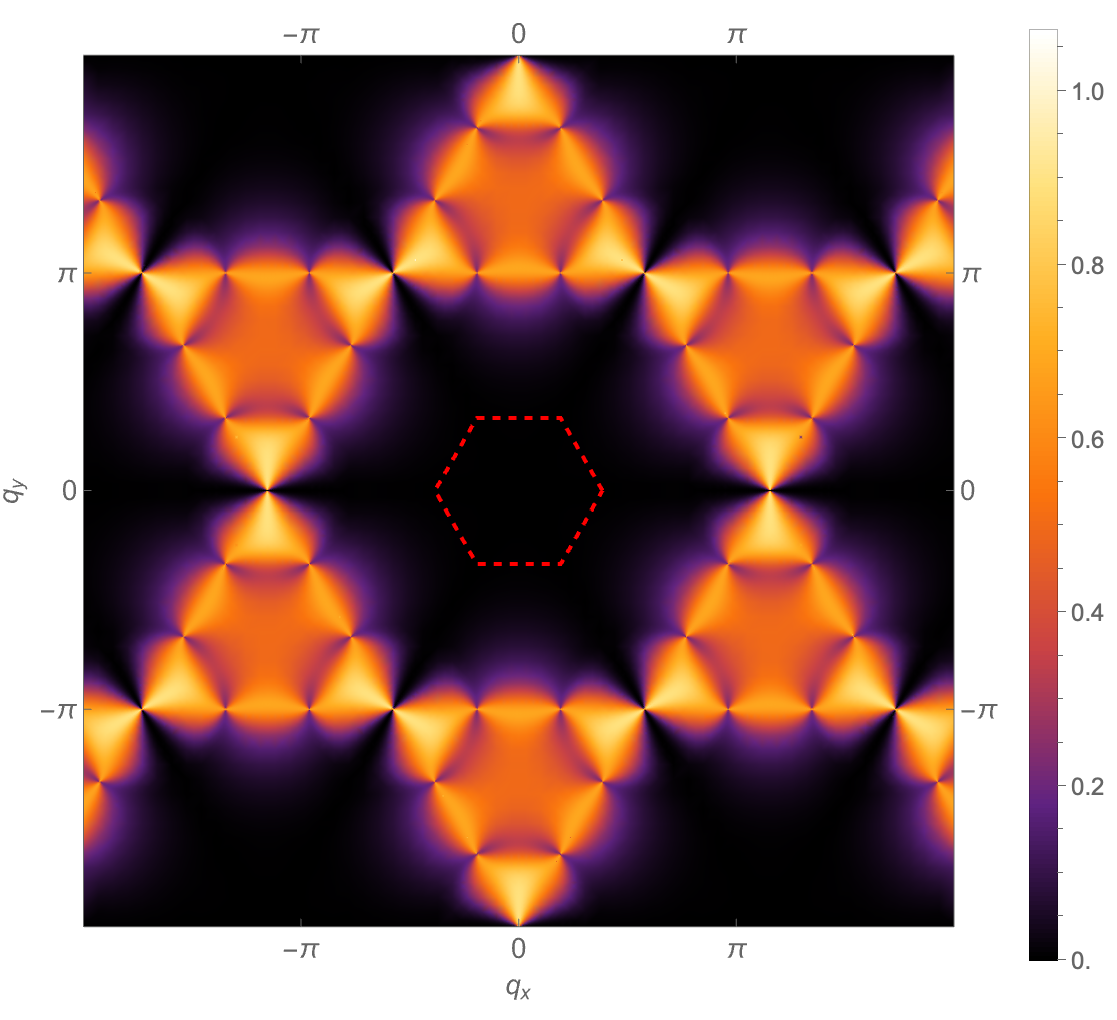} 
        \\
        \makecell{ \vspace{-2.2cm} \\ \textbf{3D Band}  \\ \textbf{Structure} } & 
        \includegraphics[width=2.5cm]{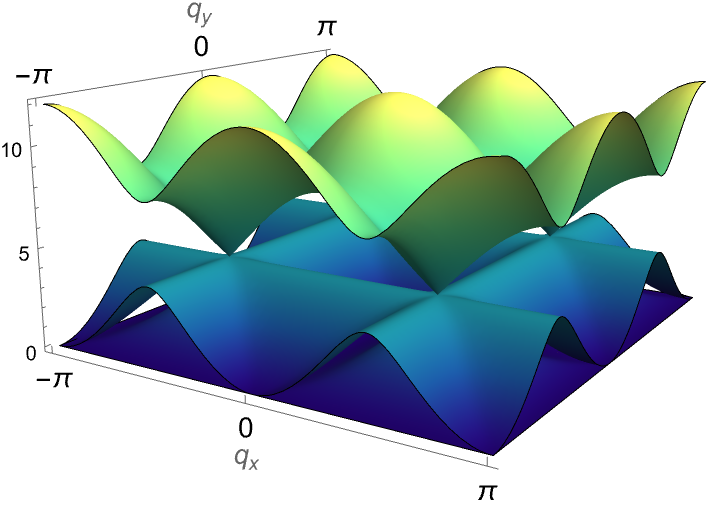} & 
        \includegraphics[width=2.5cm]{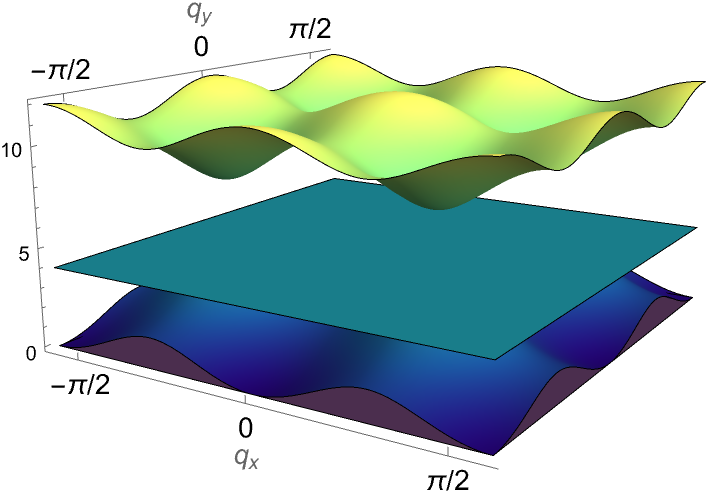} & 
        \includegraphics[width=2.5cm]{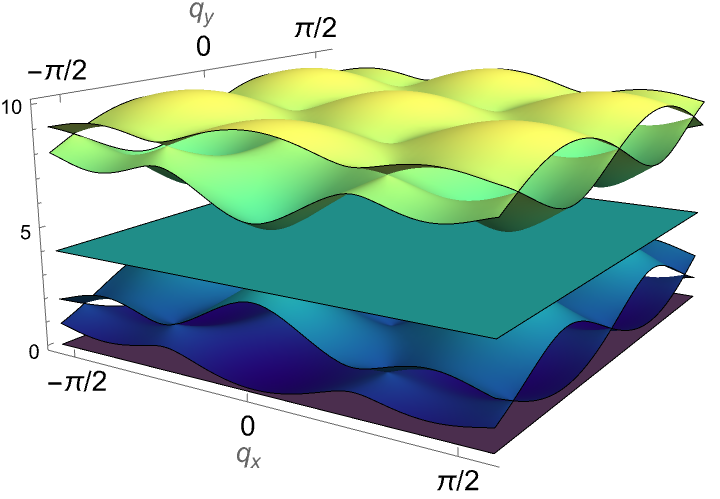} & 
        \includegraphics[width=2.5cm]{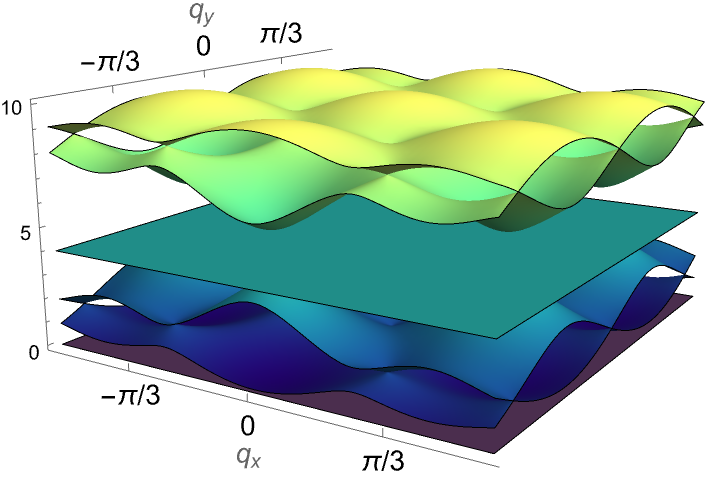} & 
        \includegraphics[width=2.5cm]{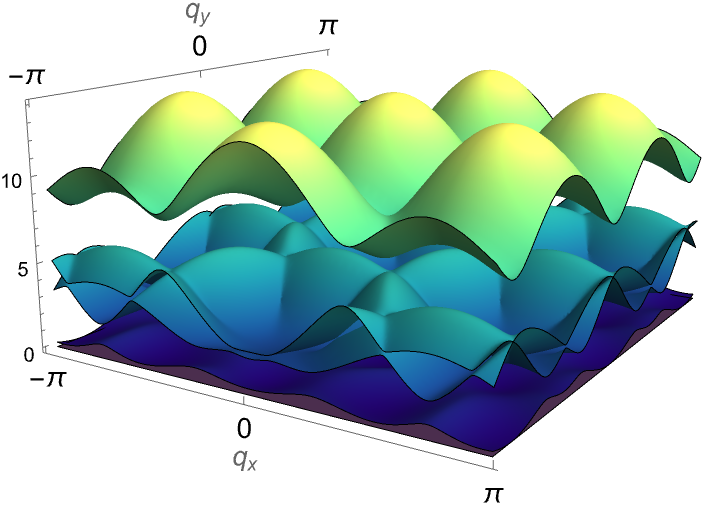} & 
        \includegraphics[width=2.5cm]{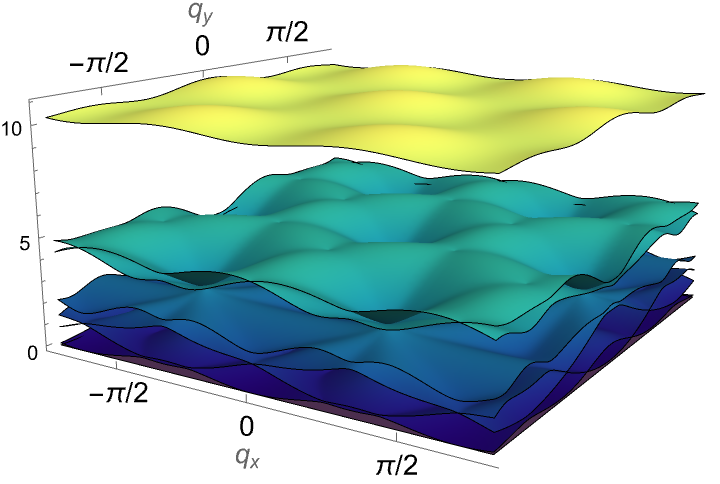} 
        \\

        \makecell{ \vspace{-2.cm} \\ \textbf{Band}  \\ \textbf{Structure} } & 
        \includegraphics[width=2.5cm]{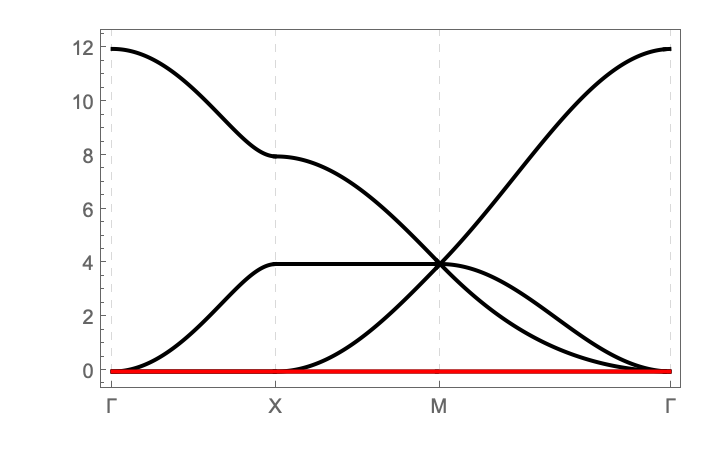} & 
        \includegraphics[width=2.5cm]{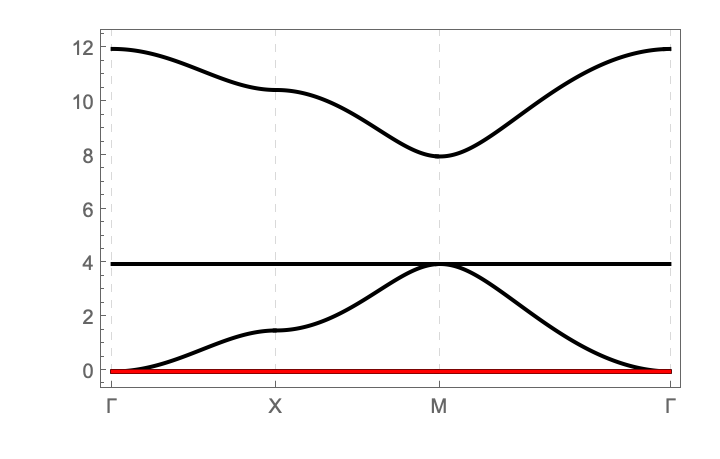} & 
        \includegraphics[width=2.5cm]{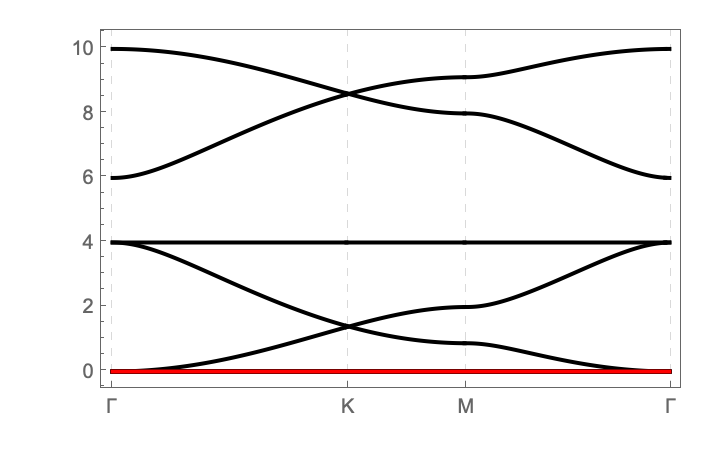} & 
        \includegraphics[width=2.5cm]{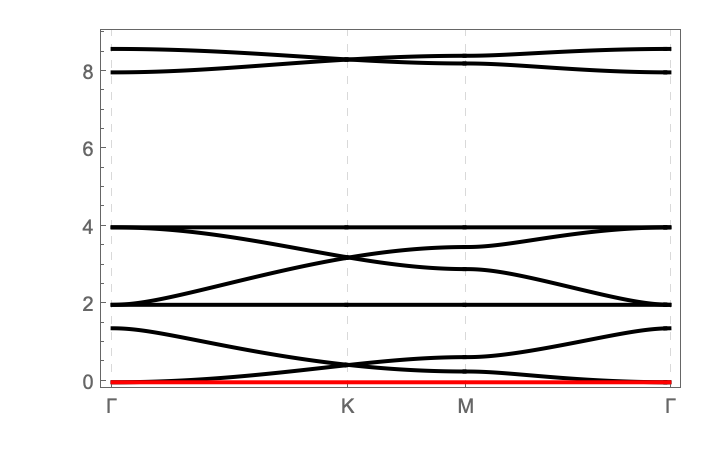} & 
        \includegraphics[width=2.5cm]{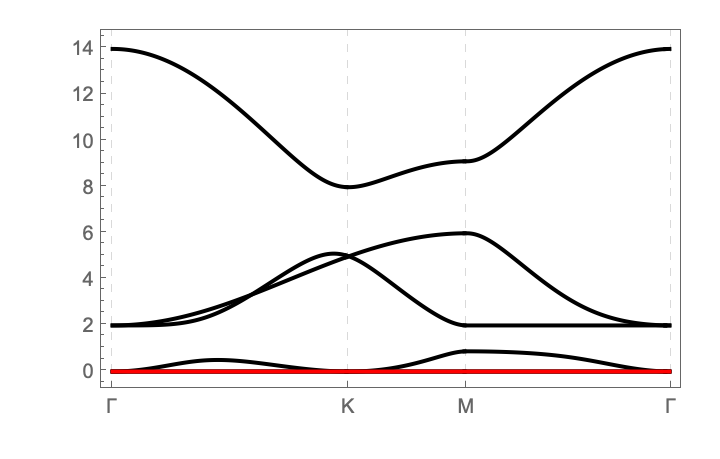} & 
        \includegraphics[width=2.5cm]{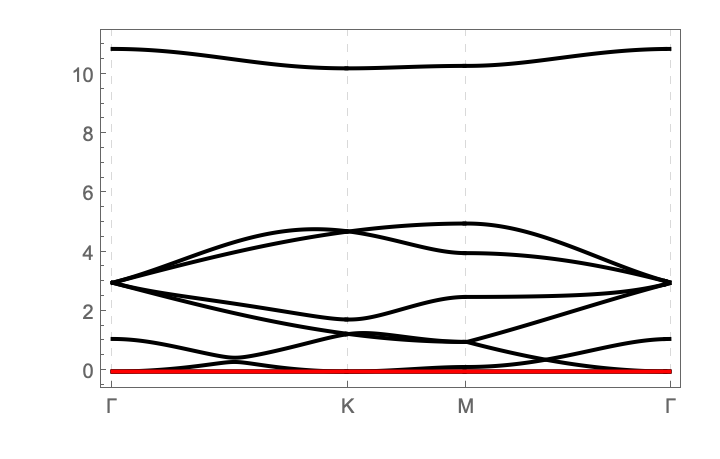} 
        \\
        
        \textbf{$n_s$}     & 4 & 8 & 12 & 15 & 6 & 9 \\
        \textbf{$n_c$}     & 3 & 3 & 5  & 8  & 4 & 7 \\
        \textbf{$n_\text{b.f.b}$} & 1 & 5 & 7  & 7  & 2 & 2 \\
        \textbf{$n_\text{i.f.b}$} & 0 & 1 & 1  & 1+1  & 0 & 0 \\
        \hline
    \end{tabular}
    }
    \caption{Bond and vertex decorated 2D systems. Here, cluster-bonds connect cluster-vertices to form usual 2D lattices. The number of bottom flat bands $n_\text{b.f.b}$ can be well estimated for these systems as $n_s - n_c$ where $n_s$ is given for clusters hosting only spins located on vertices. For lattices based on diamond or cracker bonds, both composite and mono-block cases are indicated (main vs. parentheses). In general, these lattices display a higher ratio of flat bands to sublattices than the bond-only decorated lattices of Table~\ref{tab: 2d decorated lattices}, making them theoretically stronger candidates for classical spin liquids. 
    As for bond decorated systems, vertex and bond decorated systems often appear to host intermediate flat bands located inside their spectrum. As these may be degenerate their number is indicated as a sum over the number of distinct set of intermediate flat bands.
    }
    \label{tab: 2d cluster-links + cluster-bonds lattices S}
\end{table*}

\subsection{Vertex-decorated systems}

If each vertex of the parent lattice is replaced by a cluster, one may suppress the bonds of the original lattice altogether to directly connect the cluster-vertices. These clusters can be connected through their corners or their edges/faces, leading to two main alternative constructions.
\begin{itemize}
    \item \emph{Corner-sharing clusters.}  
    To link the clusters through their corners requires the vertices of the parent lattice to be replaced by clusters having as many corners as the coordination number $z_p$ of the parent lattice. This way each link of the parent lattice is now replaced by a contact between two neighboring clusters corners. Following this procedure starting form the honeycomb lattice lead for example to build a kagome lattice, while starting from a square lattice corresponds to draw a checkerboard lattice, see Table.~\ref{tab: 2d cluster systems}. 
    In 3D the scheme is similar, the cubic lattice leading to the octochlore lattice while the diamond lattice produces the pyrochlore lattice, see Table.~\ref{tab: 3d cluster systems}.
    \item \emph{Face-sharing clusters.}  
    If the vertices are replaced by clusters having as many faces as the parent coordination number, clusters can be connected through their faces. The simplest examples are the triangular lattice that give the hexagonal cluster lattice, see Table.~\ref{tab: 2d cluster systems}, or the honeycomb lattice which leads to the triangular lattice.
\end{itemize}
These two main schemes can also be mixed to produce cluster lattices with different types of contacts between neighboring clusters, as for the square hexagonal lattice of Table. \ref{tab: 2d cluster systems} or the quadrupahedral lattice of Table. \ref{tab: 3d cluster systems}.

Making a vertex-decorated lattice does not however require to suppress the parent system links. These can be either conserved or replaced by cluster-bonds, leading to produce bond and vertex-decorated systems.

\subsection{Bond and vertex-decorated systems}

These bond and vertex-decorated systems can also be divided into two mains categories.
\begin{itemize}
    \item \emph{Corner-sharing clusters.} 
    The first option is to replace parent system vertices by clusters having as many corners as the coordination number $z_p$ of the parent lattice. In this case the cluster-vertices can be either connected using simple bonds or cluster bonds. 
    The effective number of sites per cluster-vertex is then
    \begin{equation}
        n_\text{c.v} = \Omega_\text{c.v} + z_p \frac{\Omega_\text{c.l}-2}{2},
    \end{equation}
    where $\Omega_\text{c.v}$ and $\Omega_\text{c.l}$ denote the number of sites in a cluster-vertex and cluster-bond, and $z_p$ is the coordination number of the parent lattice. The subtraction of 2 accounts for the two sites at the ends of each cluster-bond (already counted in the vertices), while the division by 2 reflects the fact that each bond is shared between two vertices. The total number of sublattices is then
    \begin{equation}
    n_s = n_s^p \times n_\text{c.v} = n_s^p \left( \Omega_\text{c.v} + z_p \frac{\Omega_\text{c.l}-2}{2} \right),
    \end{equation}
    where $n_s^p$ is the number of sublattices of the parent lattice.
        
    \item \emph{Face-sharing clusters.}  Using clusters with a number of faces equal to the parent coordination number allows this time to join cluster-vertices through their edges, using cluster-bonds that terminate in parallel faces. This can be achieved with, for example, rectangular blocks or double triangles forming a “butterfly”, see Fig.~\ref{fig: large cluster-links}. 
    \begin{figure}[ht]
        \centering
        \includegraphics[height=0.65cm]{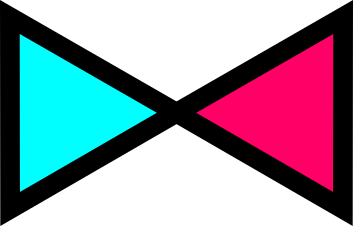} \hspace{0.2 mm}
        \includegraphics[height=0.65cm]{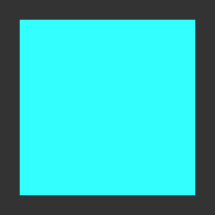} \hspace{0.2 mm}
        \includegraphics[height=0.65cm]{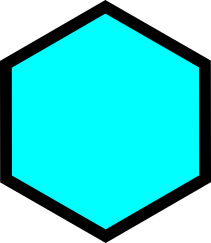}  \hspace{0.2 mm}
        \includegraphics[height=0.65cm]{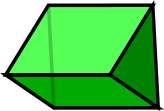}  \hspace{0.2 mm}
        \includegraphics[height=0.65cm]{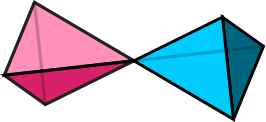}  \hspace{0.2 mm}
        \includegraphics[height=0.65cm]{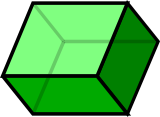}  \hspace{0.2 mm}
        \includegraphics[height=0.65cm]{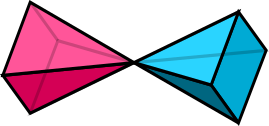}  \hspace{0.2 mm}
        \caption{Examples of large cluster-bonds: butterfly, rectangle, hexagon, triangular prism, double-tetrahedron, rectangular prism, double-pyramid.}
        \label{fig: large cluster-links}
    \end{figure}
    A honeycomb lattice decorated with triangular vertices and square bonds gives the ruby lattice (Table~\ref{tab: 2d cluster systems}). In this case, the number of sublattices becomes
    \begin{equation}
        n_s = n_s^p \left(\Omega_\text{c.v} + z_p \frac{\Omega_\text{c.l}-4}{2}\right).
        \label{Eq : bumber of sublattices for decorated systems}
    \end{equation}
    For example, decorating a hexagonal lattice ($z_p=3$, $n_s^p=2$) with hexagonal vertices ($\Omega_\text{c.v}=6$) and square bonds ($\Omega_\text{c.l}=4$) yields $n_s = 12$, see Table \ref{tab: 2d cluster-links + cluster-bonds lattices S} for the corresponding lattice scheme.
\end{itemize}

Both types of bond and vertex decorated lattices typically exhibit more zero modes than bond-decorated lattices, thanks to the extra degrees of freedom introduced by cluster-vertices. This makes them particularly promising candidates for classical spin liquids, despite their increased structural complexity. The same methodology can be applied in three dimensions: one may retain simple vertices or replace them with 3D clusters, and connect them via corners (simple or cluster-bonds) or via faces (3D cluster-bonds ending with compatible facets).\\

These construction methods provide a versatile framework for generating a wide range of cluster Hamiltonians. They make it possible to design good candidates for classical spin liquids, and in principle allow for the engineering of systems that host features such as high-rank pinch points~\cite{Davier_2023, Yan_2024_long} or pinch lines~\cite{Benton2016, Yan_2024_long, Davier_2025_pinch_lines} by enlarging the clusters, as already discussed in the literature. Furthermore, the study of the band structure of these systems reveal that a frequent feature of the decorated cluster lattice is the presence of of flat bands appearing not at the bottom of the spectrum but among the upper set of dispersive bands as it can be seen in the many band structures presented in the Tables \ref{tab: 2d cluster systems} to \ref{tab: 2d cluster-links + cluster-bonds lattices}. This feature, shared by the well studied ruby lattice, seems not to have been discussed in the literature. We therefore devote the remaining of this work to the explanation of the origin and consequences of the existence of such in-spectrum flat bands.

\section{Additional bottom flat bands for bond decorated systems and non singular band touchings}
\label{sec: additional_bfb}

\begin{figure}[h!]
    \centering
    \includegraphics[width=0.85\linewidth]{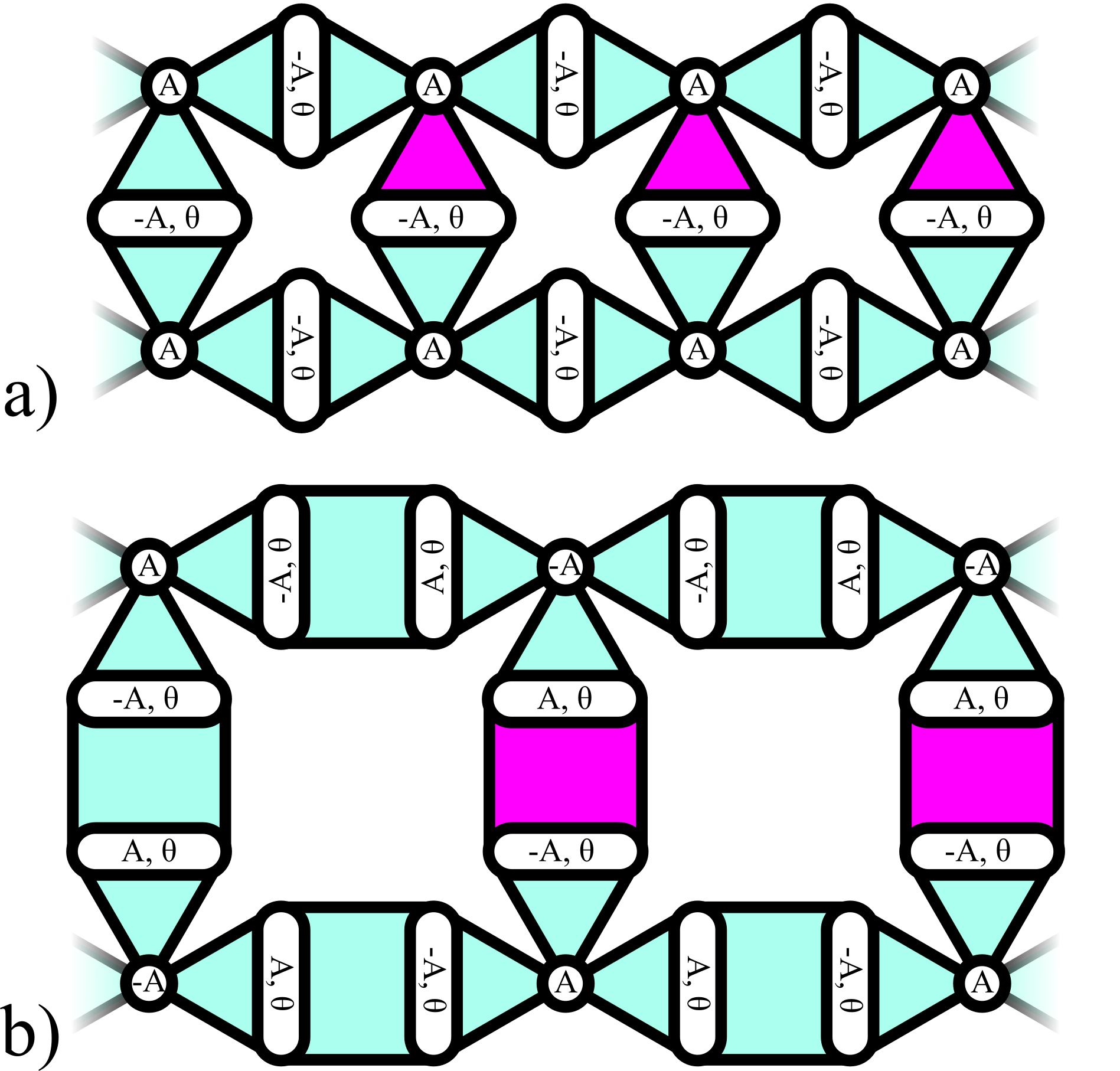}
    \caption{Real-space interpretation of the emergence of additional bottom flat band(s) in bond-decorated systems for Heisenberg spins. The notation $-A, \theta$ refers to a pair of spins which sum equals the vector $-A$, with a free angular degree of freedom $\theta$ representing the free rotation of this pair of spins around the direction given by $A$. Propagating the (anti)ferromagnetic order of the corner spins by successively imposing the local ground-state constraints on the blue clusters automatically enforces the constraints on the pink clusters. For the two systems shown, this results in one constraint per unit cell (in the thermodynamic limit) being trivially satisfied, which produces a single additional flat band. Applying the same construction to triangular (kagome) bond-decorated systems leads to two (three) trivially satisfied constraints per unit cell, explaining the presence of two (three) additional bottom flat bands. }
    \label{fig:Additional_flat_band}
\end{figure}

Before to dig in the study of in-spectrum flat bands, it is worth to first look at the presence of additional flat bands located at the bottom of the spectrum compared to what is expected from a naive counting based on the number of distinct cluster types (see Table~\ref{tab: 2d decorated lattices}). In reciprocal space, this corresponds to the fact that the subspace associated with dispersive bands has a smaller dimension than anticipated. This subspace describes constrained modes with finite energy and is therefore expected to have a dimension equal to the number of independent local constraints per unit cell.
In generic situations, the number of such constraints equals the number of clusters per unit cell, denoted by $n_c$. Consequently, observing more non-bottom flat bands than the difference $n_s - n_c$, where $n_s$ is the number of degrees of freedom per unit cell, indicates that some of the cluster constraints are not independent but are instead trivially satisfied.
This phenomenon can be understood in real space by explicitly constructing ground-state configurations for Heisenberg spins ($\mathcal{N} = 3$). One may start by fixing the state of a spin located on an arbitrarily chosen vertex of the parent lattice. As illustrated in Fig.~\ref{fig:Additional_flat_band}, imposing the vanishing of the local constraint on all clusters forming a horizontal chain enforces either a ferromagnetic or antiferromagnetic ordering of the spins belonging to the corresponding sublattice along this one-dimensional chain. Repeating the same step-by-step construction for vertical bond clusters propagates this ordering in the orthogonal direction.
Once this global ordering pattern is established, imposing the constraint $\bm{\mathcal{C}}_X = 0$ on all but one clusters among a vertical bond automatically ensures that the remaining constraint is satisfied as well. In the thermodynamic limit, this implies that one local ground-state condition per vertical bond is redundant. Equivalently, one constraint per unit cell is trivially fulfilled (the same reasoning applies upon exchanging the horizontal and vertical directions). As a result, the effective number of independent constraints per unit cell is reduced by one, which directly leads to the appearance of an additional flat band at the bottom of the spectrum.

This extra flat band has important consequences for the structure factor. In particular, it enables nontrivial band touching between the dispersive bands and the flat-band manifold, rendering the system gapless while simultaneously suppressing pinch-point singularities in the structure factor, as reported in Table~\ref{tab: 2d decorated lattices}. The underlying reason is that the Gram-matrix argument of Ref.~\cite{Davier_2025_interacting_clusters} no longer applies, since one of the Fourier transforms of the constrainers can always be expressed as a linear combination of the others.
An alternative viewpoint is obtained by considering the dimensionality of the band-touching manifold. When this dimension exceeds the spatial dimension minus two, the low-energy expansion around the contact manifold is at most one-dimensional and does not generally imply the emergence of an effective Gauss law and therefore of pinch points \cite{Davier_2023, Yan_2024_long}. 
This explains the absence of pinch points in two-dimensional systems exhibiting contact lines, such as the square kagome lattice (Table~\ref{tab: 2d cluster systems}) or the square–square–square decorated lattice (Table~\ref{tab: 2d cluster-links + cluster-bonds lattices S}). These one-dimensional contact manifolds, referred to as \emph{degeneracy lines} \cite{Davier_2023}, are associated with correlated one-dimensional strings in real space and do not give rise to Coulomb phases \footnote{Such degeneracy lines can coexist with singular band-touching points associated with pinch points in the structure factor, as illustrated in Ref.~\cite{Davier_2023} for the generalized checkerboard model. These features correspond to distinct forms of long-range spin--spin correlations: degeneracy lines produce correlations that extend along a single spatial direction, whereas pinch points are associated with isotropic algebraic correlations. Both originate from the same underlying constraint structure, namely the need to assemble local cluster ground-state configurations satisfying $\mathcal{C}=0$ in order to construct global ground states spanning the entire system.}.
In bond-decorated systems, by contrast, the band touching forms a full contact plane, which merges with the set of conventional flat bands. This leads to an unusual situation in which the band structure alone appears to indicate the presence of contact points—and hence pinch points—whereas none actually occur. The reason is that these contact points involve a hybridization between the lowest dispersive-band eigenvector and the accidental flat-band eigenvector, rather than with those of the conventional flat bands responsible for Coulomb-phase physics. 
A more intuitive understanding of this result can be obtained by examining the typical ground-state configurations shown in Fig.~\ref{fig:Additional_flat_band}. One observes that the degrees of freedom allowed to fluctuate within the ground-state manifold—parametrized by the local angular variables $\theta$—are completely independent from one bond to the next. As a consequence, correlations cannot propagate through the system, in contrast to what occurs in a Coulomb phase. This behavior corresponds to special lattice geometries in which the interpenetration of the constraint-carrying clusters suppresses any configurational propagation, thereby preventing the emergence of long-range correlations.

We now turn to the general study of in-spectrum flat bands, within which the phenomena described in the present section naturally emerge as a specific realization.

\section{Additional in-spectrum flat bands and graph connectivity}
\label{sec: additional_ifb_connectivity}

\subsection{Notion of connectivity matrices}

The band spectra of decorated lattices often display \emph{additional} flat bands embedded in the continuum, i.e., above the set of bottom flat bands (see Tables~\ref{tab: 2d decorated lattices}--\ref{tab: 2d cluster-links + cluster-bonds lattices}). Their origin can be understood naturally in terms of connectivity (adjacency) matrices~\cite{Mizoguchi_2018, Davier_2025_interacting_clusters, Roychowdhury_2024}.

Any bilinear classical spin Hamiltonian can be written as
\begin{equation}
    \mathcal{H} = \sum_{i,j} H_{ij}\,\mathbf{S}_i \cdot \mathbf{S}_j ,
    \label{Eq: H(1) connectivity matrix}
\end{equation}
where $H$ is the connectivity matrix of the lattice graph: $H_{ij}\neq 0$ only when vertices $i$ and $j$ are connected, and its value encodes the corresponding interaction strength.
For cluster systems, the underlying graph is built from fully connected clusters. This implies a factorized form for $H$~\cite{Davier_2025_interacting_clusters},
\begin{equation}
    H = h^{v \leftarrow c}\, h^{c \leftarrow v},
    \label{Eq: connectivity matrix expression for H(1)}
\end{equation}
where $h^{v \leftarrow c}$ is a rectangular connectivity matrix encoding connections between vertices and cluster centers, with $h^{v \leftarrow c}=\left(h^{c \leftarrow v}\right)^t$. Concretely, if $i$ denotes a site belonging to cluster $n$ of type $X$, one may define
\begin{equation}
    h^{v \leftarrow c}_{i,n} \equiv \gamma^{X}_{i}, 
\end{equation}
so that Eq.~\eqref{Eq: connectivity matrix expression for H(1)} follows directly from the cluster-Hamiltonian structure in Eq.~(\ref{Eq: general cluster H}) as the constrainer associated with cluster $n$ of type $X$ reads
\begin{equation}
    \bm{\mathcal{C}}_{n,X} = \sum_i h^{c \leftarrow v}_{n,i}\,\mathbf{S}_i.
\end{equation}

Because the cluster structure is translationally invariant, it is convenient to work in reciprocal space. We define the Fourier transform of a spin on sublattice $\mu\in\{1,\dots,n_s\}$ as
\begin{equation}
    \begin{split}
        \mathbf{S}_\mu(\mathbf{q}) &= \sum_{i}  \mathbf{S}_{\mu,i} \,e^{-i (\mathbf{R}_i + \mathbf{r}_\mu) \cdot \mathbf{q}}, \\ 
         \mathbf{S}_{\mu,i} &= \frac{1}{N_{u.c}}\sum_{\mathbf{q}}  \mathbf{S}_\mu(\mathbf{q}) \,e^{i (\mathbf{R}_i + \mathbf{r}_\mu) \cdot \mathbf{q}},
    \end{split}
    \label{Eq: Spin components Fourier transform}
\end{equation}
where $\mathbf{R}_i$ is the unit-cell position and $\mathbf{r}_\mu$ is the sublattice offset within the unit cell. Translation invariance implies $H_{\mu,\nu}(\mathbf{R}_i,\mathbf{R}_j)=H_{\mu,\nu}(\mathbf{R}_j-\mathbf{R}_i)$, and the Hamiltonian becomes 
\begin{equation}
    \mathcal{H} = \frac{1}{N_{u.c}}\sum_\mathbf{q} \sum_{\mu,\nu} H_{\mu \nu}(\mathbf{q}) \,\mathbf{S}_\mu(\mathbf{q}) \cdot \mathbf{S}_\nu(-\mathbf{q}),
\end{equation}
in momentum space~\cite{Mizoguchi_2018, Davier_2025_interacting_clusters}, with
\begin{equation}
    H_{\mu\nu}(\mathbf{q}) = \sum_{j} H_{(0,\mu),(j,\nu)}\,e^{i(- \mathbf{R}_j + \mathbf{r}_\mu-\mathbf{r}_\nu)\cdot\mathbf{q}}.
\end{equation}
The factorization in Eq.~\eqref{Eq: connectivity matrix expression for H(1)} carries over to reciprocal space:
\begin{equation}
    H(\mathbf{q}) = h^{v \leftarrow c}(\mathbf{q})\, h^{c \leftarrow v}(\mathbf{q}),
\end{equation}
where the $n_s\times n_c$ matrix $h^{v \leftarrow c}(\mathbf{q})$ is the Fourier transform of the vertex--cluster connectivity matrix,
\begin{equation}
\begin{split}
    h^{v \leftarrow c}_{\mu, X}(\mathbf{q})
    &= \sum_{j} h^{v \leftarrow c}_{(0,\mu),(j,X)}\,e^{i(- \mathbf{R}_j + \mathbf{r}_\mu -\mathbf{r}_X^c)\cdot\mathbf{q}} \\
    &= \sum_{k} h^{v \leftarrow c}_{(k,\mu),(0,X)}\,e^{i(\mathbf{R}_k + \mathbf{r}_\mu -\mathbf{r}_X^c)\cdot\mathbf{q}} \\
    &= \left[h^{c \leftarrow v}(\mathbf{q})\right]^\dagger_{\mu, X},
\end{split}
    \label{Eq: h fourier transform}
\end{equation}
and $\mathbf{r}_X^c$ denotes the position of the cluster center (type $X$) within the unit cell. Here $n_s$ is the number of sublattices of the original lattice, and $n_c$ is the number of cluster types per unit cell~\cite{Davier_2025_interacting_clusters}. Equivalently, $n_c$ is the number of sublattices of the \emph{bidual} lattice obtained by placing sites at cluster centers and connecting sites associated with touching clusters (see Table~\ref{tab: bidual_lattices}).

\begin{table*}[ht]
    \centering
    \begin{tabular}{c c c c c c c}
        \hline
        \hline
        \makecell{\textbf{Parent}\\ \textbf{cluster} \\ \textbf{lattice}} &
        Ruby & \makecell{Bond\\decorated\\square} & \makecell{Bond\\decorated\\honeycomb} &
        \makecell{Bond\\decorated\\kagome} & \makecell{Bond\\decorated\\triangular} \\
        &
        \includegraphics[width= 0.17 \linewidth]{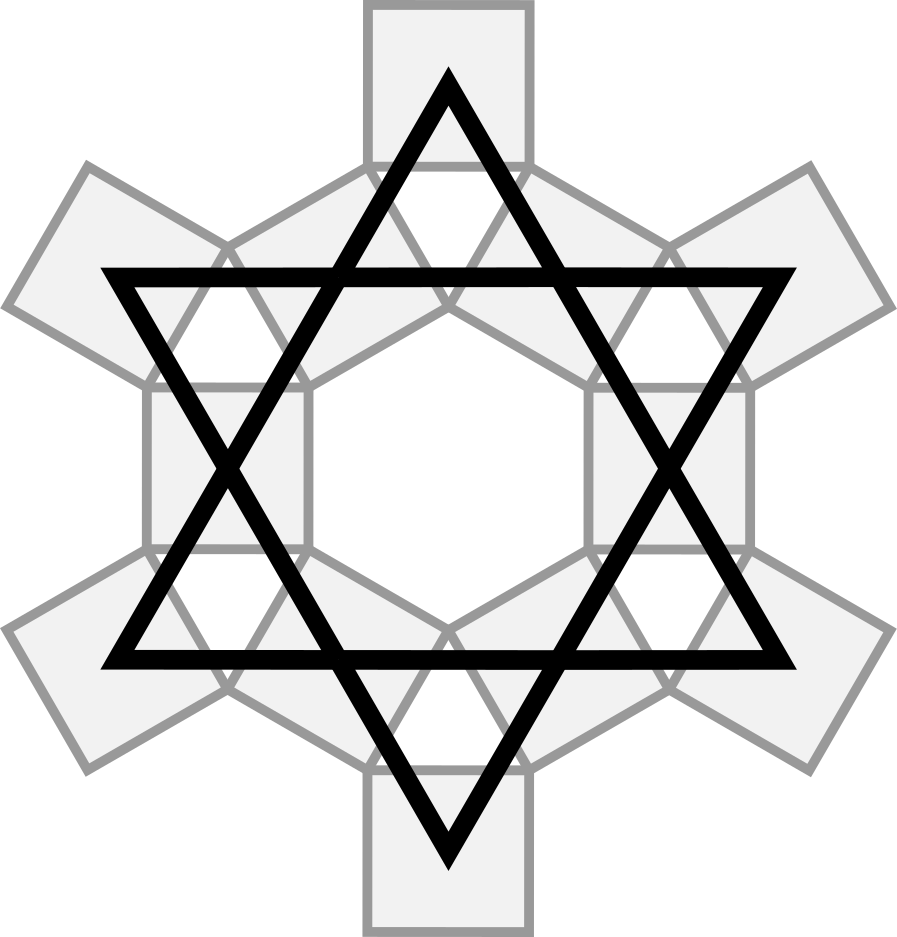} &
        \includegraphics[width= 0.18 \linewidth]{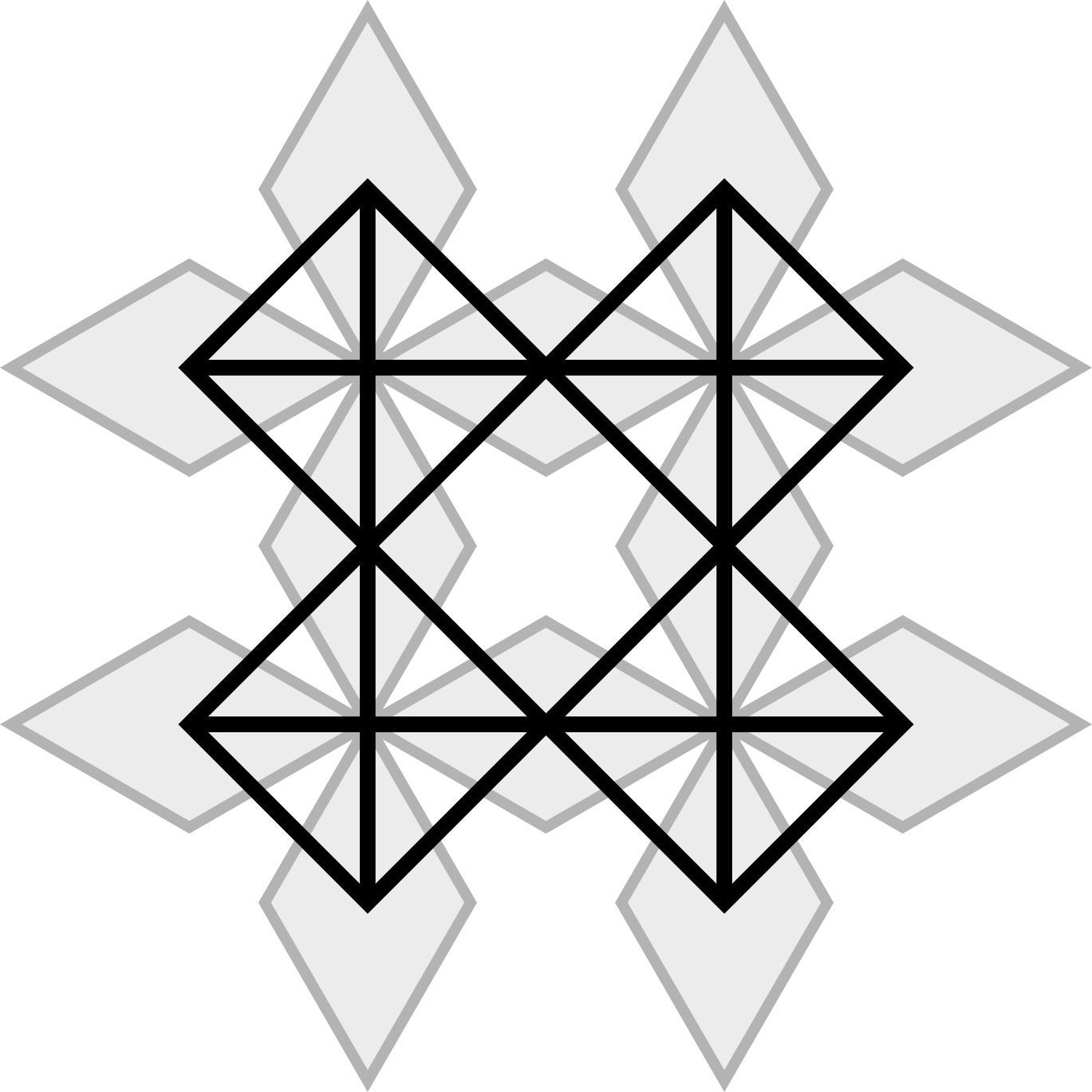} &
        \includegraphics[width= 0.16 \linewidth]{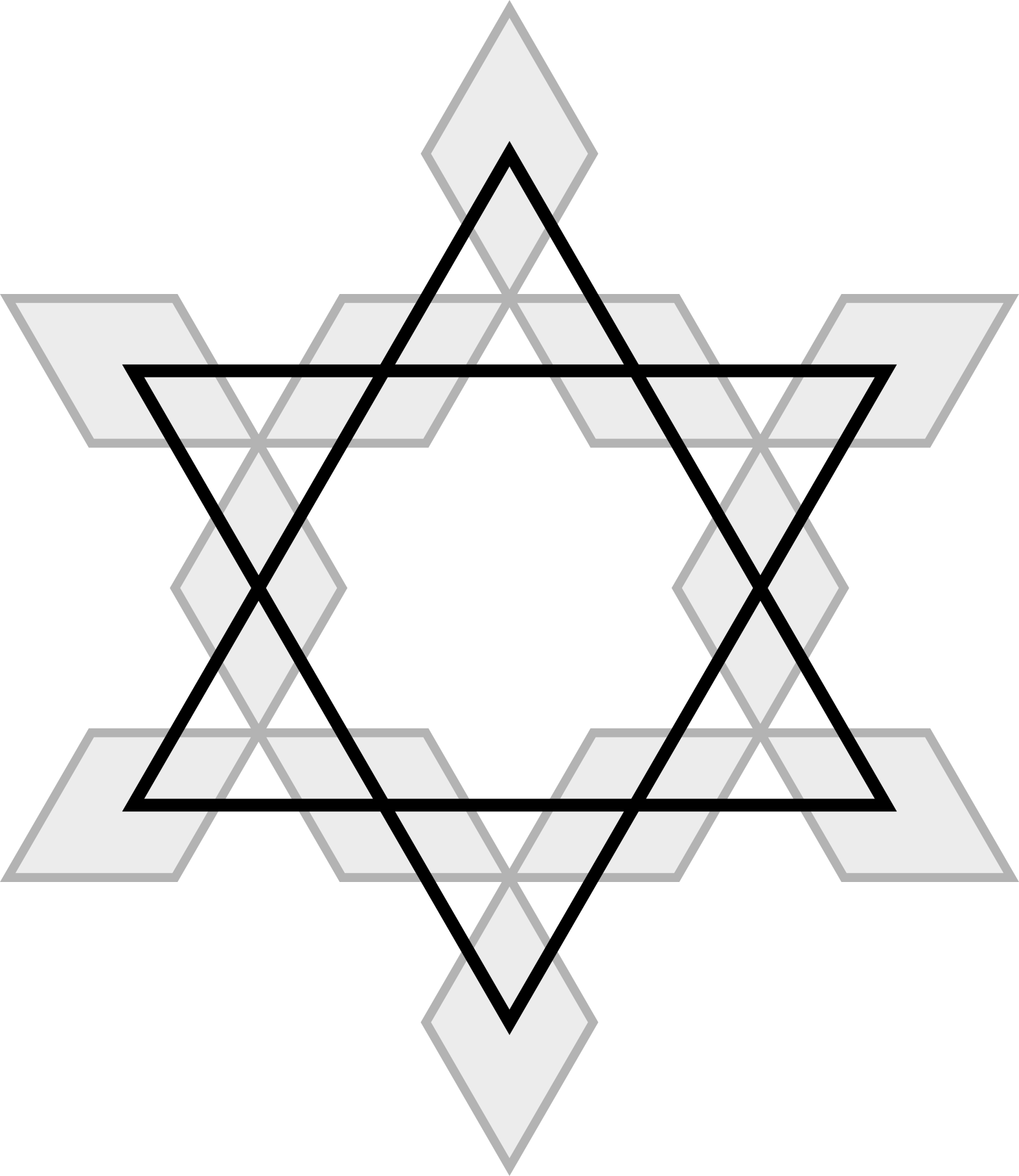} &
        \includegraphics[width= 0.2 \linewidth]{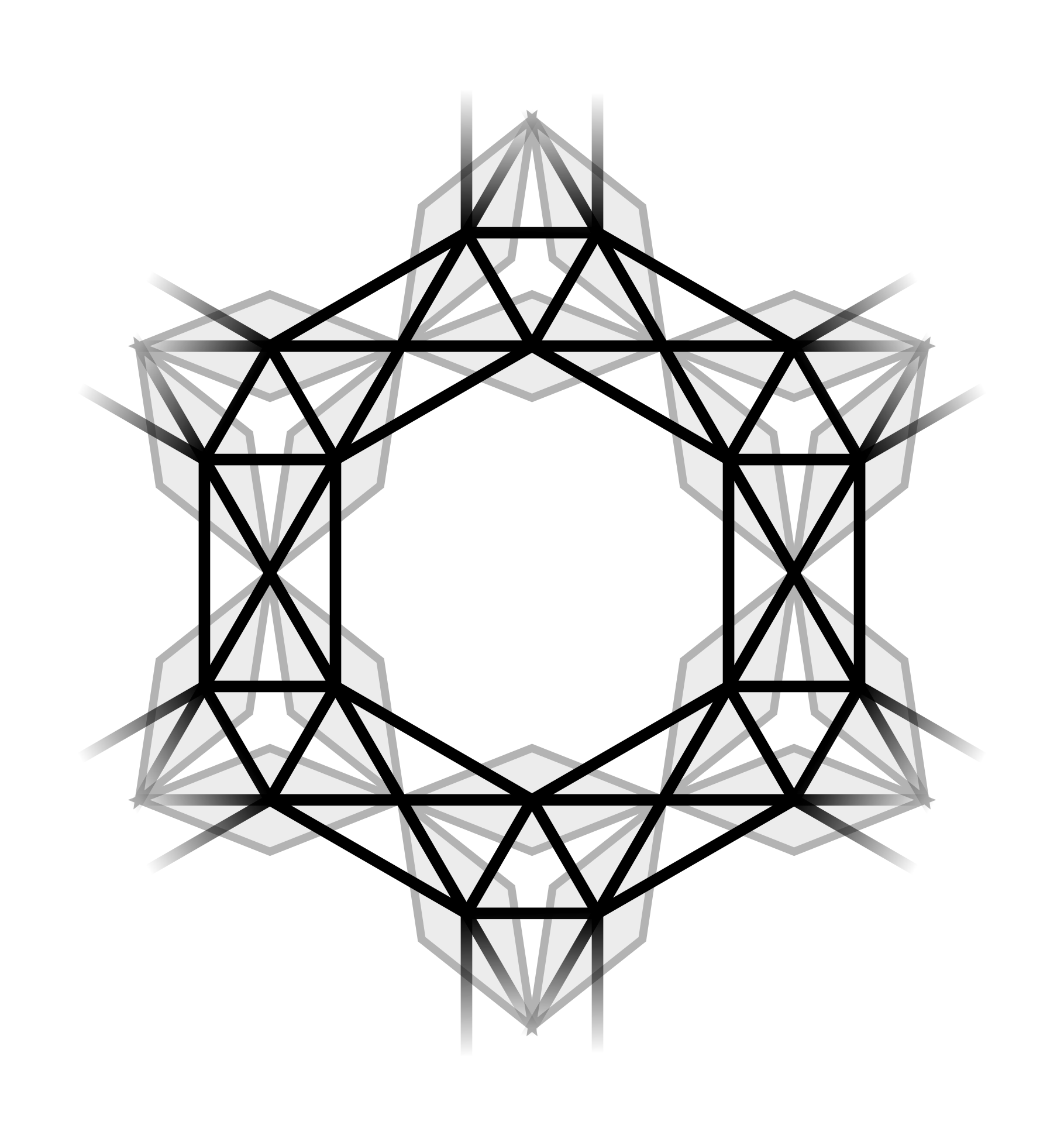} &
        \includegraphics[width= 0.16 \linewidth]{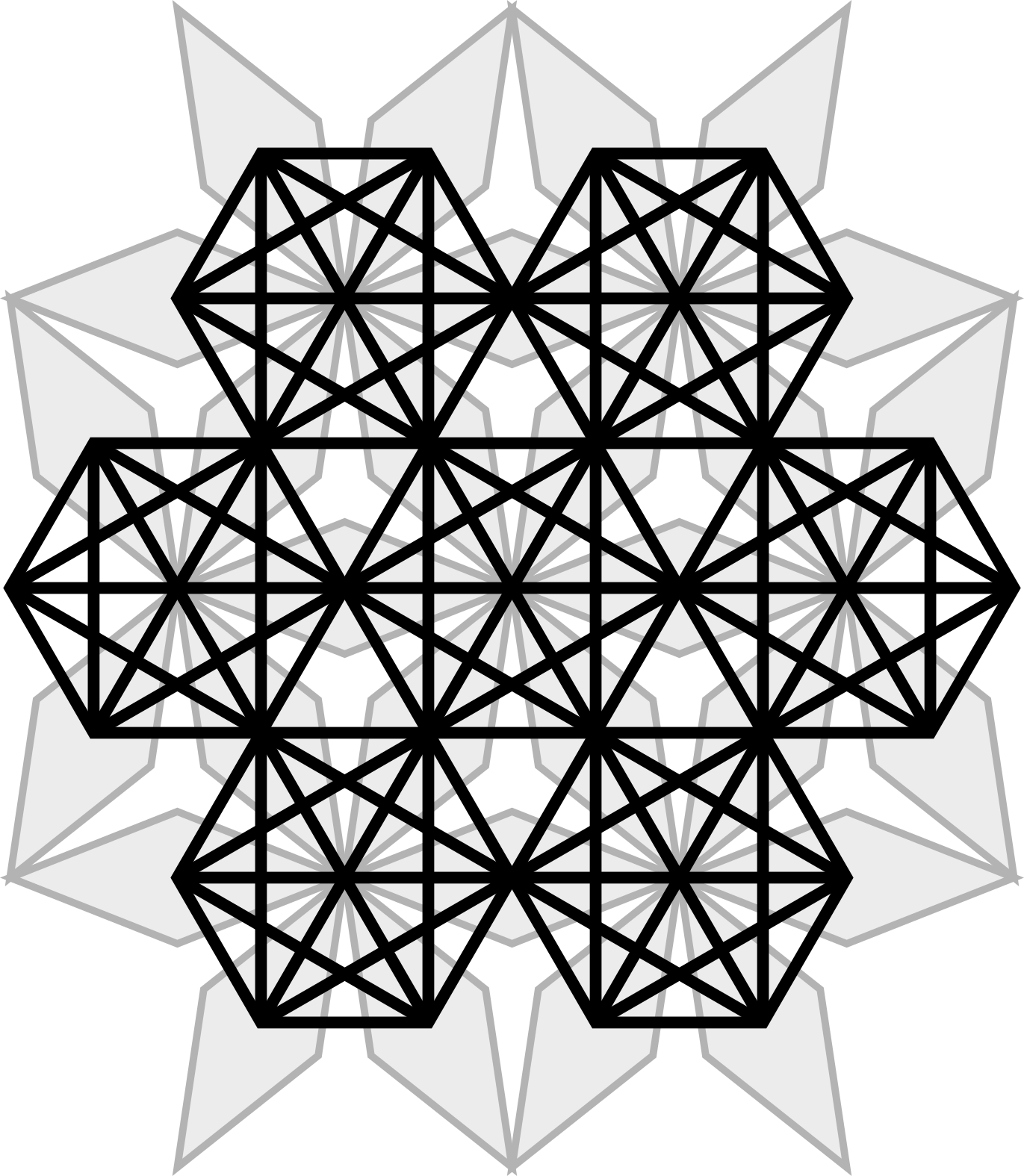} \\

        \makecell{\textbf{Bidual}\\ \textbf{lattice}} &
        Kagome & Checkerboard & Kagome & Ruby & \makecell{Hexagonal\\Kagome} \\
        \hline

        $\Xi^{(c)}$ & 4 & 4 & 4 & 4 & 4 \\
        $\Omega^{(bd)}$ & 2 & 2 & 2 & 2 & 2 \\
        $n_c$ & 3 & 2 & 3 & 6 & 3 \\
        $\mathfrak{n_c}$ & 2 & 1 & 2 & 3 & 1 \\
        \hline
        \hline
    \end{tabular}
    \caption{ Cluster lattices (represented in gray) admitting bidual lattices (black lines) which are themselves cluster lattices. Note that for monocracker bond decorated systems $\Xi^{(c)} = 6$ while $\Omega^{(bd)}$ remains identical because the bidual lattice is the same.
    }
    \label{tab: bidual_lattices}
\end{table*}

A key consequence of Eq.~\eqref{Eq: connectivity matrix expression for H(1)} is purely algebraic. Since $H(\mathbf{q})=h(\mathbf{q})h^\dagger(\mathbf{q})$ with $h(\mathbf{q})$ of size $n_s\times n_c$, the rank of $H(\mathbf{q})$ is at most $n_c$. Therefore, for each $\mathbf{q}$, $H(\mathbf{q})$ has at least $n_s-n_c$ \emph{exact zero modes}, which form $n_s-n_c$ flat bands. Moreover, the remaining (generally dispersive) eigenvalues of $H(\mathbf{q})$ coincide with those of the $n_c\times n_c$ matrix
\begin{equation}
    H^{\mathrm{bd}}(\mathbf{q}) \equiv h^{c \leftarrow v}(\mathbf{q})\,h^{v \leftarrow c}(\mathbf{q}) ,
\end{equation}
i.e., with the spectrum of the bidual connectivity problem. Consequently, any \emph{additional} flat bands appearing \emph{inside} the spectrum of $H(\mathbf{q})$ must originate from flat bands of $H^{\mathrm{bd}}(\mathbf{q})$.

In the remainder of this section, we work exclusively with connectivity matrices in reciprocal space and omit the explicit $\mathbf{q}$ dependence for notational simplicity. We also focus on the case $\gamma_i^X=1$, which reduces the adjacency matrices $h^{c \leftarrow v},\,h^{v \leftarrow c}$ to purely geometric (0/1) connectivity and simplifies both the discussion and the graphical representations. Most arguments are expected to extend to the general case $\gamma_i^X\neq 1$.

\subsection{Case of gapped flat bands}

We first consider in-spectrum flat bands that are gapped and lie immediately above the bottom flat bands. This situation is realized, for instance, in the ruby lattice (see Table. \ref{tab: 2d cluster systems}) and in monocluster bond-decorated systems (see Table. \ref{tab: 2d monobond decorated lattices}). For these systems, the product of cluster connectivity matrices can be written \cite{Davier_2025_interacting_clusters} as
\begin{equation}
    h^{c \leftarrow v} h^{v \leftarrow c} = \Xi^{(c)}\, I^{c \leftarrow c} + g^{c \leftarrow c},
    \label{Eq: hcvhvc}
\end{equation}
where $I^{c \leftarrow c}$ is the identity matrix, $\Xi^{(c)}$ is the number of spins per cluster, and $g^{c \leftarrow c}$ is the connectivity matrix linking the centers of clusters that share a common site. Note that, geometrically, the product $h^{c \leftarrow v} h^{v \leftarrow c}$ should be understood as the connectivity matrix encoding all paths that connect two cluster centers via a vertex site, including paths that return a cluster center to itself; these self-returning paths are responsible for the identity term.

We now consider the case where $g^{c \leftarrow c}$ itself encodes a cluster system graph, which is a common feature of monoblock bond decorated systems, see Table~\ref{tab: bidual_lattices}. In this case, it can be written as
\begin{equation}
    g^{c \leftarrow c} = \mathfrak{h}^{c \leftarrow \mathfrak{c}} \mathfrak{h}^{\mathfrak{c} \leftarrow c}
    - \Omega^{(bd)}\, I^{c \leftarrow c},
    \label{Eq:gcc meta h}
\end{equation}
where $\mathfrak{h}^{c \leftarrow \mathfrak{c}}$ connects the cluster centers to the centers of the clusters they themselves form, which we refer to as \emph{meta-centers} and \emph{meta-clusters}. This decomposition follows from the fact that the product $\mathfrak{h}^{c \leftarrow \mathfrak{c}} \mathfrak{h}^{\mathfrak{c} \leftarrow c}$ accounts both for connections between cluster centers mediated by a meta-center and for self-return paths, the latter giving rise to the identity term. Therefore $\Omega^{(bd)}$ is the number of clusters between which a site of the bidual lattice is shared.
The matrix $\mathfrak{h}^{c \leftarrow \mathfrak{c}}(\mathbf{q})$ has dimension $n_c \times \mathfrak{n}_c$, where $\mathfrak{n}_c$ is the number of sublattices of the bidual lattice of the bidual lattice of the parent system. As a consequence, its kernel has dimension $n_c - \mathfrak{n}_c$, which is associated with flat bands. This implies that $g^{c \leftarrow c}(\mathbf{q})$ also possesses $n_c - \mathfrak{n}_c$ flat bands at the bottom of its spectrum, located at energy $-\Omega^{(bd)}$. 
Finally, re-expressing Eq.~\eqref{Eq: hcvhvc} as
\begin{equation}
    h^{c \leftarrow v} h^{v \leftarrow c}
    = \left( \Xi^{(c)} - \Omega^{(bd)} \right) I^{c \leftarrow c}
    + \mathfrak{h}^{c \leftarrow \mathfrak{c}} \mathfrak{h}^{\mathfrak{c} \leftarrow c}
    \label{Eq:additional flat bands}
\end{equation}
shows that the matrix $h^{c \leftarrow v} h^{v \leftarrow c}$ must admit the same number of flat bands $n_c - \mathfrak{n}_c$, now located at energy $\Xi^{(c)} - \Omega^{(bd)}$. Inspection of the band structures of monocluster bond-decorated systems (see Table~\ref{tab: 2d monobond decorated lattices}) confirms that the energies of the in-spectrum flat bands are indeed given by this difference. The degeneracies also match exactly the value $n_c - \mathfrak{n}_c$, i.e., the difference between the number of cluster types in the cluster lattice and in the meta-cluster lattice (see Table~\ref{tab: bidual_lattices}).
A particularly illustrative example is the bond-decorated kagome lattice. Its bidual lattice is the ruby lattice, which itself admits a cluster bidual structure. This hierarchical organization results in the presence of two distinct sets of upper flat bands in the spectrum, as shown in Table~\ref{tab: 2d decorated lattices}.

This discussion in fact extends beyond the mere presence of flat bands. Since the structure of the dispersive bands is encoded in the bidual Hamiltonian $ H^{\mathrm{bd}} = h^{c \leftarrow v} h^{v \leftarrow c}$, which can be rewritten as in Eq.~(\ref{Eq:additional flat bands}), the full spectrum of the cluster system—up to the bottom flat bands—is determined by the spectrum of the dual cluster-lattice Hamiltonian $\mathfrak{h}^{c \leftarrow \mathfrak{c}} \mathfrak{h}^{\mathfrak{c} \leftarrow c}$, rigidly shifted by an energy $\Xi^{(c)} - \Omega^{(bd)}$.
This correspondence can be verified directly by comparing the spectra of monocluster bond-decorated systems (Table~\ref{tab: 2d monobond decorated lattices}) with those of their bidual lattices (Table~\ref{tab: 2d cluster systems}). More generally, it demonstrates that the dispersive part of the spectrum is entirely governed by the structure of the bidual lattice, irrespective of whether this lattice itself admits a cluster representation. This explains, for instance, why the hexagonal and kagome--hexagonal lattices listed in Table~\ref{tab: 2d cluster systems} share the same dispersive band, namely that of the triangular lattice. Indeed, in both cases the connectivity matrix carrying the spectrum can be recast as
\begin{equation}
    h^{c \leftarrow v} h^{v \leftarrow c}
    = \Xi^{(c)} I^{c \leftarrow c} + \Omega\, g^{c \leftarrow c}_\text{triangular},
\end{equation}
where $\Omega$ denotes the number of sites shared by two neighboring clusters, equal to \(2\) for the hexagonal cluster lattice and to \(1\) for the kagome--hexagonal lattice, while $\Xi^{(c)} = 6$ for both lattices.

\subsection{General origin of the flat bands}

The above reasoning, however, does not account for situations in which in-spectrum flat bands are surrounded by dispersive bands. This is the case for composite cluster bond-decorated systems (see Tables~\ref{tab: 2d decorated lattices}, \ref{tab: 2d cluster-links + cluster-bonds lattices S} and \ref{tab: 2d cluster-links + cluster-bonds lattices}). The derivation is then more involved, and we illustrate it through a simple example: the bidiamond-decorated square lattice.

\subsubsection{Introductory example}

\begin{figure}[h]
    \centering
    \includegraphics[width=\linewidth]{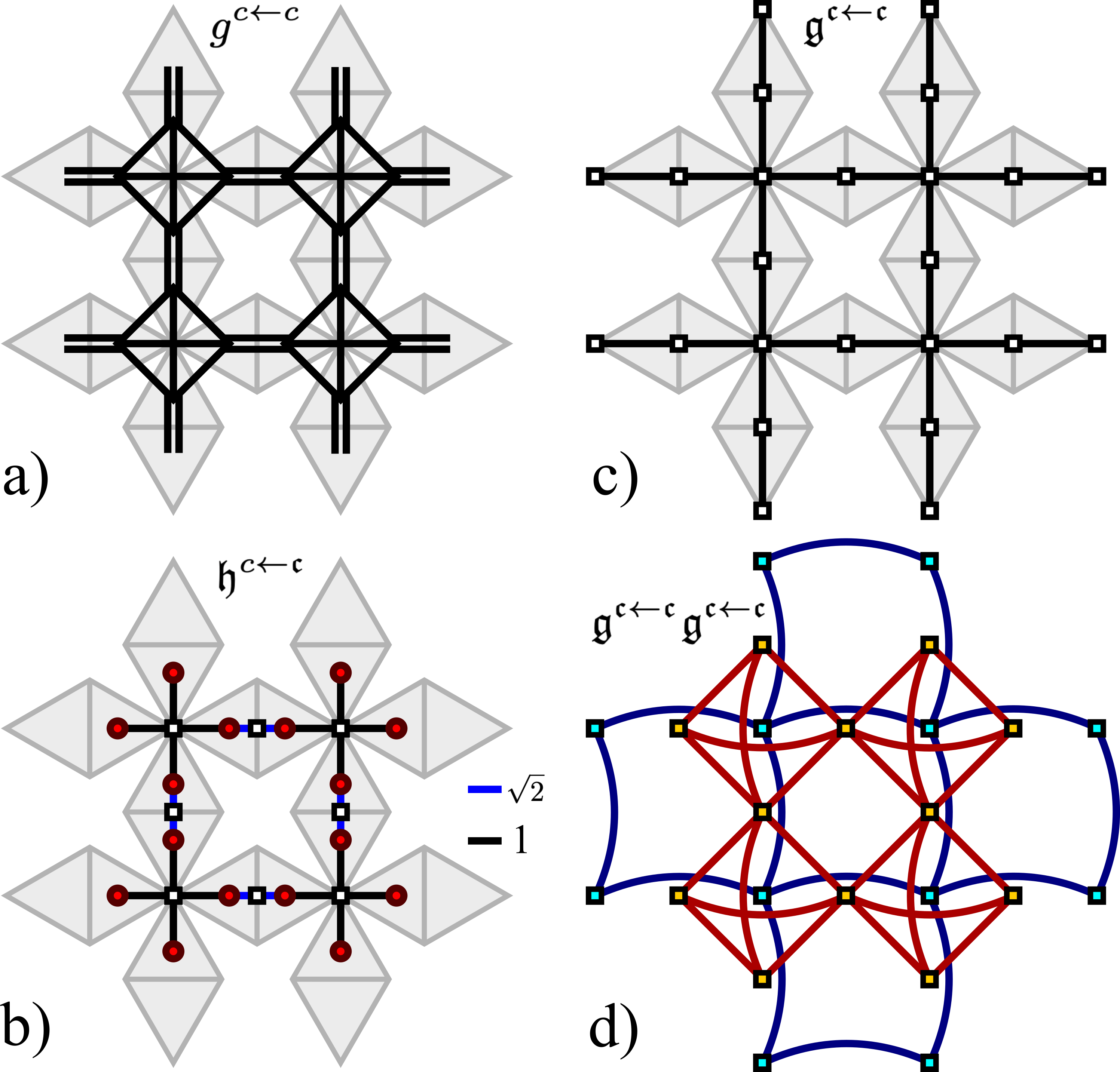}
    \caption{
    (a) Dual lattice associated with the connectivity matrix $g^{c \leftarrow c}$.
    (b) Cluster-connectivity matrix $\mathfrak{h}^{c \leftarrow \mathfrak{c}}$ linking the centers of bidiamond clusters to the centers of the meta-clusters they form.
    (c) Lattice associated with the connectivity matrix $\mathfrak{g}^{\mathfrak{c} \leftarrow \mathfrak{c}}$ obtained from
    $\mathfrak{h}^{\mathfrak{c} \leftarrow c} \mathfrak{h}^{c \leftarrow \mathfrak{c}} = \mathfrak{h}^{ \sqbox0{black} \, \circbox0{black}} \mathfrak{h}^{\circbox0{black} \, \sqbox0{black}}$.
    (d) Decoupled square and checkerboard lattices obtained by squaring $\mathfrak{g}^{\mathfrak{c} \leftarrow \mathfrak{c}}$.}
    \label{fig:bidiamond_dual_construction}
\end{figure}

In this system, the product $h^{c \leftarrow v} h^{v \leftarrow c}$ yields a connectivity matrix $g^{c \leftarrow c}$ whose associated graph resembles that of the checkerboard lattice, but with additional bonds linking neighboring squares, as shown in Fig.~\ref{fig:bidiamond_dual_construction}(a). These additional bonds carry weight~2, as they correspond to two distinct paths $c \rightarrow v \rightarrow c$, reflecting the fact that a bidiamond consists of two triangles sharing a pair of sites.
As in the previous cases, this lattice can itself be interpreted as a cluster lattice, whose clusters are squares connected by two distinct types of bonds. Consequently, the connectivity matrix $g^{c \leftarrow c}$ can again be expressed in the form of Eq.~(\ref{Eq:gcc meta h}). Care must be taken here, however, because the initial cluster system involves both corner-sharing and bond-sharing clusters. As a result, the entries of the matrix $\mathfrak{h}^{c \leftarrow \mathfrak{c}}$ are no longer restricted to values $0$ or $1$.
Instead, the coefficients must satisfy
\begin{equation}
    \sum_{\mathfrak{k}}
    \mathfrak{h}^{c \leftarrow \mathfrak{c}}_{m \mathfrak{k}}
    \mathfrak{h}^{\mathfrak{c} \leftarrow c}_{\mathfrak{k} n}
    = N_{m \cap n},
\end{equation}
where $N_{m \cap n}$ denotes the number of sites shared by clusters $m$ and $n$. This condition is fulfilled by the weights shown in Fig.~\ref{fig:bidiamond_dual_construction}(b).
In this situation, the coefficient $\Omega^{(bd)}$ must be computed explicitly by summing all two-step paths that return to the initial site $n$,
\begin{equation}
    \Omega^{(bd)} =
    \sum_{\mathfrak{k} \in \langle n \rangle}
    \mathfrak{h}^{c \leftarrow \mathfrak{c}}_{n \mathfrak{k}}
    \mathfrak{h}^{\mathfrak{c} \leftarrow c}_{\mathfrak{k} n},
\end{equation}
which yields $\Omega^{(bd)} = 1 + 2 = 3$ for the bidiamond-decorated square lattice.
As before, the spectrum is therefore composed of a set of flat bands followed by the spectrum of the matrix
$\mathfrak{h}^{c \leftarrow \mathfrak{c}} \mathfrak{h}^{\mathfrak{c} \leftarrow c}$, shifted by $\Xi^{(c)} - \Omega^{(bd)}$. In the present case, the elementary clusters are triangles, so $\Xi^{(c)} = 3$, and since $\Omega^{(bd)} = 3$ the shift vanishes. The flat bands associated with the meta-cluster lattice therefore merge with those of the original cluster system, in agreement with the additional bottom flat bands reported in Table~\ref{tab: 2d monobond decorated lattices}. This completes the real-space interpretation given previously.

We now turn to the existence of flat bands located higher in the spectrum. Such bands must arise within the three-band spectrum of the matrix
$\mathfrak{h}^{\mathfrak{c} \leftarrow c} \mathfrak{h}^{c \leftarrow \mathfrak{c}}$.
To elucidate their origin, it is useful to note that this adjacency matrix can be rewritten as
\begin{equation}
    \mathfrak{h}^{\mathfrak{c} \leftarrow c} \mathfrak{h}^{c \leftarrow \mathfrak{c}}
    = \sqrt{2}\,\mathfrak{g}^{\mathfrak{c} \leftarrow \mathfrak{c}}
    + 4\, I^{\mathfrak{c} \leftarrow \mathfrak{c}},
    \label{eq hh gcc}
\end{equation}
where $\mathfrak{g}^{\mathfrak{c} \leftarrow \mathfrak{c}}$ is the connectivity matrix of a square lattice decorated by additional sites located at the midpoints of the bonds (called a Lieb lattice), with all bonds having unit weight, as illustrated in Fig.~\ref{fig:bidiamond_dual_construction}(c).
Because this lattice is bipartite, squaring this connectivity matrix connects only second neighbors and the resulting graph decomposes into two uncoupled lattices: a simple square lattice and a checkerboard lattice (including self-interaction terms), as shown in Fig.~\ref{fig:bidiamond_dual_construction}(d). Schematically, the resulting connectivity matrix can be expressed in the two decoupled subspaces as
\begin{equation}
    \mathfrak{g}^{\mathfrak{c} \leftarrow \mathfrak{c}}
    \mathfrak{g}^{\mathfrak{c} \leftarrow \mathfrak{c}}
    \sim
    \begin{pmatrix}
        H_{\text{ckb}} & 0 \\
        0 & 2 H_{\text{square}} + 4 I_1^{\mathfrak{c} \leftarrow \mathfrak{c}}
    \end{pmatrix},
\end{equation}
where $I_1^{\mathfrak{c} \leftarrow \mathfrak{c}}$ denotes the identity matrix restricted to the square-lattice subspace.

The spectrum of $\mathfrak{g}^{\mathfrak{c} \leftarrow \mathfrak{c}} \mathfrak{g}^{\mathfrak{c} \leftarrow \mathfrak{c}}$ thus consists of the dispersive band of the square lattice shifted by~4, together with the full spectrum of the checkerboard lattice. Since the checkerboard lattice possesses a flat band, this implies that $\mathfrak{g}^{\mathfrak{c} \leftarrow \mathfrak{c}}$ itself admits a flat band at the bottom of its spectrum.
Moreover, the dispersive band associated with
$2 H_{\text{square}} + 4 I^{\mathfrak{c} \leftarrow \mathfrak{c}}$
coincides exactly with the dispersive band of the checkerboard lattice,
\begin{equation}
    \begin{split}
        \lambda_{\text{ckb}}(\mathbf{q})
        &= 4 \left[ \cos^2(q_x a) + \cos^2(q_y a) \right] \\
        &= 4 + 2 \left[ \cos(2 q_x a) + \cos(2 q_y a) \right],
    \end{split}
\end{equation}
where $a$ is the lattice spacing of the decorated square lattice in Fig.~\ref{fig:bidiamond_dual_construction}(c).
As a result, the spectrum of
$\mathfrak{g}^{\mathfrak{c} \leftarrow \mathfrak{c}}
\mathfrak{g}^{\mathfrak{c} \leftarrow \mathfrak{c}}$
is identical to that of the checkerboard lattice, except that the dispersive band is doubled. Writing Eq.~(\ref{eq hh gcc}) as
\begin{equation}
    \left(
    \mathfrak{h}^{\mathfrak{c} \leftarrow c}
    \mathfrak{h}^{c \leftarrow \mathfrak{c}}
    - 4 I^{\mathfrak{c} \leftarrow \mathfrak{c}}
    \right)^2
    = 2\,
    \mathfrak{g}^{\mathfrak{c} \leftarrow \mathfrak{c}}
    \mathfrak{g}^{\mathfrak{c} \leftarrow \mathfrak{c}},
\end{equation}
it follows that the matrix
$\mathfrak{h}^{\mathfrak{c} \leftarrow c}
\mathfrak{h}^{c \leftarrow \mathfrak{c}}$
must possess a flat band at energy~4. This flat band is surrounded by two dispersive bands with dispersions
$\pm \sqrt{2\,\lambda_{\text{ckb}}(\mathbf{q})}$.
Finally, since $\Xi^{(c)} - \Omega^{(bd)} = 0$ in the present case, Eq.~(\ref{Eq:additional flat bands}) implies that a flat band must appear at energy~4 in the spectrum of the bidiamond-decorated square lattice, in perfect agreement with Table~\ref{tab: 2d decorated lattices}. The symmetry of the dispersive bands about this flat band follows directly from the square-root structure of the dispersion.

The existence of this additional flat band is therefore ultimately tied to the cluster lattice obtained by linking the centers of the bidiamonds, which forms a checkerboard lattice. This establishes a direct connection between monocluster-decorated and composite cluster-decorated systems: in both cases, the flat bands originate from an emergent meta-cluster structure built by connecting bond centers. 

\subsubsection{Generalization}

From a more general perspective, applying the same construction to any bidiamond-decorated system amounts to generating, upon linking the sites of the associated meta-cluster system, a particular class of bipartite lattices. By construction, the resulting bipartite lattice is the parent lattice decorated with diamond bonds, supplemented with additional sites located at the midpoints of each bond. Since this lattice is always bipartite, one may partition its sites into two sets, $\mathrm{I}$ and $\mathrm{II}$, in which case the associated connectivity matrix can be written as \cite{Sutherland_1986_FB, Roychowdhury_2024}
\begin{equation}
    \mathfrak{g}^{\mathfrak{c} \leftarrow \mathfrak{c}}
    =
    \begin{pmatrix}
        0 & A_{\mathrm{I\text{-}II}} \\
        A_{\mathrm{II\text{-}I}} & 0
    \end{pmatrix},
\end{equation}
where $A_{\mathrm{I\text{-}II}}$ and $A_{\mathrm{II\text{-}I}}$ are adjacency matrices connecting sites of type $\mathrm{I}$ to sites of type $\mathrm{II}$, and conversely. Squaring $\mathfrak{g}^{\mathfrak{c} \leftarrow \mathfrak{c}}$ then always yields a block-diagonal matrix,
\begin{equation}
    \begin{split}
        \left(\mathfrak{g}^{\mathfrak{c} \leftarrow \mathfrak{c}}\right)^2
        =
        \begin{pmatrix}
            A_{\mathrm{I\text{-}II}} A_{\mathrm{II\text{-}I}} & 0 \\
            0 & A_{\mathrm{II\text{-}I}} A_{\mathrm{I\text{-}II}}
        \end{pmatrix} 
        =
        \begin{pmatrix}
            A_{\mathrm{I}} & 0 \\
            0 & A_{\mathrm{II}}
        \end{pmatrix}.
    \end{split}
\end{equation}
In general, the numbers of sublattices of types $\mathrm{I}$ and $\mathrm{II}$ differ, so that $A_{\mathrm{I\text{-}II}}$ and $A_{\mathrm{II\text{-}I}}$ are rectangular matrices with a nontrivial kernel. Standard singular-value arguments then imply that $A_{\mathrm{I}}$ and $A_{\mathrm{II}}$ share the same spectrum, up to a set of flat bands whose degeneracy is fixed by the dimension mismatch \cite{Roychowdhury_2024}. As a consequence, the spectrum of $\mathfrak{g}^{\mathfrak{c} \leftarrow \mathfrak{c}}$ generically contains flat band(s) at zero energy, surrounded symmetrically by dispersive bands determined by the smallest of the two blocks, as illustrated in Ref.~\cite{Sutherland_1986_FB, Roychowdhury_2024}.
This reasoning is expected to extend to the case where the bond cluster contains more than two elementary clusters, and to the case where the clusters composing the bond are not all equivalent (e.g., ``cracker'' bonds). In these situations as well, one observes in-spectrum flat bands surrounded by dispersive bands (see Tables~\ref{tab: 2d decorated lattices}). For larger bonds, one may again rewrite
\begin{equation}
    h^{c \leftarrow v} h^{v \leftarrow c}
    = \mathfrak{h}^{c \leftarrow \mathfrak{c}} \mathfrak{h}^{\mathfrak{c} \leftarrow c},
\end{equation}
since $\mathfrak{h}^{c \leftarrow \mathfrak{c}}$ can be naturally built such that for each cluster $\Xi^{(c)} = \Omega^{(bd)}$ as we did for the case of the bidiamond square decorated lattice, see Fig. \ref{fig:bidiamond_dual_construction}b). As the meta-cluster description involves fewer cluster types than the original cluster lattice, $\mathfrak{n}_c < n_c$ and this provides a general explanation for why the number of bottom flat bands is $n_s - \mathfrak{n}_c$ and therefore larger than what would be expected from the cluster counting of the original description. Determining the additional in-spectrum flat bands is then system dependent, but it typically relies on rewriting the nontrivial spectral part of $h^{c \leftarrow v} h^{v \leftarrow c}$ as a polynomial in a matrix $X$ that inherits a nontrivial kernel from an underlying cluster-structure argument. 

\subsubsection{Inequivalent identity shifts}

A subtle point concerns the identity contributions generated by the product $h^{c \leftarrow v} h^{v \leftarrow c}$, as they can slightly modify the polynomial construction needed to expose the kernel responsible for flat bands. We illustrate this with the square lattice decorated with square bonds and octagonal vertices (s--s--o). For this system, the nontrivial spectrum is determined by studying
\begin{equation}
   \begin{split}
        h^{c \leftarrow v} h^{v \leftarrow c}
       &= 4 I_s + 8 I_o + 2 H_{\text{d-s}} \\
       &= 4 I + 2\left(2 I_o + H_{\text{d-s}}\right),
       \label{Eq:sso}
   \end{split}
\end{equation}
where d--s denotes the decorated-square connectivity matrix, $I_s$ the identity acting on square-cluster centers $s$ subspace, and $I_o$ the identity acting octagon centers $o$ subspace. Since the identity terms act with different coefficients on the inequivalent cluster-center subspaces, the relevant matrix to analyze is
$X \equiv 2 I_o + H_{\text{d-s}}$. In the basis separating square-cluster centers $s$ (dimension $2$) and octagon centers $o$ (dimension $1$), it reads
\begin{equation}
    X =
    \begin{pmatrix}
        0 & A_{\mathrm{so}} \\
        A_{\mathrm{os}} & \zeta
    \end{pmatrix},
    \qquad
    \zeta = 2.
\end{equation}
This structure generally implies that
\begin{equation}
    X^2 - \zeta X
    =
    \begin{pmatrix}
        A_{\mathrm{ss}} & 0 \\
        0 & A_{\mathrm{oo}}
    \end{pmatrix}
\end{equation}
is block diagonal. Its spectrum coincides with that of the checkerboard lattice, with the dispersive band being twofold degenerate, as stated above. In particular, this matrix has a one-dimensional kernel, corresponding to a zero-eigenvalue flat band. Therefore, $X$ must possess a flat band at eigenvalue $0$ or $\zeta$.
Here the nontrivial kernel resides in the $s$ sector (there are two square clusters per unit cell for a single octagonal cluster), so there exists $\mathbf{u}_s$ such that $A_{\mathrm{ss}} \mathbf{u}_s = 0$, which in turn implies $A_{\mathrm{os}} \mathbf{u}_s = 0$. Hence,
\begin{equation}
    X
    \begin{pmatrix}
        \mathbf{u}_s \\ 0
    \end{pmatrix}
    =
    \begin{pmatrix}
        0 \\
        A_{\mathrm{os}} \mathbf{u}_s
    \end{pmatrix}
    =
    \begin{pmatrix}
        0 \\ 0
    \end{pmatrix},
\end{equation}
showing that the corresponding flat band lies at eigenvalue $0$ in the spectrum of $X$. This is in fact generic: whenever the identity shift $\zeta$ acts in the sector of smallest dimension, the flat-band eigenvalue is pinned to $0$.
The remaining spectrum of $X$ consists of two dispersive bands with dispersions
\begin{equation}
    \Lambda_\pm = 1 \pm \sqrt{1 + \lambda_{\text{ckb}}^2}
\end{equation}
as $\zeta = 2$.
Accordingly, Expressing (\ref{Eq:sso}) as $h^{c \leftarrow v} h^{v \leftarrow c} = 4I + 2X$ , the nontrivial spectrum of the s--s--o lattice is composed of a flat band at energy $4$, together with two dispersive bands
\begin{equation}
    \lambda_\pm
    = 4 + 2 \Lambda_\pm
    = 4 + 2\left(1 \pm \sqrt{1 + \lambda_{\text{ckb}}^2}\right),
\end{equation}
in perfect agreement with the band structure shown in Table~\ref{tab: 2d cluster-links + cluster-bonds lattices S}.

This approach naturally extends to systems for which the product $h^{c \leftarrow v} h^{v \leftarrow c}$ generates distinct identity terms acting on inequivalent cluster-center subspaces. In particular, it provides a straightforward route to determining the band structures of the monocluster versions of the bond- and vertex-decorated systems presented in Tables~\ref{tab: 2d cluster-links + cluster-bonds lattices S} and \ref{tab: 2d cluster-links + cluster-bonds lattices}. For more intricate constructions—such as systems featuring simultaneous vertex decoration and composite bond decoration—analogous procedures can still be devised, but they become increasingly involved as the decoration complexity grows, as illustrated by the square–bidiamond–square lattice discussed in Appendix~\ref{Appendix: square diamond square fb}.

Having elucidated the origin of flat bands via a formal algebraic derivation, we now turn to two complementary approaches that offer a more intuitive understanding of the physics underlying in-spectrum flat bands.

\section{Interpretation of the flat bands in terms of compact localized states}
\label{sec: CLS}

\begin{figure}[h]
    \centering
    \includegraphics[width=0.95\linewidth]{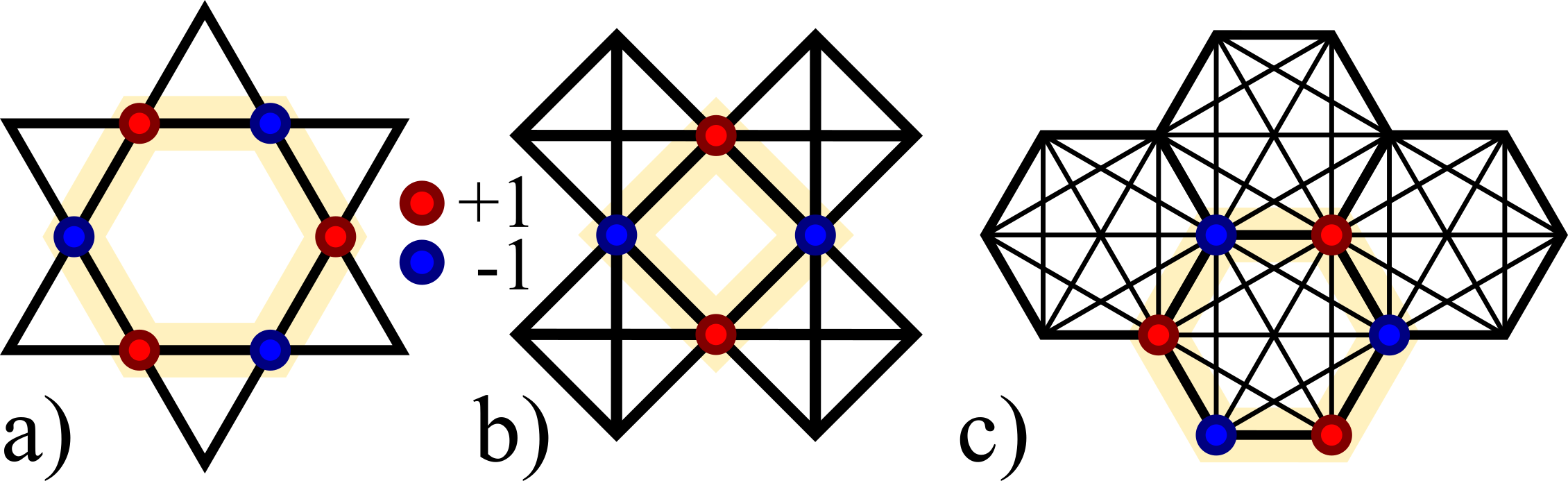}
    \caption{Compact localized states (CLS) for (a) kagome, (b) checkerboard, and (c) hexagonal cluster lattices. For each state, acting with the hopping Hamiltonian~\eqref{eq: H hopping} reproduces the same state with eigenvalue zero (once the onsite term is chosen with the coefficients corresponding to the cluster model).}
    \label{fig:CLS_simple_clusters}
\end{figure}

\begin{figure}[h]
    \centering
    \includegraphics[width=0.9\linewidth]{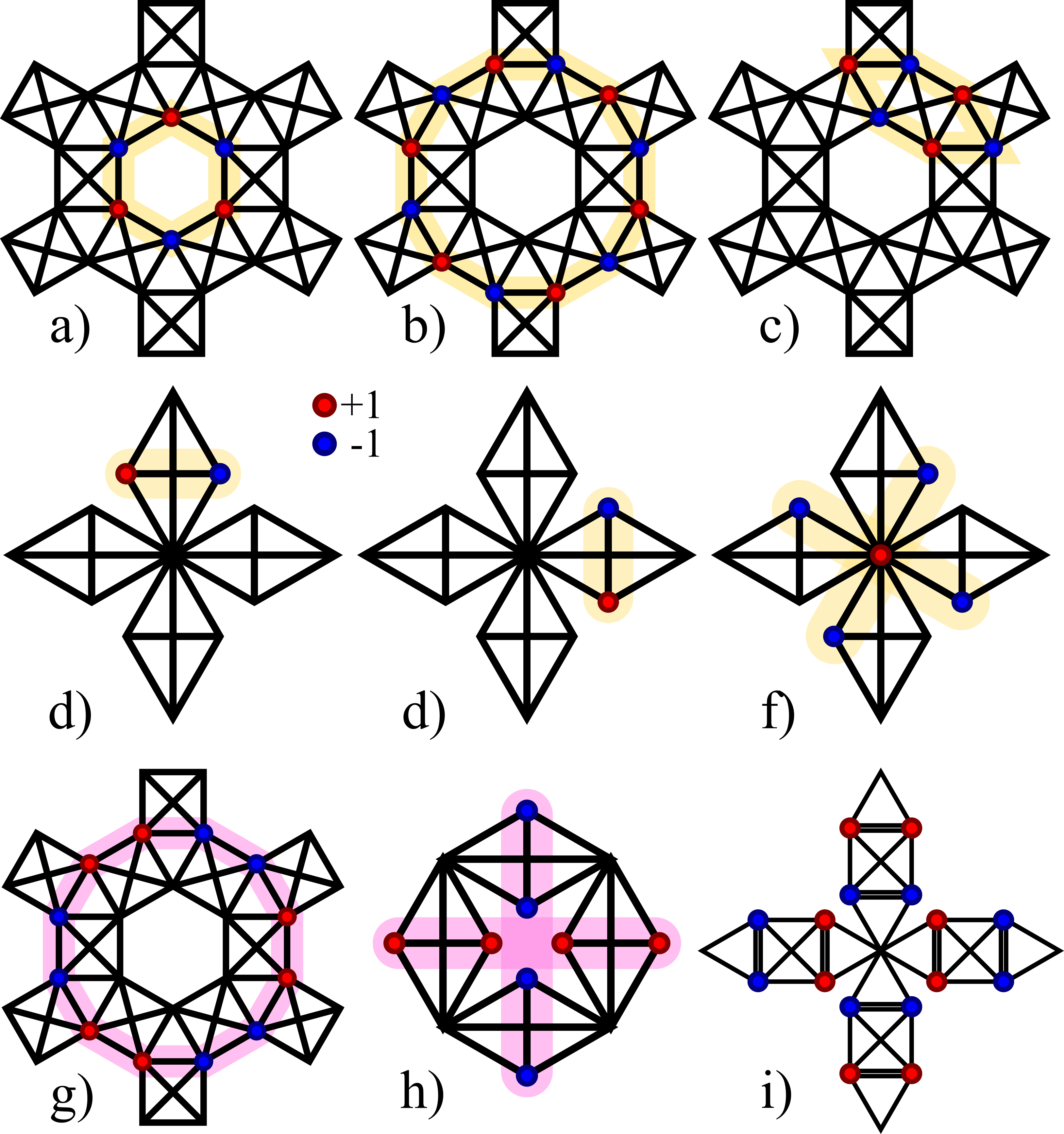}
    \caption{
    Compact localized states for the (a--c,g) ruby lattice and the (d--f,h) monodiamond-decorated square lattice. In both systems, the presence of three degenerate bottom flat bands implies three distinct families of zero-energy CLS (yellow), built as closed loops with alternating weights and yielding perfect cancellations under \eqref{eq: H hopping}. Finite-energy CLS (magenta) correspond instead to pairs of equal-sign amplitudes, alternating from one cluster to the next so that destructive interference still suppresses leakage outside the support. For the ruby lattice, the onsite term multiplies each nonzero amplitude by $z=2$, giving a flat band at energy $2$. For the monodiamond-decorated square lattice, the corresponding rescaling is $z+t=1+1=2$, again producing a flat band at energy $2$. (i) Example of a CLS associated with the flat band at energy $2$ in cracker-decorated systems.}
    \label{fig:CLS_ruby_and_diamond}
\end{figure}

The previous interpretation of in-spectrum flat bands can be complemented by an alternative picture, which also relies on the emergence of meta-cluster structures built from the original clusters. This interpretation is based on the general fact that flat bands are often associated with the existence of compact localized states (CLS) of an appropriate hopping Hamiltonian \cite{Kim_Rhim_2023_flat_bands, Yan_2024_long, Rhim_2019_flat_bands, Sutherland_1986_FB, Dias_2015_FB, Morales-Inostroza_2016_FB, Miyahara_2005_FB, Graf_Piechon_2021_flat_bands}. A CLS is a finite-support eigenstate, constructed as a superposition of flat-band Bloch eigenstates\cite{Kim_Rhim_2023_flat_bands}.
Such states can be obtained by considering the tight-binding model that maps exactly onto the spin cluster Hamiltonian. For simplicity, we again restrict to the case $\gamma_X^i = 1$ for all clusters $X$ and all sites $i$. The corresponding hopping Hamiltonian can be written as
\begin{equation}
    \mathcal{H}
    = \sum_{i \neq j} t_{ij}\; c^\dagger_i c_j
    + \sum_i z_i\; c^\dagger_i c_i ,
    \label{eq: H hopping}
\end{equation}
where the hopping amplitude $t_{ij}$ equals the number of clusters containing the link $i\!-\!j$, while the onsite term weight $z_i$ equals the number of clusters containing site $i$. For example, for the kagome cluster model one has $t=1$ and $z=2$, whereas for the hexagonal cluster model one has $t_i=1$ for bonds inside a cluster, $t_d=1$ for bonds delimiting clusters, and $z=3$ (see Table~\ref{tab: 2d cluster systems}).
In this framework, a CLS is an eigenstate of \eqref{eq: H hopping} with nonzero amplitudes on only a finite set of neighboring sites. Constructing such a state amounts to choosing amplitudes such that, upon acting with \eqref{eq: H hopping}, all single-hop processes interfere destructively on sites outside the initial support. In addition, the amplitudes on the support must be reproduced up to a global prefactor, which fixes the flat-band energy.
A generic way to generate zero-energy CLS on many cluster lattices is to consider closed loops with alternating coefficients $\pm 1$, arranged so that each crossed cluster contains pairs of opposite amplitudes. This construction is illustrated for the kagome, checkerboard, and hexagonal lattices in Fig.~\ref{fig:CLS_simple_clusters}, and for ruby lattice and bi-diamond square lattice on Fig. \ref{fig:CLS_ruby_and_diamond}(a-f). In these examples, having two opposite weights per cluster enforces local cancellations, yielding CLS associated with a zero-energy flat band.

For systems exhibiting in-spectrum flat bands, the relevant CLS can be constructed by mimicking the zero-energy CLS of the meta-cluster lattice obtained by linking the centers of the decorated bonds that are in contact. For instance, in the ruby lattice, one can use the kagome CLS pattern of Fig.~\ref{fig:CLS_simple_clusters}(a) to build the finite-energy CLS shown in Fig.~\ref{fig:CLS_ruby_and_diamond}(g). Acting with Hamiltonian \eqref{eq: H hopping} on this state cancels all hopping contributions by destructive interference, while the onsite term multiplies each nonzero amplitude by $z$. This produces an eigenvalue $z=2$, consistent with the previous connectivity-matrix analysis and with the band structure in Table~\ref{tab: 2d cluster systems} (although the two derivations are formally different).
For bond-decorated systems, compact localized states associated with excited flat bands can be constructed in a largely systematic manner. One places groups of sites with equal amplitudes at the midpoints of the decorated bonds, and chooses the relative signs of these groups according to the CLS pattern of the corresponding meta-cluster lattice. This is illustrated for the monodiamond-decorated square lattice in Fig.~\ref{fig:CLS_ruby_and_diamond}(h). In such a construction, acting with \eqref{eq: H hopping} rescales each nonzero amplitude by a factor $z + (\alpha-1)t$, where $\alpha$ denotes the number of nonzero amplitudes per cluster involved in the local pattern. For monodiamond-decorated systems one has $z=1$ and $t=1$, which yields a flat band at energy $2$; for monocracker-decorated systems the same reasoning gives a flat band at energy $4$, in agreement with the connectivity-matrix derivation.
The construction is analogous for bidiamond-decorated lattices, except that the effective parameters are $t=2$ and $z=2$, leading to an excited flat band at energy $4$. Note that the zero-energy CLS shown in Fig.~\ref{fig:CLS_ruby_and_diamond}(f), and related CLS in other decorated systems, are no longer eigenstates of \eqref{eq: H hopping} in the bidiamond case. This explains why, for bidiamond-decorated systems, the number of bottom flat bands is tied to the number of inequivalent cluster bonds.

For cracker-decorated systems, the same construction instead yields an eigenvalue
$z + t_1 + 2 t_2 = 2 + 2 + 2 = 6$. Since these systems also exhibit a flat band at energy $2$, there must exist a second family of compact localized states. One such state is obtained by placing two oppositely signed pairs on each bond, as shown in Fig.~\ref{fig:CLS_ruby_and_diamond}(i).  

\begin{figure}[h]
    \centering
    \includegraphics[width=0.9\linewidth]{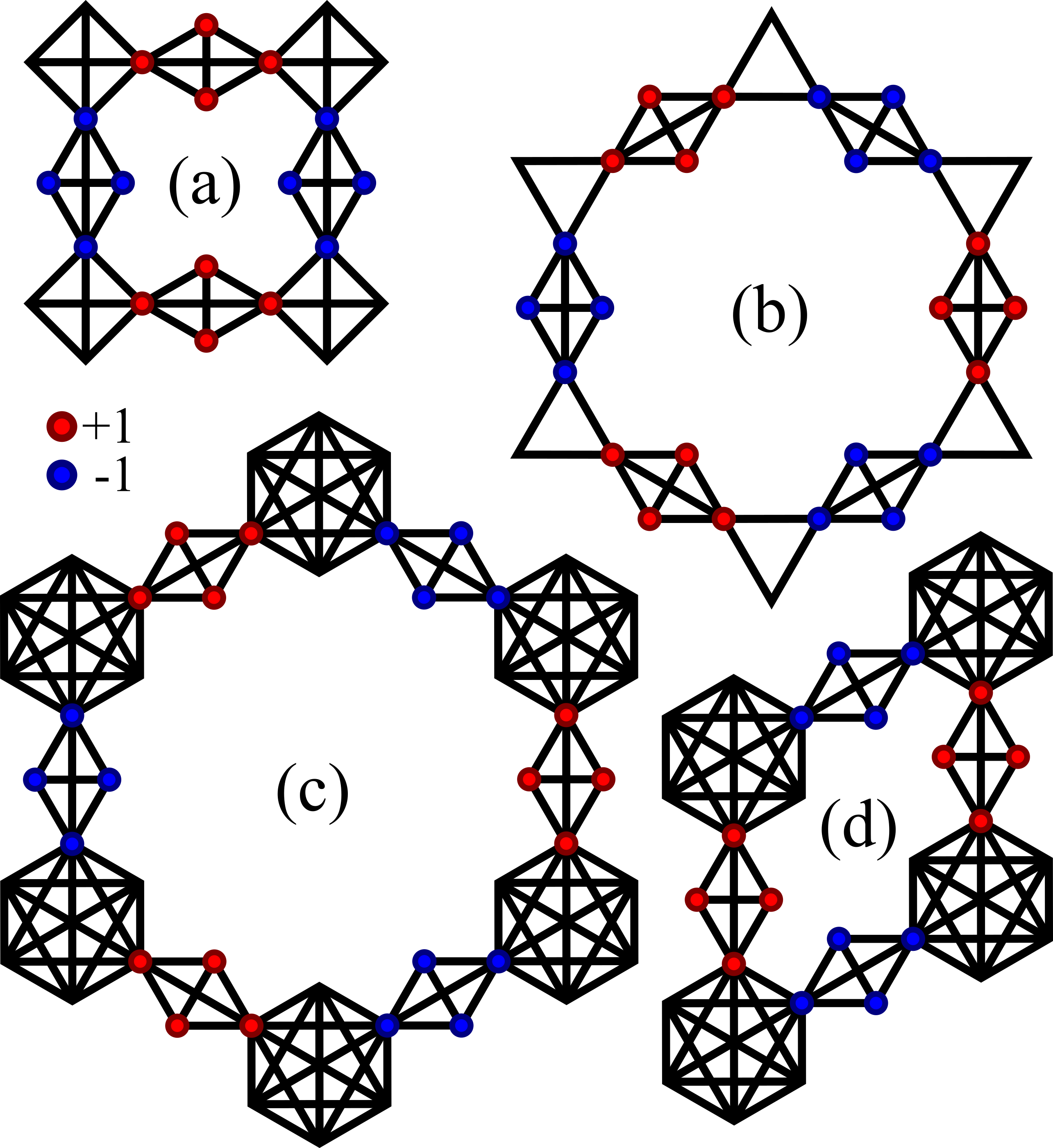}
    \caption{
    Finite energy compact localized states for vertex and monobond decorated systems. (a) Square - monodiamond -square, (b) Hexagonal - monodiamond - triangle, (c) Hexagonal - monodiamond - hexagon and (d) Triangular - monodiamond - Hexagon.}
    \label{fig:CLS_decorated}
\end{figure}

For vertex-decorated systems with monoblock-decorated bonds, an analogous construction can also be readily implemented by placing groups of equal amplitudes on each bond and alternating their signs so as to form a closed ring containing an even number of bonds (see Fig.~\ref{fig:CLS_decorated}). This produces flat bands whose energies are equal to the number of sites per bond, as confirmed by the band structures reported in Table~\ref{tab: 2d cluster-links + cluster-bonds lattices}.  

Overall, regardless of the specific decoration, in-spectrum flat bands appears to systematically reflect the presence of an emergent meta-cluster lattice encapsulated in the structure of the cluster system itself. The corresponding flat-band energies are then set by the internal structure of the vertices and bond clusters used in the decoration. Regardless of the framework adopted, the construction of the appropriate matrix polynomial or of the corresponding finite-energy compact localized states responsible for the in-spectrum flat bands rapidly becomes increasingly intricate as the structure of the decorated lattices grows in complexity. In particular, this task proves especially challenging for vertex-decorated systems with composite bonds.

\section{Additional in-spectrum flat bands: notion of meta-spin}
\label{sec: meta_spin}

A natural interpretation of the above results on in-spectrum flat bands can be obtained by introducing explicit cluster--cluster interactions in the Hamiltonian,
\begin{equation}
    \mathcal{H}
    = \alpha \sum_n \lvert \bm{\mathcal{C}}_n \rvert^2
    + 2\eta \sum_{\langle m,n \rangle}
    \bm{\mathcal{C}}_m \cdot \bm{\mathcal{C}}_n ,
    \label{Eq: general H(2)}
\end{equation}
where the additional term couples neighboring clusters. Using the cluster--cluster connectivity matrix $g^{c \leftarrow c}$, this Hamiltonian can be rewritten as
\begin{equation}
    \mathcal{H}
    = \alpha \sum_n \lvert \bm{\mathcal{C}}_n \rvert^2
    + \eta \sum_{m,n}
    g^{c \leftarrow c}_{mn}\,
    \bm{\mathcal{C}}_m \cdot \bm{\mathcal{C}}_n .
\end{equation}
Systems exhibiting a gapped in-spectrum flat band are precisely those for which the lattice formed by the connected clusters itself constitutes a meta-cluster system. In such cases, the connectivity matrix $g^{c \leftarrow c}$ can be expressed using Eq.~(\ref{Eq:gcc meta h}), which allows the Hamiltonian to be recast as
\begin{equation}
    \mathcal{H}
    =
    \left( \alpha - \eta \Omega^{(bd)} \right)
    \sum_n \lvert \bm{\mathcal{C}}_n \rvert^2
    + \eta \sum_{\mathfrak{n}}
    \lvert \bm{\mathfrak{C}}_{\mathfrak{n}} \rvert^2 ,
\end{equation}
where
\begin{equation}
    \bm{\mathfrak{C}}_{\mathfrak{n}}
    = \sum_{n \in \mathfrak{n}} \bm{\mathcal{C}}_n
\end{equation}
defines a \emph{meta-constrainer} associated with the meta-clusters.
When $\alpha - \eta \Omega^{(bd)} > 0$, the ground state coincides with that of the original cluster Hamiltonian, characterized by vanishing constrainers everywhere. By contrast, when $\alpha - \eta \Omega^{(bd)} < 0$, the energy is minimized by maximizing the norms of the individual constrainers while simultaneously enforcing $\bm{\mathfrak{C}}_{\mathfrak{n}} = 0$ on every meta-cluster. In this regime, the constrainers $\bm{\mathcal{C}}_n$ effectively behave as classical \emph{meta-spins} interacting on a meta-cluster lattice. It is therefore natural to recover Coulomb-phase physics, accompanied by pinch-point singularities in the structure factor, upon increasing the inter-cluster coupling $\eta$ beyond a critical value. The pinch-points are then expected to be the ones of the meta-cluster system, which are encoded in the gapped flat bands. 

\begin{figure*}
    \centering
    \includegraphics[width=0.95\linewidth]{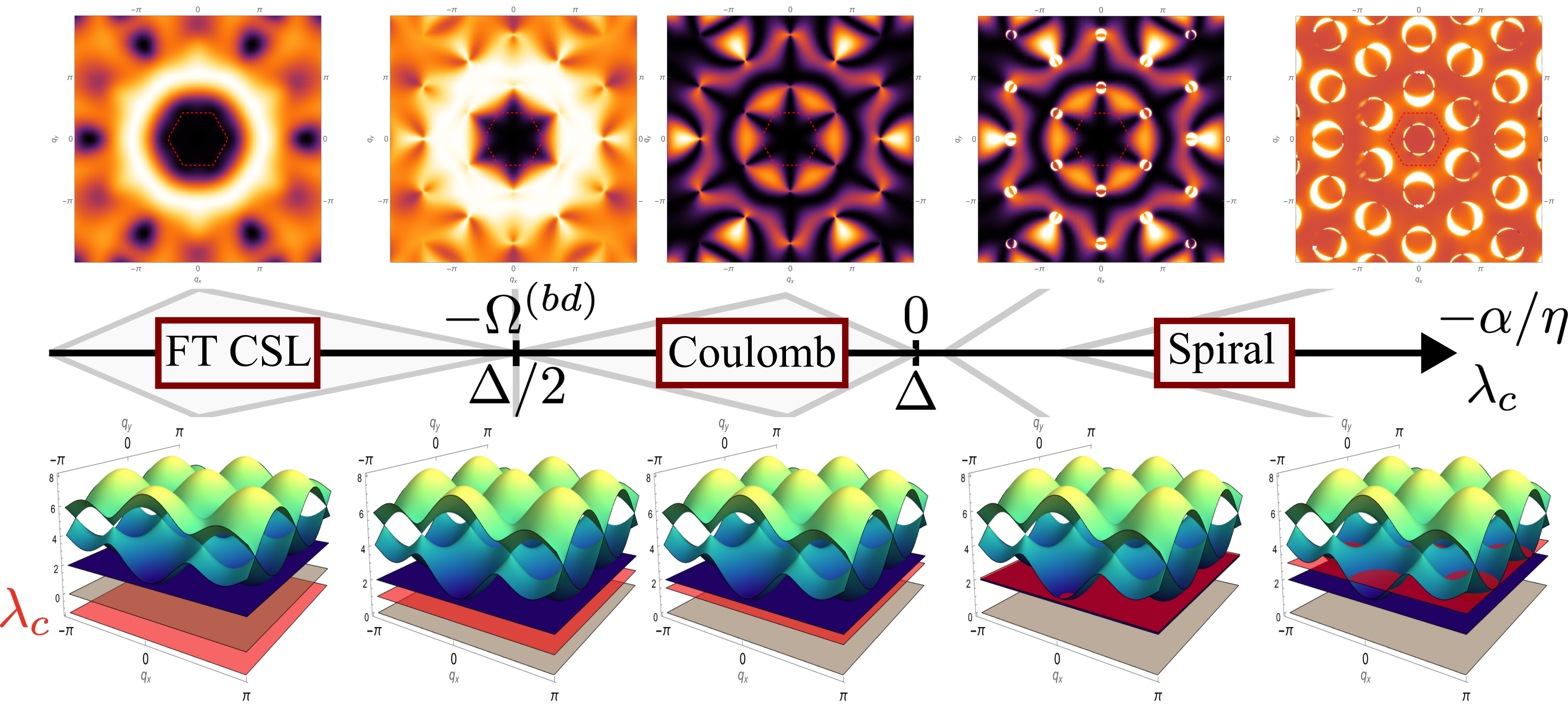}
    \caption{Phase diagram for the interacting Ruby lattice given as a function of the ratio $-\alpha/\eta$ and equivalently as a function of the minima $\lambda_c$ of the the polynomial (\ref{eq: Polynomial}). The top raw depicts characteristic structure factors obtained using Self Consistent Gaussian approximation \cite{Garanin_SCGA} (with $T = 10^{-3}$ in units of $\gamma = 1$) using Heisenberg spins, enforcing self consistently the spin norm constraint averaged over all spins, see Appendix \ref{Appendix : SCGA} for details. The structure factors show pinch points in the intermediary range $-\Omega^{(bd)} \leq - \alpha / \eta \leq \Xi^{(c)}-2 \Omega^{(bd)} = 0$ indicating a Coulomb phase. The structure factors next present half moons for $-\alpha/\eta \gtrsim 0$ revealing the transition into a spiral CSL phase \cite{Davier_2025_interacting_clusters}. The bottom row illustrates the position of the plane of energy $\lambda_c$ with respect with the non interacting ruby lattice band structure, their intersection determining the ground state manifold.}
    \label{fig:Interacting_ruby_phase_diag}
\end{figure*}

This picture can be further precised using the connectivity-matrix formalism. The adjacency matrix of the interacting cluster system can be written, using Eq.~(\ref{Eq: hcvhvc}), as
\begin{equation}
    \begin{split}
        H^{(2)}
        &=
        \alpha\, h^{c \leftarrow v} h^{v \leftarrow c}
        + \eta\, h^{c \leftarrow v} g^{c \leftarrow c} h^{v \leftarrow c} \\
        &=
        \left( \alpha - \eta \Xi^{(c)} \right)
        h^{c \leftarrow v} h^{v \leftarrow c}
        + \eta \left( h^{c \leftarrow v} h^{v \leftarrow c} \right)^2 .
    \end{split}
\end{equation}
As a consequence, the interacting band structure can be obtained directly from the noninteracting one $\lambda_\mu(\mathbf{q})$ as \cite{Davier_2025_interacting_clusters}
\begin{equation}
    \Lambda_\mu(\mathbf{q})
    =
    \left( \alpha - \eta \Xi^{(c)} \right)
    \lambda_\mu(\mathbf{q})
    + \eta \left( \lambda_\mu(\mathbf{q}) \right)^2 .
\end{equation}
Recalling that the gap of the cluster band structure is given by
$\Delta = \Xi^{(c)} - \Omega^{(bd)}$, the critical value of the ratio $-\alpha/\eta$ at which the gapped flat band becomes the ground state is \cite{Davier_2025_interacting_clusters}
\begin{equation}
    -\frac{\alpha_c}{\eta_c}
    = \Delta - \Xi^{(c)}
    = -\Omega^{(bd)},
\end{equation}
in agreement with the above qualitative energetic considerations. Introducing the polynomial relating the interacting and noninteracting spectra,
\begin{equation}
    P(\lambda) =
    \left( \alpha - \eta \Xi^{(c)} \right) \lambda
    + \eta \lambda^2 ,
    \label{eq: Polynomial}
\end{equation}
one sees that the abscissa of its minimum,
\begin{equation}
    \lambda_c =
    \frac{1}{2} \left( \Xi^{(c)} - \frac{\alpha}{\eta} \right),
\end{equation}
controls the nature of the ground state. If $\lambda_c$ remains smaller than the gap
$\Delta = \Xi^{(c)} - \Omega^{(bd)}$, that is if
\begin{equation}
    -\frac{\alpha}{\eta}
    <
    \Xi^{(c)} - 2 \Omega^{(bd)},
\end{equation}
the system remains in a Coulomb phase analogous to that of the meta-cluster system. Beyond this threshold, the ground state corresponds to the intersection between the plane at energy $\lambda_c$ and the dispersive part of the band structure of the noninteracting cluster model \cite{Davier_2025_interacting_clusters}, see Fig. \ref{fig:Interacting_ruby_phase_diag} for an illustration. In this regime, the ground-state manifold has codimension one: its dimension is $d-1$, corresponding in $d=2$ to a set of lines. The system is therefore expected to realize a spiral spin liquid \cite{Yan_2022_Spiral}.

Altogether, mono-cluster bond-decorated systems thus form a family of classical spin liquids undergoing a sequence of transitions from a fragile topological CSL (FT-CSL) to a Coulomb CSL, followed by a transition to a spiral CSL, as illustrated for the ruby lattice in Fig.~\ref{fig:Interacting_ruby_phase_diag}. At the special point $-\alpha/\eta = -\Omega^{(bd)}$, corresponding to the transition from the FT-CSL phase to the Coulomb CSL, the number of bottom flat bands equals the sum of the bottom flat bands of the non interacting cluster system and the flat bands of the meta-cluster system. As a result, the degeneracy of the ground-state manifold is maximal. The corresponding low temperature structure factor is therefore expected to be a superposition of the FT-CSL and Coulomb CSL ones, as depicted for the ruby lattice in Fig.~\ref{fig:Interacting_ruby_phase_diag}. Note that in the spiral CSL phase the ground-state manifold is no longer spanned by a set of flat bands, but instead collapses onto a set of lines. Consequently, the structure factor cannot be obtained from the flat-band projector used in the preceding sections; it can nevertheless be computed at low temperature within the self-consistent Gaussian approximation described in Appendix~\ref{Appendix : SCGA}.

The dispersion relations further impose that the nature and location of pinch points in the Coulomb phase of the interacting cluster model coincide with those of the associated meta cluster model. However, the resulting structure factors are expected to differ, since the eigenvectors associated with the flat bands are not the same in the interacting cluster model and in the meta-cluster system. In particular, they live in spaces of different dimensionality, reflecting the different numbers of sublattices involved.

For other types of decorated systems featuring flat bands that are not gapless but instead embedded within the dispersive part of the band structure, two main scenarios may arise. If the initial cluster spin liquid is gapless but exhibits nonsingular band touchings, as in composite cluster-decorated systems, increasing the ratio $-\alpha/\eta$ may drive a sequence of transitions
\[
\text{FT} \;\to\; \text{spiral} \;\to\; \text{Coulomb} \;\to\; \text{spiral},
\]
the CSL regime being determined by the intersection between the non interacting band structure and the plane at energy $\lambda_c$. By contrast, if the band touching in the non interacting cluster system are singular, the succession of phases is instead expected to alternate between Coulomb and spiral phases.

Altogether, this picture confirms that flat bands are generically associated with spin-liquid physics in classical spin systems, even though in-spectrum flat bands signatures are expected to remain difficult to access in realistic systems, where only the vicinity of the ground-state manifold can typically be explored.

\section{Discussion}

We have presented a \emph{unified} and \emph{tunable} framework for designing interacting spin systems within the class of cluster models. Because the assembly rules are simple, the associated design space is remarkably large, making this approach particularly well suited for the systematic generation and exploration of classical spin liquids.

A particularly effective strategy consists in decorating a standard parent lattice—on its bonds and/or vertices—with symmetry-compatible clusters. By varying the internal structure of the clusters and the decoration pattern, a wide variety of cluster lattices can be generated from a single parent system. The counting criterion introduced in Eq.~(\ref{Eq: Criterion Classical Spin liquid}) then provides a rapid and practical diagnostic for identifying constructions likely to host classical spin liquids, turning the framework into an efficient tool for model design. The examples presented here illustrate the simplest realizations of this idea, and Tables~\ref{tab: 2d cluster systems} to \ref{tab: 2d decorated lattices with inter-clusters} serve as a basic catalog of relatively simple cluster systems. Many further families—including mixed bond–vertex decorations and straightforward three-dimensional generalizations—can be constructed following the same principles.

Beyond this initial classification, the additional analysis presented here highlights several nontrivial features specific to decorated cluster systems. In particular, bond- and vertex-decorated lattices generically host additional flat bands—either at the bottom of the spectrum or embedded within the dispersive bands—whose origin can be traced to emergent meta-cluster structures. Adding further cluster-cluster interactions, these flat bands can strongly influence the nature of the associated spin-liquid phases, leading to rich phenomenology that includes Coulomb phases, FT-CSLs and spiral CSLs regimes.

Finally, combining decoration with cluster extension provides a versatile route to engineer targeted reciprocal-space features—most notably higher-rank pinch points, pinch lines, and extended band-touching manifolds—within a single, controllable framework.

Altogether, this study reveals a number of particularly intriguing candidate systems that clearly merit further investigation beyond the large $\mathcal{N}$ approximation. In particular, numerical approaches such as classical Monte Carlo simulations would be valuable to assess the stability of the predicted spin-liquid phases, to characterize their finite-temperature behavior, and to determine which of the features identified here survive in more realistic settings.

A notable strength of the present framework lies in the breadth of decorated models it generates, which makes experimental realizations conceivable. Classical spin liquids are stabilized by an extensive entropy and can therefore persist over a finite temperature range, even when microscopic parameters are not finely tuned. This robustness enhances the chances that some of the proposed models could be realized in real materials. Importantly, the constructions considered here rely solely on standard isotropic Heisenberg exchange interactions between spins, without requiring multi-spin couplings, which are typically difficult to engineer experimentally. Moreover, in materials with significant spin–orbit coupling, the cluster-based approach can be naturally generalized to include anisotropic interactions. Note that the construction principles presented here apply equally to Heisenberg, XY, and Ising spin degrees of freedom. Beyond solid-state realizations, synthetic quantum platforms offer particularly promising routes toward experimental implementation. In particular, recent progress in Rydberg-atom arrays \cite{shauss2015, scholl2021, ebadi2021, chen2023} has demonstrated a high degree of control over lattice geometry, interaction range, and dimensionality, making them well suited for engineering two-dimensional cluster systems. Such platforms may therefore provide a flexible setting in which several of the decorated cluster models introduced in this work could be realized and explored experimentally.

\textit{Acknowledgements:} 
The author acknowledges financial support from Grant No. ANR-23-CE30-0038-01 and warmly thanks Ludovic Jaubert for insightful discussions and valuable advice.

\appendix

\section{Large-$\mathcal{N}$ approximation}
\label{Appendix:Large N}

In the large-$\mathcal{N}$ approach~\cite{Henley2005, Garanin_LargeN, Canals_2001_LargeN}, one promotes the classical spins to $\mathcal{N}$-component vectors and enforces the fixed-length constraint only on average (via a Lagrange multiplier), which becomes exact in the $\mathcal{N}\to\infty$ limit. At zero temperature, the correlations are then governed by the lowest-energy modes of the quadratic Hamiltonian, i.e., by the ground-state subspace of the interaction matrix in reciprocal space.

For cluster systems, the Fourier-space interaction matrix can be constructed conveniently from the factorization
\begin{equation}
    H(\mathbf{q}) = h^{v \leftarrow c}(\mathbf{q})\, h^{c \leftarrow v}(\mathbf{q}),
    \label{Eq:H Appendix}
\end{equation}
where $h^{v \leftarrow c}(\mathbf{q})$ is a rectangular vertex--cluster incidence matrix,  with $h^{c \leftarrow v}(\mathbf{q})=[h^{v \leftarrow c}(\mathbf{q})]^\dagger$. For a cluster of type $X$, the matrix elements of $h^{v \leftarrow c}(\mathbf{q})$ are obtained by summing over the sites of that cluster that belong to a given sublattice $\mu$,
\begin{equation}
    \big[h^{v \leftarrow c}(\mathbf{q})\big]_{\mu X}
    = \sum_{i\in X \cap \mu} \gamma_i^X \, e^{i \mathbf{q} \cdot \mathbf{r}_i},
\end{equation}
where $\mathbf{r}_i$ denotes the position of site $i$ (within the chosen reference cell and with respect to an arbitrary origin, consistently used throughout).

Once $H(\mathbf{q})$ is built, the zero-temperature structure factor can be expressed~\cite{Henley2005} in terms of the projector $P_{\mathrm{G.S.}}(\mathbf{q})$ onto the ground-state manifold,
\begin{equation}
    \mathcal{S}(\mathbf{q}) = \frac{1}{N_{u.c}}\sum_{\mu, \nu} \langle \mathbf{S}_\mu(\mathbf{q}) \cdot \mathbf{S}_\nu(-\mathbf{q}) \rangle 
    \propto \sum_{\mu, \nu} P_\text{G.S.}^{\mu\nu}(\mathbf{q}).
    \label{Eq : Projector Pi at zero temperature}
\end{equation}
Because of the factorized form~\eqref{Eq:H Appendix}, the ground-state subspace corresponds to the kernel of $h^{c \leftarrow v}(\mathbf{q})$, i.e., to the subspace orthogonal to the rows of $h^{c \leftarrow v}(\mathbf{q})$. The corresponding projector can be written as~\cite{Henley2005}
\begin{equation}
    P_\text{G.S.}(\mathbf{q}) = I - h^{v \leftarrow c} \left[h^{c \leftarrow v} h^{v \leftarrow c}\right]^{-1}h^{c \leftarrow v}.
\end{equation}
where the inverse acts on the $n_c\times n_c$ matrix $h^{c \leftarrow v} h^{v \leftarrow c}$.

For numerical purposes, it can alternatively be more efficient to compute an orthonormal basis of ground-state eigenvectors $\bm{\Psi}^{\mathrm{G.S.}}_n(\mathbf{q})$ and build the projector as
\begin{equation}
    P_{\mathrm{G.S.}}(\mathbf{q})
    = \sum_{n\in\mathrm{G.S.}} \bm{\Psi}^{\mathrm{G.S.}}_n(\mathbf{q})
    \big[\bm{\Psi}^{\mathrm{G.S.}}_n(\mathbf{q})\big]^\dagger.
\end{equation}

\section{Self Consistent Gaussian Approximation}
\label{Appendix : SCGA}

Consider any isotropic Heisenberg Hamiltonian that can be expressed as
\begin{equation}
    \begin{split}
        \mathcal{H} &= \frac{1}{2}\sum_{i,j} H_{i,j} \mathbf{S}_i \cdot \mathbf{S}_j \\ 
        &=\frac{1}{2N}\sum_\mathbf{q} \sum_{\mu, \nu} H_{\mu \nu}(\mathbf{q}) \mathbf{S}_\mu(\mathbf{q}) \cdot \mathbf{S}_\nu(-\mathbf{q})
    \end{split}
\end{equation}
with $N$ the number of unit cells. Further consider there are $n_s$ sublattices, and denote the $n_s$ eigenvalues and eigenvectors from the Hamiltonian $H(\mathbf{q})$ as
\begin{equation}
    \varepsilon_\kappa(\mathbf{q}), \qquad \bm{\psi}_\kappa(\mathbf{q}).
\end{equation}
The self consistent Gaussian approximation\cite{Garanin_SCGA} corresponds to take into account the spin length constraint $|\mathbf{S}_i| = 1$ only in average, enforcing only the constraint $\langle |\mathbf{S}_i| \rangle = 1$. Using that the three spin components are equivalent, the constraint that is really enforced in practice is
\begin{equation}
    \langle (S_i ^\alpha)^2 \rangle = \frac{1}{3}.
\end{equation}
This can be done using a single Lagrange multiplier $\lambda$ to enforce the constraint in the Hamiltonian, while enforcing the hard constraint for every spin would amount to introduce $Nn_s$ Lagrange multipliers. Also shifting the energy origin by defining $\varepsilon_0 =$ Min$(\varepsilon_\mu)$ and considering the positive semi definite matrix $K(\mathbf{q}) = H(\mathbf{q}) - \varepsilon_0 I$, the Hamiltonian can finally be expressed as 
\begin{equation}
    \mathcal{H} = \frac{1}{2N}\sum_\mathbf{q} \left( K_{\mu \nu}(\mathbf{q}) + \frac{\lambda}{\beta} \delta_{\mu \nu} \right) \mathbf{S}_\mu(\mathbf{q}) \cdot \mathbf{S}_\nu(\mathbf{q}).
\end{equation}
The eigenvalues of $K$ are equal to the ones of $H$ but shifted of $\varepsilon_0$ and are therefore positive, while its eigenvectors are simply identical. For a semi-positive definite matrix $A$ the general formula
\begin{equation}
    \frac{\int \prod_i  dx_i \; x_\mu x_\nu e^{-\frac{1}{2} \mathbf{x}^t A \mathbf{x}}}{\int \prod_i  dx_i e^{-\frac{1}{2} \mathbf{x}^t A \mathbf{x}}} = \left(A^{-1}\right)_{\mu \nu}
\end{equation}
can be used to write that, defining $M = \beta K + \lambda I$,
\begin{equation*}
    \begin{split}
        \langle S^\alpha_\mu(-\mathbf{q})  S^\alpha_\nu(\mathbf{q}) \rangle &= \frac{\int \prod_{i,}  dS_i^\alpha \; S^\alpha_\mu S_\nu^\alpha e^{-\frac{ 1}{2 N} \sum_{\mathbf{q}'}\sum_{\eta, \xi} M_{\eta \xi} S_\eta^\alpha  S_\xi^\alpha }} 
    {\int \prod_{i}  dS_i^\alpha e^{-\frac{ 1}{2 N} \sum_{\mathbf{q}'}\sum_{\eta, \xi} M_{\eta \xi} S_\eta^\alpha  S_\xi^\alpha }} \\
    &= N \left(M^{-1}\right)_{\mu \nu} = N \left[ \lambda I + \beta K(\mathbf{q})\right]^{-1}_{\mu\nu}.
    \end{split}
\end{equation*}
This allows to self consistently fix the Lagrange multiplier $\lambda$ by enforcing the constraint $\langle (S_i ^\alpha)^2 = \frac{1}{3}$ (using that the three spin components are equivalent) as
\begin{equation}
    \begin{split}
        \langle (S_i ^\alpha)^2 \rangle &= \langle S_i^\alpha S_i^\alpha \rangle 
    = \frac{1}{N n_s} \sum_i\langle S_i^\alpha S_i^\alpha \rangle \\
    &= \frac{1}{N^2 n_s} \sum_\mathbf{q} \sum_{\mu = 1}^{n_s} \langle S^\alpha_\mu(-\mathbf{q})   S^\alpha_\mu(\mathbf{q}) \rangle \\
    &= \frac{1}{N n_s}\sum_\mathbf{q} \sum_{\mu = 1}^{n_s} \left[ \lambda I + \beta K(\mathbf{q})\right]^{-1}_{\mu\mu} 
    = \frac{1}{3},
    \end{split}
    \label{Eq: self eq for lambda}
\end{equation}
where we used that all spins in the system are equivalent by symmetry, and that there is a total of $N n_s$ spins in the system. This also allows to express the static structure factor as 
\begin{equation*}
    \begin{split}
        \mathcal{S}(\mathbf{q}) &= \frac{1}{N}\sum_{\mu, \nu} \langle \mathbf{S}_\mu(-\mathbf{q}) \cdot \mathbf{S}_\nu(\mathbf{q}) \rangle 
        = \frac{3}{N}\sum_{\mu, \nu} \langle S^\alpha_\mu(-\mathbf{q}) \cdot S^\alpha_\nu(\mathbf{q}) \rangle \\
        =&\; 3\sum_{\mu, \nu} \left[ \lambda I + \beta K(\mathbf{q})\right]^{-1}_{\mu\nu} 
        = 3\sum_{\kappa = 1}^{n_s}\sum_{\mu, \nu} \frac{\left[\psi_\kappa ^*(\mathbf{q})\right]_\mu \left[\psi_\kappa(\mathbf{q})\right]_\nu }{\lambda + \beta \varepsilon_\kappa(\mathbf{q})}
    \end{split}
\end{equation*}
because the three spin components are equivalent, and then the spin components can be treated as three independent variables. 

In the limit of zero temperature $\beta \to \infty$, only flat bands contributions are selected. In this case the static structure factor becomes simply proportional to the sum of the components of the projector into flat band manifold 
\begin{equation}
    \begin{split}
        \Pi_{\mu \nu} &= \sum_{\kappa = 1}^{n_\text{f.b}} \left[\psi_\kappa ^*(\mathbf{q})\right]_\mu \left[\psi_\kappa(\mathbf{q})\right]_\nu \\ 
        &= I_{\mu \nu} - \sum_{\kappa = n_\text{f.b} + 1}^{n_s} \left[\psi_\kappa ^*(\mathbf{q})\right]_\mu \left[\psi_\kappa(\mathbf{q})\right]_\nu.
    \end{split}
\end{equation}
As the manifold spanned by the dispersive bands eigenvectors $\bm{\psi}_\kappa, \; \kappa > n_\text{f.b}$ is identical to the one spanned by the constraint vectors, this projector can be re-expressed as in Eq.~(\ref{Eq : Projector Pi at zero temperature}).

\begin{table*}[ht]
    \centering
    \renewcommand{\arraystretch}{1.5} 
    \resizebox{\textwidth}{!}{
    \begin{tabular}{c c c c c c c}
        \hline    
        \textbf{Lattice} & \makecell{Square \\ diamond \\ with squares}  & \makecell{Hexagonal \\ diamond\\ with hexagons} & \makecell{Square \\ cracker \\ with octagons} & 
        \makecell{Square \\ monodiamond \\ with squares}  & \makecell{Hexagonal \\ monodiamond\\ with hexagons} & \makecell{Square \\ monocracker \\ with octagons} \\
        \hline
        \noalign{\vskip 1mm}
        \makecell{ \vspace{-2.5cm} \\ \textbf{Lattice}  \\ \textbf{Scheme} } & 
        \includegraphics[width=2.5cm]{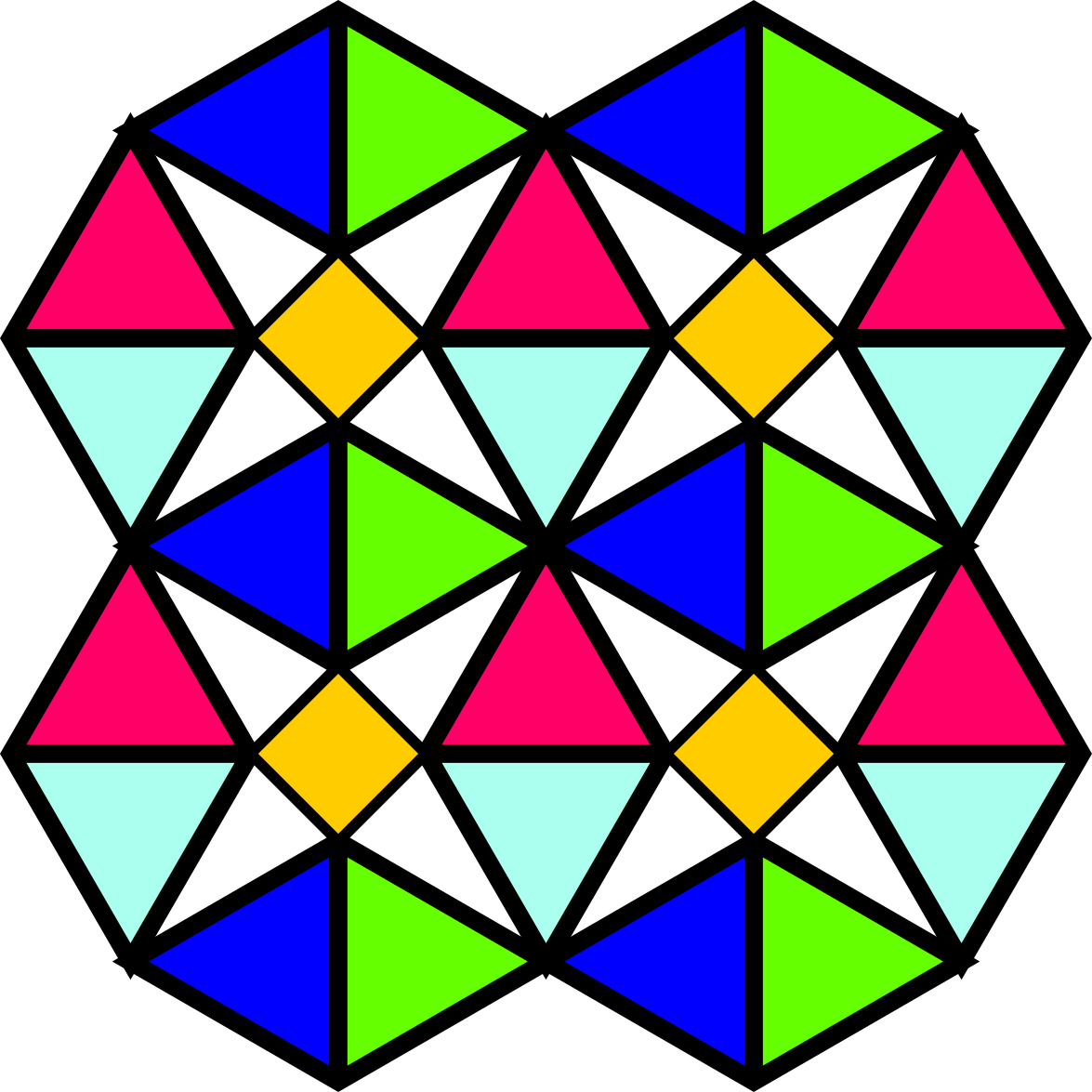} &
        \includegraphics[width=2.5cm]{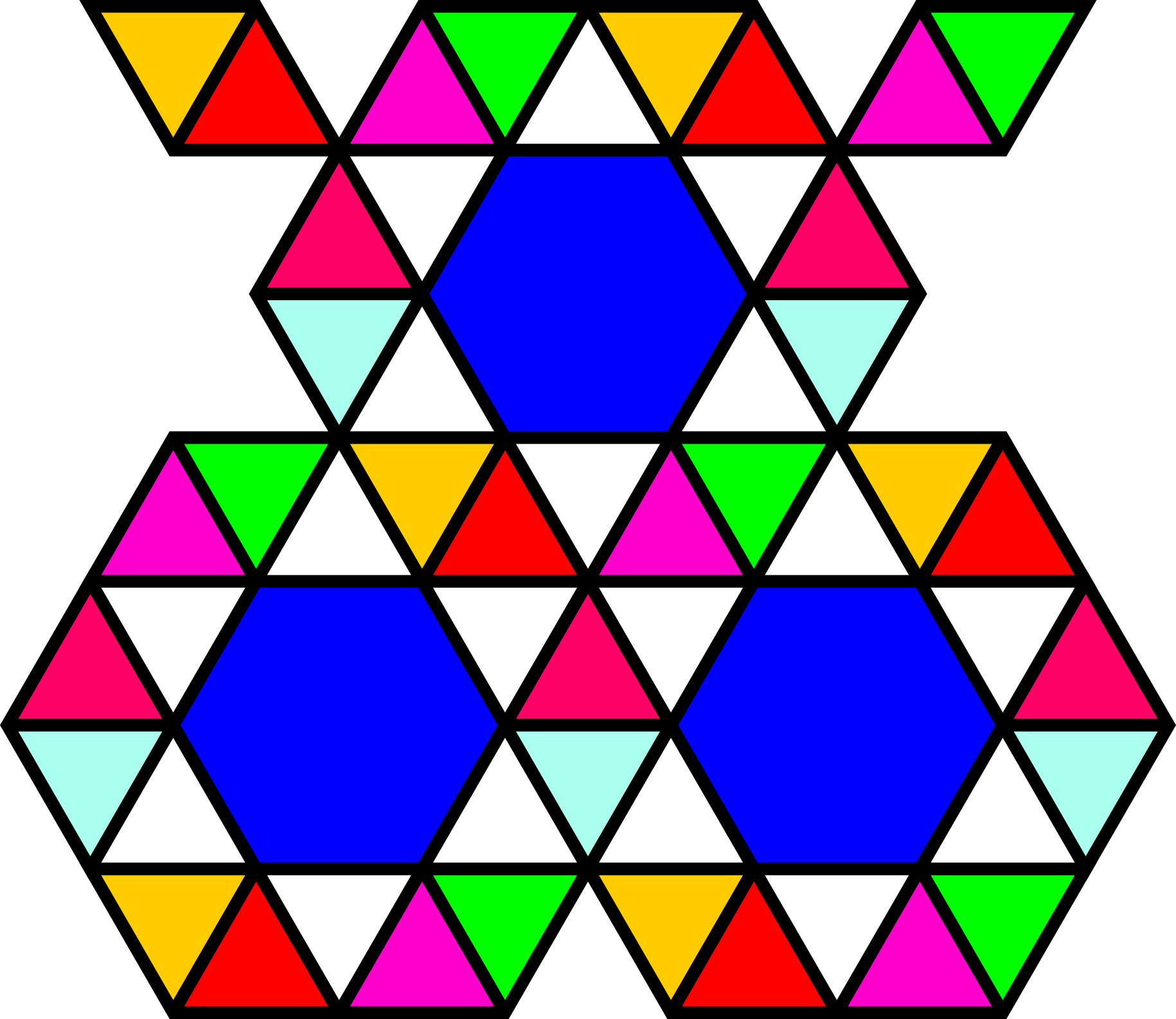} & 
        \includegraphics[width=2.5cm]{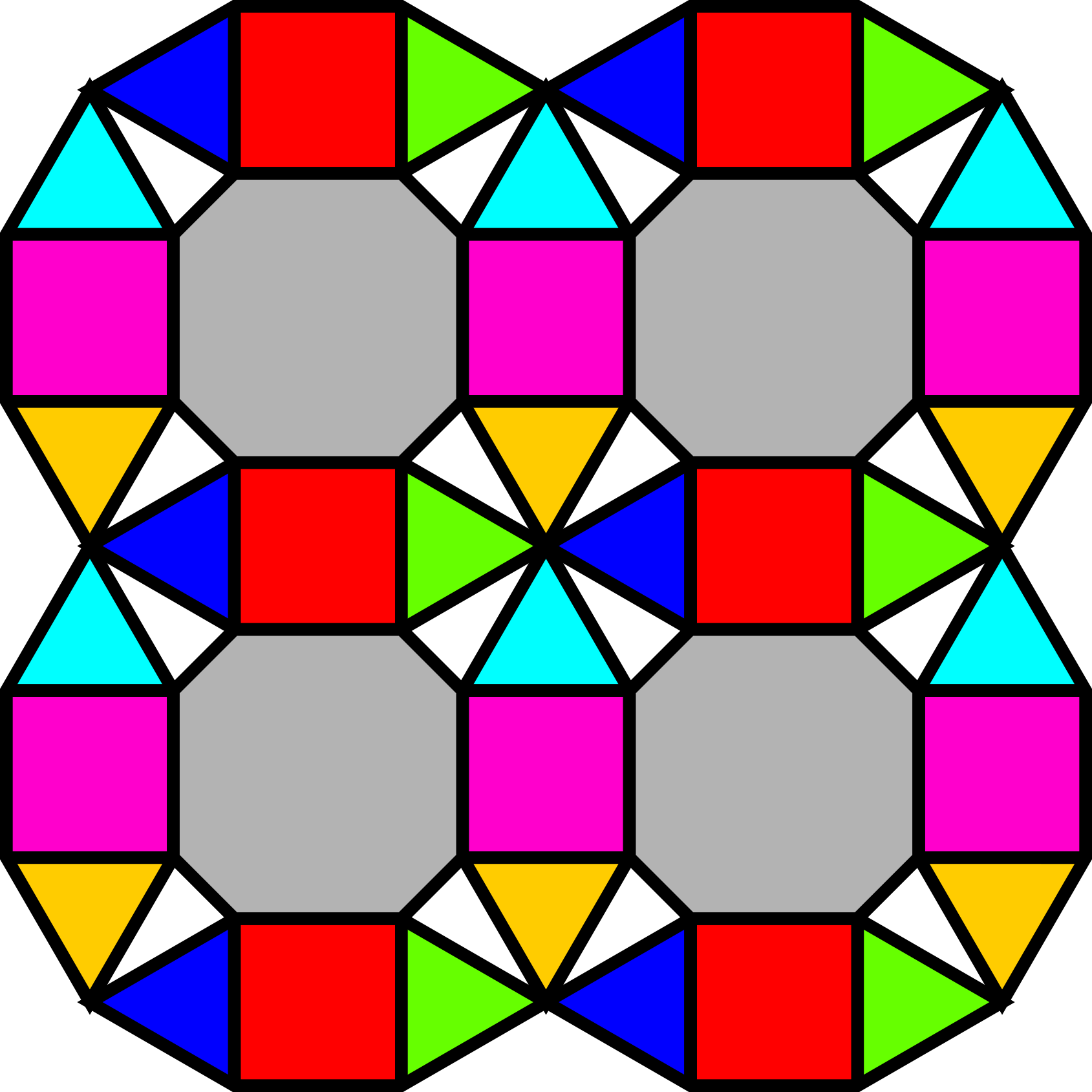} & 
        \includegraphics[width=2.5cm]{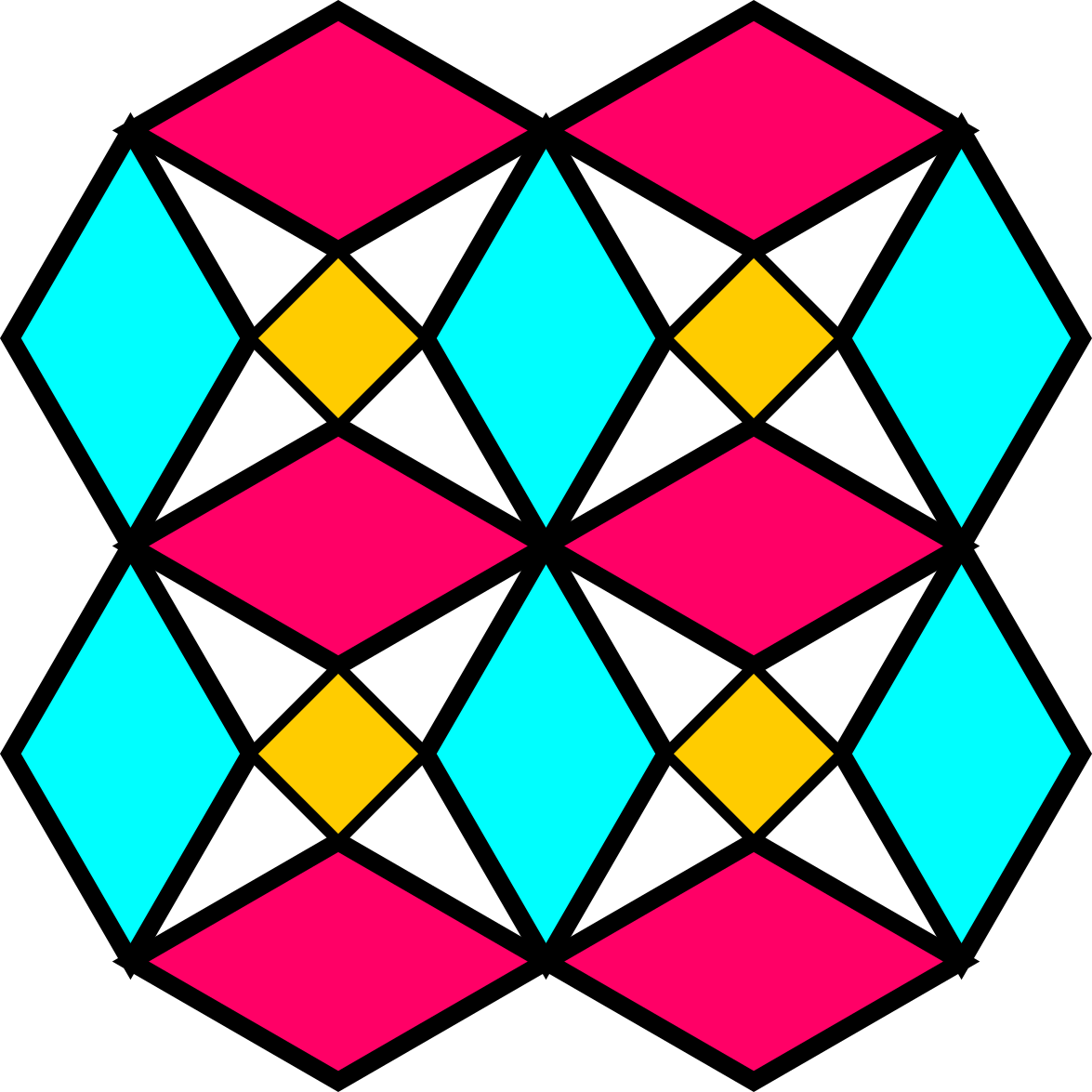} &
        \includegraphics[width=2.5cm]{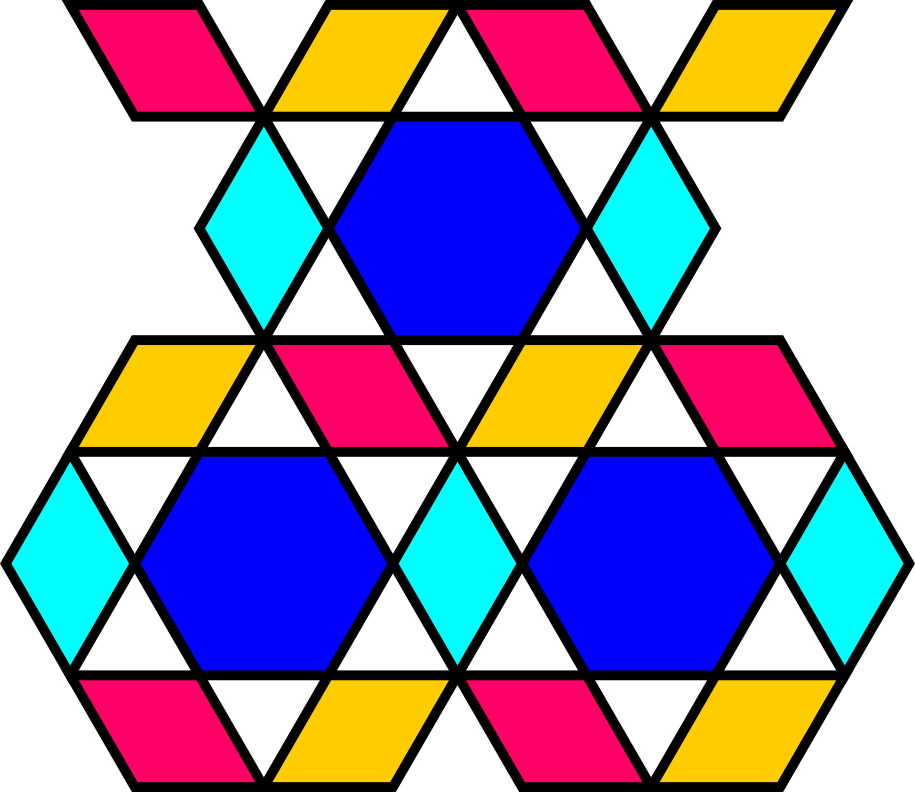} & 
        \includegraphics[width=2.5cm]{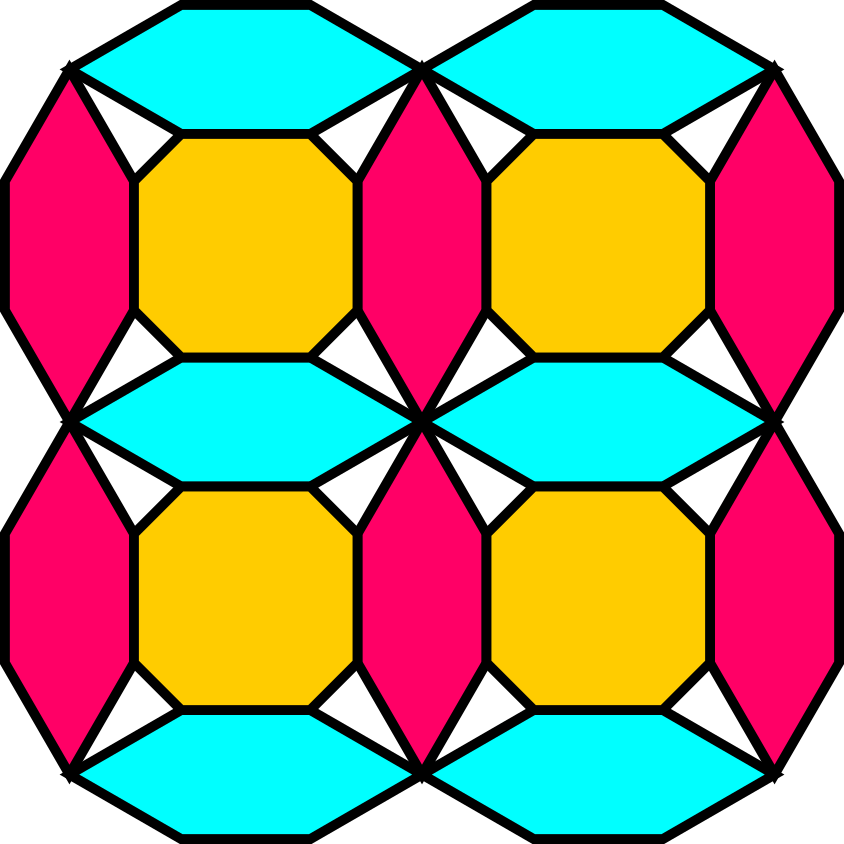} \\

        \makecell{ \vspace{-2.5cm} \\ \textbf{Structure}  \\ \textbf{Factor} } & 
        \includegraphics[width=2.5cm,trim={1cm 1cm 2cm 0},clip]{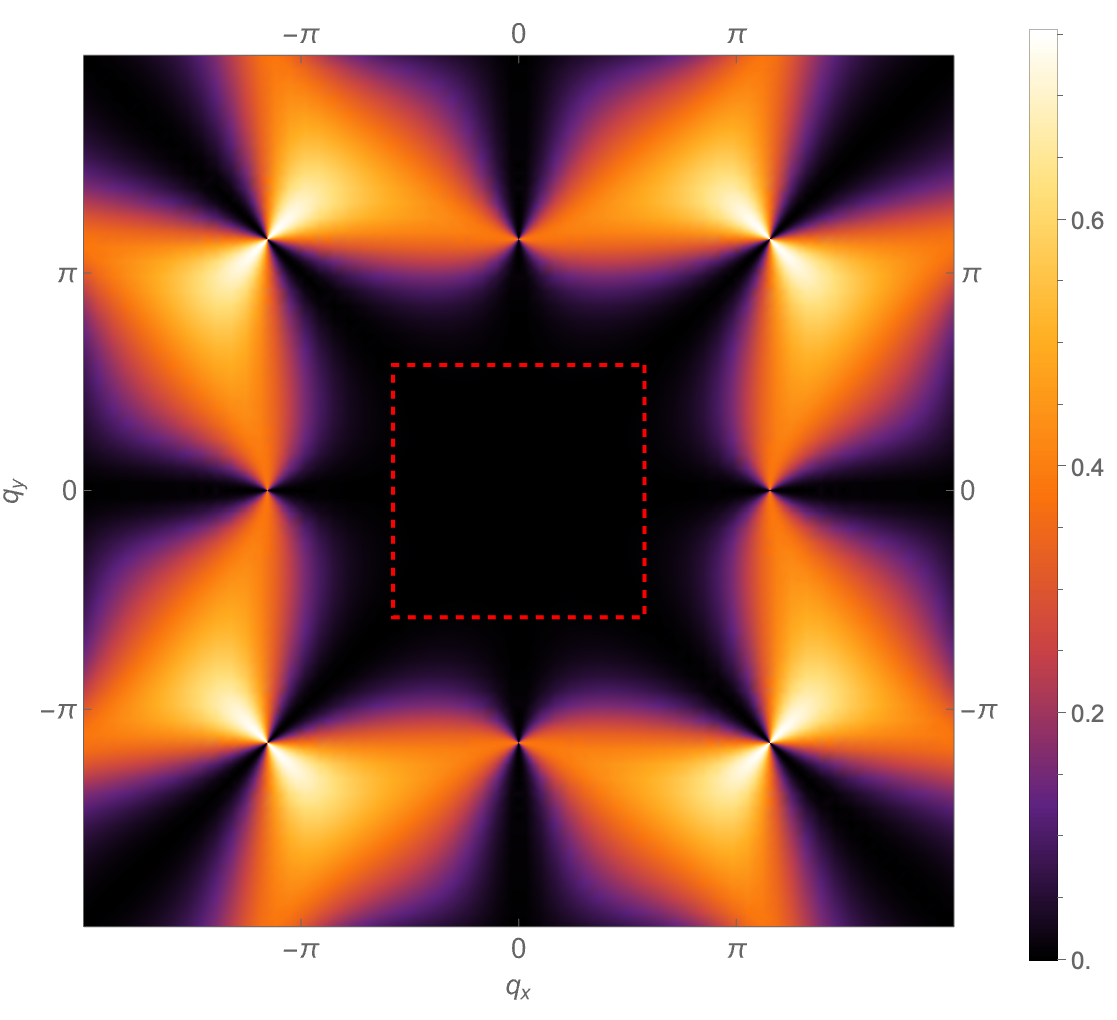} & 
        \includegraphics[width=2.5cm,trim={1cm 1cm 2cm 0},clip]{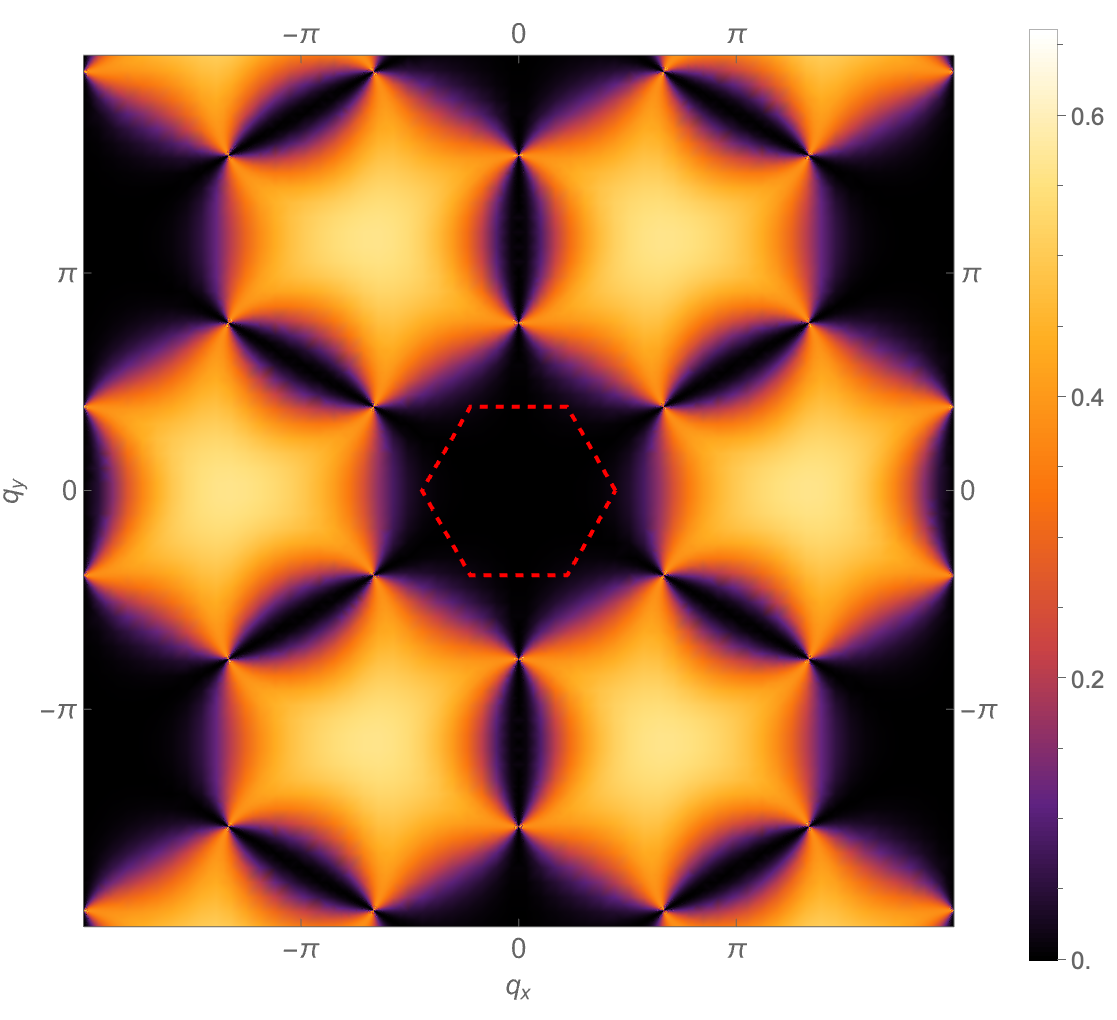} & 
        \includegraphics[width=2.5cm,trim={1cm 1cm 2cm 0},clip]{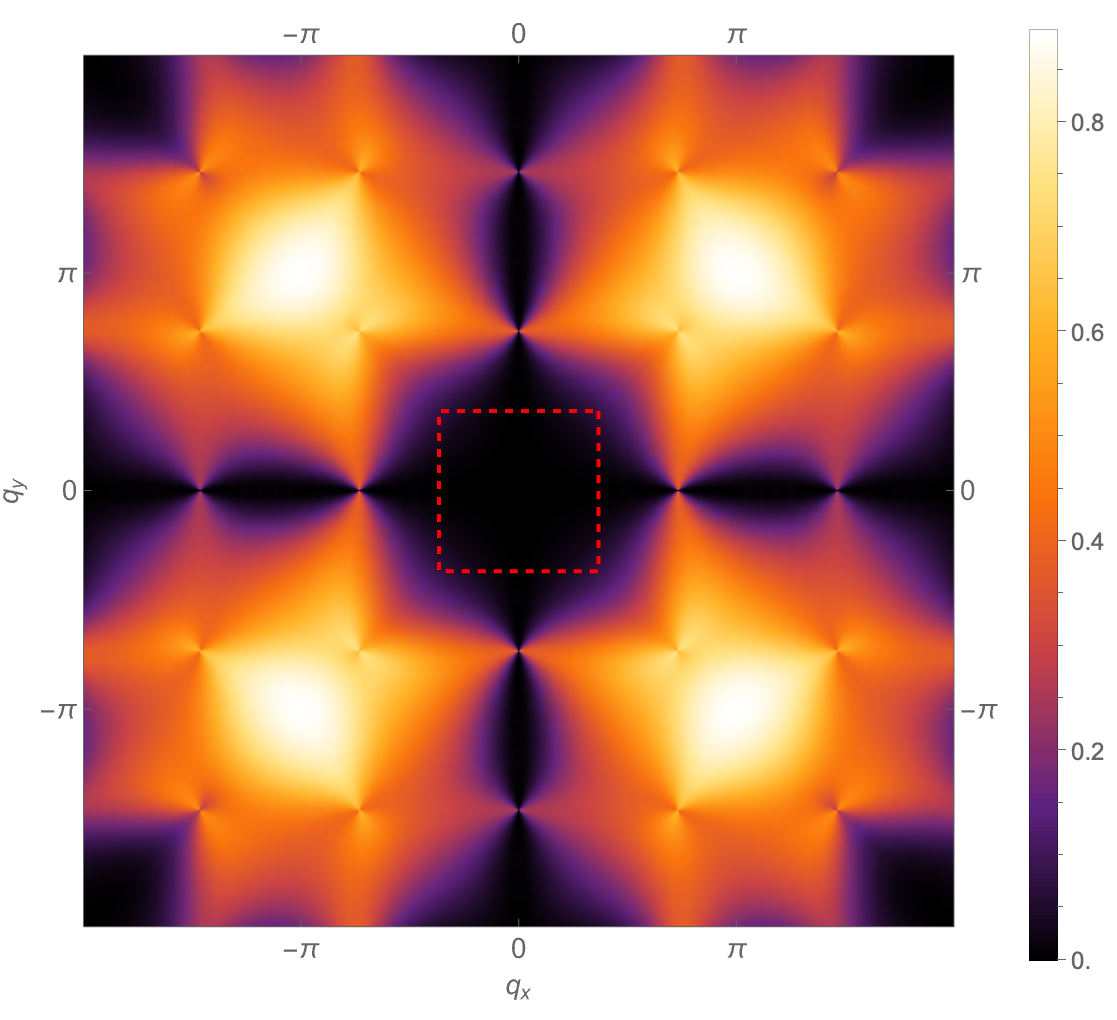} & 
        \includegraphics[width=2.5cm,trim={1cm 1cm 2cm 0},clip]{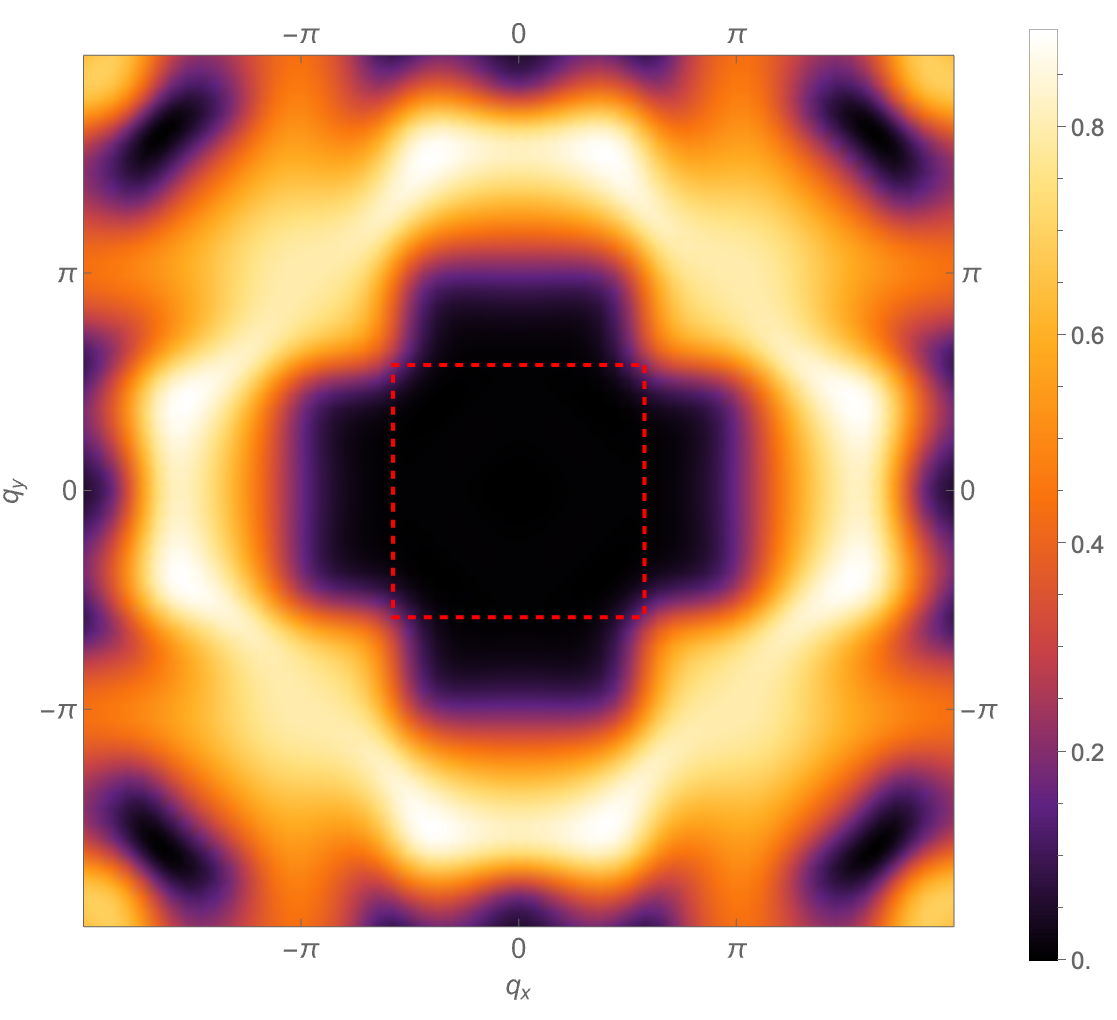} &
        \includegraphics[width=2.5cm,trim={1cm 1cm 2cm 0},clip]{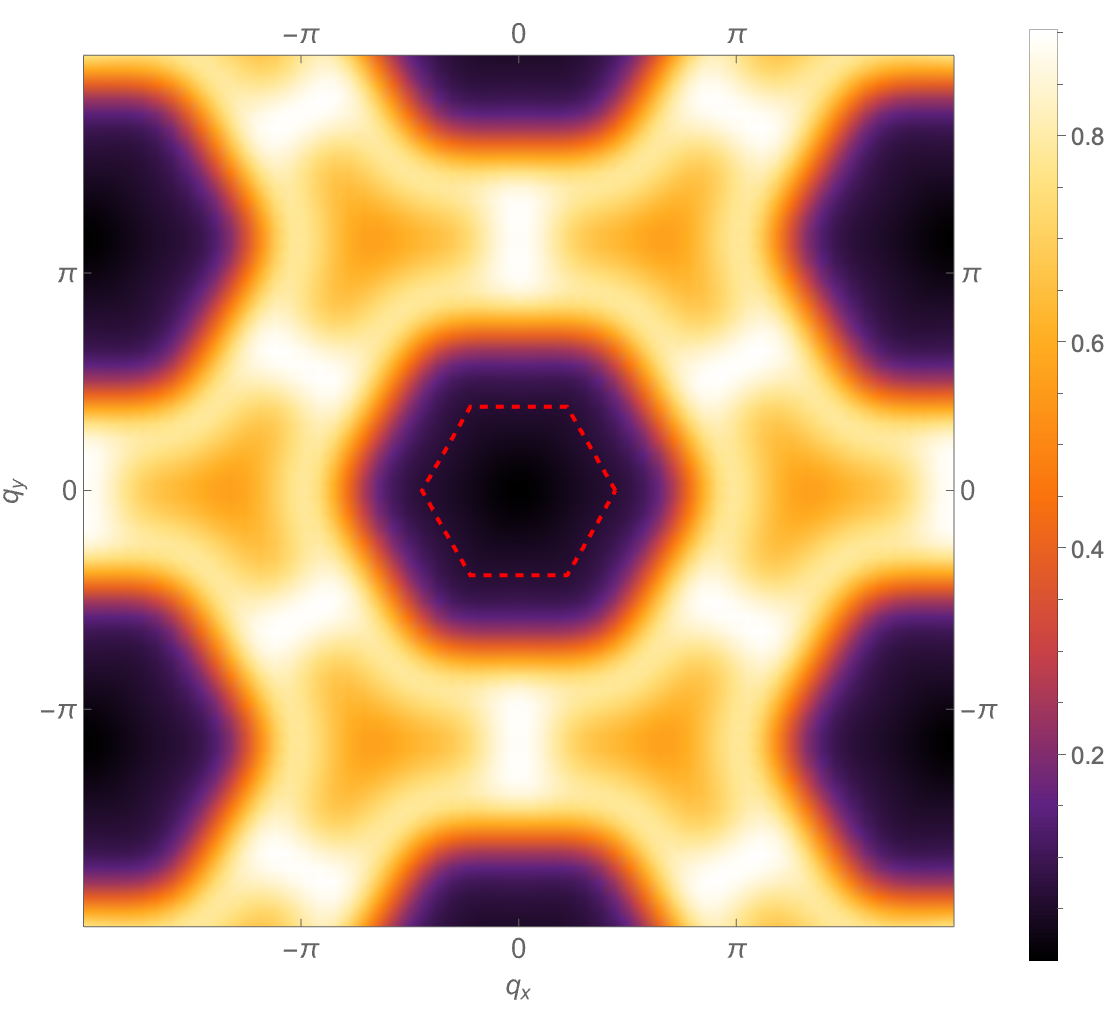} & 
        \includegraphics[width=2.5cm,trim={1cm 1cm 2cm 0},clip]{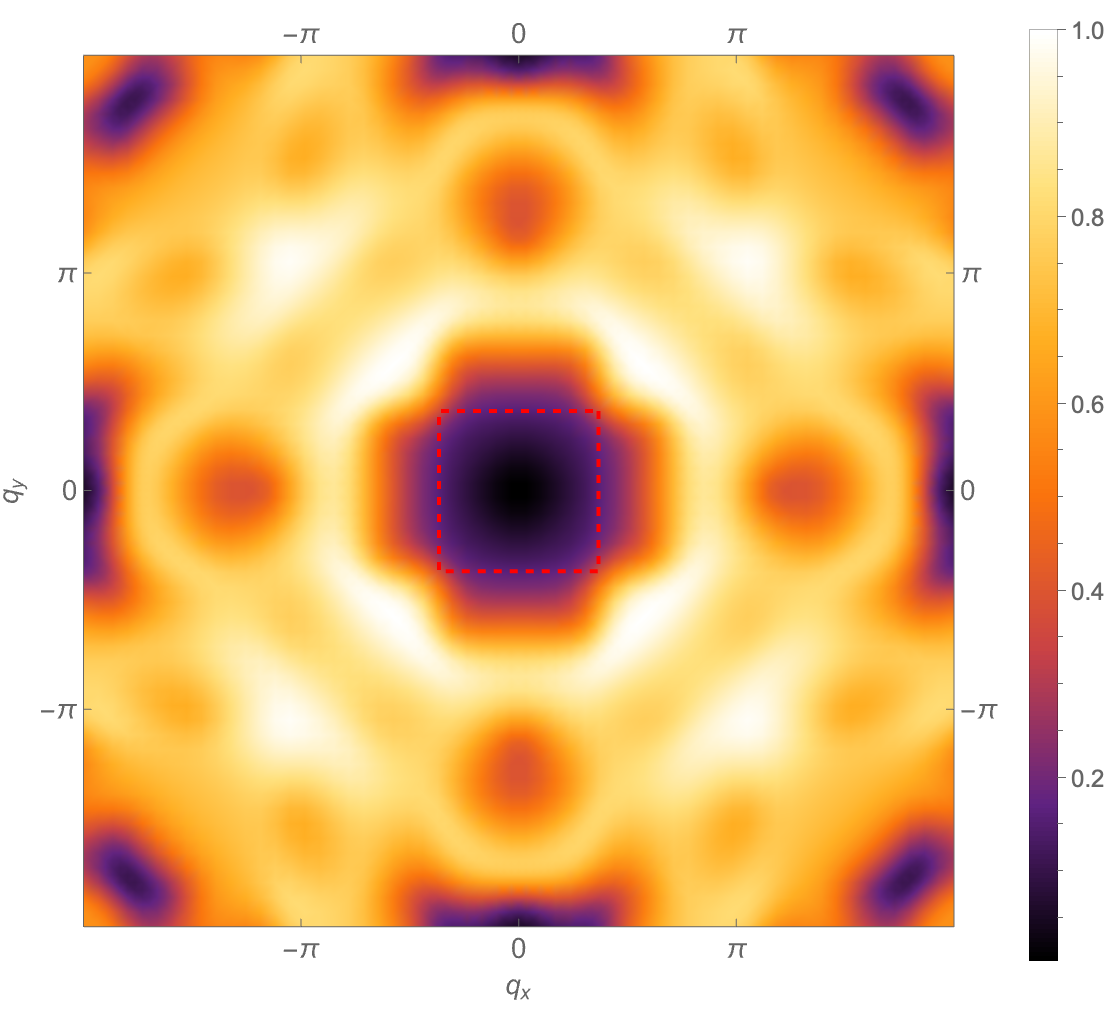} \\

        \makecell{ \vspace{-2.3cm} \\ \textbf{3D Band}  \\ \textbf{Structure} } & 
        \includegraphics[width=2.5cm]{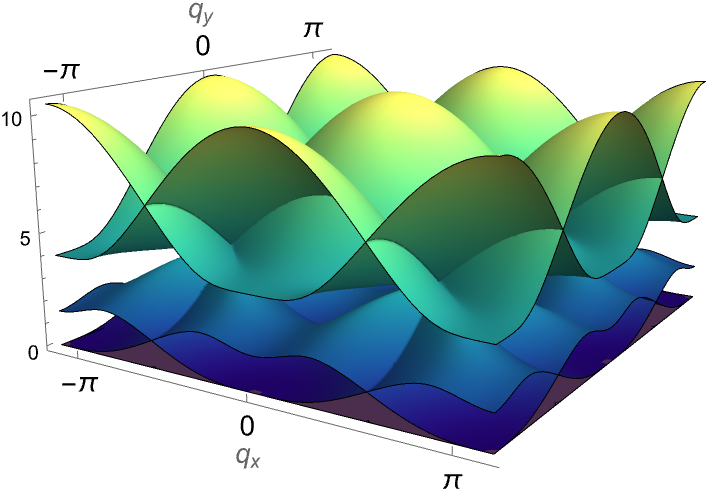} & 
        \includegraphics[width=2.5cm]{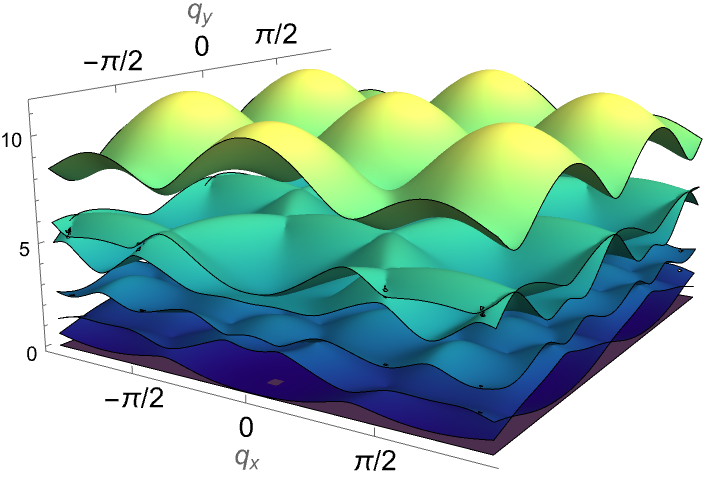} & 
        \includegraphics[width=2.5cm]{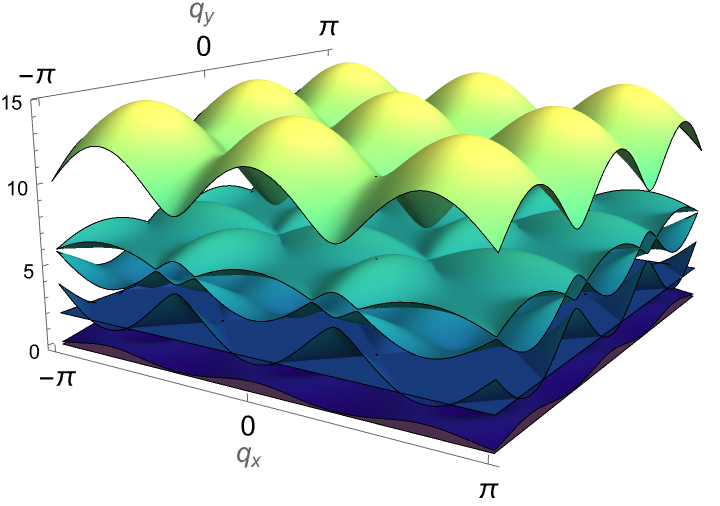} &
        \includegraphics[width=2.5cm]{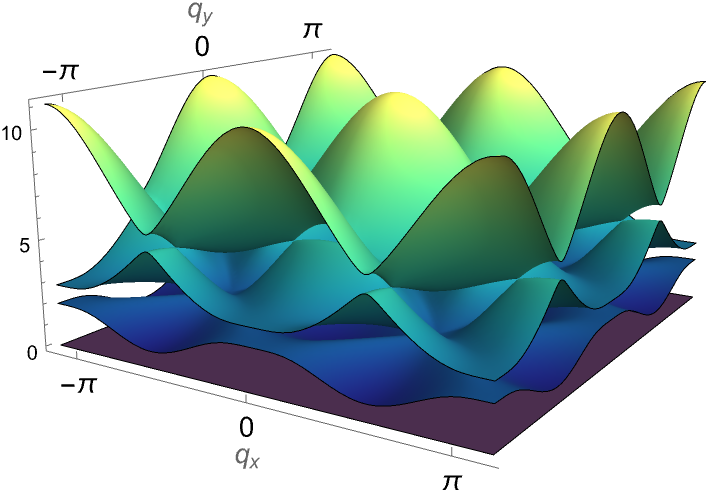} &
        \includegraphics[width=2.5cm]{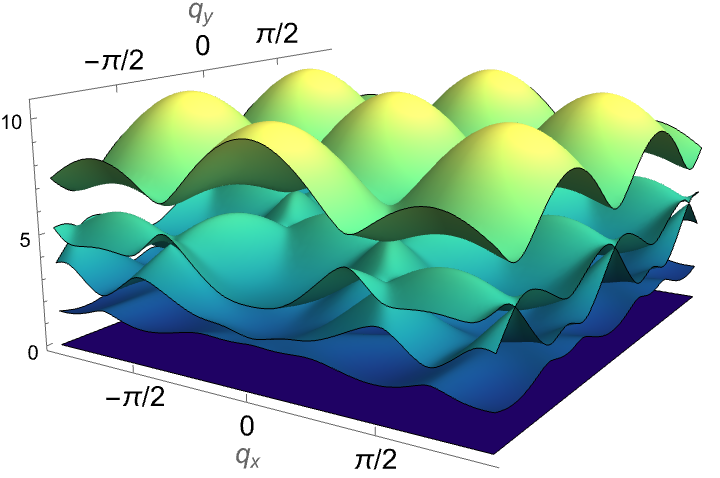} & 
        \includegraphics[width=2.5cm]{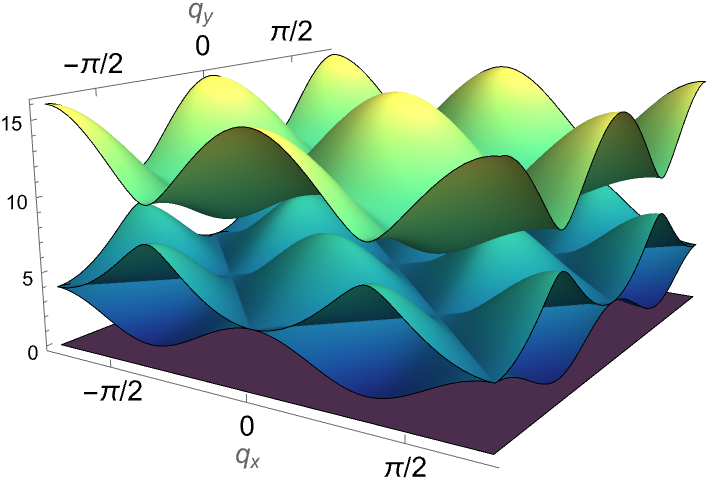} \\

        \makecell{ \vspace{-2.1cm} \\ \textbf{Band}  \\ \textbf{Structure} } & 
        \includegraphics[width=2.5cm]{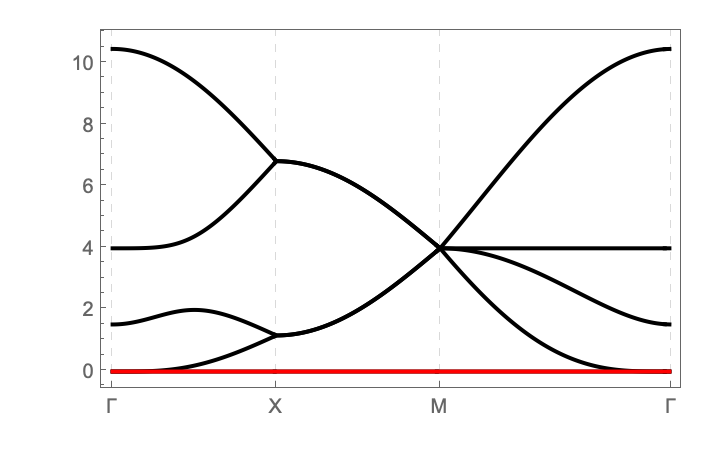} & 
        \includegraphics[width=2.5cm]{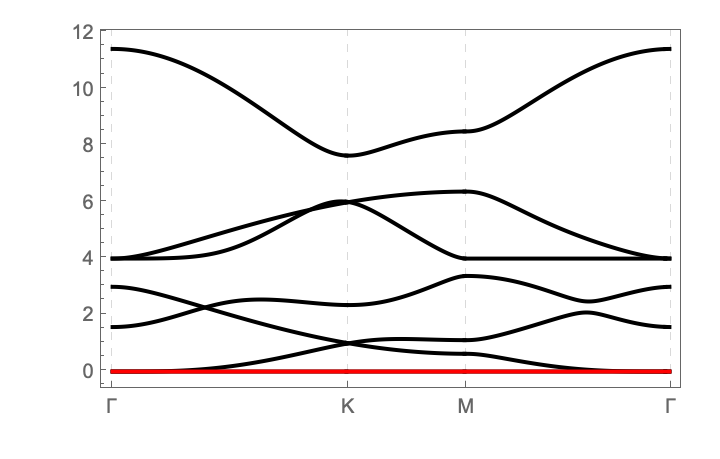} & 
        \includegraphics[width=2.5cm]{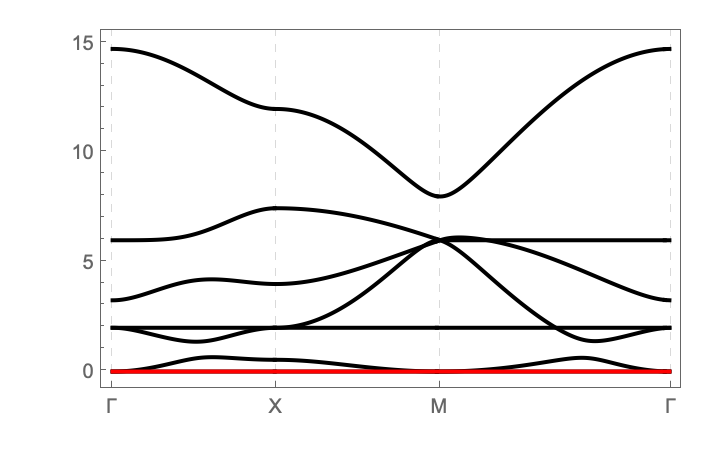} &
        \includegraphics[width=2.5cm]{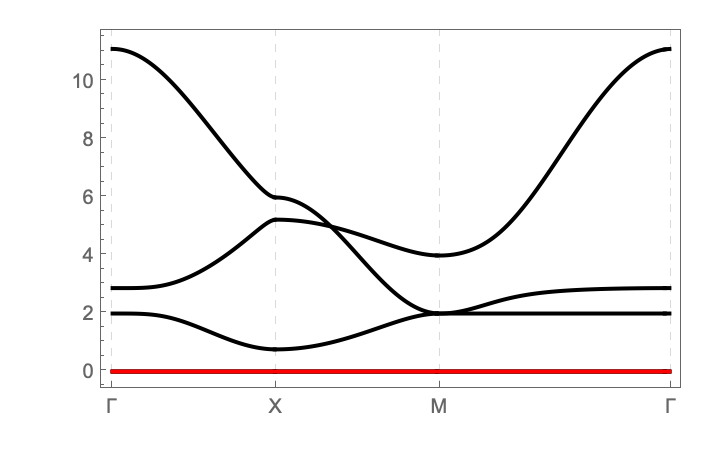} &
        \includegraphics[width=2.5cm]{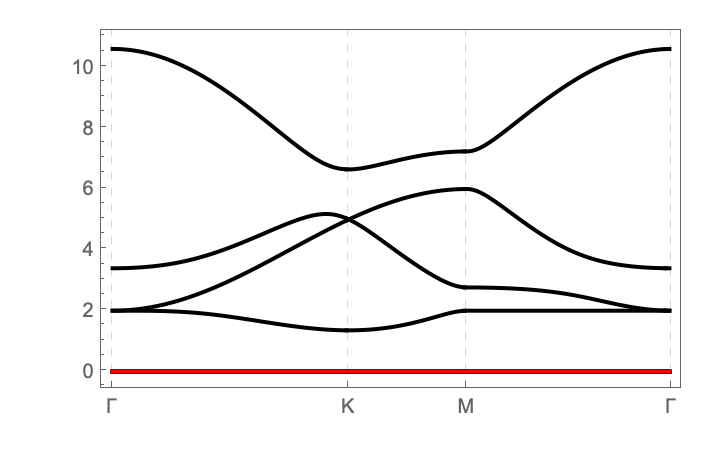} & 
        \includegraphics[width=2.5cm]{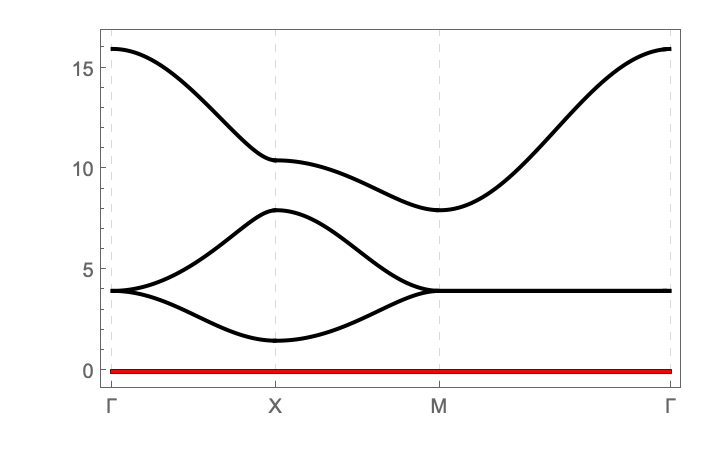} \\
        
        \textbf{$n_s$}     & 5 & 8 & 9 & 5 & 8 & 9 \\
        \textbf{$n_c$}     & 5 & 7 & 7  & 3 & 4 & 3 \\
        \textbf{$n_\text{b.f.b}$} & \textcolor{red}{1} & \textcolor{red}{2} & \textcolor{red}{3} & 2  & 4 & 6 \\
        \hline
    \end{tabular}}
    \caption{Bond-decorated systems in two dimensions, with additional inter-clusters fitting within the parent lattice plaquettes. The number of bottom flat bands $n_\text{b.f.b}$ again does not match the difference $n_s-n_c$ for composite bonds decorated systems. The plots and counting are performed assuming spins sit only on cluster vertices. The addition of inter-clusters destroys the intermediate flat bands, and restores the singular nature of the band touchings in gapless bond-decorated systems. Each structure factor and band structure are given in inverse units of the first neighbor distance among bond clusters.    }
    \label{tab: 2d decorated lattices with inter-clusters}
\end{table*}

\section{Square diamond square flat bands}
\label{Appendix: square diamond square fb}

For the square diamond square lattice, the non trivial part of spectrum is governed by the matrix
\begin{equation}
   \begin{split}
        h^{c \leftarrow v} h^{v \leftarrow c} &= 4I^{ss} + 3I^{tt} + 
    \begin{pmatrix}
        2 (A^{to}A^{ot}-I^{tt}) & A^{ts} \\
        A^{st} & 0 \\
    \end{pmatrix} \\
    &= I + \begin{pmatrix}
        2 A^{to}A^{ot} & A^{ts} \\
        A^{st} & 3I^{ss} \\
    \end{pmatrix} 
   \end{split}
\end{equation}
with $s$ labeling square cluster centers, $t$ triangular cluster centers and $o$ referring to virtual sites located at the middle of the bidiamond bonds. Setting
\begin{equation}
    X = \begin{pmatrix}
        0 & A^{ts} \\
        A^{st} & 3 \\
    \end{pmatrix}
\end{equation}
allows to obtain
\begin{equation}
    X^2 - 3X = \begin{pmatrix}
        A^{ts}A^{st} & 0 \\
        0 & A^{st}A^{ts} \\
    \end{pmatrix}
\end{equation}
where there are four $t$ indices (sublattices), two $o$ indices and one $s$ index. Therefore $A^{to}A^{ot}$ has a kernel of dimension 2, and $A^{ts}A^{st}$ has a kernel of dimension 3. This suggests that the subpart of $h^{c \leftarrow v} h^{v \leftarrow c}$ living in the $t$ subspace has a kernel of dimension at least 1, resulting in a flat band located at energy 1 in the spectrum. Next, noticing that $A^{st}A^{ts}$ is simply four times the identity over square centers subspace, this generates another flat band located at energy $1+4 = 5$, therefore in perfect agreement with the spectrum presented in Table. \ref{tab: 2d cluster-links + cluster-bonds lattices}.

\begin{table*}[ht]
    \centering
    \renewcommand{\arraystretch}{1.5} 
    \resizebox{\textwidth}{!}{
    \begin{tabular}{c c c c c c c}
        \hline    
        \makecell{\textbf{2D parent}\\ \textbf{lattice}} & Square & Square & Hexagonal & Hexagonal & Hexagonal & Triangular \\
        \textbf{Cluster-bond} & diamond & Cracker & Diamond & Cracker & Diamond & Diamond \\
        \textbf{Cluster-vertex} & Square & square & Triangle & Triangle & Hexagon & Hexagon \\
        \hline
        \noalign{\vskip 1mm}
        \makecell{ \vspace{-2.5cm} \\ \textbf{Composite} \\ \textbf{Cluster-bond} \\ \textbf{Lattice}  \\ \textbf{Scheme} } & 
        \includegraphics[height=2.5cm]{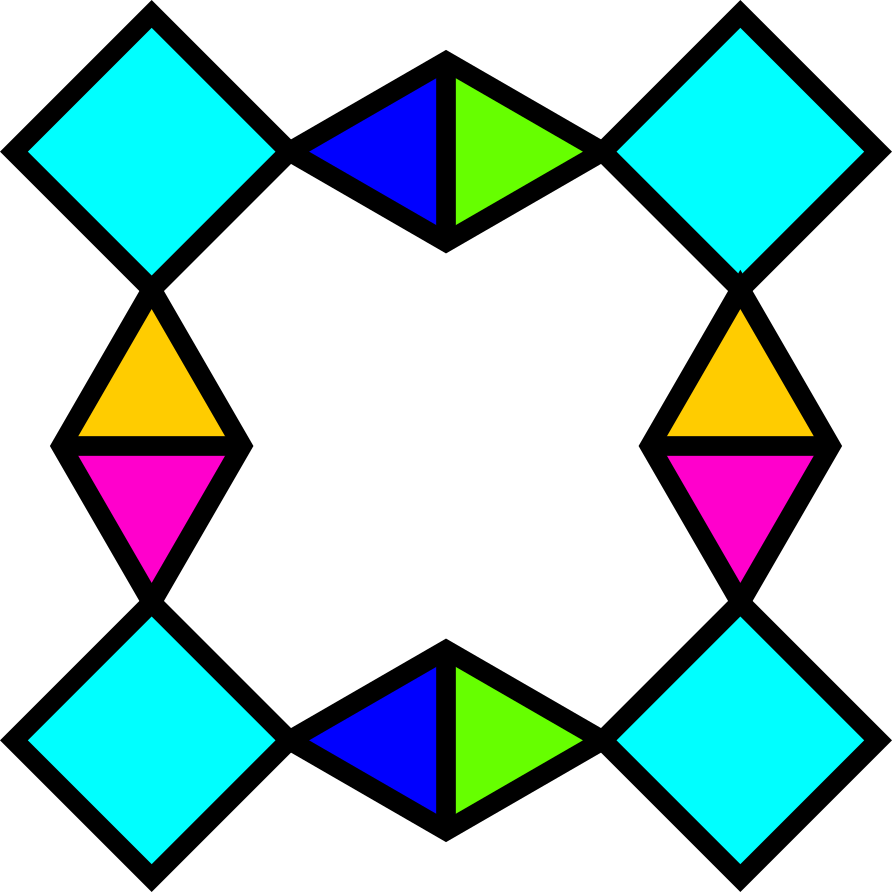} & 
        \includegraphics[height=2.5cm]{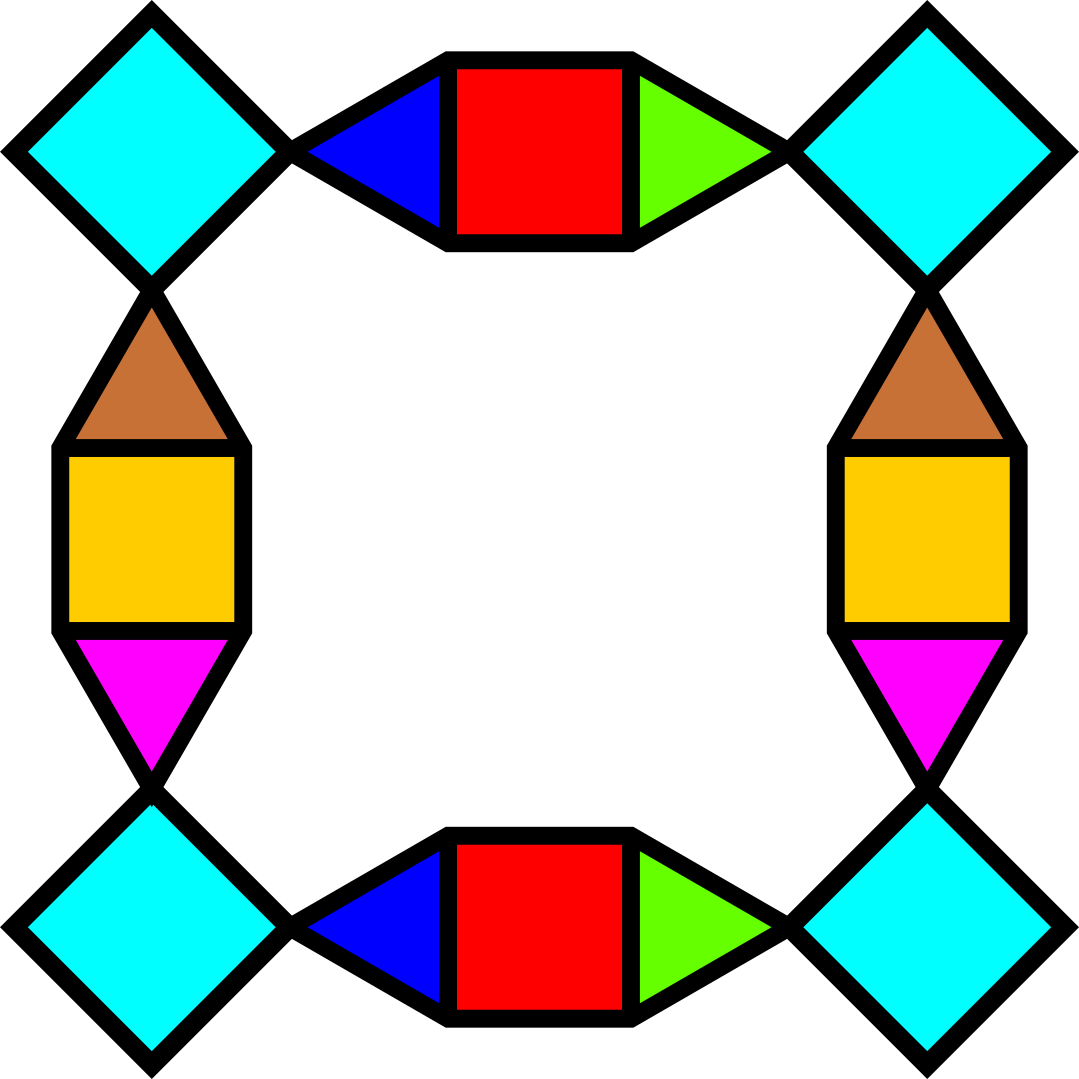} & 
        \includegraphics[height=2.5cm]{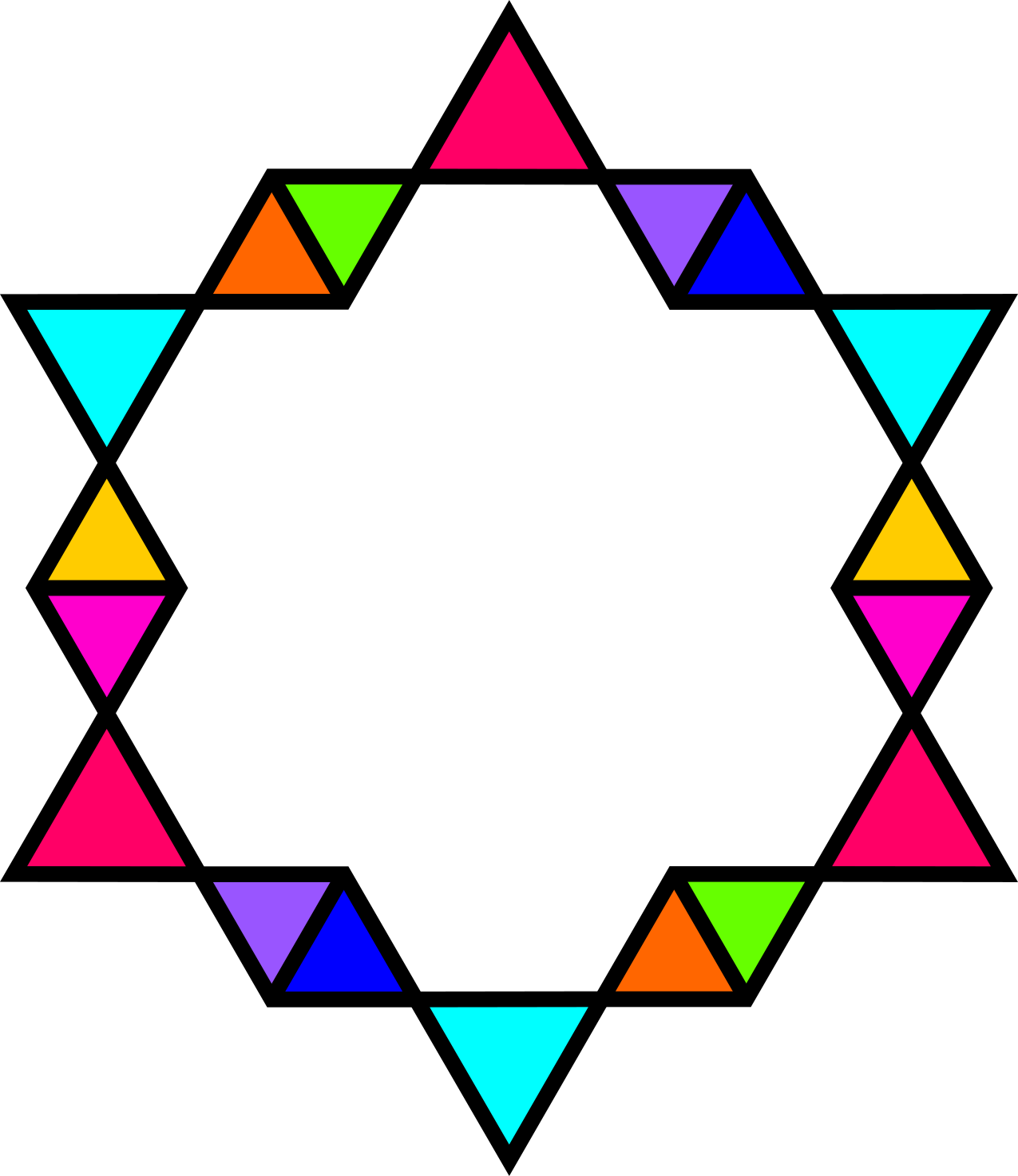} & 
        \includegraphics[height=2.5cm]{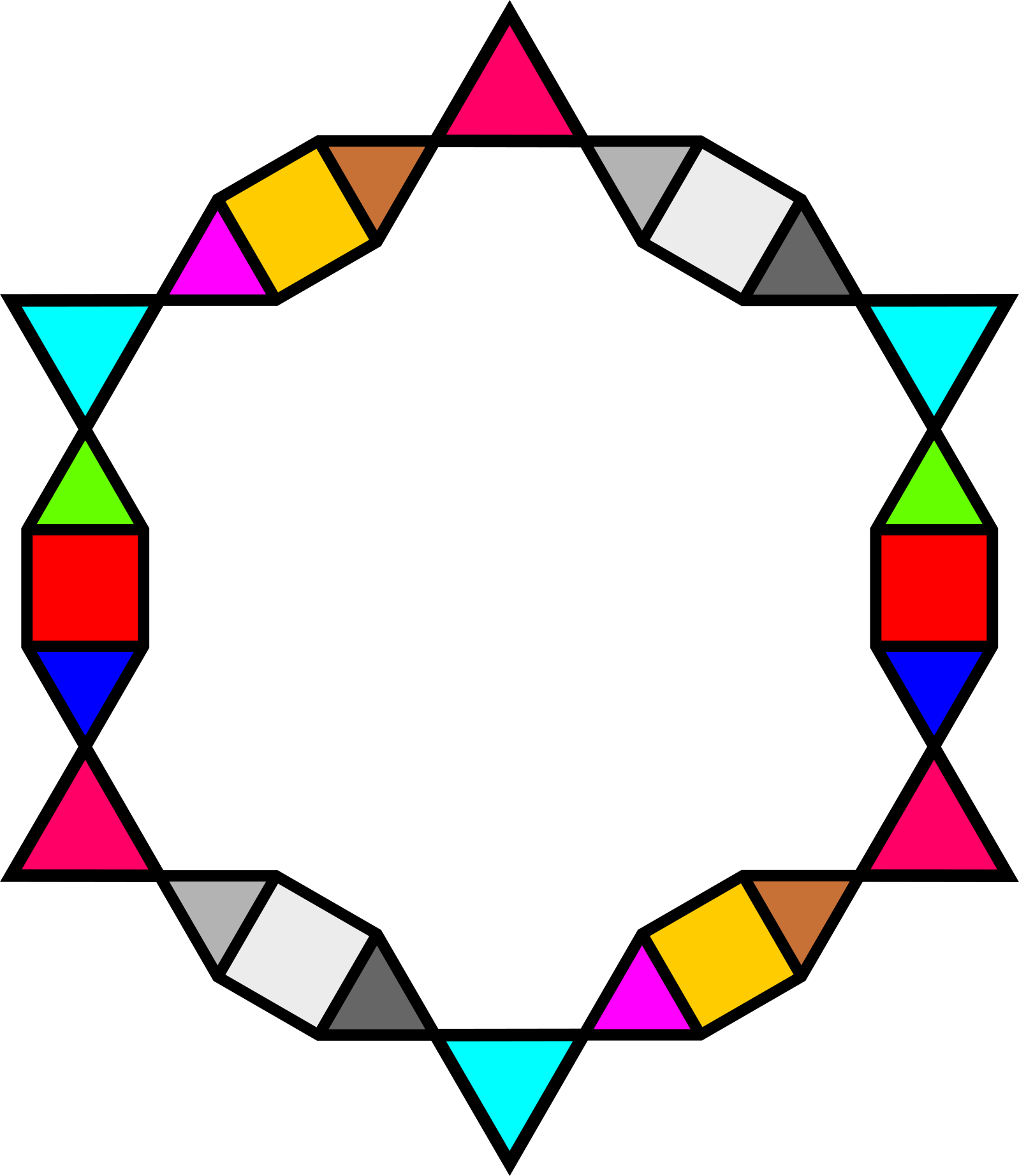} & 
        \includegraphics[height=2.5cm]{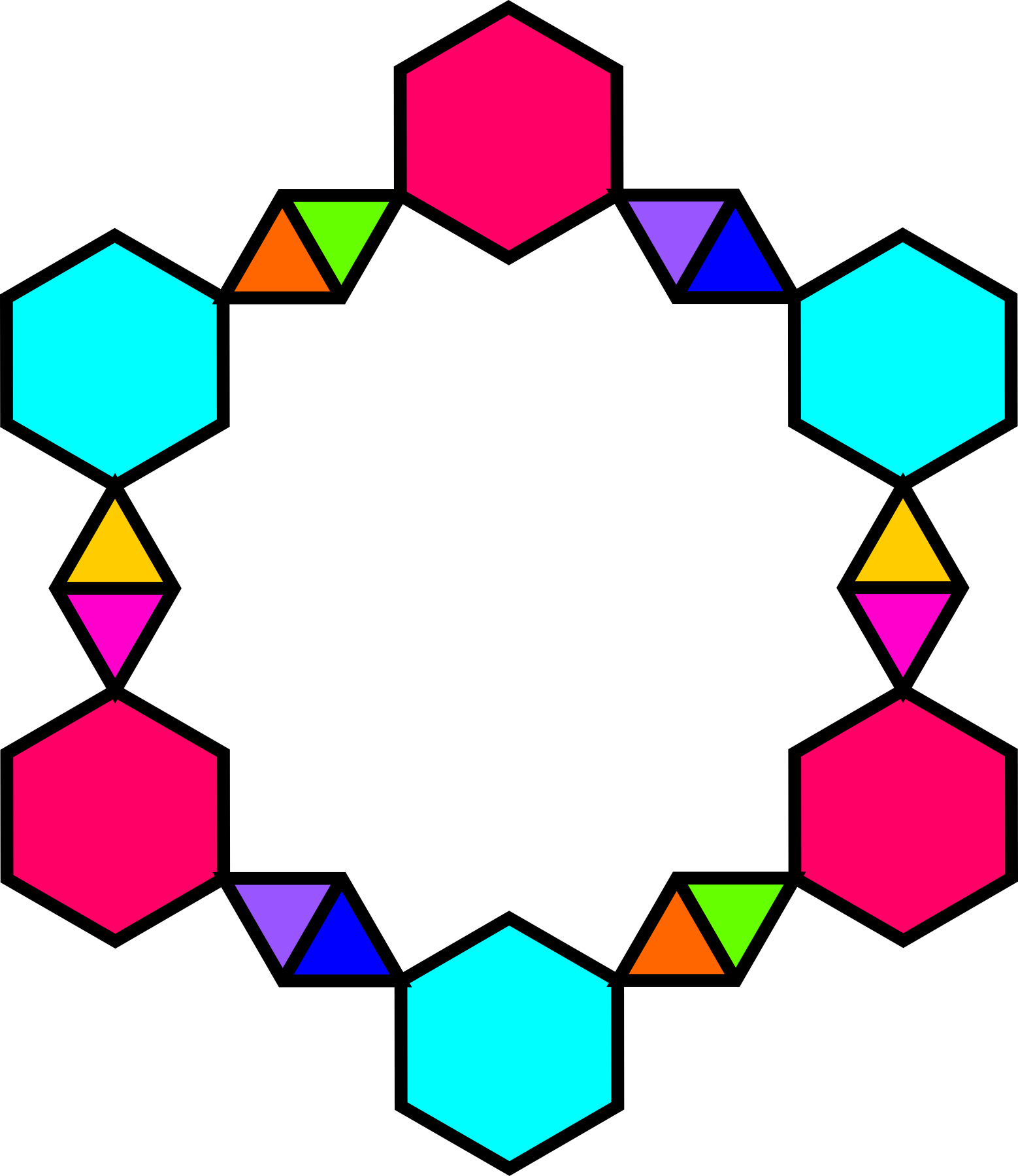} & 
        \includegraphics[width=2.5cm]{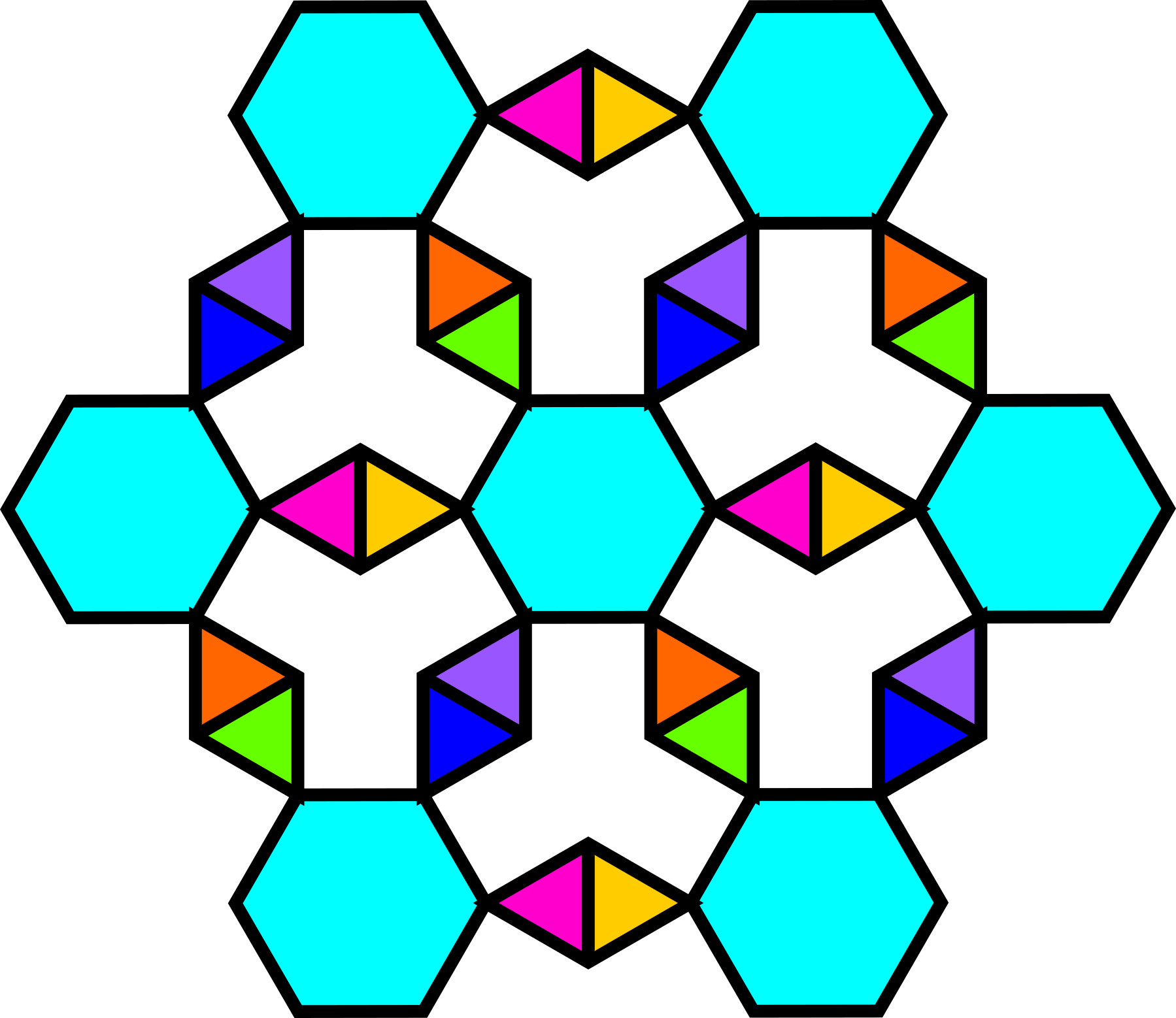} 
        \\

        \hline
        \noalign{\vskip 1mm}

        \makecell{ \vspace{-2.5cm} \\ \textbf{Monoblock} \\ \textbf{Cluster-bond} \\ \textbf{Structure}  \\ \textbf{Factor} } & 
        \includegraphics[width=2.5cm,trim={1cm 1cm 2cm 0},clip]{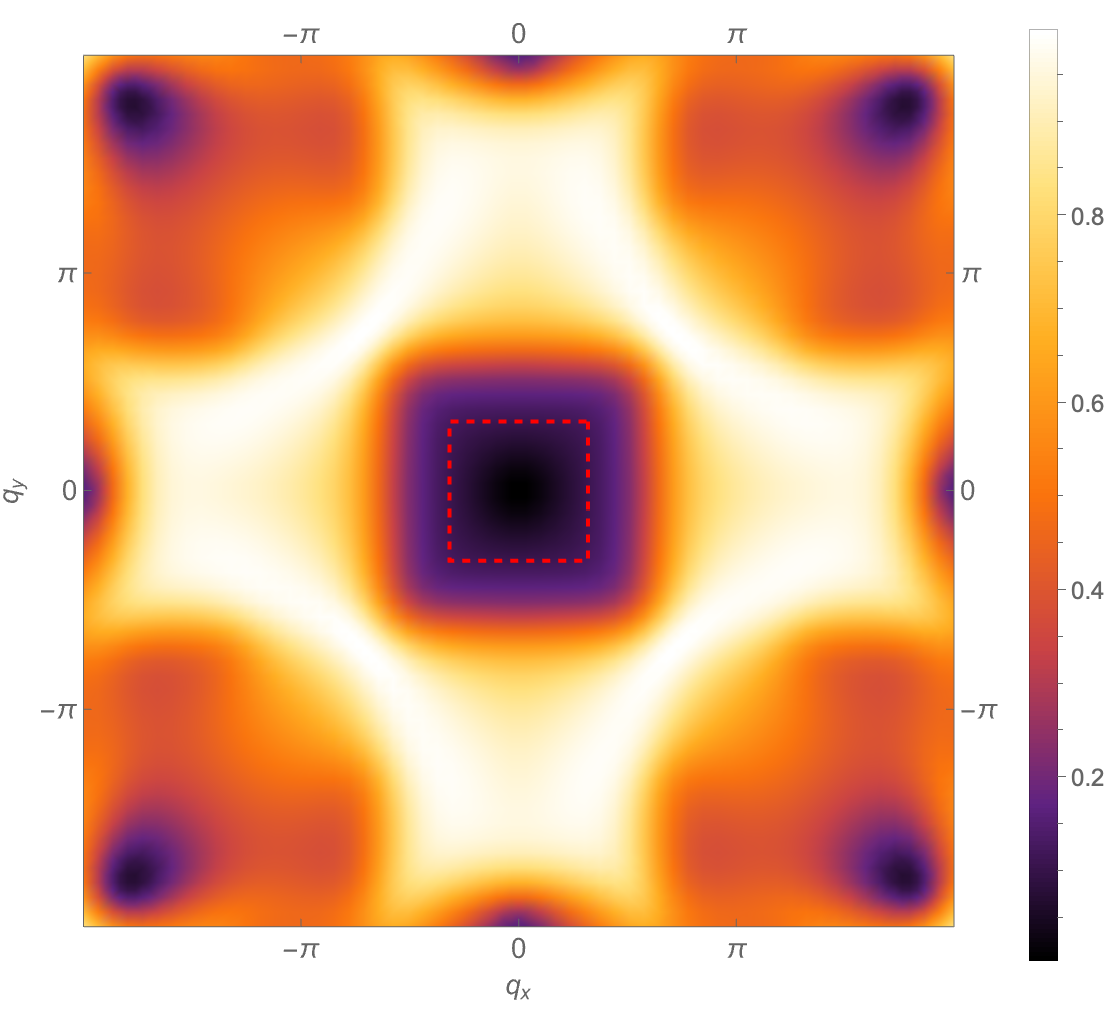} & 
        \includegraphics[width=2.5cm,trim={1cm 1cm 2cm 0},clip]{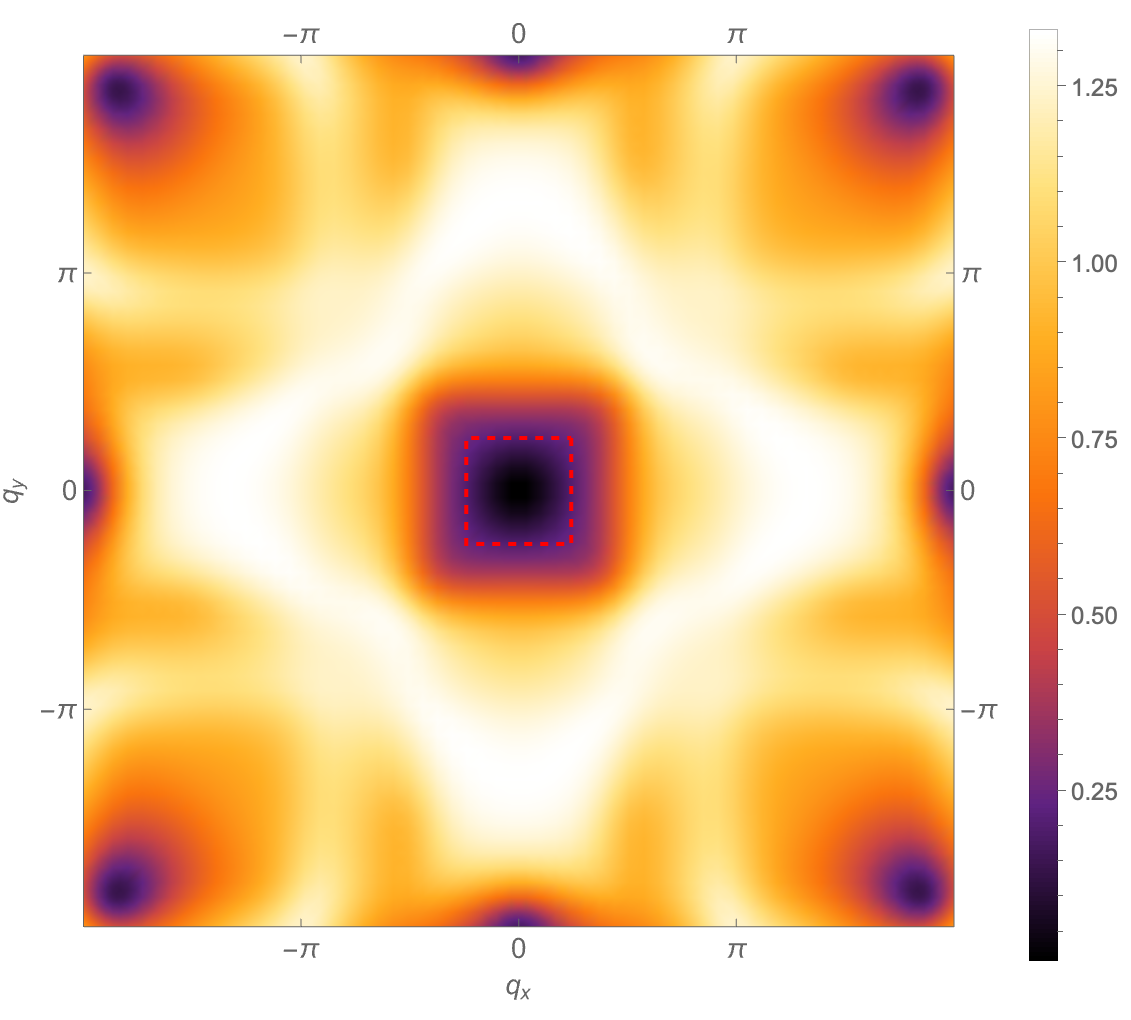} & 
        \includegraphics[width=2.5cm,trim={1cm 1cm 2cm 0},clip]{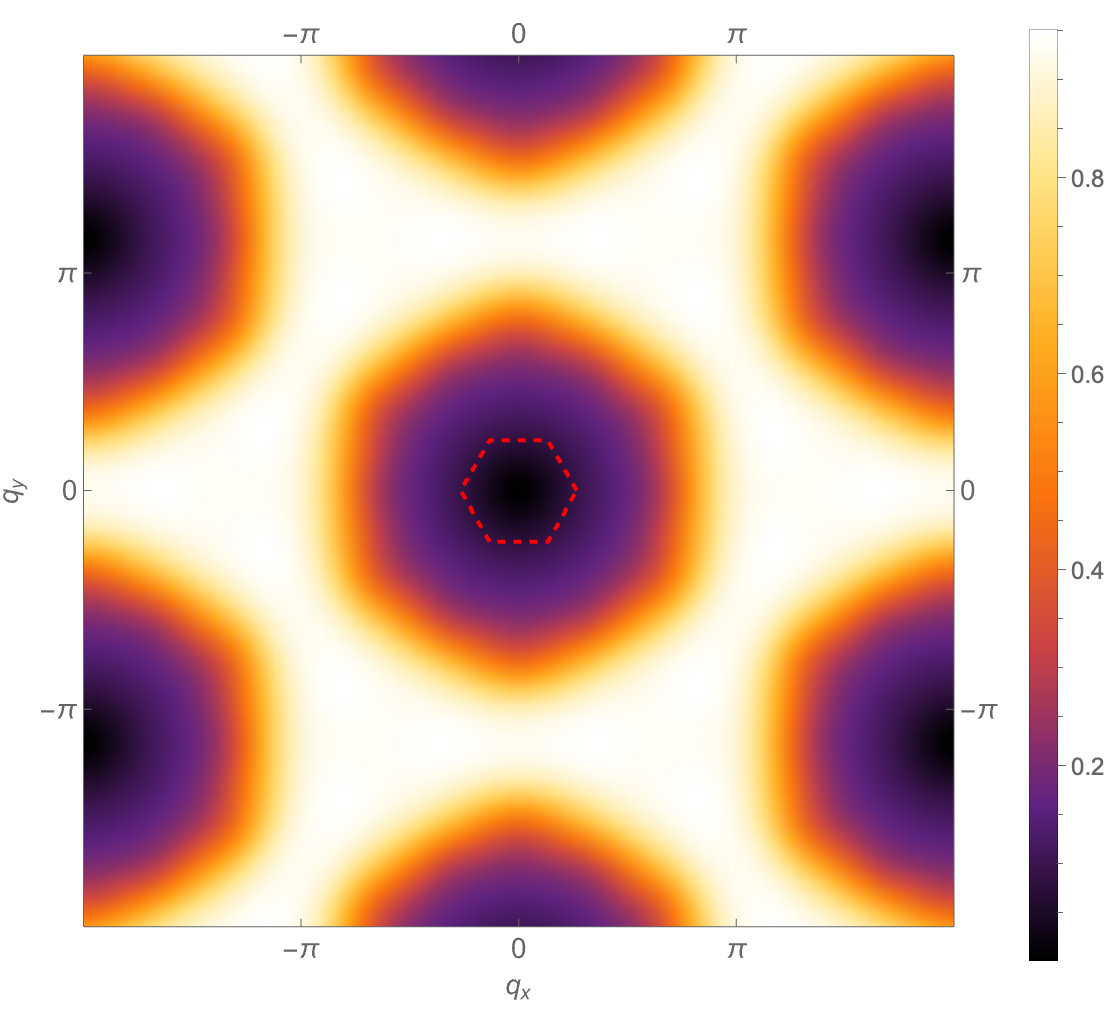} & 
        \includegraphics[width=2.5cm,trim={1cm 1cm 2cm 0},clip]{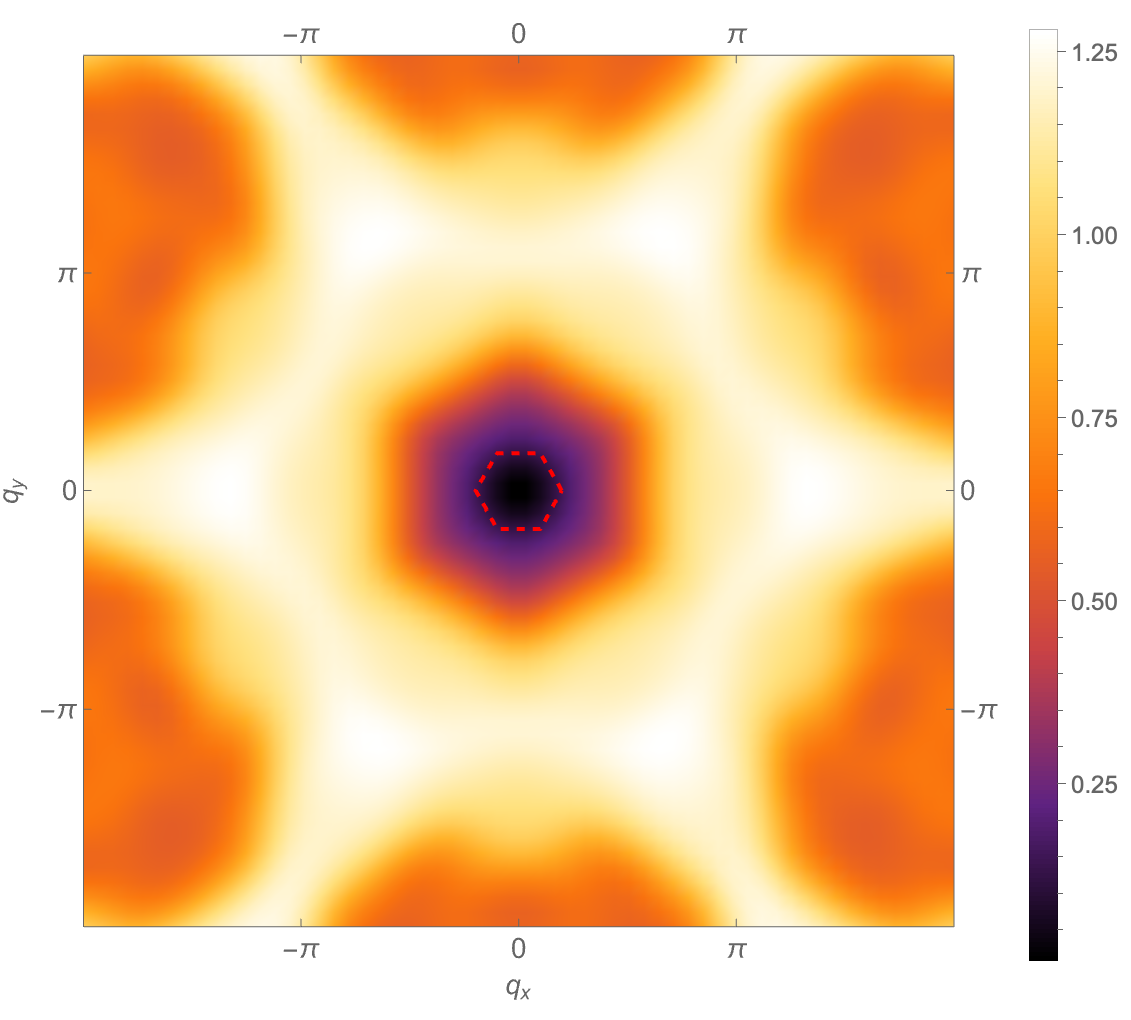} & 
        \includegraphics[width=2.5cm,trim={1cm 1cm 2cm 0},clip]{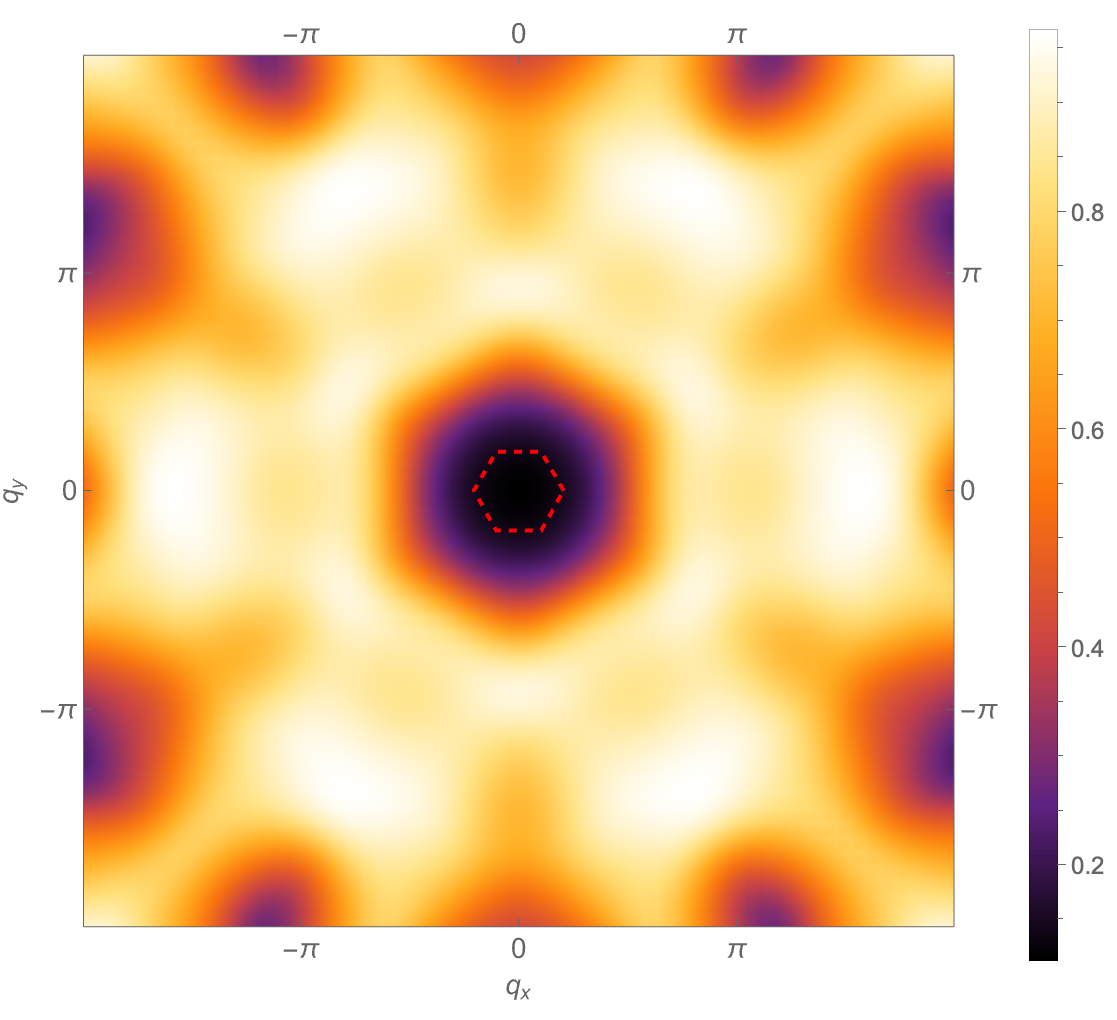} & 
        \includegraphics[width=2.5cm,trim={1cm 1cm 2cm 0},clip]{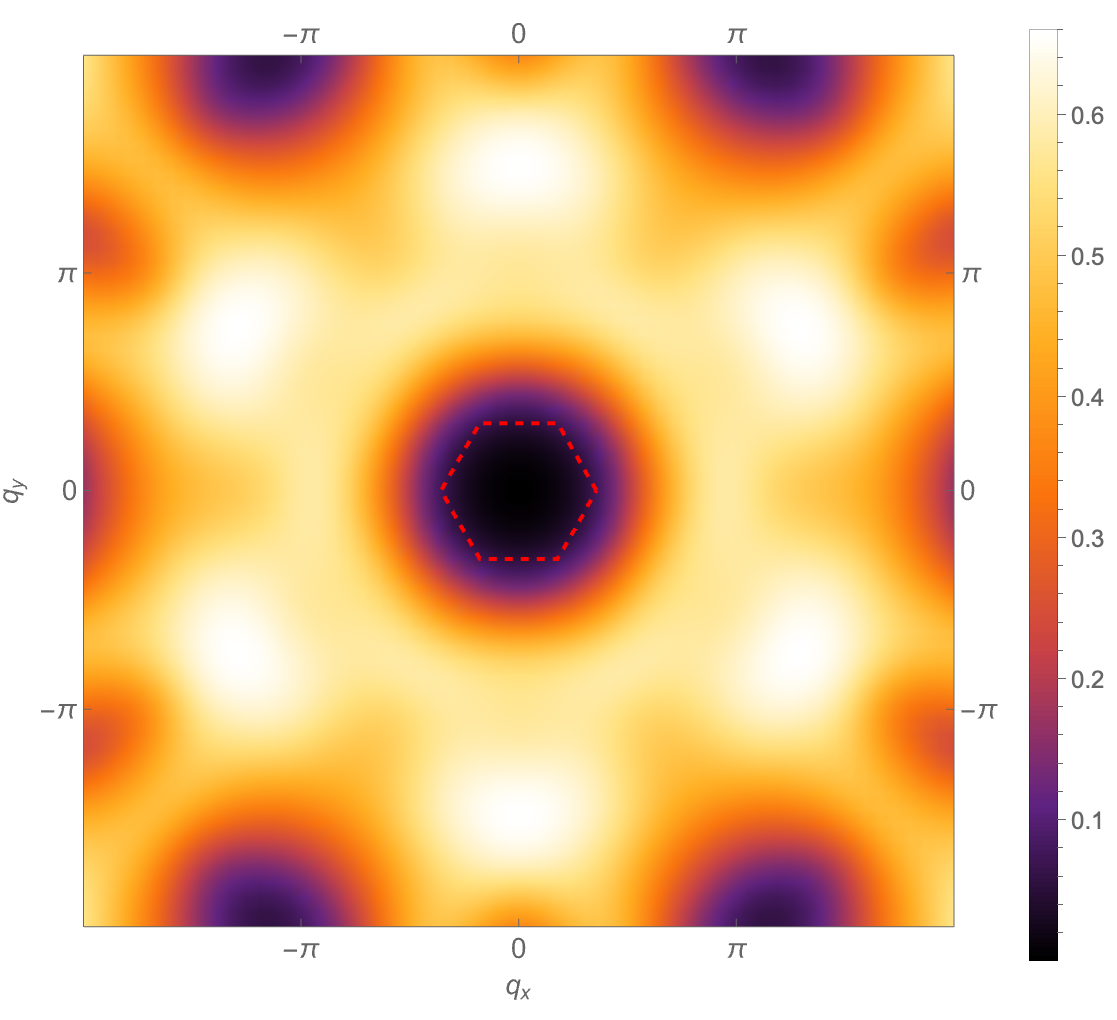} 
        \\

        \makecell{ \vspace{-2.3cm} \\ \textbf{Monoblock} \\ \textbf{Cluster-bond} \\ \textbf{3D Band}  \\ \textbf{Structure} } & 
        \includegraphics[width=2.5cm]{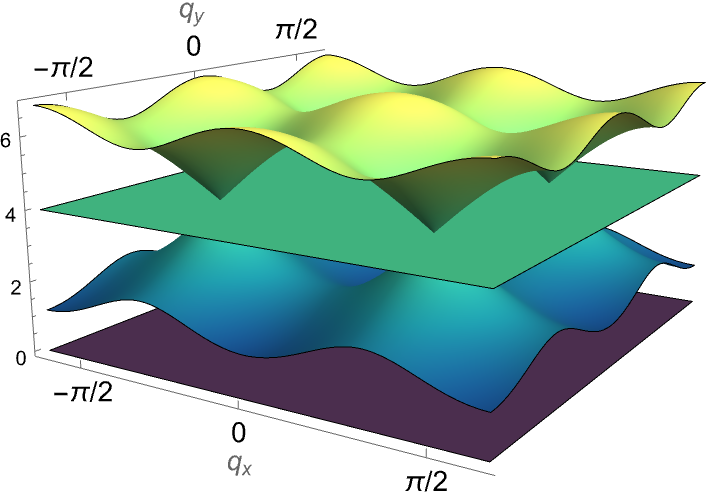} & 
        \includegraphics[width=2.5cm]{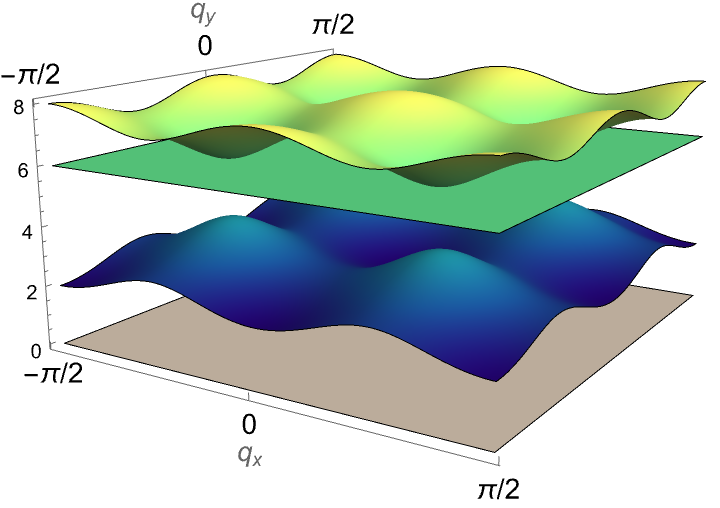} & 
        \includegraphics[width=2.5cm]{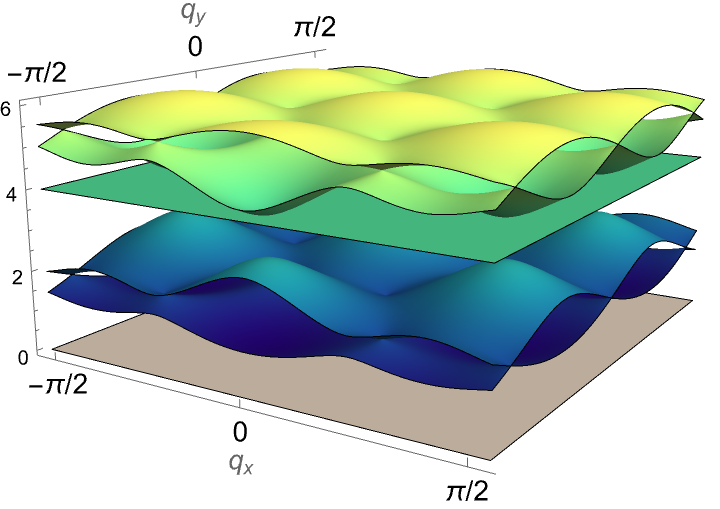} & 
        \includegraphics[width=2.5cm]{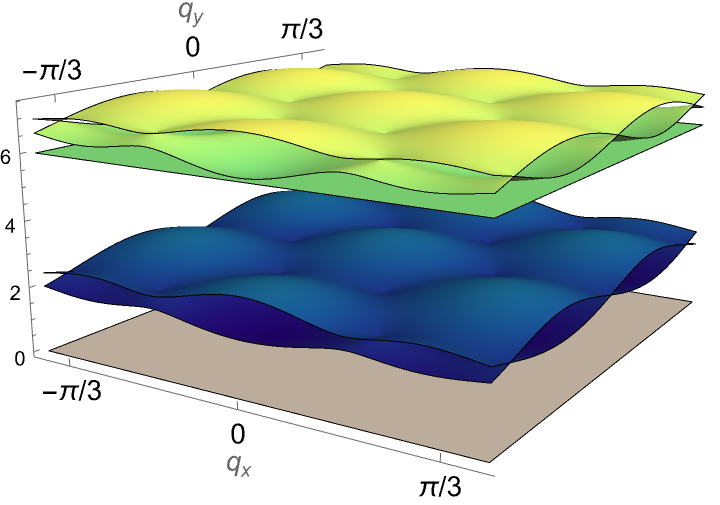} & 
        \includegraphics[width=2.5cm]{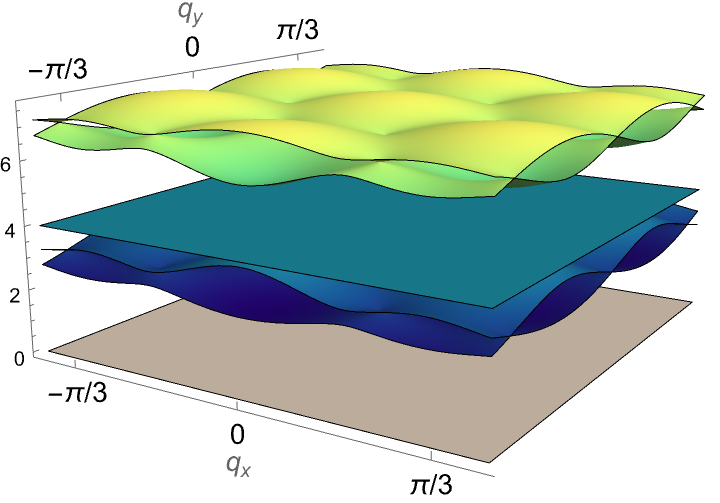} & 
        \includegraphics[width=2.5cm]{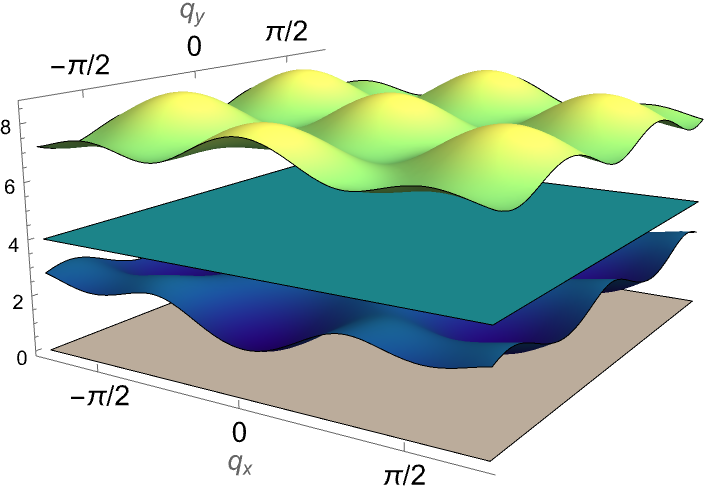} 
        \\

        \makecell{ \vspace{-2cm} \\ \textbf{Monoblock} \\ \textbf{Cluster-bond} \\ \textbf{Band} \\ \textbf{Structure} } & 
        \includegraphics[width=2.5cm]{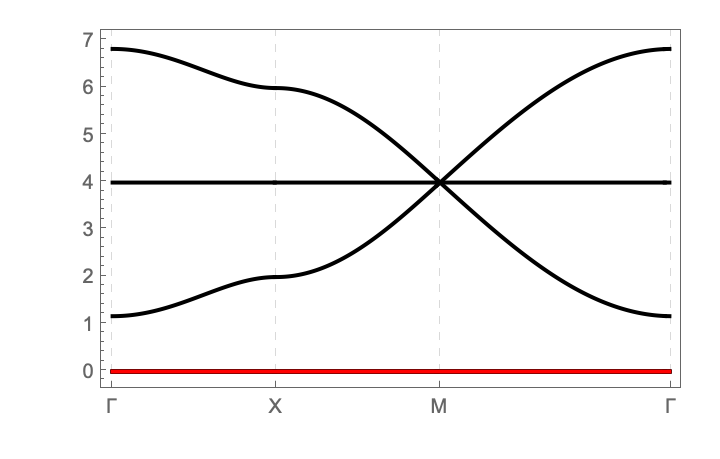} & 
        \includegraphics[width=2.5cm]{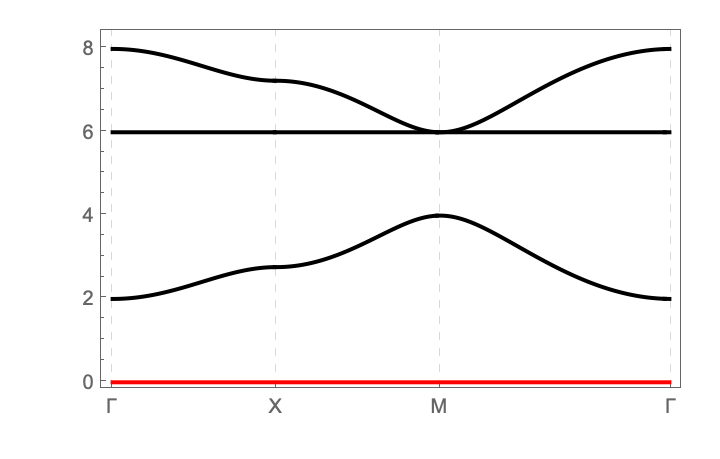} & 
        \includegraphics[width=2.5cm]{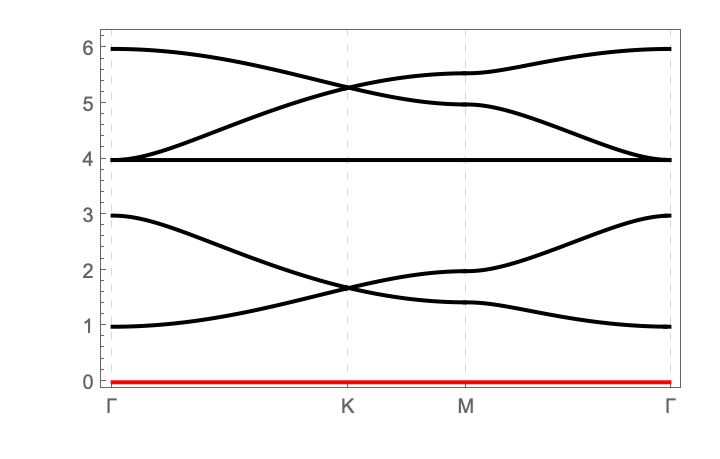} & 
        \includegraphics[width=2.5cm]{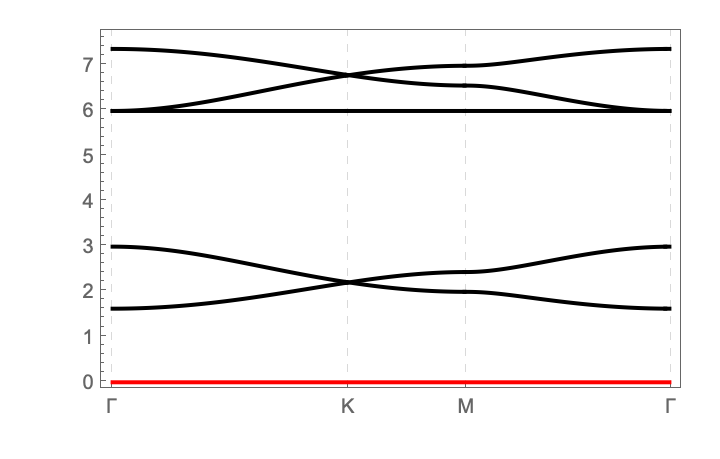} & 
        \includegraphics[width=2.5cm]{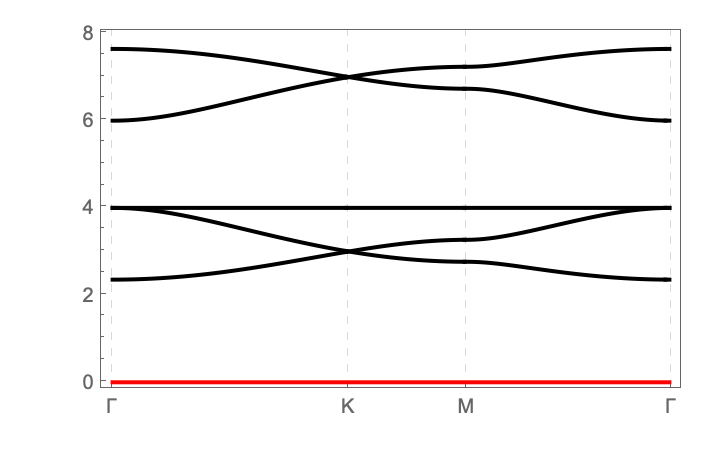} & 
        \includegraphics[width=2.5cm]{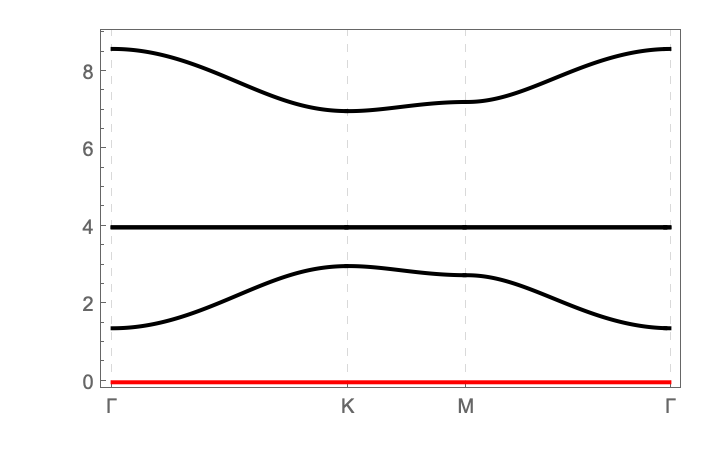} 
        \\

        \textbf{$n_s$}     & 8    & 12    & 12    & 18     & 18     & 12 \\
        \textbf{$n_c$}     & 3 & 3  & 5  & 5  & 5   & 4 \\
        \textbf{$n_\text{b.f.b}$} & 5 & 9  & 7  & 13  & 13 & 8 \\
        \textbf{$n_\text{i.f.b}$} & 1 & 1  & 1  & 1  & 1 & 2 \\

        \hline
        \noalign{\vskip 1mm}

        \makecell{ \vspace{-2.5cm} \\ \textbf{Composite} \\ \textbf{Cluster-bond} \\ \textbf{Structure}  \\ \textbf{Factor} } & 
        \includegraphics[width=2.5cm,trim={1cm 1cm 2cm 0},clip]{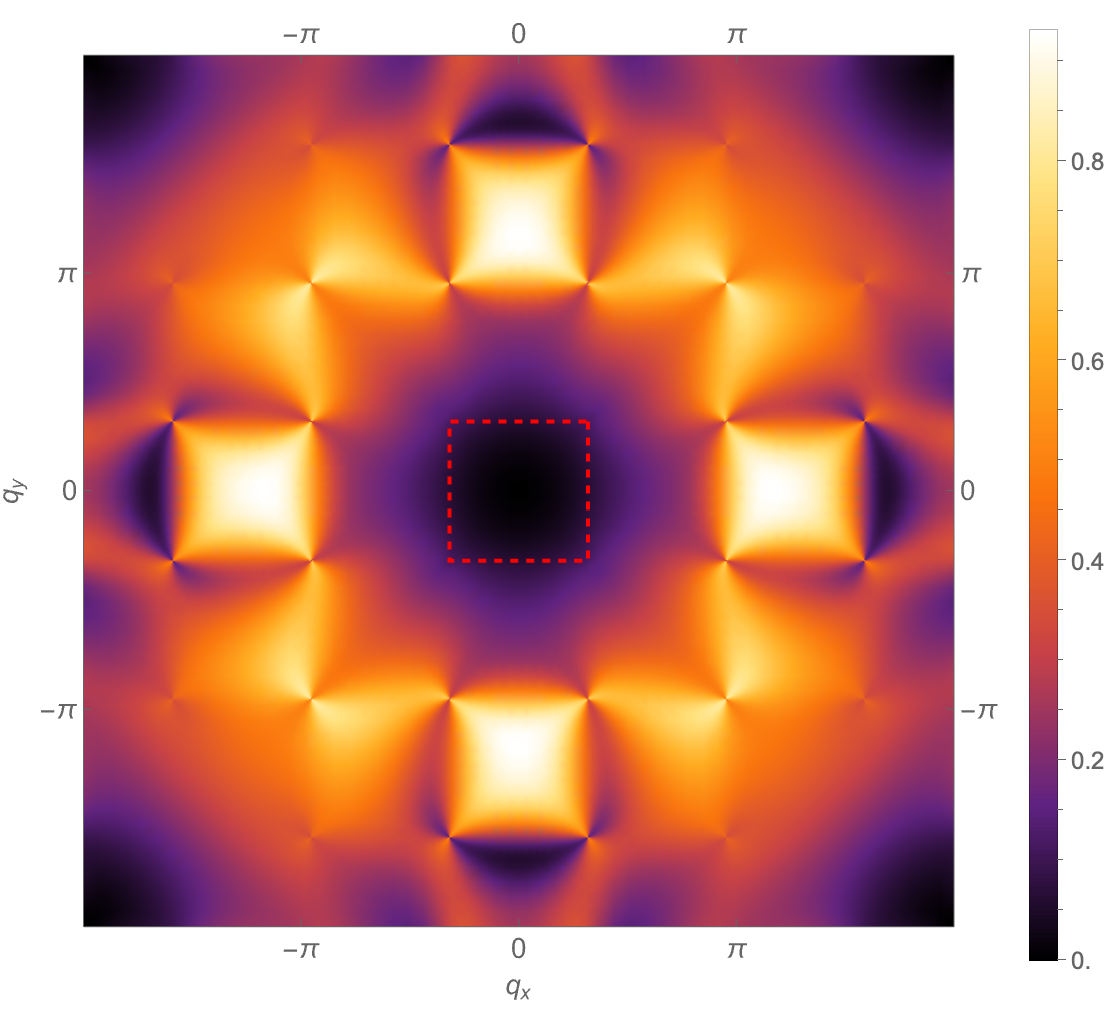} & 
        \includegraphics[width=2.5cm,trim={1cm 1cm 2cm 0},clip]{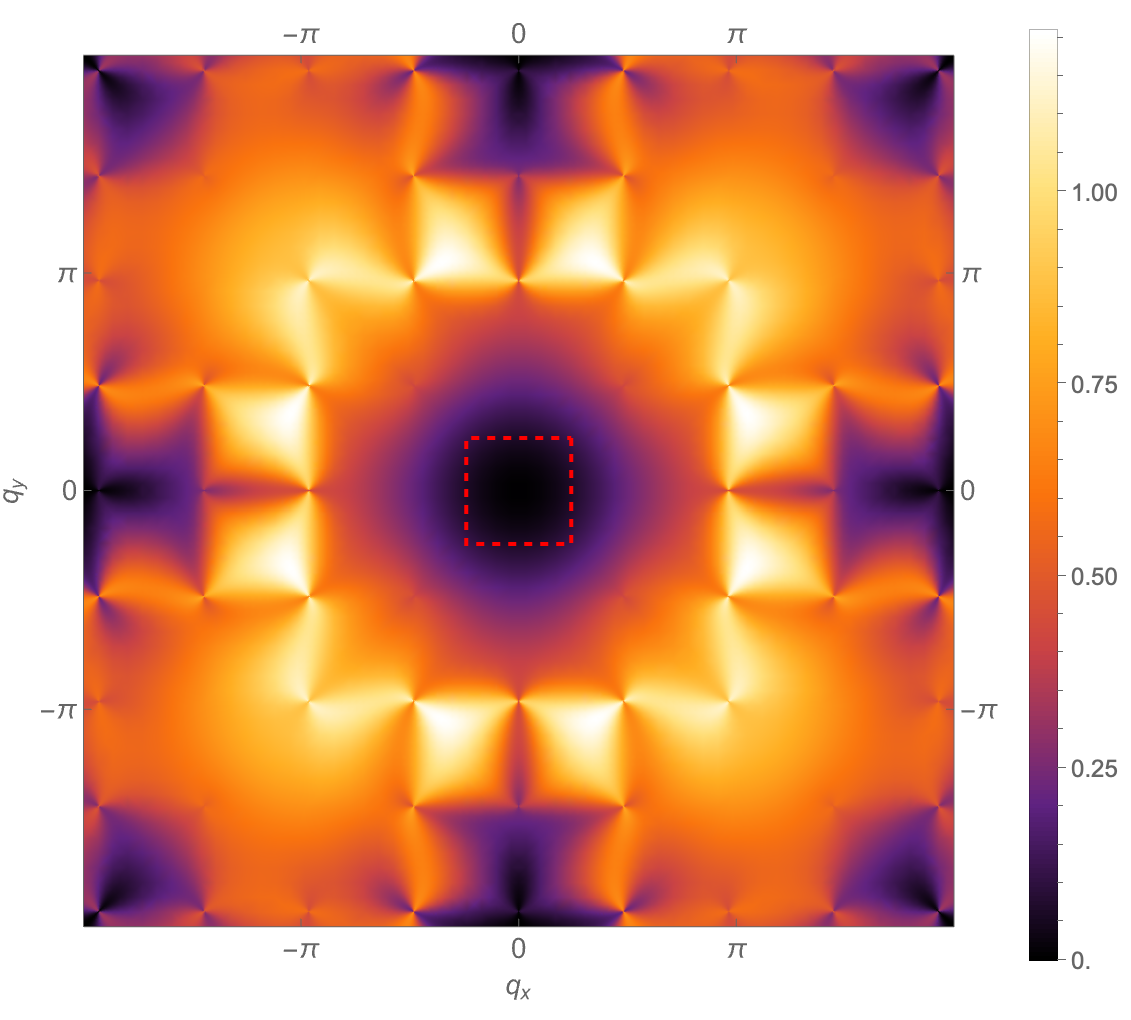} & 
        \includegraphics[width=2.5cm,trim={1cm 1cm 2cm 0},clip]{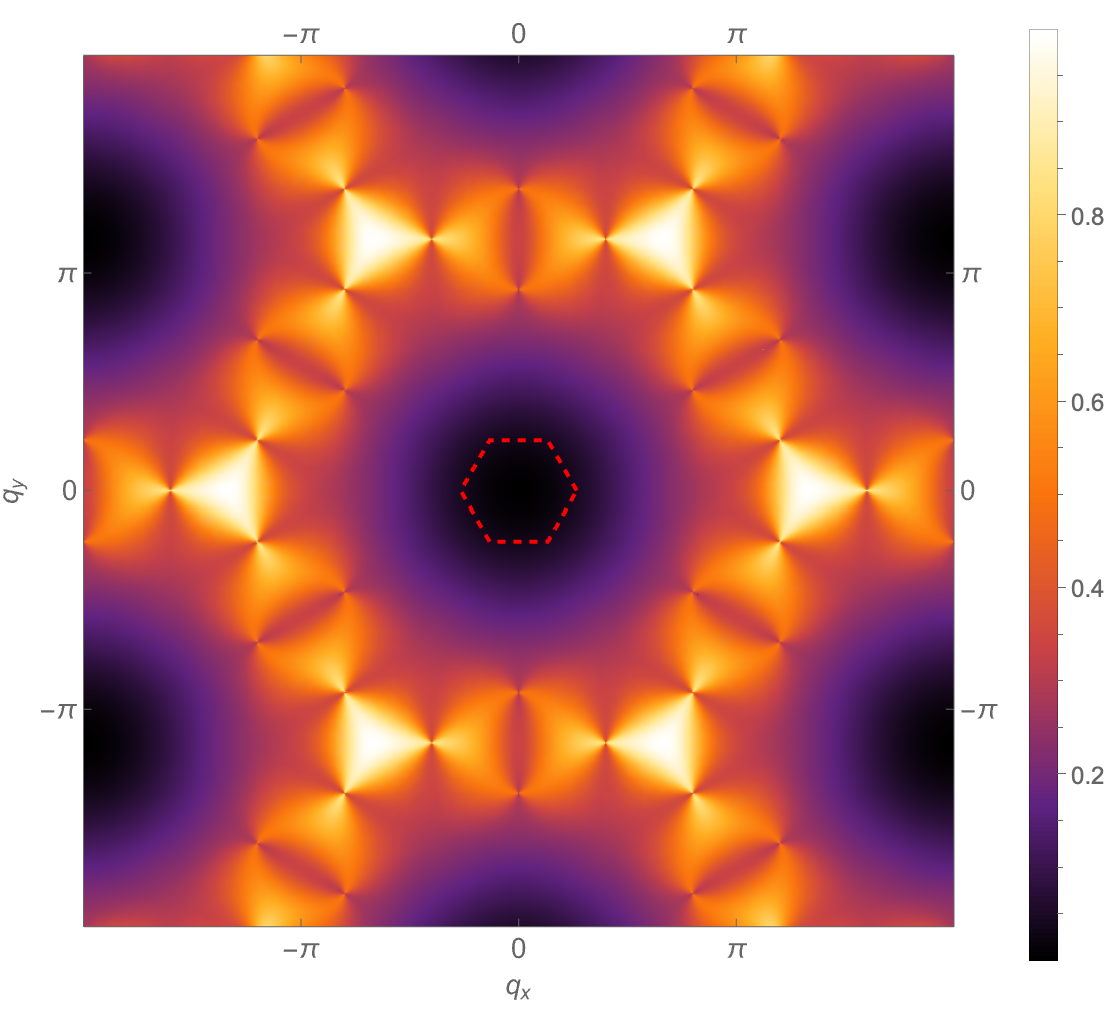} & 
        \includegraphics[width=2.5cm,trim={1cm 1cm 2cm 0},clip]{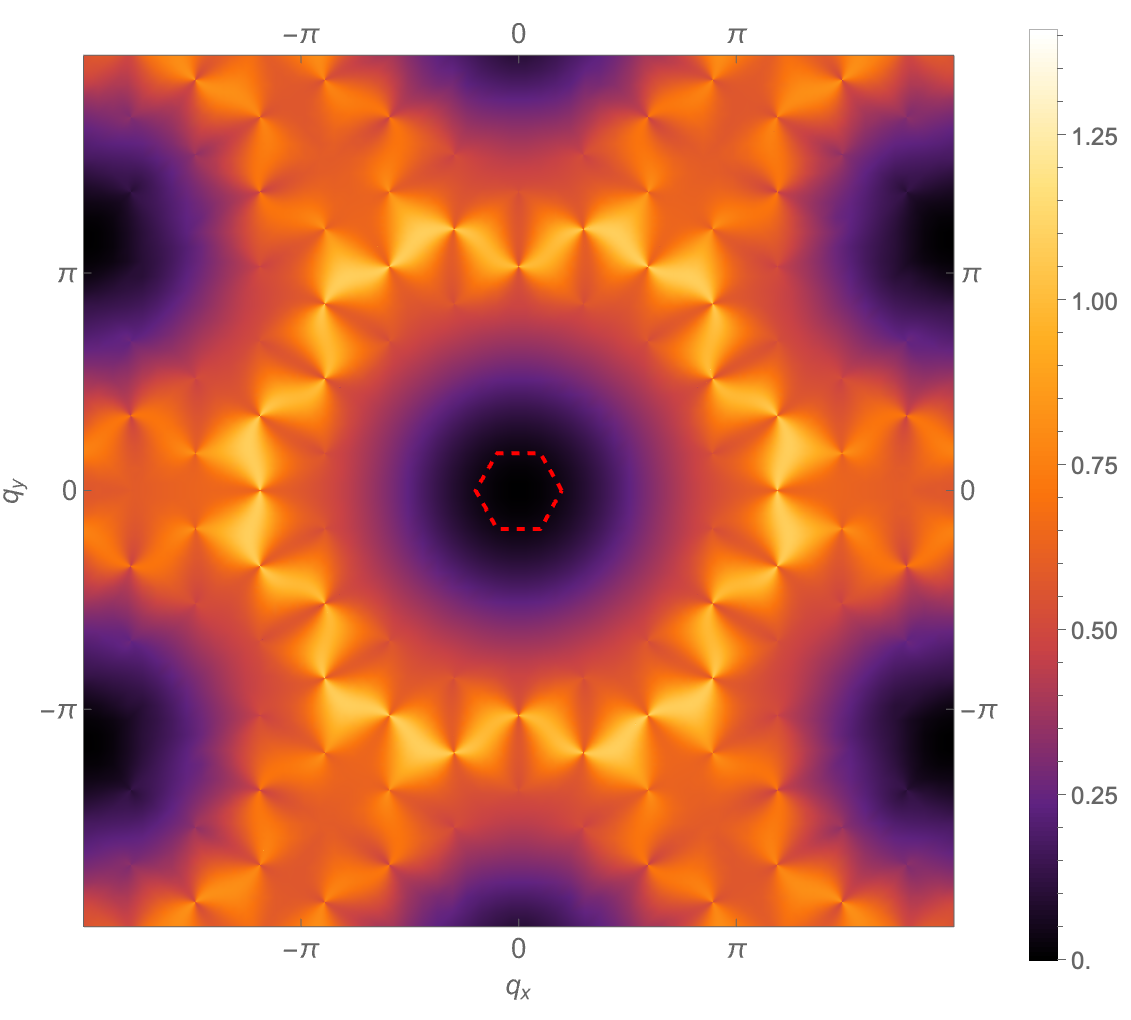} & 
        \includegraphics[width=2.5cm,trim={1cm 1cm 2cm 0},clip]{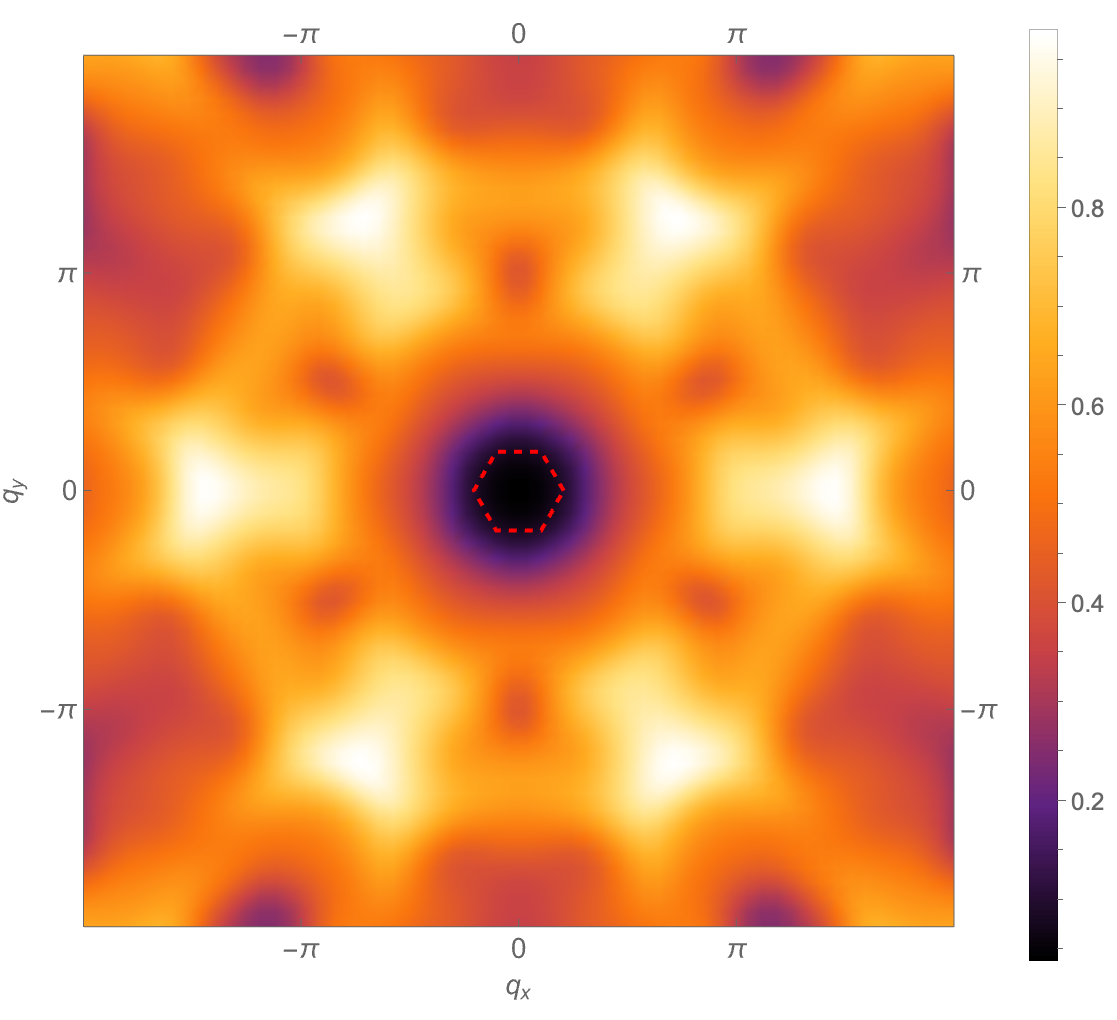} & 
        \includegraphics[width=2.5cm,trim={1cm 1cm 2cm 0},clip]{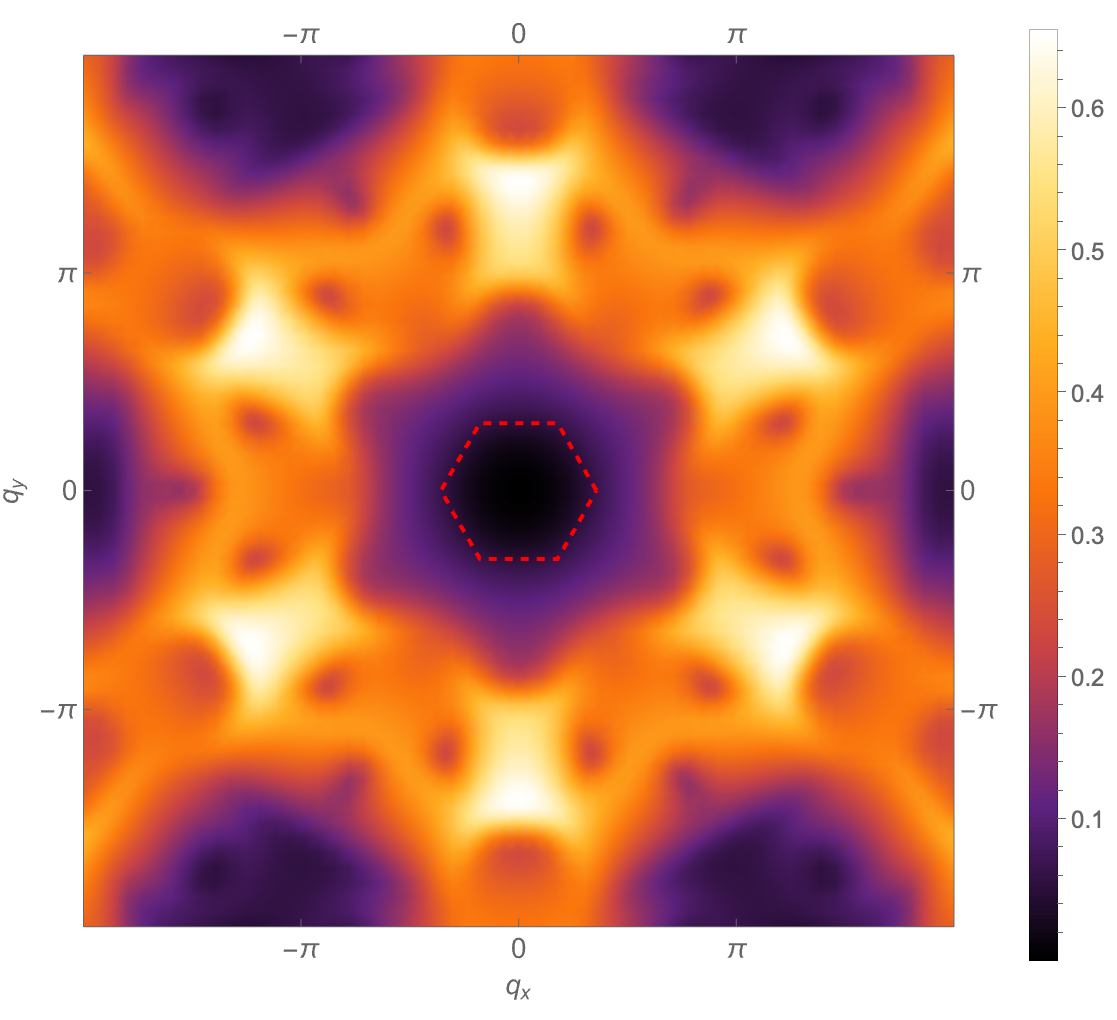} 
        \\

        \makecell{ \vspace{-2.3cm} \\ \textbf{Composite} \\ \textbf{Cluster-bond} \\ \textbf{3D Band}  \\ \textbf{Structure} } & 
        \includegraphics[width=2.5cm]{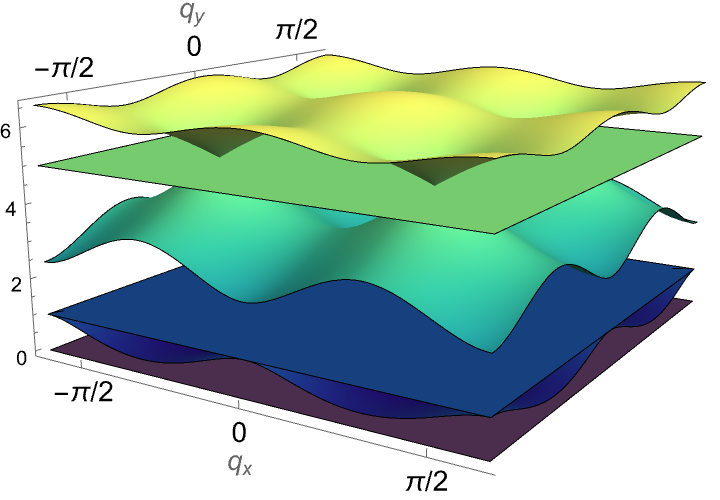} & 
        \includegraphics[width=2.5cm]{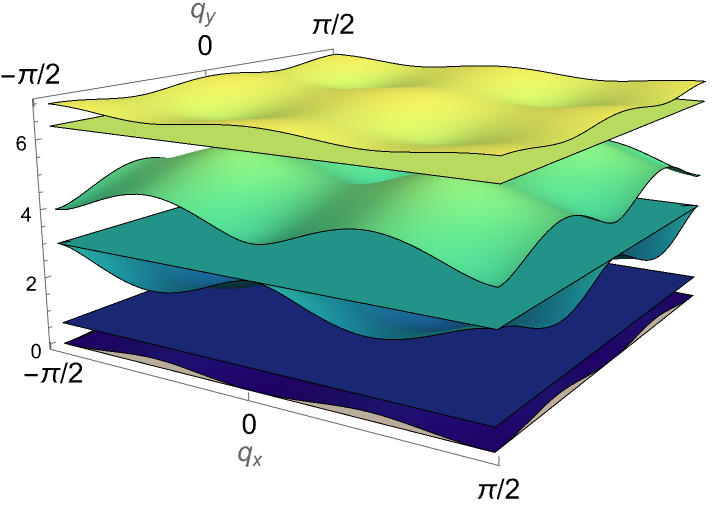} & 
        \includegraphics[width=2.5cm]{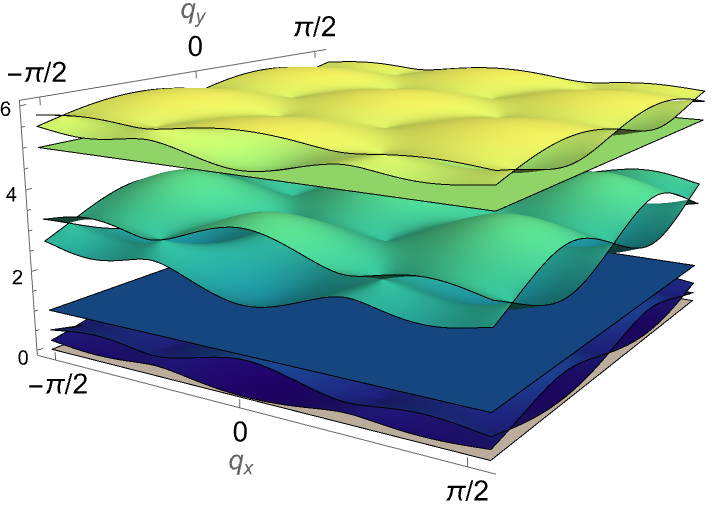} & 
        \includegraphics[width=2.5cm]{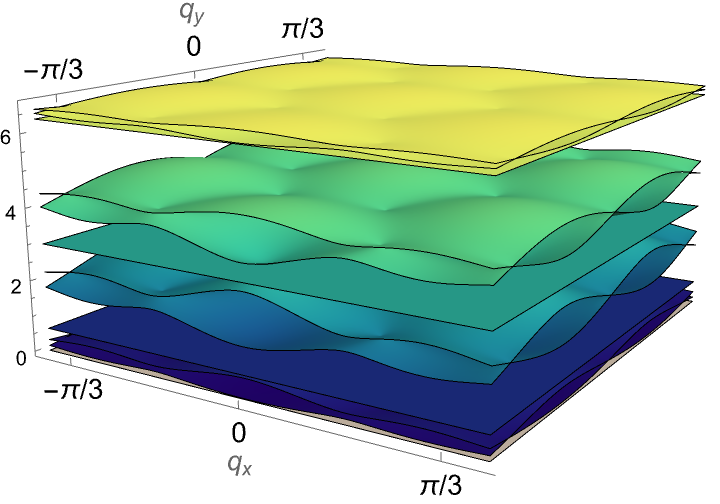} & 
        \includegraphics[width=2.5cm]{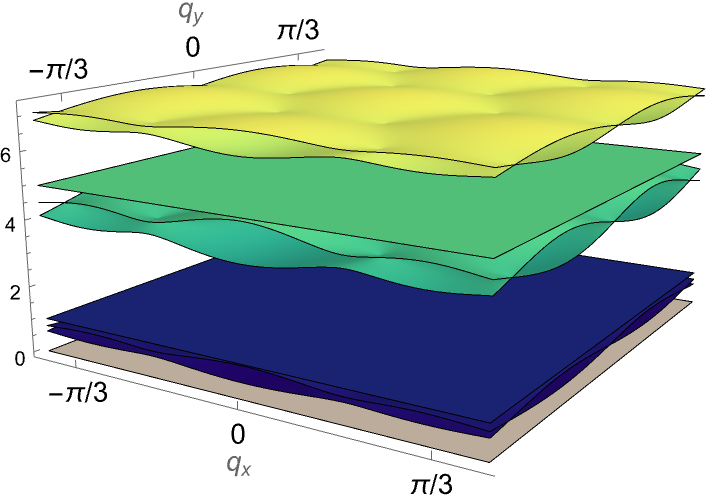} & 
        \includegraphics[width=2.5cm]{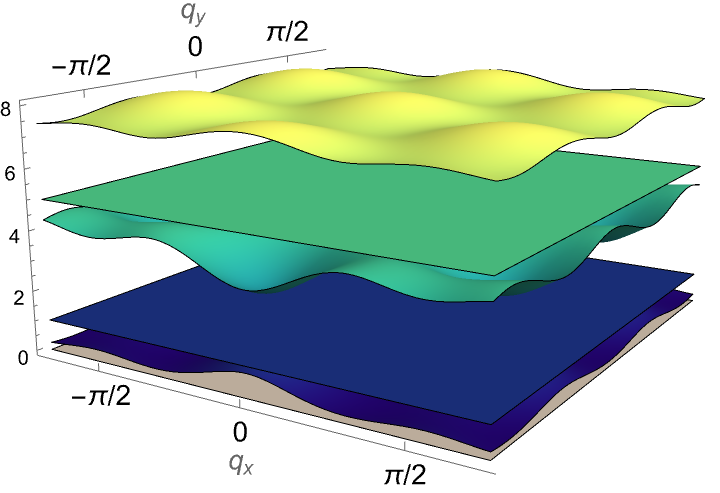} 
        \\

        \makecell{ \vspace{-2cm} \\ \textbf{Composite} \\ \textbf{Cluster-bond} \\ \textbf{Band}  \\ \textbf{Structure} } & 
        \includegraphics[width=2.5cm]{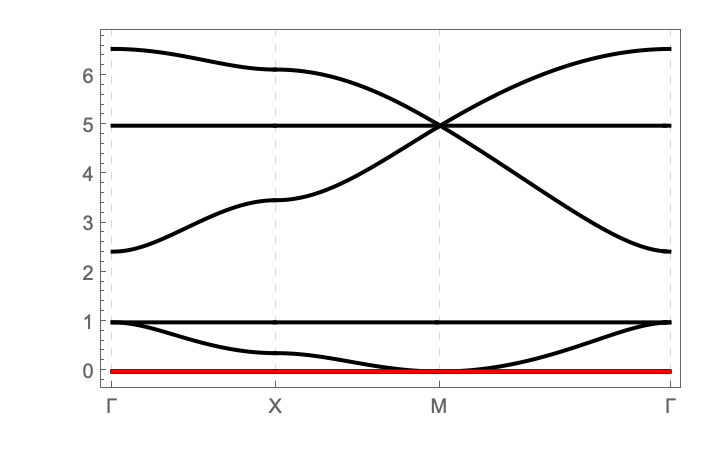} & 
        \includegraphics[width=2.5cm]{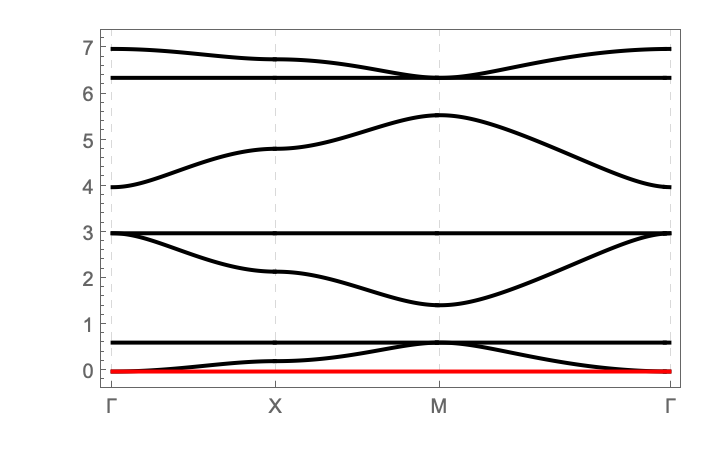} & 
        \includegraphics[width=2.5cm]{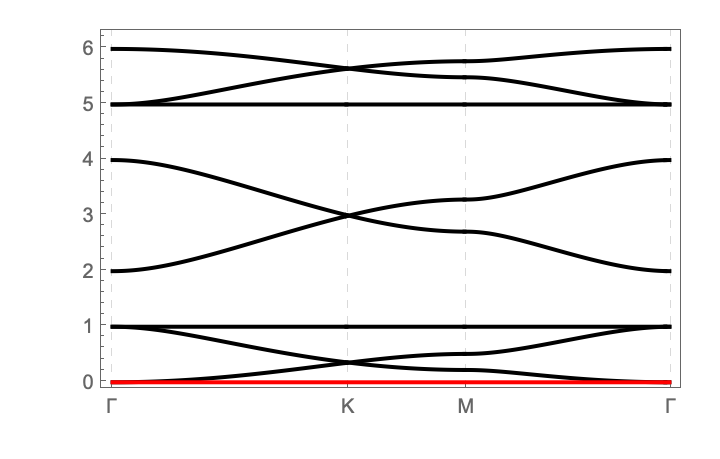} & 
        \includegraphics[width=2.5cm]{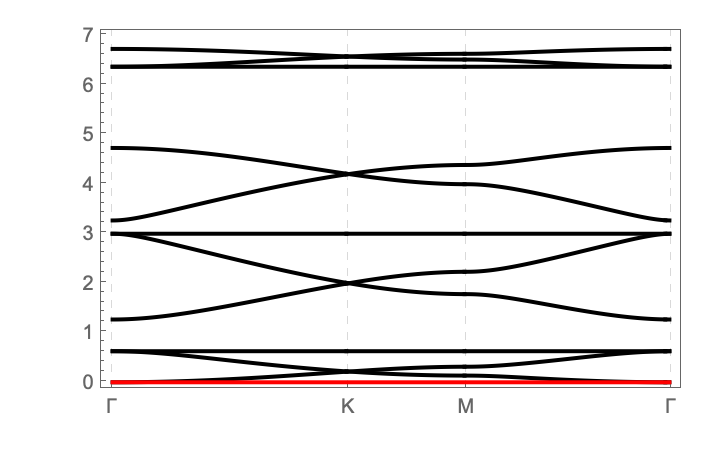} & 
        \includegraphics[width=2.5cm]{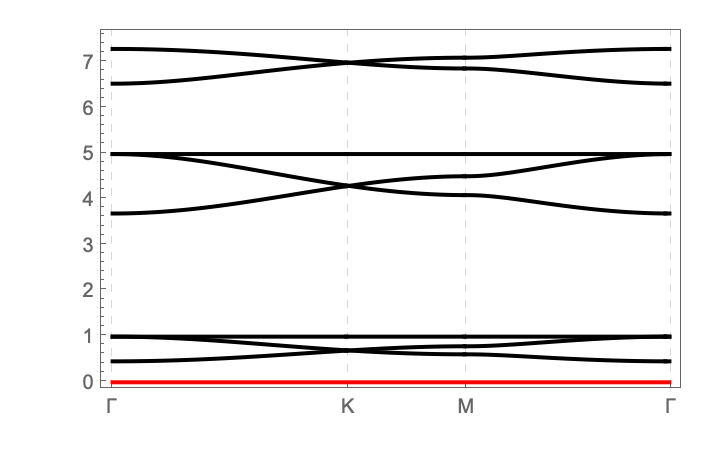} & 
        \includegraphics[width=2.5cm]{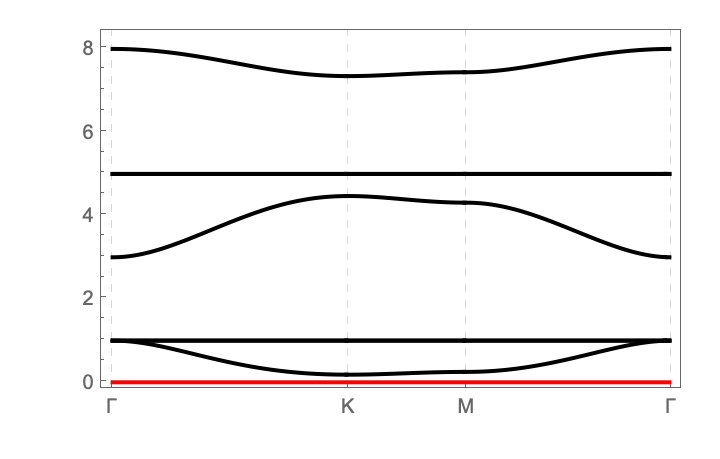} 
        \\
        
        \textbf{$n_s$}     & 8    & 12    & 12    & 18     & 18     & 12 \\
        \textbf{$n_c$}     & 5 & 7  & 8  & 11  & 8   & 7 \\
        \textbf{$n_\text{b.f.b}$} & 3 & 5  & 4  & 7  & 10 & 5 \\
        \textbf{$n_\text{i.f.b}$} & 1+1 & 1+1+1  & 1+1  & 1+1+1  & 1+1 & 2+2 \\
        \hline
    \end{tabular}
    }
    \caption{Bond and vertex decorated 2D systems. Here, cluster-bonds connect cluster-vertices to form usual 2D lattices. The number of bottom flat bands $n_\text{b.f.b}$ can be well estimated for these systems as $n_s - n_c$ where $n_s$ is given for clusters hosting only spins located on vertices. Both composite and mono-block band structure are depicted. Monoblock cluster-bond systems appears to be always gapped contrary to the case of composite cluster-bonds, revealing the contact point and therefore the pinch points are due to the internal structure of the cluster-bonds.
    As for bond decorated systems, vertex and bond decorated systems often appear to host intermediate flat bands located inside their spectrum. As these may be degenerate their number is indicated as a sum over the number of distinct set of intermediate flat bands.
    }
    \label{tab: 2d cluster-links + cluster-bonds lattices}
\end{table*}

\FloatBarrier


\bibliography{Biblio}

\end{document}